\documentclass{emulateapj}

\usepackage{natbib}
\usepackage{color}

\slugcomment{Received October 13, 2014; accepted December 10, 2014}

\shorttitle{Ground-based Pa$\alpha$ Narrow-band Imaging of Local Luminous Infrared Galaxies I}
\shortauthors{Tateuchi et al.}

\begin{document}

\title{Ground-based Pa$\alpha$ Narrow-band Imaging of Local Luminous Infrared Galaxies I:\\ Star Formation Rates and Surface Densities}

\author{Ken Tateuchi\altaffilmark{1}, Masahiro Konishi\altaffilmark{1}, Kentaro Motohara\altaffilmark{1}, Hidenori Takahashi\altaffilmark{1}, Natsuko Mitani Kato\altaffilmark{1}, Yutaro\\
Kitagawa\altaffilmark{1}, Soya Todo\altaffilmark{1}, Koji Toshikawa\altaffilmark{1}, Shigeyuki Sako\altaffilmark{1}, Yuka K. Uchimoto\altaffilmark{1,3}, Ryou Ohsawa\altaffilmark{1}, Kentaro\\
Asano\altaffilmark{1}, Yoshifusa Ita\altaffilmark{3}, Takafumi Kamizuka\altaffilmark{1}, Shinya Komugi\altaffilmark{4}, Shintaro Koshida\altaffilmark{5}, Sho Manabe\altaffilmark{7},\\
Tomohiko Nakamura\altaffilmark{1,2}, Asami Nakashima\altaffilmark{2,6}, Kazushi Okada\altaffilmark{1}, Toshinobu Takagi\altaffilmark{8}, Toshihiko\\
Tanab\'{e}\altaffilmark{1}, Mizuho Uchiyama\altaffilmark{1}, Tsutomu Aoki\altaffilmark{9}, Mamoru Doi\altaffilmark{1,10}, Toshihiro Handa\altaffilmark{11},\\
Kimiaki Kawara\altaffilmark{1}, Kotaro Kohno\altaffilmark{1,10}, Takeo Minezaki\altaffilmark{1}, Takashi Miyata\altaffilmark{1},\\
Tomoki Morokuma\altaffilmark{1}, Takeo Soyano\altaffilmark{9}, Yoichi Tamura\altaffilmark{1}, Masuo\\
Tanaka\altaffilmark{1}, Ken'ichi Tarusawa\altaffilmark{9}, Yuzuru Yoshii\altaffilmark{1}
 }

\affil{$^{1}$ Institute of Astronomy, Graduate School of Science, The University of Tokyo, 2-21-1 Osawa, Mitaka, Tokyo 181-0015, Japan}
%\email{tateuchi@ioa.s.u-tokyo.ac.jp}
\affil{$^{2}$ Department of Astronomy, Graduate School of Science, The University of Tokyo, 7-3-1 Hongo, Bunkyo-ku, Tokyo 113-0033, Japan}
\affil{$^{3}$ Astronomical Institute, Tohoku University, 6-3 Aoba, Aramaki, Aoba-ku, Sendai, Miyagi 980-8578, Japan}
\affil{$^{4}$ Division of Liberal Arts, Kogakuin University, 2665-1, Hachioji, Tokyo 192-0015, Japan}
\affil{$^{5}$ Subaru Telescope, National Astronomical Observatory of Japan, Hilo, HI 96720, USA}
\affil{$^{6}$ National Astronomical Observatory, Mitaka, Tokyo 181-8588, Japan}
\affil{$^{7}$ Department of Earth and Planetary Sciences, Kobe University, 657-8501 Kobe, Japan}
\affil{$^{8}$ Institute of Space and Astronautical Science, Japan Aerospace Exploration Agency, 3-1-1 Yoshinodai, Sagamihara, Kanagawa 229-8510, Japan}
\affil{$^{9}$ Kiso Observatory, Institute of Astronomy, The University of Tokyo, 10762-30, Mitake, Kiso-machi, Kiso-gun, Nagano 397-0101, Japan}
\affil{$^{10}$ Research Center for the Early Universe (RESCEU), University of Tokyo, 7-3-1, Hongo, Bunkyo, Tokyo 113-0033, Japan}
\affil{$^{11}$ Faculty of Science, Kagoshima University, Korimoto 1-21-24, Kagoshima 890-8580, Japan}

%%%==========================================================================%%%
%%%   ABSTRACT
%%%==========================================================================%%%
\begin{abstract}
Luminous infrared galaxies (LIRGs) are enshrouded by a large amount of dust, produced by their active star formation, and it is difficult to measure their activity in the optical wavelength. We have carried out Pa$\alpha$ narrow-band imaging observations of 38 nearby star-forming galaxies including 33 LIRGs listed in $IRAS$ RBGS catalog with the Atacama Near InfraRed camera (ANIR) on the University of Tokyo Atacama Observatory (TAO) 1.0 m telescope (miniTAO). Star formation rates (SFRs) estimated from the Pa$\alpha$ fluxes, corrected for dust extinction using the Balmer Decrement Method (typically $A_V$ $\sim$ 4.3 mag), show a good correlation with those from the bolometric infrared luminosity of $IRAS$ data within a scatter of 0.27 dex. This suggests that the correction of dust extinction for Pa$\alpha$ flux is sufficient in our sample.
We measure the physical sizes and the surface density of infrared luminosities ($\Sigma_{L(\mathrm{IR})}$) and SFR ($\Sigma_{\mathrm{SFR}}$) of star-forming region for individual galaxies, and find that most of the galaxies follow a sequence of local ultra luminous or luminous infrared galaxies (U/LIRGs) on the $L(\mathrm{IR})$-$\Sigma_{L(\mathrm{IR})}$ and SFR-$\Sigma_{\mathrm{SFR}}$ plane. We confirm that a transition of the sequence from normal galaxies to U/LIRGs is seen at $L(\mathrm{IR})=8\times10^{10}$ $L_{\sun}$. Also, we find that there is a large scatter in physical size, different from those of normal galaxies or ULIRGs. Considering the fact that most of U/LIRGs are merging or interacting galaxies, this scatter may be caused by strong external factors or differences of their merging stage.
\end{abstract}

\keywords{galaxies:interactions --- galaxies:starburst --- H{\sc ii} regions --- stars:formation --- infrared: galaxies }

%%%==========================================================================%%%
%%%   Section 1 Introduction
%%%==========================================================================%%%
\section{Introduction}
In recent years, many large deep cosmological surveys have been performed in various wavelengths, including ultraviolet, visible, infrared and submillimeter. These surveys have revealed that the star formation rate (SFR) density of the universe (cosmic SFR density, cSFRD) increases with redshift, and peaks at $1 < z < 3$ \citep[e.g.,][]{ruj10,2006ApJ...651..142H}. It is known that the cSFRD at the high-redshift universe is dominated by bright infrared galaxies; ULIRGs (Ultra Luminous Infrared Galaxies; $L(\mathrm{IR})$ $\equiv$ $L$ (8--1000 $\mu$m) $\geq$ 10$^{12}$ $L_{\sun}$) and LIRGs (Luminous Infrared Galaxies; $L(\mathrm{IR})$ $\equiv$ 10$^{11}$--10$^{12}$ $L_{\sun}$) dominate 80\% of total star formation activities at $z$ $\sim$ 1 \citep[e.g.,][]{2007ApJ...660...97C,2010A&A...514A...6G}, and these galaxies are in the starburst sequence which have high star formation efficiency ($\mathrm{SFE}=\mathrm{SFR}\ (M_{\sun}\ \mathrm{yr}^{-1})/M_{\mathrm{gas}}\ (M_{\sun})$) of around 10$^{-8}$ yr$^{-1}$ \citep[e.g.,][]{2012ApJ...745..190L,2010ApJ...714L.118D} in contrast of 10$^{-9}$ yr$^{-1}$ for normal galaxies. In order to probe the detailed properties of these galaxies, and to understand how they have been formed and evolved, nearby ULIRGs and LIRGs are ideal laboratories because they can be spatially resolved.

In the local universe, normal galaxies show extended star-forming regions over a few kilo-parsecs along the spiral arms, while distributions of star-forming regions of LIRGs at the high infrared luminosity end ($\log(L(\mathrm{IR})/L_{\sun})$ = 11.8--12.0) and ULIRGs, which are considered to be in the starburst sequence, become very compact and concentrated at central regions (e.g., \citealt{2000AJ....119..509S,2010ApJ...723..993D,2011ApJ...726...93R} (R11); \citealt{2012ApJ...744....2A}). However, the relationship between star formation activities and spatial distribution of star-forming regions, and the mechanism of starburst is still an open question. Some simulations of galaxy formation suggest that the central concentration of stars and gases is formed by interacting/merging systems, which accumulates dense gas clouds and triggers starburst \citep{1996ApJ...471..115B}. However, there have been not enough observational studies to reveal the relations.

\begin{figure}
\begin{center}
\plotone{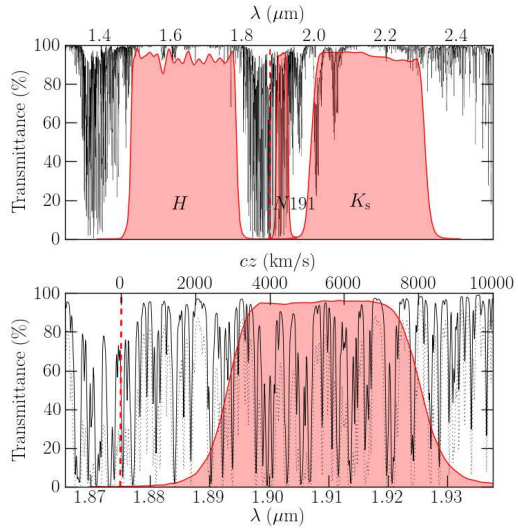}
\caption{Top: atmospheric transmittance in the wavelength range containing the $H$- and the $K_\mathrm{s}$-band. The bold line represents the difference of the median value of the transmittance at the TAO site (altitude of 5,640 m, PWV = 0.5 mm) and that at the VLT site (2,600 m, PWV = 2.0 mm), both calculated by ATRAN \citep{Lord...1992}. The shaded area with red shows the filter transmittance curve of the $H$-band, $N191$, and $K_\mathrm{s}$-band filters. Bottom: atmospheric transmittance around the $N191$ filter. The bold line corresponds to that at the TAO site, and the dotted line at the VLT site. The overlaid thick red area shows the transmittance of the $N191$ filter. The dashed vertical red line in the both plots represents the wavelength of the rest-frame Pa$\alpha$ (1.8751 $\mu$m).}\label{fig:transmodel}
\end{center}
\end{figure}

To understand the detailed mechanism of starburst activities in U/LIRGs, the star formation rate is one of the most important parameters. Many indicators for estimation of star formation rate, such as X-ray, ultraviolet (UV), H$\alpha$, mid-infrared (MIR), and far-infrared (FIR) emission have been used. Hydrogen recombination lines, which is emitted from the current star-forming regions within 10 Myr \citep{1996ARA&A..34..749S}, are direct tracers of star-forming region. Especially, optical hydrogen recombination lines such as H$\alpha$ and H$\beta$ are usually used because they can be observed easily. However, U/LIRGs are affected by a large amount of dust (extinction of $A_V\sim 3$ mag for LIRGs; \citealp{2006ApJ...650..835A}, and $A_V > 10$ mag for ULIRGs; \citealp{2013arXiv1304.0894P}), typically associated with the regions of active star formation. Therefore, those lines are easily attenuated by the dust. Wherein the hydrogen Pa$\alpha$ emission line (1.8751 $\mu$m) is a good tracer of the dusty star-forming region because of its insensitivity to the dust-extinction \citep{2006ApJ...650..835A} and being the strongest emission line in the near-infrared wavelength range (NIR, $\lambda$ $\sim$ 0.9--2.5 $\mu$m), which can reach higher spatial resolution easily than in the MIR and FIR. However, because of poor atmospheric transmission around the wavelength of Pa$\alpha$ emission line (Figure \ref{fig:transmodel}) due to absorptions mainly by water vapor, no Pa$\alpha$ imaging from a ground-based telescope is reported so far, although there are some spectroscopic observations of Pa$\alpha$ in redshifted galaxies \citep[e.g.,][]{1996ApJ...462..163H,1998ApJ...494L.155F,1999ApJ...525L..85M,2010ApJ...724..386K}. To overcome these difficulties, it is necessary to observe the emission line by either a space-borne facilities such as the Near Infrared Camera and Multi-object Spectrometer (NICMOS) on $Hubble\ Space\ Telescope$ ($HST$) \citep[e.g.,][]{2001AJ....122.3017S,2006ApJ...650..835A,2013ApJ...772...27L} or facilities built at sites with low PWV. However, some researchers have pointed out that $HST$/NICMOS (already decommissioned in 2010) may be insensitive to diffuse Paschen-alpha emission due to its intrinsic high angular resolution \citep[e.g.,][]{2006ApJ...650..835A,2007ApJ...666..870C,2007ApJ...671..333K,2009ApJ...692..556R}.

Therefore, we have been carrying out Pa$\alpha$ narrow-band imaging observations \citep{2012PKAS...27..297T,2013ASPC..476..301T} with Atacama Near InfraRed camera (ANIR; \citealt{2008SPIE.7014E..94M}, Konishi et al. 2014; K14), on the University of Tokyo Atacama Observatory (TAO, \citealt{2010SPIE.7733E...6Y}) 1.0m telescope (miniTAO; \citealt{2010SPIE.7733E.163M}) installed at the summit of Co. Chajnantor (5640m altitude) in northern Chile to understand distributions of star-forming region and properties of dust-enshrouded infrared galaxies in the local universe. Thanks to the high altitude and the extremely low water vapor content of the site we can stably observe Pa$\alpha$ emission line has been observationally confirmed \citep{2010SPIE.7735E.120M,2011RMxAC..41...83M,2013PASJ...65...55T}.

In this paper, we describe our sample of luminous infrared galaxies and the observation procedure in $\S$ 2, the method of data reduction and flux calibration in $\S$ 3, and the derived Pa$\alpha$ flux and properties of individual galaxies in $\S$ 4. In $\S$ 5, we evaluate the selection bias due to our luminosity-limited sample, the effects of dust-extinction, the relationship between SFRs estimated from Pa$\alpha$ and those from FIR fluxes, and the properties of surface densities of SFR. Then, we summarize them in $\S$ 6.

Throughout this paper, we use a $\Lambda$-CDM cosmology with $\Omega_{m}$ = 0.3, $\Omega_{\Lambda}$ = 0.7, and $H_0$ = 70 $\rm km$ $\rm s^{-1}$ $\rm Mpc^{-1}$.

%-------------
%-------------

%%%==========================================================================%%%
%%%   Section 2 Observations
%%%==========================================================================%%%
\section{Observations}

%%%---< Subsection 2.1 The Sample >
\subsection{The Sample}
We have selected the target galaxies from the $IRAS$ Revised Bright Galaxy Sample (RBGS : \citealt{2003AJ....126.1607S}). The location of the observatory limits the declination of the targets to be $<$ 30$^{\circ}$. The wavelength range of the $N191$ narrow-band filter limits the recession velocity to be 2800 km s$^{-1}$ -- 8100 km s$^{-1}$, corresponding to the distances of 46.6 Mpc -- 109.6 Mpc. From these conditions, the number of observable galaxies in the RBGS is 151, and we have observed 38 galaxies out of them at random. The selected galaxies are listed in Table \ref{tbl-sample}. In our sample, the bolometric infrared luminosity ($L(\mathrm{IR})$) ranges between $4.5\times10^{10}$ $L_{\sun}$ and $6.5\times10^{11}$ $L_{\sun}$, which ranges from the high luminosity end of normal galaxies to LIRGs.

%%%___ Subsection 2.2 observation with ANIR ___________________________
\subsection{Observation with Atacama Near Infrared Camera}
The observations have been carried out using Atacama NIR camera (ANIR; \citealt{2008SPIE.7014E..94M}, K14) installed

\clearpage
\onecolumngrid

%-------------
%\begin{turnpage}
%\begin{landscape}
\begin{deluxetable}{ccccccccccc}
%\tabletypesize{\normalsize}
\tabletypesize{\tiny}
%\rotate
\tablecolumns{10}
\tablecaption{Sample of Local Luminous Infrared Galaxies\label{tbl-sample}}
\tablewidth{0pt}
%\tablewidth{\columnwidth}
\tablehead{
\colhead{ID} & \colhead{Galaxy} & \colhead{IRAS}& \colhead{R.A.}& \colhead{Dec.}& \colhead{$cz$}& \colhead{Dist.}& \colhead{log$L_{\rm IR}$}& \colhead{Spectral}& \colhead{Obs.}\\
 & \colhead{Name} & \colhead{Name}& \colhead{(J2000)}& \colhead{(J2000)}& \colhead{(km s$^{-1}$)}& \colhead{(Mpc)}& \colhead{($L_{\mathrm{\odot}}$)}& \colhead{Class} & \colhead{Date}\\
\colhead{(1)} & \colhead{(2)} & \colhead{(3)}& \colhead{(4)}& \colhead{(5)}& \colhead{(6)}& \colhead{(7)}& \colhead{(8)}& \colhead{(9)} & \colhead{(10)}
}
\startdata
1	& NGC 23\dotfill				& F00073$+$2538	& 00 09 55.1	& $+$25 55 37	
& 4536	& 65.6	& 11.13	& H{\sc ii}       & 2009-10-26    \\
2	& NGC 34\dotfill				& F00085$-$1223	& 00 11 06.6	& $-$12 06 27	
& 5931	& 86.0	& 11.52	& Sy2       & 2011-10-22   \\
3	& NGC 232\dotfill				& F00402$-$2349	& 00 42 46.5	& $-$23 33 31	
& 6047	& 87.7	& 11.39	& H{\sc ii}       & 2009-10-21    \\
4	& IC 1623A/B\dotfill			& F01053$-$1746	& 01 07 46.3	& $-$17 30 32	
& 6028	& 87.4	& 11.74	& H{\sc ii}       & 2009-10-17    \\
5	& ESO 244-G012\dotfill			& F01159$-$4443	& 01 18 08.6	& $-$44 27 40	
& 6866	& 99.8	& 11.48	& LINER     & 2009-10-21 \\
6	& UGC 2238\dotfill				& F02435$+$1253	& 02 46 17.0	& $+$13 05 45	
& 6436	& 93.5	& 11.33	& LINER     & 2009-10-27  \\
7	& IRAS F02437$+$2122\dotfill	& F02437$+$2122	& 02 46 38.3	& $+$21 35 06	
& 6987	& 101.6	& 11.21	& LINER     & 2009-10-22  \\
8	& UGC 2982\dotfill				& F04097$+$0525	& 04 12 22.4	& $+$05 32 49	
& 5161	& 74.7	& 11.20	& H{\sc ii}       & 2011-10-19  \\
9	& NGC 1614\dotfill				& F04315$-$0840	& 04 34 00.1	& $-$08 34 46	
& 4746	& 68.6	& 11.66	& H{\sc ii}/Sy2	& 2009-10-15 \\
10	 & MCG $-$05-12-006\dotfill		& F04502$-$3304	& 04 52 06.8	& $-$32 59 24	
& 5622	& 81.5	& 11.17	& H{\sc ii}       & 2009-10-14   \\
11	& NGC 1720\dotfill				& F04569$-$0756	& 04 59 19.9	& $-$07 51 34	
& 4186	& 60.4	& 10.90	& N         & 2011-10-19 \\
12	& ESO 557-G002\dotfill			& F06295$-$1735	& 06 31 46.3	& $-$17 37 15	
& 6339	& 92.0	& 11.24	& H{\sc ii}       & 2009-10-19   \\
13	& IRAS F06592-6313\dotfill	& F06592$-$6313	& 06 59 40.3	& $-$63 17 53	
& 6882	& 100.0	& 11.20	& H{\sc ii}       & 2009-10-27  \\
14	& NGC 2342\dotfill				& 07063$+$2043	& 07 09 19.6	& $+$20 38 12	
& 5276	& 76.4	& 11.40	& H{\sc ii}       & 2009-10-23 \\
15	& ESO 320-G030\dotfill			& F11506$-$3851	& 11 53 12.0	& $-$39 07 54	
& 3232	& 46.6	& 11.28	& H{\sc ii}       & 2011-04-21  \\
16	& NGC 4922\dotfill				& F12590$+$2934	& 13 01 25.9	& $+$29 18 46	
& 7071	& 102.8	& 11.33	& LINER     & 2009-06-13 \\
17	& MCG $-$03-34-064\dotfill		& F13197$-$1627	& 13 22 23.5	& $-$16 43 34	
& 5152	& 74.6	& 11.28	& Sy1       & 2011-04-28    \\
18	& NGC 5135\dotfill				& F13229$-$2934	& 13 25 43.0	& $-$29 49 54	
& 4114	& 59.4	& 11.27	& Sy2       & 2011-04-28 \\
19	& NGC 5257/8\dotfill			& F13373$+$0105	& 13 39 54.9	& $+$00 50 07	
& 6798	& 98.8	& 11.54	& H{\sc ii}       & 2011-04-27 \\
20	& IC 4518A/B\dotfill			& F14544$-$4255	& 14 57 43.1	& $-$43 08 01	
& 4715	& 68.2	& 11.09	& Sy2       & 2011-04-24  \\
21	& IC 4687/86\dotfill			& F18093$-$5744	& 18 13 38.6	& $-$57 43 36	
& 5188	& 75.0	& 11.55	& H{\sc ii} (both)	& 2011-04-25 \\
22	& IRAS F18293-3413\dotfill	& F18293$-$3413	& 18 32 40.2	& $-$34 11 26	
& 5449	& 78.9	& 11.82	& H{\sc ii}       & 2009-06-12 \\
23	& ESO 339-G011\dotfill			& F19542$-$3804	& 19 57 37.5	& $-$37 56 10	
& 5722	& 82.9	& 11.14	& Sy2       & 2009-10-25 \\
24	& NGC 6926\dotfill				& F20304$-$0211	& 20 33 04.8	& $-$02 01 39	
& 5970	& 86.6	& 11.32	& Sy2       & 2009-06-12 \\
25	& IC 5063\dotfill				& F20482$-$5715	& 20 52 03.5	& $-$57 04 03	
& 3380	& 48.7	& 10.86	& Sy2       & 2010-10-15 \\
26	& ESO 286-G035\dotfill			& F21008$-$4347	& 21 04 11.2	& $-$43 35 34	
& 5208 & 75.4	& 11.25	& H{\sc ii}       & 2009-10-27 \\
27	& ESO 343-IG013\dotfill			& F21330$-$3846	& 21 36 10.8	& $-$38 32 38	
& 5714	& 82.8	& 11.10	& H{\sc ii}       & 2009-10-25 \\
28	& NGC 7130\dotfill				& F21453$-$3511	& 21 48 19.6	& $-$34 57 05	
& 4824	& 69.8	& 11.39	& LINER/Sy1	& 2009-10-26 \\
29	& IC 5179\dotfill				& F22132$-$3705	& 22 16 10.0	& $-$36 50 35	
& 3398	& 49.0	& 11.21	& H{\sc ii}       & 2010-10-14 \\
30	& ESO 534-G009\dotfill			& F22359$-$2606	& 22 38 40.8	& $-$25 51 05	
& 3393	& 48.9	& 10.70	& LINER     & 2010-10-07 \\
31	& NGC 7469\dotfill				& F23007$+$0836	& 23 03 15.5	& $+$08 52 25	
& 4922	& 71.2	& 11.67	& Sy1       & 2009-10-22 \\
32	& CGCG 453-062\dotfill			& F23024$+$1916	& 23 04 55.2	& $+$19 33 01	
& 7524	& 109.6	& 11.41	& LINER     & 2010-10-19 \\
33	& NGC 7591\dotfill				& F23157$+$0618 & 23 18 15.7	& $+$06 35 06	
& 4961	& 71.8	& 11.11	& LINER     & 2011-10-16  \\
34	& NGC 7678\dotfill				& F23259$+$2208	& 23 28 27.0	& $+$22 25 09	
& 3482	& 50.2	& 10.84	& H{\sc ii}       & 2010-10-20 \\
35	& MCG $-$01-60-022\dotfill		& F23394$-$0353	& 23 42 02.2	& $-$03 36 48	
& 6966	& 101.3	& 11.29	& H{\sc ii}       & 2009-10-17 \\
36	& NGC 7771\dotfill				& F23488$+$1949	& 23 51 24.7	& $+$20 06 39	
& 4336	& 62.6	& 11.42	& H{\sc ii}       & 2009-10-27 \\
37	& Mrk 0331\dotfill				& F23488$+$2018	& 23 51 26.1	& $+$20 35 08	
& 5371	& 77.8	& 11.48	& H{\sc ii}/Sy2	& 2010-10-15 \\
38	& UGC 12914/15\dotfill			& F23591$+$2312	& 00 01 40.7	& $+$23 29 37	
& 4534	& 65.5	& 10.99	& LINER     & 2010-10-09
\enddata
%% Text for table notes should follow after the \enddata but before
%% the \end{deluxetable}. Make sure there is at least one \tablenotemark
%% in the table for each \tablenotetext.
\tablecomments{Column (1): Galaxy ID in this paper. Column (2): Galaxy name. Column (3): IRA catalog ID. Column (4): Right Ascension for the epoch 2000. Column (5): Declination for the epoch 2000. Column (6): Recession velocity. Column (7): Luminosity distance based on the parameters of $\Lambda$-CDM cosmology \citep[e.g.,][]{2003ApJS..148..175S}. Column (8): Bolometric Infrared luminosities ($L_{\mathrm{IR(8-1000\mu m)}}$ $(L_{\odot})$) in \citet{2003AJ....126.1607S} corrected for the different cosmic parameters. Column (9): Classification by optical spectroscopic observation taken from \cite{1998ApJ...508..627K} and \citet{2006ApJ...650..835A}. Sy1: Seyfert 1, Sy2: Seyfert 2, LINER: LINER, H{\sc ii}: H{\sc ii} region. (10): Observation date, yyyy-mm-dd.}
%\tablenotetext{a}{Sample footnote for table~\ref{tbl-1} that was generated with the deluxetable environment}
%\tablenotetext{b}{Another sample footnote for table~\ref{tbl-1}}
\end{deluxetable}
%\end{landscape}
%\end{sidewaystable}
%\clearpage
%\end{turnpage}

%-------------

\twocolumngrid

\noindent at the Cassegrain focus of the University of Tokyo Atacama Observatory (TAO, \citealt{2010SPIE.7733E...6Y}) 1.0 m telescope (miniTAO; \citealt{2010SPIE.7733E.163M}).The observatory is located at the summit of Co. Chajnantor (5640m altitude) in northern Chile.

The value of the precipitable water vapor (PWV) is 0.1--0.7 mm (K14), and the atmospheric window opens around the wavelength of Pa$\alpha$ (1.8751 $\mu \mathrm{m}$). The median value of the PWV during the observation is 0.5 mm which is described in section 3.2. Figure \ref{fig:transmodel} shows the simulated atmospheric transmittance with the ATRAN \citep{Lord...1992} at the TAO site. It can be seen that the TAO site (altitude of 5,640 m, bold line) shows higher transmission than the Very Large Telescope (VLT) site at a lower altitude (2,600 m, dotted line), especially at the wavelength range around Pa$\alpha$.

We have carried out 5 observation runs from 2009 to 2011. 38 targets (44 individual galaxies) have been observed with the $N191$ narrow-band filter, which has the central wavelength ($\lambda_\mathrm{c}$) of 1.9105 $\mu$m with a FWHM of 0.0079 $\mu$m ($\Delta\lambda$) to cover redshifted Pa$\alpha$ line. Also, we have carried out the observations with the $H$ and $K_\mathrm{s}$ broad-band filters to obtain stellar continuum images. The profiles of these filters are also shown in Figure \ref{fig:transmodel}. Typical seeing size is $0\farcs8$ during the observations. The total integration time for each galaxy is 540 s (60 s $\times$ 9 dithering) for the $H$ and $K_\mathrm{s}$ filters, and 1620 s (180 s $\times$ 9 dithering) for the $N191$ except for UGC 12914/5 which is observed with longer integration time of 12420 s for a detailed study \citep{2012ApJ...757..138K}. As the Pa$\alpha$ emission line is strongly affected by the PWV, the observations have been carried out during nights with low PWV (about 0.5 mm; details are described in the next section).

%%%==========================================================================%%%
%%%   Section 3 Reductions
%%%==========================================================================%%%
\section{Data Reductions}

%%%---< Subsection 3.1 Data Reduction >
\subsection{Reduction Procedure}
The data are reduced using the standard IRAF software packages. In the first step of the data reduction, flat pattern and sky background are removed from raw images. A flat pattern image is made from a sky image produced by stacking all object-masked images per 
\clearpage
\onecolumngrid

%-------------
%\begin{turnpage}
%\begin{landscape}
\begin{deluxetable}{ccccccccc}
\tabletypesize{\tiny}
%\rotate
%\tabletypesize{\normalsize}
\tablecolumns{9}
\tablewidth{0pc}
%\rotate
\tablecaption{Pa$\alpha$ fluxes and derived quantities of atmospheric condition with miniTAO/ANIR. \label{table_flux}}
%\tablewidth{0pt}

\tablehead{
\colhead{ID} & \colhead{Galaxy}& \colhead{$f(\mathrm{Pa}\alpha)$}& \colhead{$\sigma(\mathrm{Pa}\alpha)_\mathrm{phot}$}& \colhead{$\sigma(\mathrm{Pa}\alpha)_\mathrm{atm}$}& \colhead{$f(\mathrm{Pa}\alpha)_\mathrm{miss}$}& \colhead{PWV}
& \colhead{$(T^{\mathrm{PWV},N191}_\mathrm{atm})^X$} & \colhead{$T_\mathrm{line}$}\\
 & \colhead{Name}& \colhead{(ergs cm$^{-2}$ s$^{-1}$)}& \colhead{(ergs cm$^{-2}$ s$^{-1}$)}& \colhead{(ergs cm$^{-2}$ s$^{-1}$)}& \colhead{(ergs cm$^{-2}$ s$^{-1}$)}& \colhead{($\mu$m)}
& \colhead{(\%)} & \colhead{(\%)}\\
\colhead{(1)} & \colhead{(2)} & \colhead{(3)}& \colhead{(4)}& \colhead{(5)}& \colhead{(6)}& \colhead{(7)}& \colhead{(8)}& \colhead{(9)}
}
\startdata
1 & NGC 23\dotfill & 2.28$\times$10$^{-13}$ & $\pm$ 7.21$\times$10$^{-16}$ & $\pm$ 2.79$\times$10$^{-14}$ & $+$ 4.33$\times$10$^{-15}$ & 629.6 & 57.3 & 51.4\\
2 & NGC 34\dotfill & 3.73$\times$10$^{-13}$ & $\pm$ 1.54$\times$10$^{-15}$ & $\pm$ 4.55$\times$10$^{-14}$ & $+$ 1.05$\times$10$^{-14}$ & 485.7 & 63.2 & 55.6\\
3 & NGC 232\dotfill & 8.66$\times$10$^{-14}$ & $\pm$ 9.29$\times$10$^{-16}$ & $\pm$ 1.06$\times$10$^{-14}$ & $+$ 4.84$\times$10$^{-15}$ & 197.3 & 74.1 & 66.3\\
4 & IC 1623A/B\dotfill & 6.58$\times$10$^{-13}$ & $\pm$ 5.81$\times$10$^{-16}$ & $\pm$ 8.03$\times$10$^{-14}$ & $+$ 2.62$\times$10$^{-15}$ & 446.3 & 64.7 & 55.4\\
5 & ESO 244-G012\dotfill & 3.63$\times$10$^{-13}$ & $\pm$ 6.88$\times$10$^{-16}$ & $\pm$ 4.43$\times$10$^{-14}$ & $+$ 4.66$\times$10$^{-15}$ & 385.6 & 67.1 & 85.6\\
6 & UGC 2238\dotfill & 3.71$\times$10$^{-13}$ & $\pm$ 1.42$\times$10$^{-15}$ & $\pm$ 4.53$\times$10$^{-14}$ & $+$ 9.88$\times$10$^{-15}$ & 576.6 & 59.2 & 60.6\\
7 & IRAS F02437$\pm$2122\dotfill & 1.99$\times$10$^{-14}$ & $\pm$ 1.26$\times$10$^{-15}$ & $\pm$ 2.42$\times$10$^{-15}$ & $+$ 5.35$\times$10$^{-15}$ & 359.7 & 68.1 & 81.4\\
8 & UGC 2982\dotfill & 4.00$\times$10$^{-13}$ & $\pm$ 7.62$\times$10$^{-16}$ & $\pm$ 4.88$\times$10$^{-14}$ & $+$ 4.00$\times$10$^{-15}$ & 82.7 & 83.4 & 80.2\\
9 & NGC 1614\dotfill & 1.06$\times$10$^{-12}$ & $\pm$ 1.18$\times$10$^{-15}$ & $\pm$ 1.29$\times$10$^{-13}$ & $+$ 8.53$\times$10$^{-15}$ & 507.5 & 62.3 & 50.1\\
10 & MCG $-$05-12-006\dotfill & 1.61$\times$10$^{-13}$ & $\pm$ 6.69$\times$10$^{-16}$ & $\pm$ 1.97$\times$10$^{-14}$ & $+$ 5.44$\times$10$^{-15}$ & 1035.1 & 48.1 & 55.5\\
11 & NGC 1720\dotfill & 7.60$\times$10$^{-14}$ & $\pm$ 3.21$\times$10$^{-16}$ & $\pm$ 9.28$\times$10$^{-15}$ & $+$ 2.04$\times$10$^{-15}$ & 74.9 & 84.3 & 82.9\\
12 & ESO 557-G002\dotfill & 8.90$\times$10$^{-13}$ & $\pm$ 5.56$\times$10$^{-16}$ & $\pm$ 1.09$\times$10$^{-14}$ & $+$ 2.89$\times$10$^{-15}$ & 435.0 & 65.2 & 61.9\\
13 & IRAS F06592-6313\dotfill & 6.17$\times$10$^{-14}$ & $\pm$ 6.52$\times$10$^{-16}$ & $\pm$ 7.53$\times$10$^{-15}$ & $+$ 3.66$\times$10$^{-15}$ & 701.3 & 55.1 & 78.4\\
14 & NGC 2342\dotfill & 2.22$\times$10$^{-13}$ & $\pm$ 7.66$\times$10$^{-16}$ & $\pm$ 2.71$\times$10$^{-14}$ & $+$ 4.46$\times$10$^{-15}$ & 1307.5 & 43.9 & 56.1\\
15 & ESO 320-G030\dotfill & 1.64$\times$10$^{-13}$ & $\pm$ 4.98$\times$10$^{-16}$ & $\pm$ 2.00$\times$10$^{-14}$ & $+$ 2.24$\times$10$^{-15}$ & 532.5 & 61.2 & 81.9\\
16 & NGC 4922\dotfill & 4.35$\times$10$^{-14}$ & $\pm$ 7.20$\times$10$^{-16}$ & $\pm$ 5.30$\times$10$^{-15}$ & $+$ 3.10$\times$10$^{-15}$ & 516.2 & 61.9 & 65.9\\
17 & MCG $-$03-34-064\dotfill & 1.12$\times$10$^{-13}$ & $\pm$ 1.41$\times$10$^{-17}$ & $\pm$ 1.37$\times$10$^{-14}$ & $+$ 3.68$\times$10$^{-16}$ & 512.2 & 62.0 & 57.8\\
18 & NGC 5135\dotfill & 4.09$\times$10$^{-13}$ & $\pm$ 6.29$\times$10$^{-16}$ & $\pm$ 4.99$\times$10$^{-14}$ & $+$ 3.28$\times$10$^{-15}$ & 759.3 & 53.4 & 39.0\\
19a & NGC 5257\dotfill & 2.27$\times$10$^{-13}$ & $\pm$ 2.28$\times$10$^{-16}$ & $\pm$ 2.76$\times$10$^{-14}$ & $+$ 1.05$\times$10$^{-15}$ & 913.5 & 50.2 & 70.5\\
19b & NGC 5258\dotfill & 1.66$\times$10$^{-13}$ & $\pm$ 2.34$\times$10$^{-16}$ & $\pm$ 2.02$\times$10$^{-14}$ & $+$ 8.62$\times$10$^{-16}$ & 913.5 & 50.2 & 70.5\\
20a & IC 4518A\dotfill & 8.11$\times$10$^{-14}$ & $\pm$ 6.30$\times$10$^{-16}$ & $\pm$ 9.89$\times$10$^{-15}$ & $+$ 2.73$\times$10$^{-15}$ & 547.2 & 60.5 & 50.6\\
20b & IC 4518B\dotfill & 5.15$\times$10$^{-14}$ & $\pm$ 1.06$\times$10$^{-15}$ & $\pm$ 6.28$\times$10$^{-15}$ & $+$ 3.31$\times$10$^{-15}$ & 547.2 & 60.5 & 50.6\\
21a & IC 4686\dotfill & 1.31$\times$10$^{-13}$ & $\pm$ 4.03$\times$10$^{-16}$ & $\pm$ 1.60$\times$10$^{-14}$ & $+$ 3.02$\times$10$^{-15}$ & 923.8 & 50.0 & 50.9\\
21b & IC 4687\dotfill & 8.04$\times$10$^{-13}$ & $\pm$ 3.75$\times$10$^{-16}$ & $\pm$ 9.80$\times$10$^{-14}$ & $+$ 1.88$\times$10$^{-15}$ & 923.8 & 50.0 & 50.9\\
21c & IC 4689\dotfill & 2.21$\times$10$^{-13}$ & $\pm$ 4.96$\times$10$^{-16}$ & $\pm$ 2.57$\times$10$^{-14}$ & $+$ 5.58$\times$10$^{-15}$ & 923.8 & 50.0 & 50.9\\
22 & IRAS F18293-3413\dotfill & 8.44$\times$10$^{-13}$ & $\pm$ 7.40$\times$10$^{-16}$ & $\pm$ 1.03$\times$10$^{-13}$ & $+$ 3.62$\times$10$^{-14}$ & 621.7 & 57.5 & 73.6\\
23 & ESO 339-G011\dotfill & 8.95$\times$10$^{-14}$ & $\pm$ 8.77$\times$10$^{-16}$ & $\pm$ 1.09$\times$10$^{-14}$ & $+$ 5.01$\times$10$^{-15}$ & 596.7 & 58.3 & 57.5\\
24 & NGC 6926\dotfill & 1.29$\times$10$^{-13}$ & $\pm$ 8.28$\times$10$^{-16}$ & $\pm$ 1.57$\times$10$^{-14}$ & $+$ 4.75$\times$10$^{-15}$ & 441.3 & 64.9 & 57.2\\
25 & IC 5063\dotfill & 9.30$\times$10$^{-14}$ & $\pm$ 3.67$\times$10$^{-16}$ & $\pm$ 1.13$\times$10$^{-14}$ & $+$ 2.76$\times$10$^{-15}$ & 350.5 & 68.4 & 86.8\\
26 & ESO 286-G035\dotfill & 2.59$\times$10$^{-13}$ & $\pm$ 7.44$\times$10$^{-16}$ & $\pm$ 3.16$\times$10$^{-14}$ & $+$ 3.92$\times$10$^{-15}$ & 526.5 & 61.4 & 64.7\\
27 & ESO 343-IG013\dotfill & 8.64$\times$10$^{-14}$ & $\pm$ 7.40$\times$10$^{-16}$ & $\pm$ 1.05$\times$10$^{-14}$ & $+$ 3.91$\times$10$^{-15}$ & 469.7 & 63.8 & 63.8\\
28 & NGC 7130\dotfill & 2.32$\times$10$^{-13}$ & $\pm$ 1.12$\times$10$^{-15}$ & $\pm$ 2.83$\times$10$^{-14}$ & $+$ 2.46$\times$10$^{-14}$ & 491.2 & 62.9 & 44.1\\
29 & IC 5179\dotfill & 5.65$\times$10$^{-13}$ & $\pm$ 2.78$\times$10$^{-16}$ & $\pm$ 6.89$\times$10$^{-14}$ & $+$ 1.17$\times$10$^{-15}$ & 501.6 & 62.5 & 82.4\\
30 & ESO 534-G009\dotfill & 2.28$\times$10$^{-14}$ & $\pm$ 8.82$\times$10$^{-16}$ & $\pm$ 2.78$\times$10$^{-15}$ & $+$ 2.92$\times$10$^{-15}$ & 255.4 & 71.5 & 88.6\\
31 & NGC 7469\dotfill & 6.11$\times$10$^{-13}$ & $\pm$ 8.18$\times$10$^{-16}$ & $\pm$ 7.45$\times$10$^{-14}$ & $+$ 4.10$\times$10$^{-15}$ & 156.7 & 77.1 & 58.9\\
32 & CGCG 453-062\dotfill & 8.59$\times$10$^{-14}$ & $\pm$ 3.44$\times$10$^{-16}$ & $\pm$ 1.05$\times$10$^{-14}$ & $+$ 1.72$\times$10$^{-15}$ & 172.5 & 76.0 & 85.4\\
33 & NGC 7591\dotfill & 1.10$\times$10$^{-13}$ & $\pm$ 6.39$\times$10$^{-16}$ & $\pm$ 1.34$\times$10$^{-14}$ & $+$ 3.05$\times$10$^{-15}$ & 37.4 & 90.4 & 81.4\\
34 & NGC 7678\dotfill & 1.13$\times$10$^{-13}$ & $\pm$ 1.08$\times$10$^{-14}$ & $\pm$ 2.38$\times$10$^{-14}$ & $+$ 1.92$\times$10$^{-15}$ & 301.2 & 70.2 & 83.0\\
35 & MCG $-$01-60-022\dotfill & 1.31$\times$10$^{-13}$ & $\pm$ 6.08$\times$10$^{-16}$ & $\pm$ 1.60$\times$10$^{-14}$ & $+$ 3.05$\times$10$^{-15}$ & 328.9 & 69.2 & 83.8\\
36a & NGC 7770\dotfill & 1.87$\times$10$^{-13}$ & $\pm$ 8.16$\times$10$^{-16}$ & $\pm$ 2.28$\times$10$^{-14}$ & $+$ 4.21$\times$10$^{-15}$ & 674.1 & 55.9 & 53.5\\
36b & NGC 7771\dotfill & 3.78$\times$10$^{-13}$ & $\pm$ 1.78$\times$10$^{-15}$ & $\pm$ 4.61$\times$10$^{-14}$ & $+$ 9.24$\times$10$^{-15}$ & 674.1 & 55.9 & 53.5\\
37 & MrK 331\dotfill & 2.96$\times$10$^{-13}$ & $\pm$ 1.06$\times$10$^{-15}$ & $\pm$ 3.60$\times$10$^{-14}$ & $+$ 5.62$\times$10$^{-15}$ & 377.1 & 67.4 & 81.2\\
38a & UGC 12914\dotfill & 1.23$\times$10$^{-13}$ & $\pm$ 2.58$\times$10$^{-16}$ & $\pm$ 1.51$\times$10$^{-14}$ & $+$ 1.41$\times$10$^{-15}$ & 364.9 & 67.9 & 62.8\\
38b & UGC 12915\dotfill & 7.49$\times$10$^{-14}$ & $\pm$ 1.72$\times$10$^{-15}$ & $\pm$ 9.14$\times$10$^{-15}$ & $+$ 1.07$\times$10$^{-15}$ & 364.9 & 67.9 & 62.8
\enddata
%% Text for table notes should follow after the \enddata but before
%% the \end{deluxetable}. Make sure there is at least one \tablenotemark
%% in the table for each \tablenotetext.
\tablecomments{Column (1): Galaxy ID in this paper. Column (2): Galaxy name. Column (3): Observed Pa$\alpha$ total flux corrected for atmospheric absorption. Column (4): 1$\sigma$ photometric error of Pa$\alpha$ flux. Column (5): Estimated error by the atmospheric absorption correction of the Pa$\alpha$ flux. Column (6): 5$\sigma$ error by continuum subtraction of the Pa$\alpha$ flux. Column (7): Estimated PWV using the calibration method described in K14, submitted. Column (8): Estimated effective atmospheric transmittance within the $N191$ filter. Column (9): Estimated the effective line transmittance.}
%\tablenotetext{a}{Sample footnote for table~\ref{tbl-1} that was generated with the deluxetable environment}
%\tablenotetext{b}{Another sample footnote for table~\ref{tbl-1}}
\end{deluxetable}
%\end{landscape}
%\end{sidewaystable}
%\clearpage
%\end{turnpage}

%-------------

\twocolumngrid

\noindent observation run for each filter. The skybackground is removed using a self-sky image which is made by stacking object-masked images in the same dither sequence. We do not correct the image distortions because it is negligible on ANIR. Each image is matched with World Coordinate System (WCS) using position of 2MASS stars \citep{2006AJ....131.1163S}. Then, these images are shifted according to WCS and co-added. In the second step, the flux scale of the combined image is calibrated by 2MASS stars. Comparing stars in the image with those in the 2MASS catalog, zero-point magnitude and system efficiency are derived. Because reference magnitudes of the $N191$ filter are not available we derived them by interpolating the $H$- and $K_\mathrm{s}$- band magnitudes in the 2MASS catalog. Details of this flux calibration procedure are described in K14, arguing that the interpolating calibration technique produces negligible ($<$ 0.01 mag) systematic error. The final $H$, $K_\mathrm{s}$ and $N191$ images are convolved with a Gaussian function to match the Point Spread Function (PSF) to the worst among the images (typical spatial resolution of the convolved images is 0\farcs9). Then, a continuum image is made by interpolating between $H$ and $K_\mathrm{s}$ images, and we derive a Pa$\alpha$ line image by subtracting the continuum image from the $N191$ image.

\clearpage
%-------------
\begin{figure}
\begin{center}
\epsscale{1.0}
\plotone{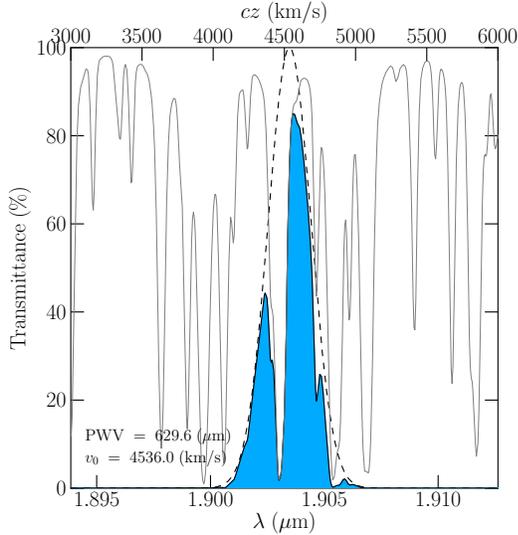}
\end{center}
\caption{An example of Pa$\alpha$ emission line we assume a redshifted Pa$\alpha$ emission line of NGC 0023, convolved with atmospheric transmittance curve, are shown with the thick solid line. The solid-thin line represents the atmospheric transmittance at PWV = 629.6$\ (\mu \mathrm{m})$ calculated by ATRAN. The dashed line represents an intrinsic Pa$\alpha$ line profile without atmospheric absorption. The recession velocity $v_0$ is set to be 4536 ({km~s$^{-1}$}), and $\sigma$ of the intrinsic profile is assumed to be 150 ({km~s$^{-1}$}).}\label{fig:lineprofile}
\end{figure}
%-------------

%%%---< Subsection 3.2 Pa$\alpha$ Flux Calibration >
\subsection{Pa$\alpha$ Flux Calibration}
As there are many atmospheric absorption features within the wavelength range of the $N191$ filter whose strength vary temporally with the PWV, it is difficult to calibrate the emission-line flux accurately (Figure \ref{fig:transmodel}). To recover the intrinsic Pa$\alpha$ fluxes accurately, we have used the following relation:

\begin{eqnarray}
F_{\mathrm{int}}^l &=& \frac{(T^{\mathrm{PWV},N191}_{\mathrm{atm}})^X}{T_{\mathrm{line}}}(f_{\mathrm{cal}}^{N191}-f_{\mathrm{cal}}^{H-K_\mathrm{s}})\Delta\lambda, \label{eq:fintfinal}
\end{eqnarray}

\noindent where $F_{\mathrm{int}}^l$ is intrinsic Pa$\alpha$ flux, $T^{\mathrm{PWV},N191}_\mathrm{atm}$ is averaged atmospheric transmittance within the wavelength of $N191$ filter toward the zenith (airmass $=$ 1), $X$ is airmass, $T_\mathrm{line}$ is effective atmospheric transmittance at the wavelength of redshifted Pa$\alpha$ emission line, $f_\mathrm{cal}^{N191}$ is observed flux-calibrated flux of the $N191$ filter, $f_\mathrm{cal}^{H-K_\mathrm{s}}$ is continuum flux obtained by interpolating $H$ and $K_\mathrm{s}$ broad-band images, and $\Delta\lambda$ is FWHM of the $N191$ narrow-band filter. Detailed derivation of Equation (\ref{eq:fintfinal}) is described in Appendix. To obtain the Pa$\alpha$ flux, we have to estimate $(T^{\mathrm{PWV},N191}_\mathrm{atm})^X$ and $T_\mathrm{line}$.

The factor $(T^{\mathrm{PWV},N191}_\mathrm{atm})^X$ is obtained using the system efficiency of the $H$, $N191$ and $K_\mathrm{s}$ image (K14; \citealt{2012SPIE.8446E..7DT}). Then, a PWV is derived from ATRAN model and $(T^{\mathrm{PWV},N191}_\mathrm{atm})^X$. The factor $T_\mathrm{line}$ is estimated from a model transmittance curve assuming the Pa$\alpha$ emission line profile with the PWV obtained above (details of the estimation is discussed in K14). In a real galaxy, width of an emission line is broadened by more than $\sigma$ = 100 {km~s$^{-1}$} due to velocity dispersion of their emission line clouds. To incorporate this effect, we assume the shape of the emission line to be a Gaussian profile with $\sigma$ = 150 {km~s$^{-1}$} for all the galaxies (Figure \ref{fig:lineprofile}), which shows the least difference between the Pa$\alpha$ fluxes obtained by the above method and those from narrow-band images by $HST$/NICMOS \citep{2012SPIE.8446E..7DT}. Although there is no Pa$\alpha$ line profile measured in local LIRGs, velocity dispersion of $\sigma$ = 150 {km~s$^{-1}$} is reasonable because the value is consistent with those derived for H$\alpha$ lines of local LIRGs \citep{2009A&A...506.1541A}. Dispersion of differences in the fluxes between those obtained from $HST$/NICMOS and from our method is 12.2\% \citep{2012SPIE.8446E..7DT}.

Figure \ref{fig:lineprofile} shows the assumed intrinsic emission-line profile (dashed line), atmospheric transmittance (solid line), and estimated emission-line profile affected by the atmospheric absorption (dark-shaded area). The factor $T_\mathrm{line}$ within the bandpass [$\lambda_1$, $\lambda_2$] of the $N191$ filter is then calculated as follows;
\begin{eqnarray}
T_\mathrm{line} = \frac{\int^{\lambda_2}_{\lambda_1}\exp\biggl[-\biggl(\frac{\lambda-(1+z)\lambda_{\mathrm{Pa}\alpha}}{\sqrt{2}\sigma}\biggl)^2\biggl] T_\mathrm{atm}(\lambda)d\lambda}{\int^{\lambda_2}_{\lambda_1}\exp\biggl[-\biggl(\frac{\lambda-(1+z)\lambda_{\mathrm{Pa}\alpha}}{\sqrt{2}\sigma}\biggl)^2\biggl]d\lambda}, 
\end{eqnarray}

\noindent where $z$ is the redshift obtained from $IRAS$ RBGS catalog \citep{2003AJ....126.1607S}, $T_\mathrm{atm}(\lambda)$ represents the model atmospheric transmittance curve, and $\lambda_{\mathrm{Pa}\alpha}$ is the intrinsic wavelength of Pa$\alpha$ (1.8751 $\mu$m).

In Table \ref{table_flux}, calculated Pa$\alpha$ fluxes as well as $(T^{\mathrm{PWV},N191}_\mathrm{atm})^X$ and $T_\mathrm{line}$ estimated by our method are listed. The median value of the PWV is 0.5 mm during our observation.

%%%==========================================================================%%%
%%%   Section 4. Results
%%%==========================================================================%%%
\section{Results}

%%%---< Subsection 4.1 Final Images >
\subsection{Final Images}
In Figure \ref{fig:imageobj}, all of our 44 individual galaxies in 38 systems are shown. The continuum images made by interpolating the $H$ and $K_\mathrm{s}$ images are shown on the left side and the Pa$\alpha$ line images, which is derived by subtracting continuum image from the $N191$ image, on the right.

%%%---< Subsection 4.2 Pa Flux >
\subsection{Pa$\alpha$ Flux}
To estimate a total Pa$\alpha$ flux of a galaxy, we first estimate how extended emission-line distributions are. Because there are large diversity in Pa$\alpha$ morphology, we use isophotal photometry technique for our image, where 5$\sigma$ isophotal area above the sky background level for Pa$\alpha$ image is used. The 1$\sigma$ noise level is measured in each Pa$\alpha$ image convolved with a Gaussian function with a FWHM of 8 pixels to reduce the noise level, and the 5$\sigma$ area is defined in this convolved line image.

The results are shown in Table \ref{table_flux} and Figure \ref{fig:imageobj}. Photometric uncertainties, $\sigma(\mathrm{Pa}\alpha)_\mathrm{phot}$, are defined as 1$\sigma$ noise level measured by applying many apertures having the same area size as that used for the measurement of the total Pa$\alpha$ flux at blank sky positions. These values are not so large ($\sim$ 0.5\% on average). The $\sigma(\mathrm{Pa}\alpha)_\mathrm{atm}$ is an error due to our Pa$\alpha$ correction method mentioned in Section 3 and set to be 12.2\% \citep{2012SPIE.8446E..7DT}. We use the combination of them,
\begin{eqnarray}
\sigma_\mathrm{total} = \sqrt{\mathstrut \sigma(\mathrm{Pa}\alpha)_\mathrm{phot}^{2} + \sigma(\mathrm{Pa}\alpha)_\mathrm{atm}^{2}},
\end{eqnarray}
\clearpage
\onecolumngrid

%-------------
\begin{figure}[t]
 \epsscale{.80}
  \begin{center}
   \plottwo{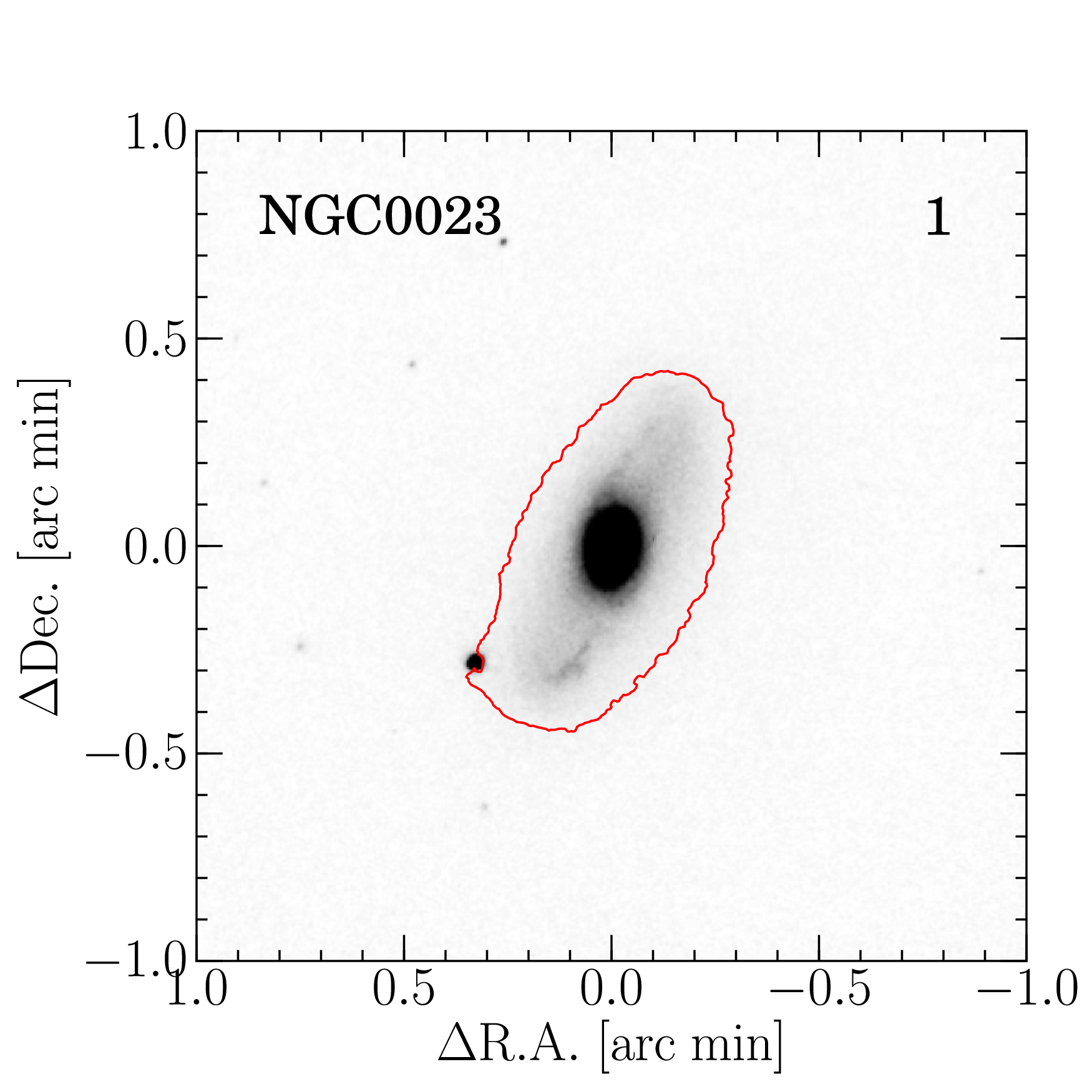}{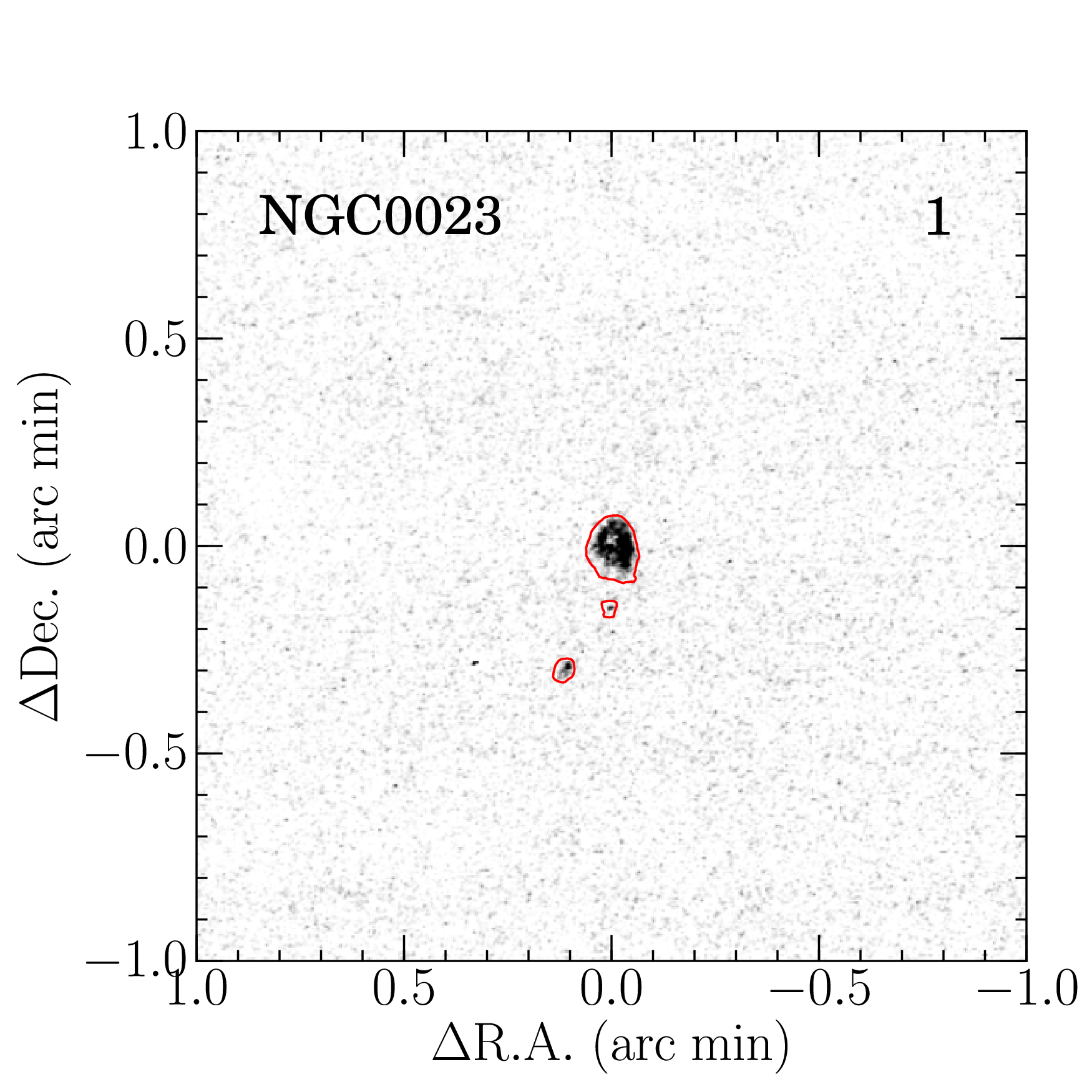}\\
   \plottwo{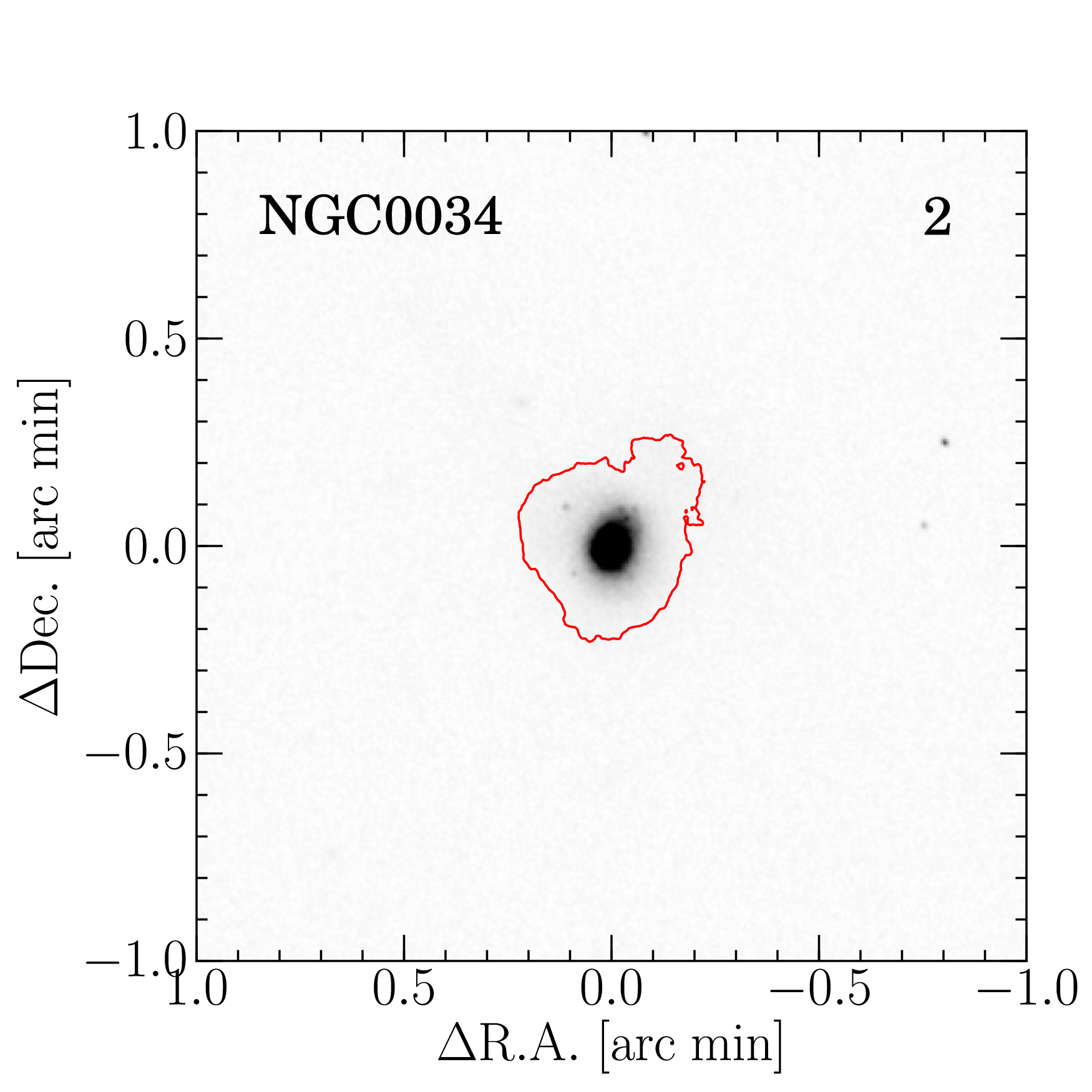}{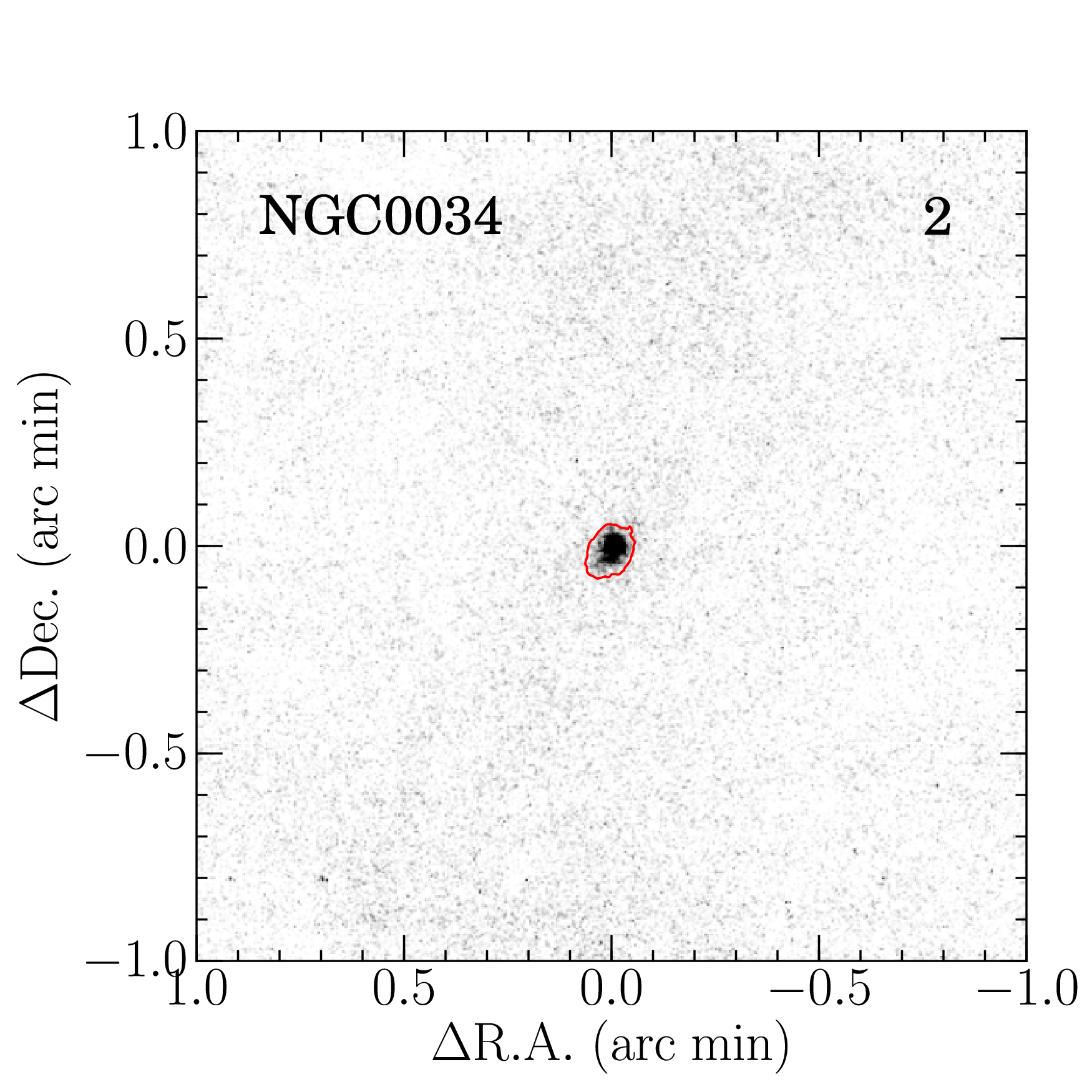}\\
   \plottwo{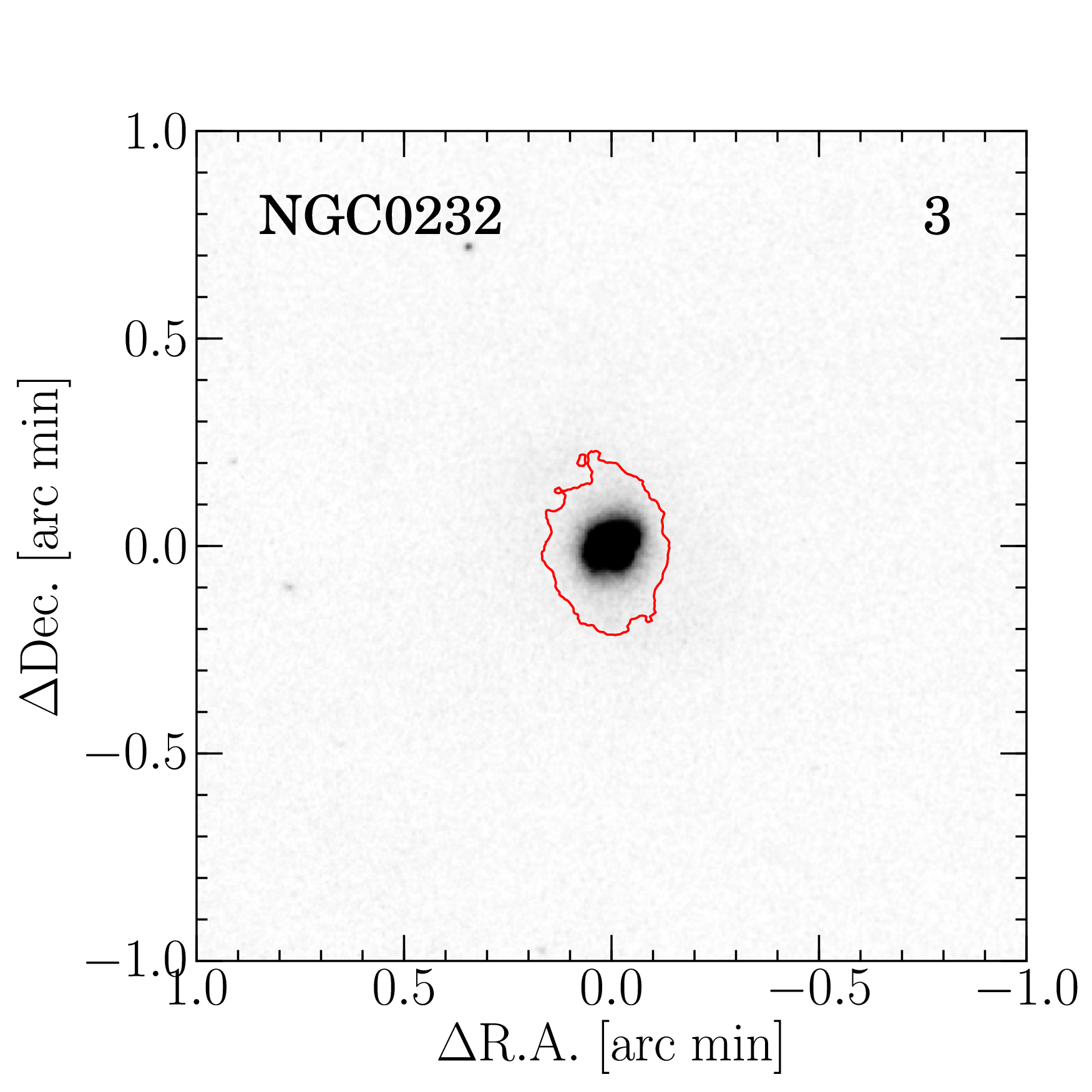}{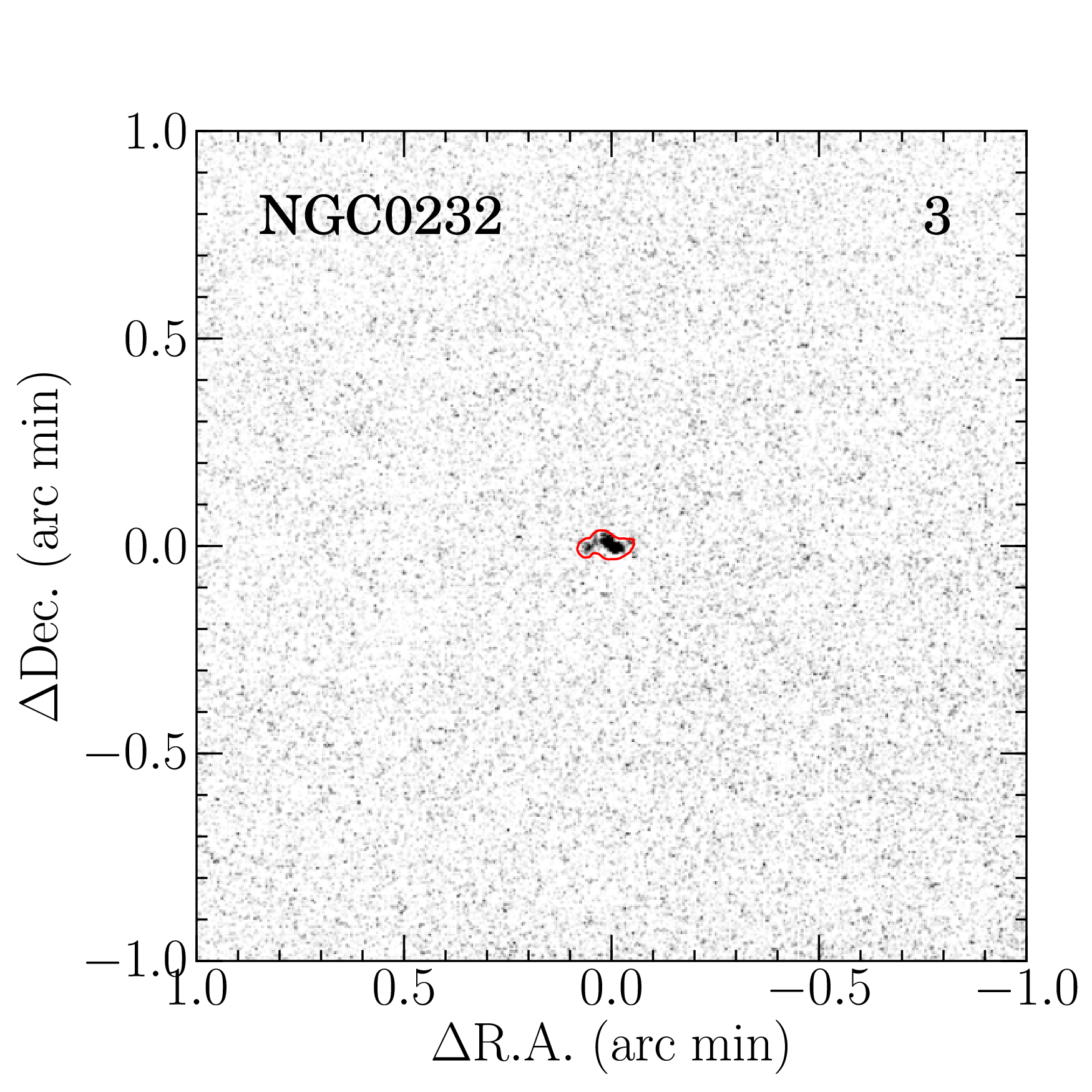}\\
   \end{center}
   \caption{Continuum and Pa$\alpha$ line images of our sample of 44 individual LIRGs in 38 systems observed with miniTAO/ANIR. The continuum images are listed on the left side and Pa$\alpha$ line images on the right. The name of a galaxy is shown at the top-left corner and ID number at the top-right corner in each figure. The solid red lines represent 10$\sigma$ surface brightness level for the continuum image and 5$\sigma$ for the Pa$\alpha$ line image measured on convolved images. The 1$\sigma$ levels are calculated on the convolved images (see text).}\label{fig:imageobj}
 \end{figure}
%-------------

\twocolumngrid
\noindent as the total uncertainties on the measurement of the Pa$\alpha$ flux.

Our results may miss any diffuse Pa$\alpha$ emission whose surface brightness is lower than the 5$\sigma$ threshold. In the $HST$/NICMOS images, missing the diffuse Pa$\alpha$ flux is also pointed out by \cite{2006ApJ...650..835A}. In order to evaluate the maximum missing flux of Pa$\alpha$ line emission ($f(\mathrm{Pa}\alpha)_\mathrm{miss}$), we assume the diffuse emission is spread over an aperture, whose area is defined by 10$\sigma$ isophote for each continuum image. Then, the amount of the missing diffuse component is estimated to be 1.8\% on average at a maximum. 
%5.9\%

\clearpage
\onecolumngrid

\setcounter{figure}{2}
\begin{figure}[htb]
 \epsscale{0.8}
  \begin{center}
   \plottwo{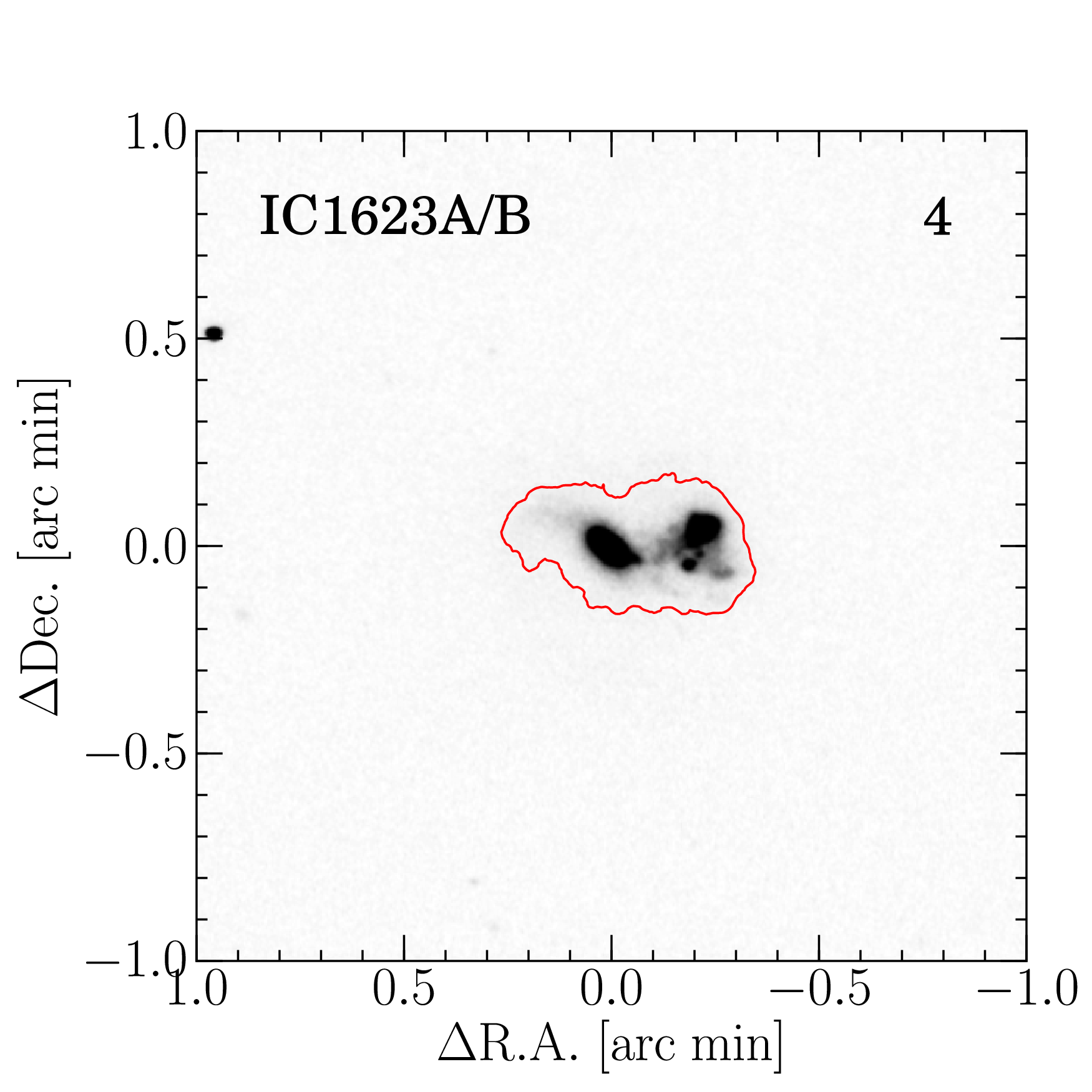}{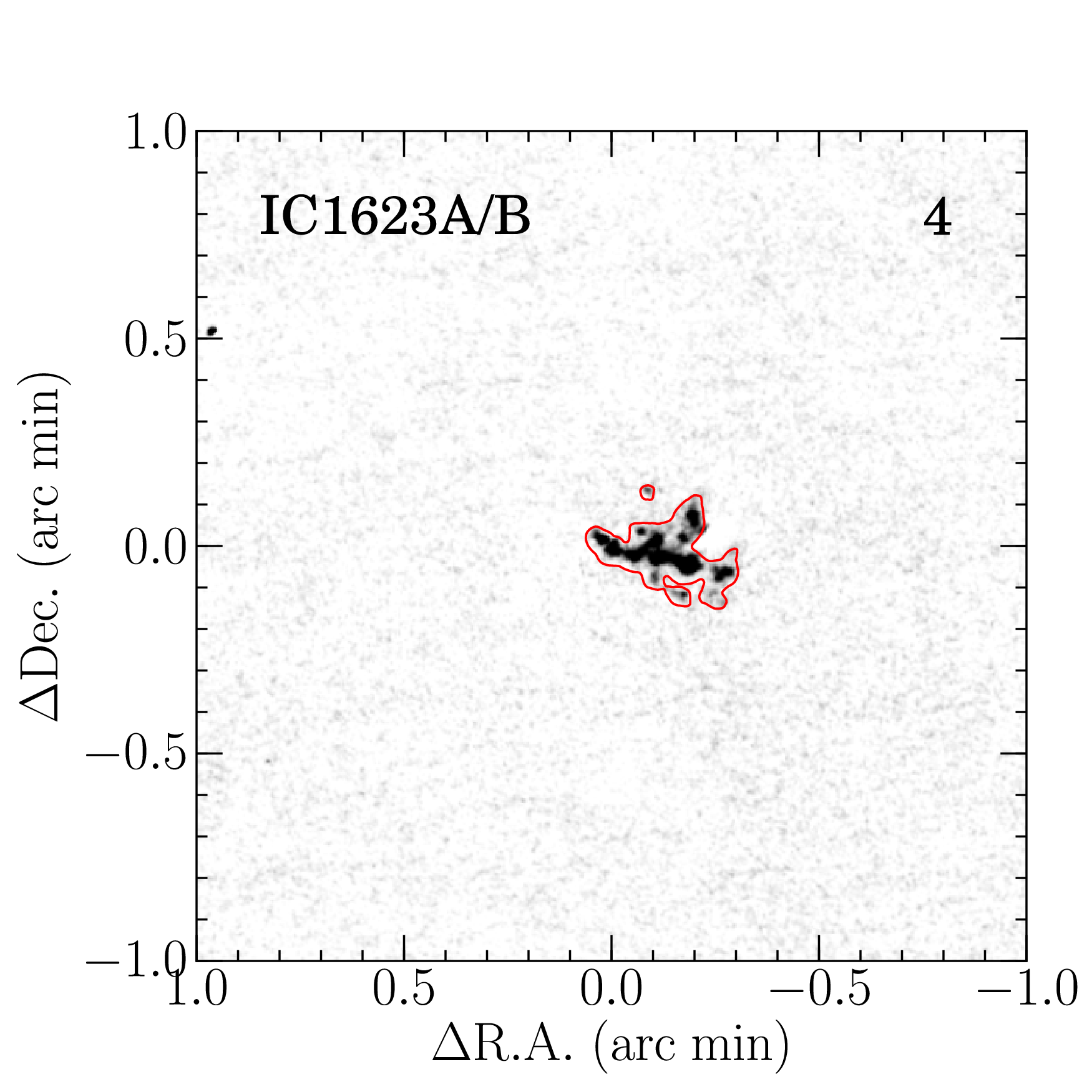}\\
   \plottwo{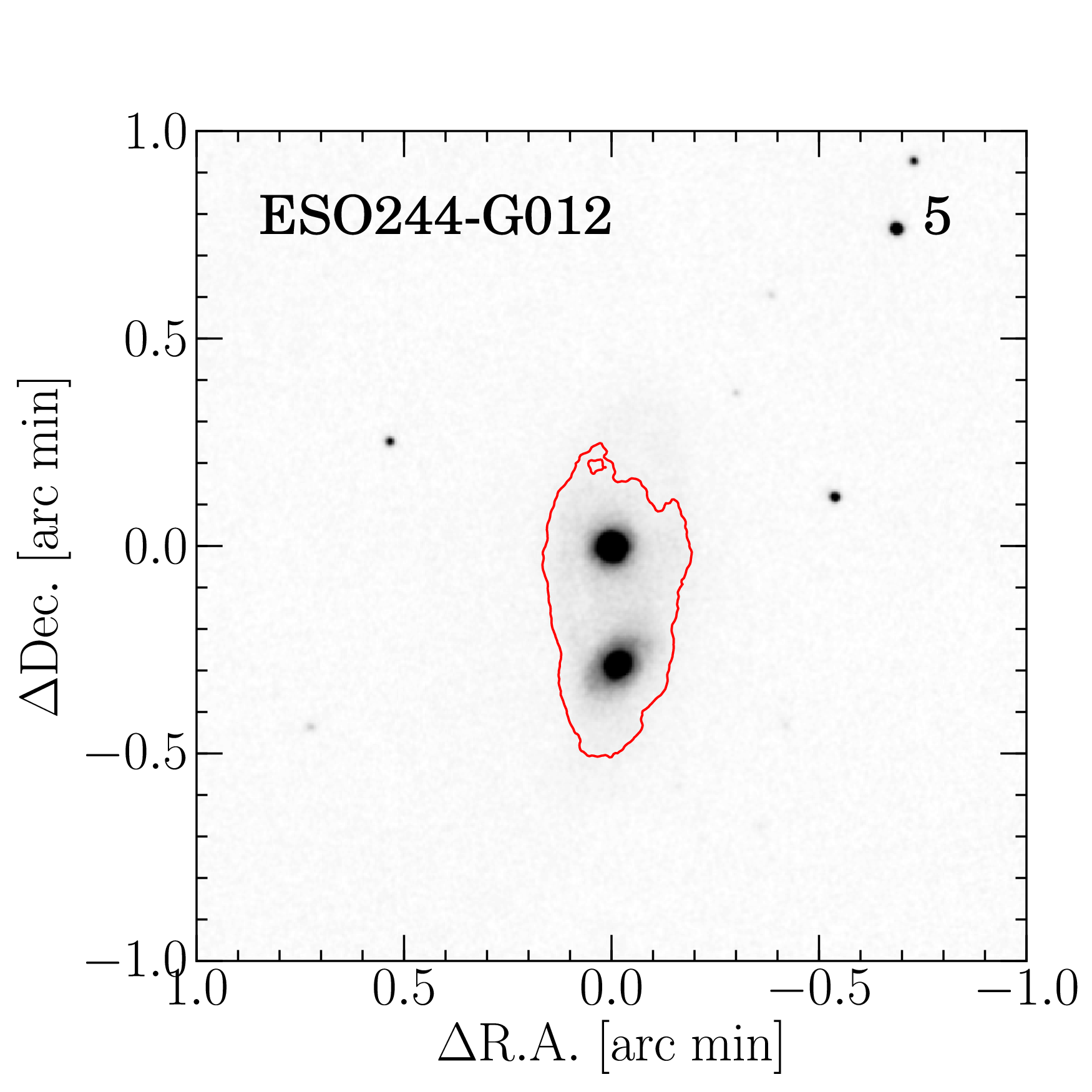}{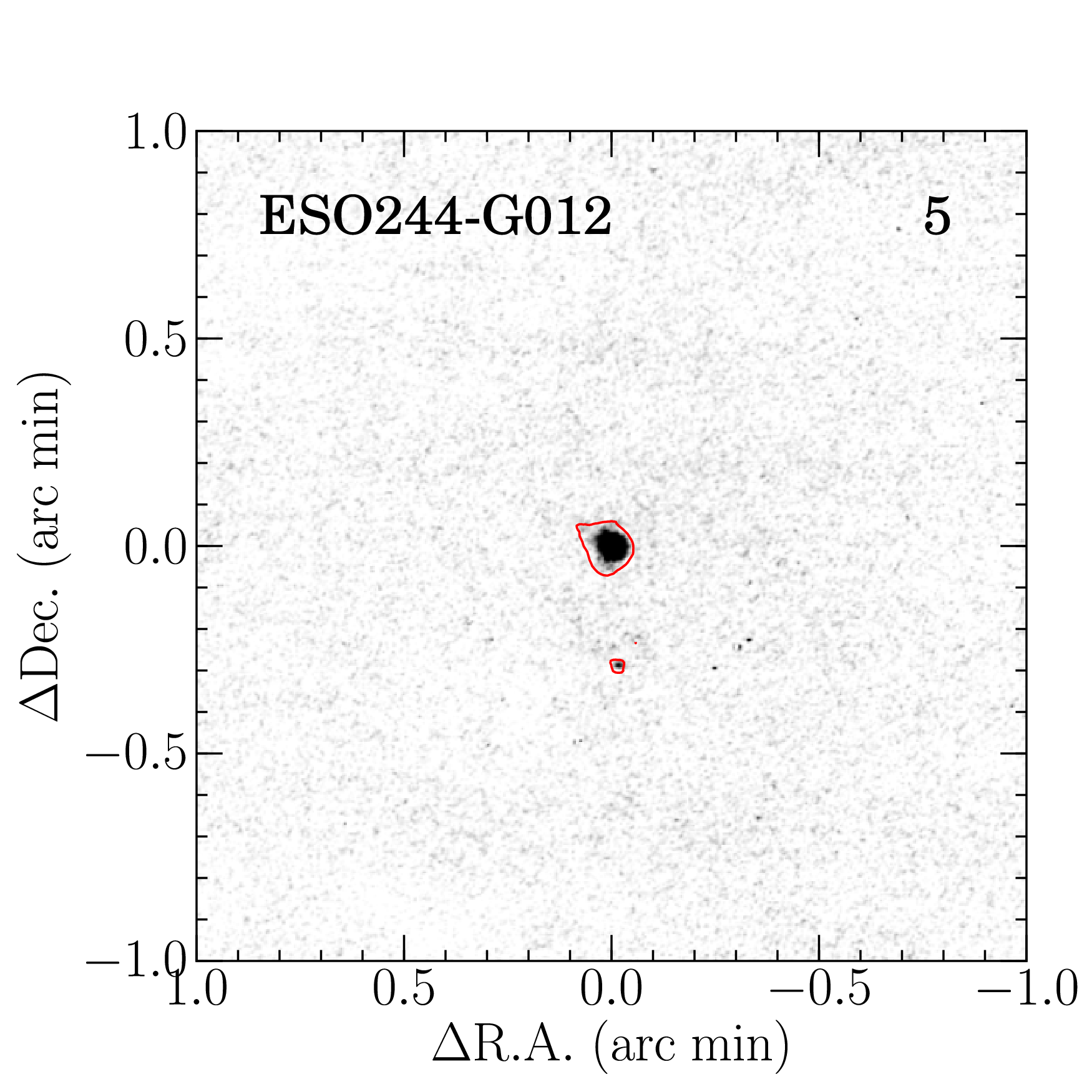}\\
   \plottwo{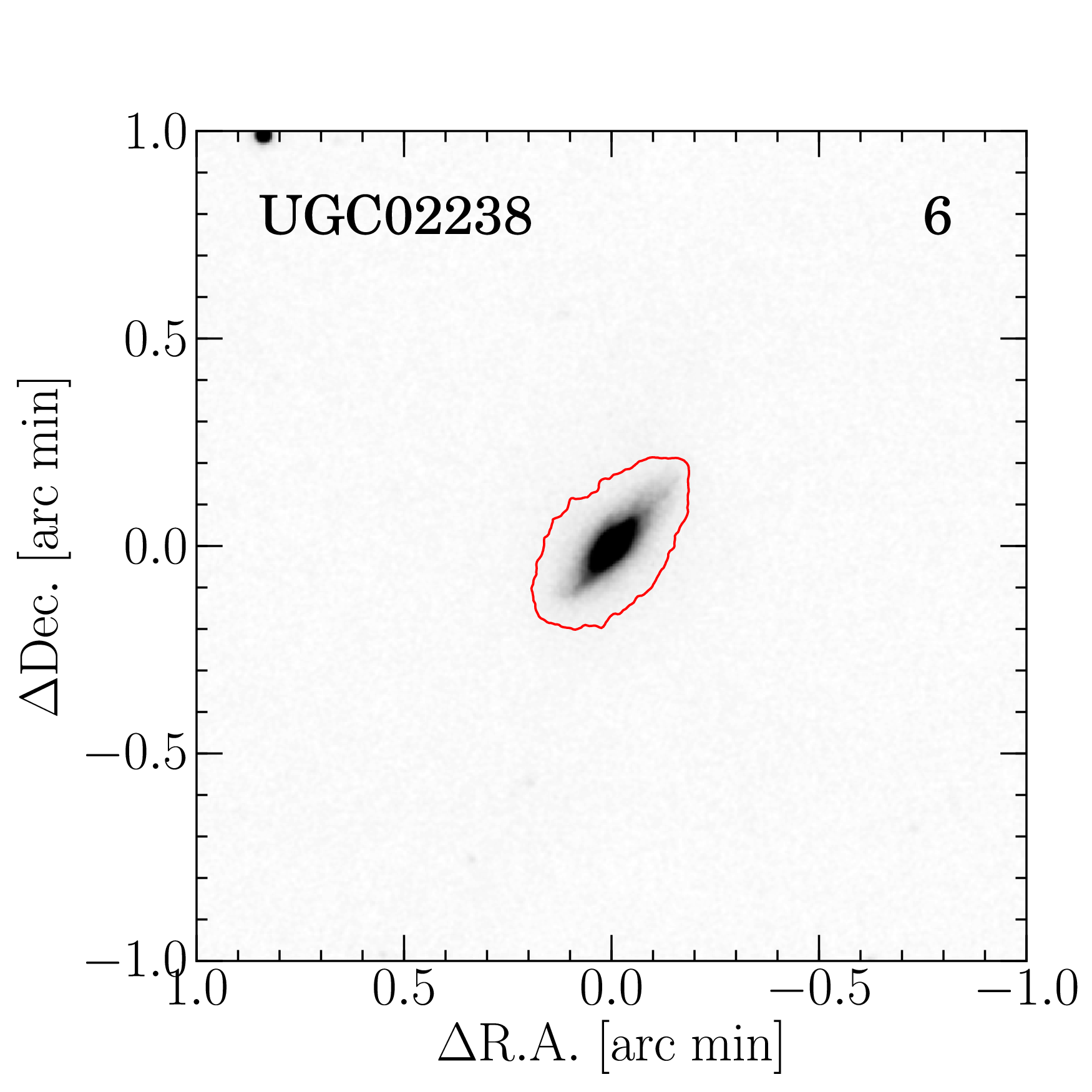}{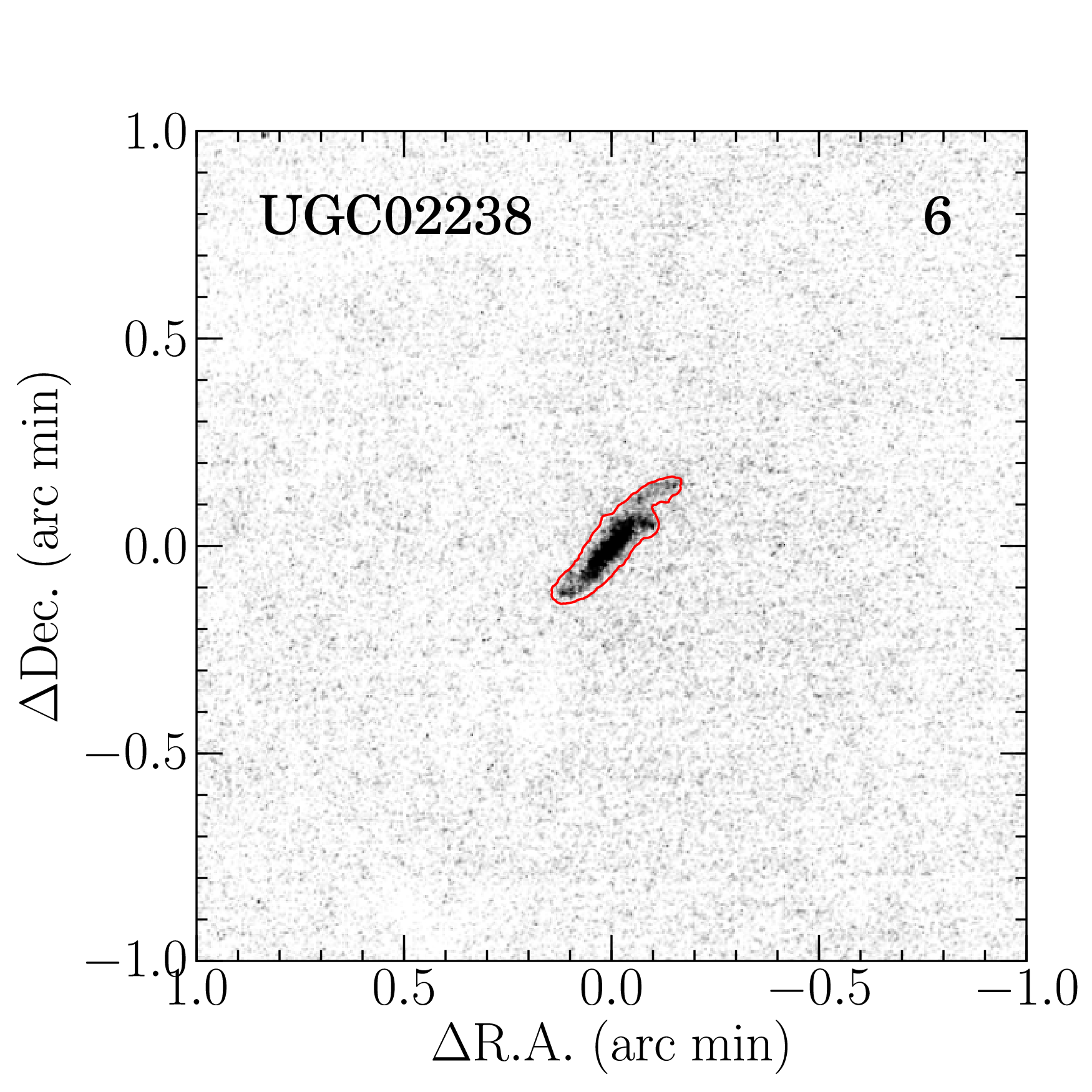}\\
   \end{center}
   \caption{Continued}
 \end{figure}

\twocolumngrid

%%%---< Subsection 4.3 Notes on Individual Objects >
\subsection{Notes on Individual Objects}
Following are notes on the individual galaxies, where the number at the top represents our sample ID. Continuum and $N191$ images are shown in Figure \ref{fig:imageobj}.\\

\indent {\bf 1. NGC 23 (IRAS F00073$+$253; Mrk 545):}
This is a paired galaxy with NGC 26 (\citealt{2006ApJ...650..835A}) at a distance of 9$\farcm$1. It is a barred spiral (Sa; HyperLeda\footnote{database for physics of galaxies; http://leda.univ-lyon1.fr \citep{2003A&A...412...45P}}) classified as an H{\sc ii} galaxy by a long-slit spectroscopic study \citep{1995ApJS...98..171V}. X-ray emission is not detected by Swift/BAT \citep{2013ApJ...765L..26K}. A ring starburst region are detected at the center of the galaxy. In addition to this structure, we find an extended Pa$\alpha$ emission-line region along the southern spiral arm which is located outside the field of view of the $HST$/NICMOS observation.
\clearpage
\onecolumngrid

\setcounter{figure}{2}
\begin{figure}[htb]
 \epsscale{0.8}
  \begin{center}
   \plottwo{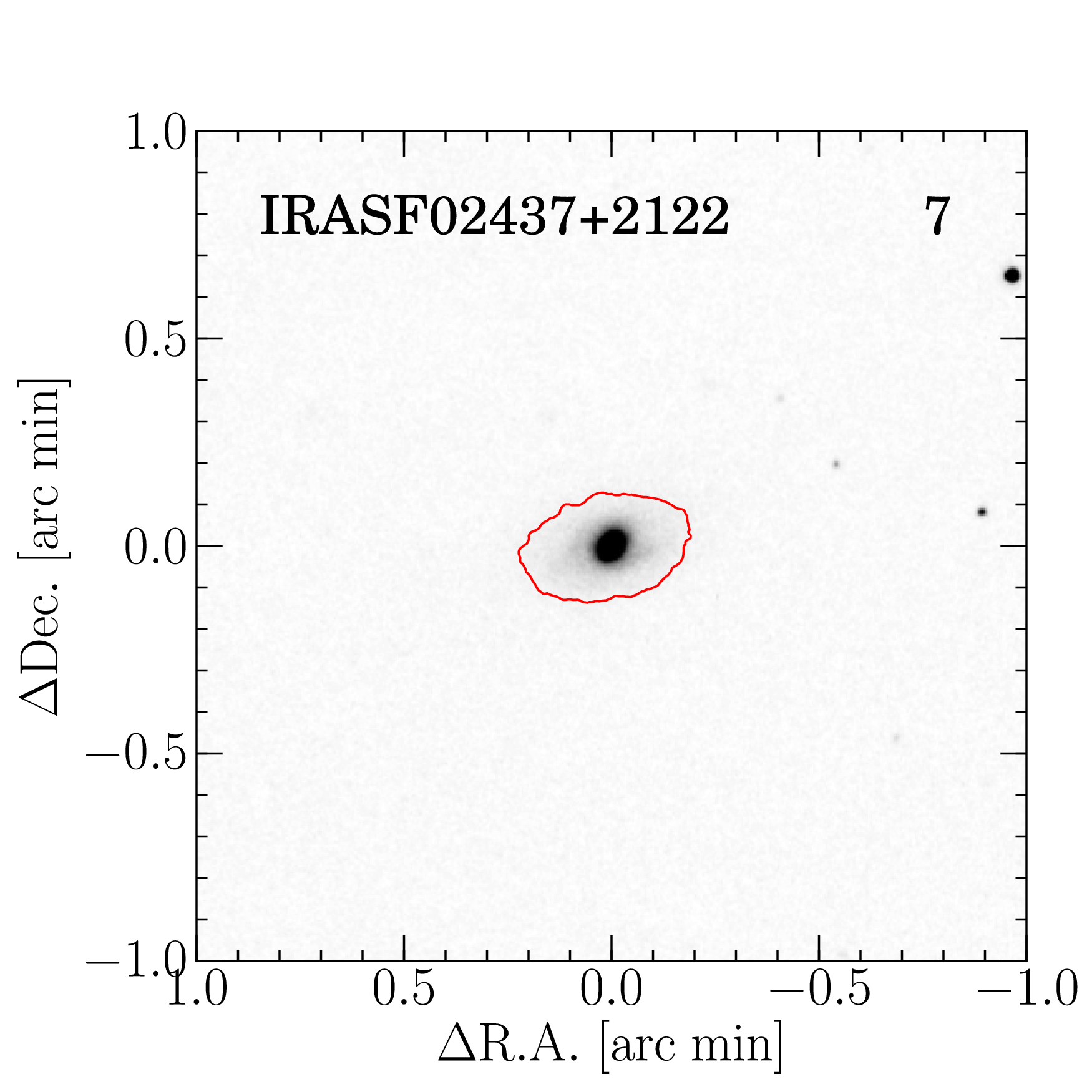}{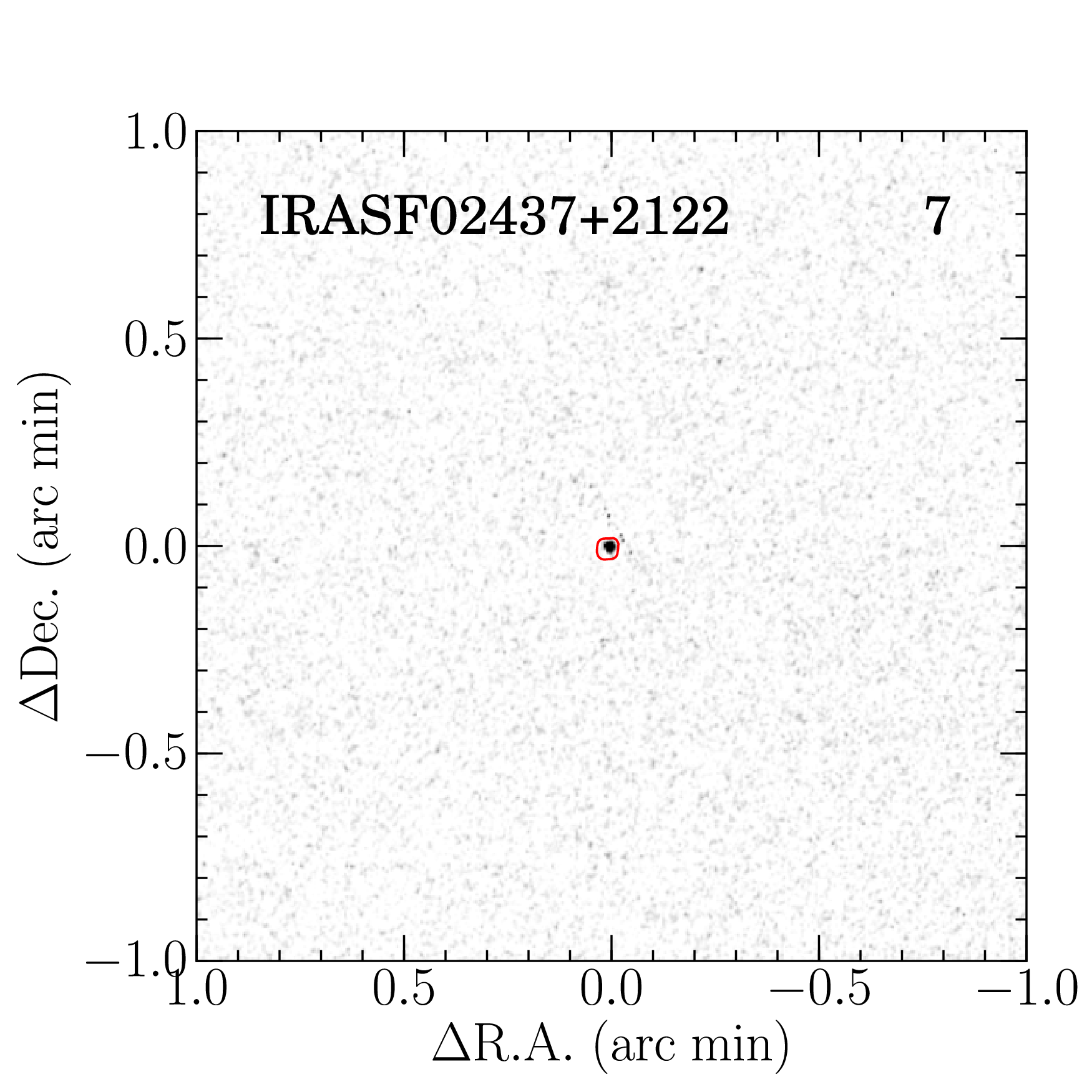}\\
   \plottwo{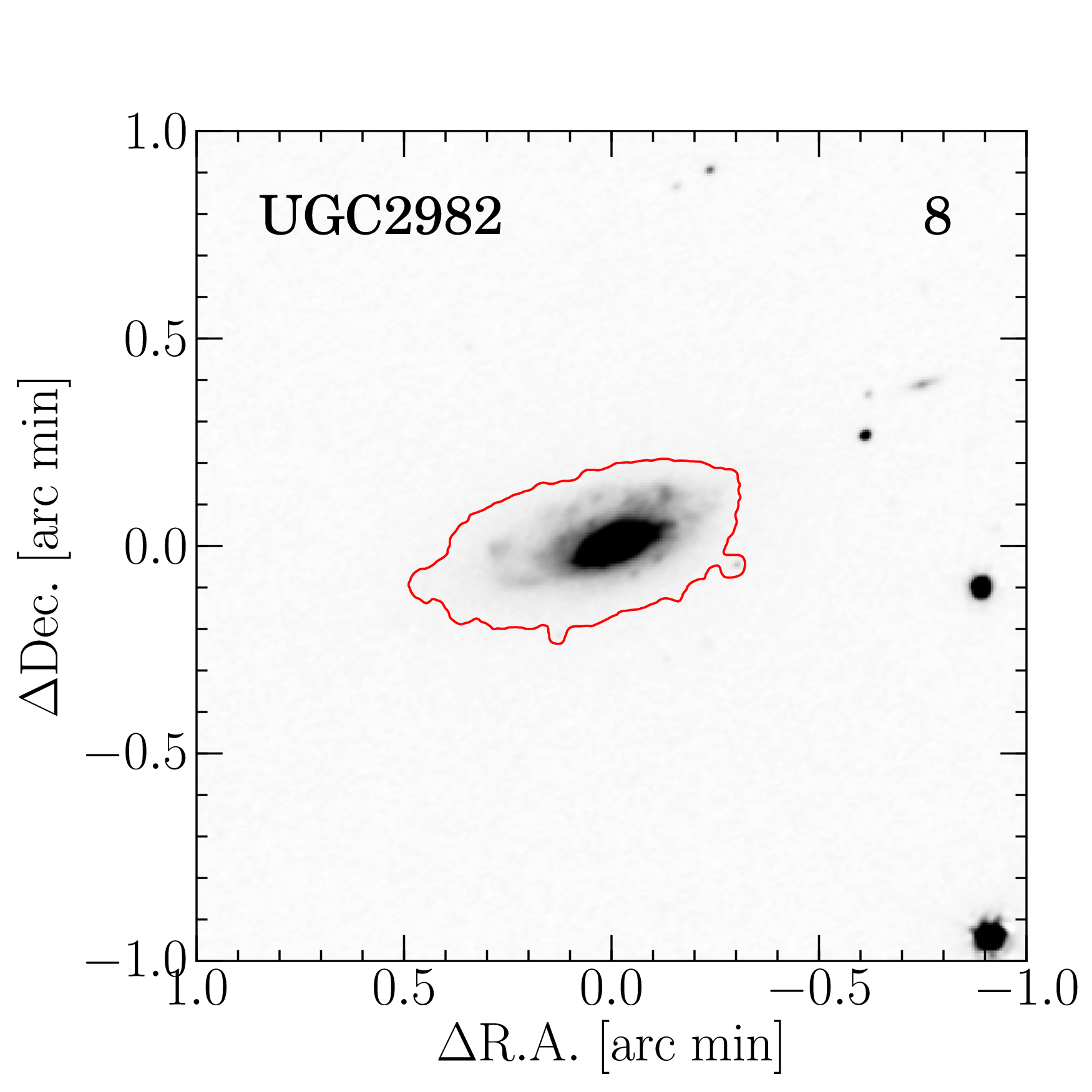}{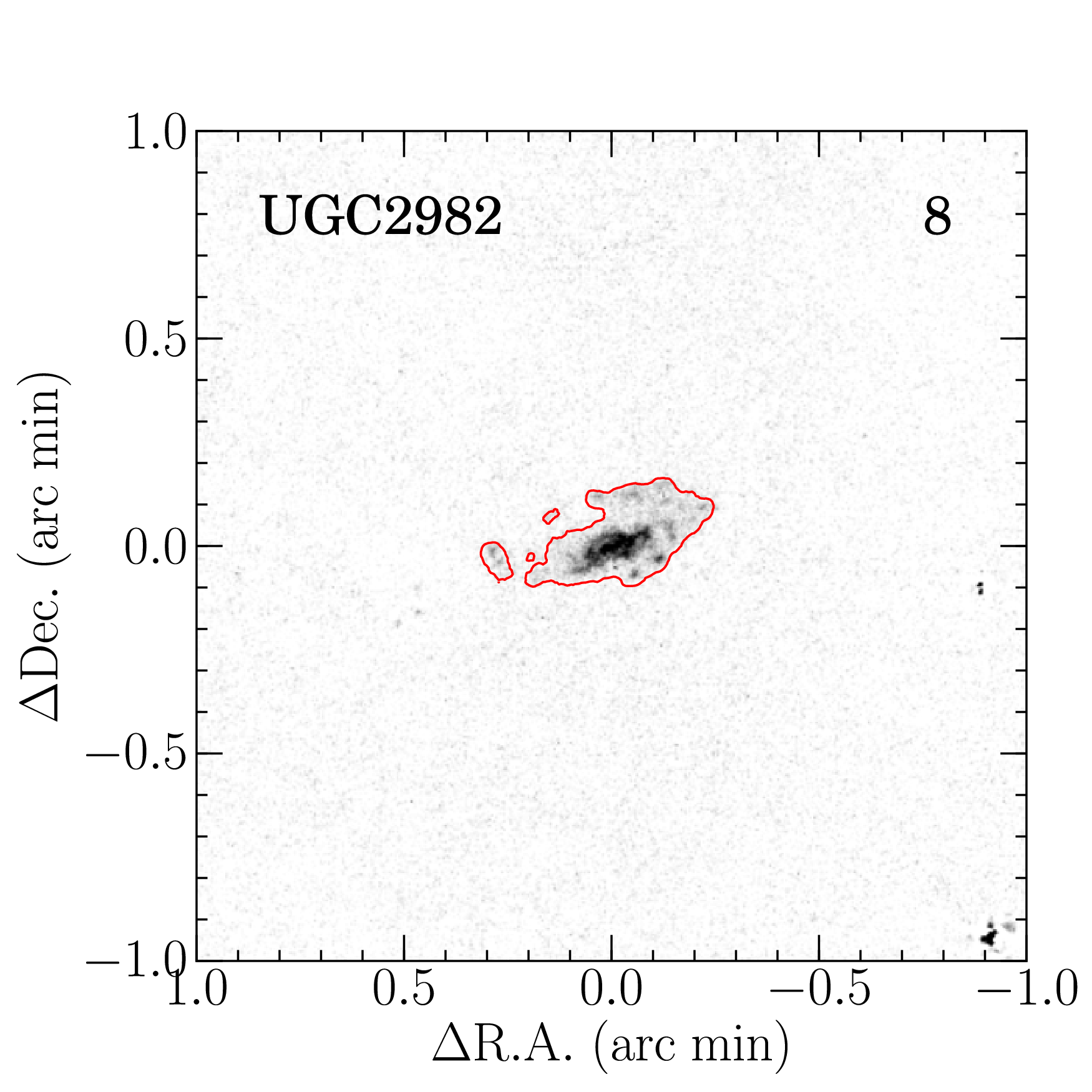}\\
   \plottwo{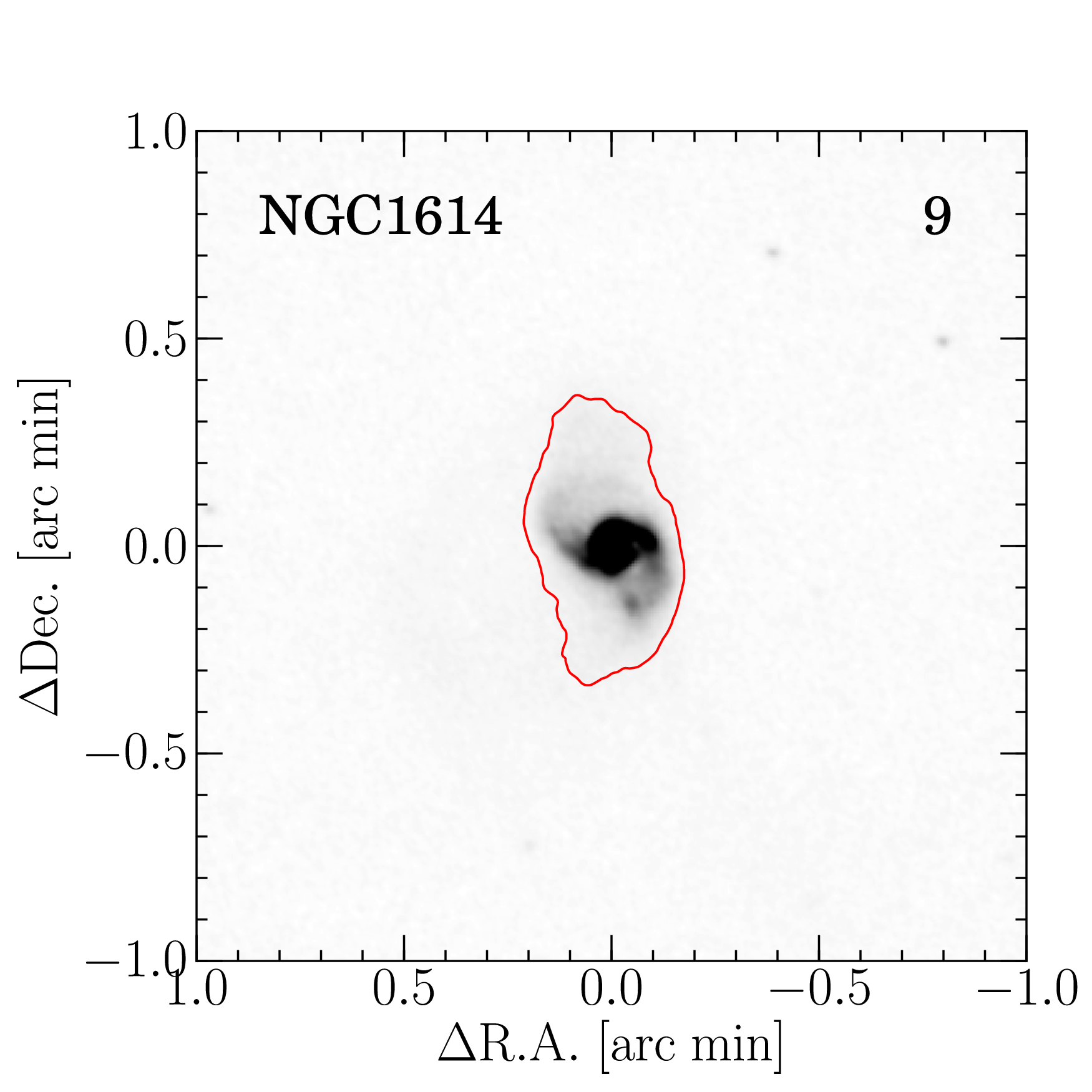}{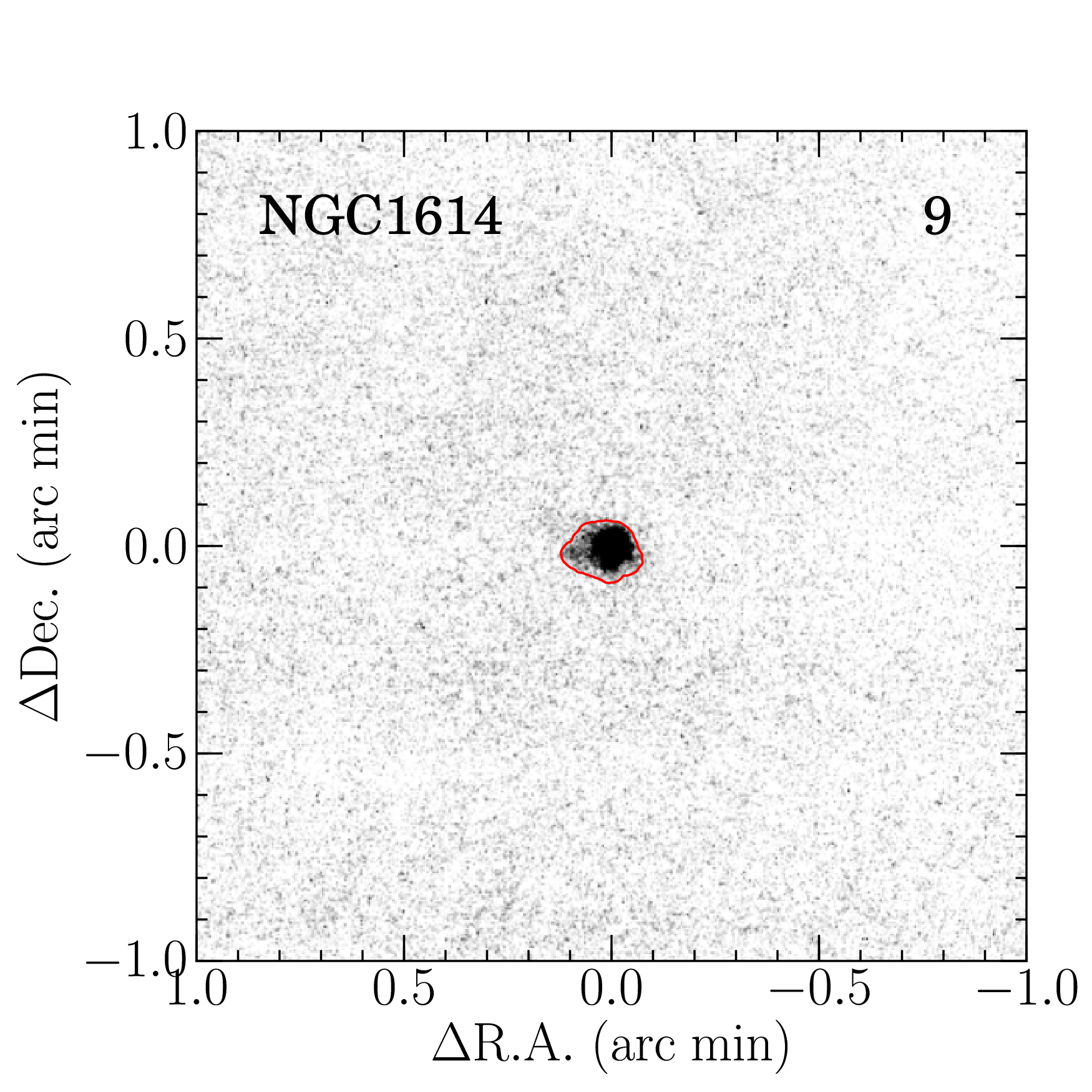}\\
   \end{center}
   \caption{Continued}
 \end{figure}

\twocolumngrid

\indent {\bf 2. NGC 34 (IRAS F00085$-$1223; Mrk 938; VV 850):}
This galaxy, hosting a strong starburst and a weak AGN, as evidenced by its optical, infrared, radio, and X-ray properties, is known as a remnant of unequal gas-rich merger \citep[e.g.,][]{2010AJ....140.1965F,2007AJ....133.2132S}. It is an S0/a (HyperLeda) and classified as a Seyfert 2 by a long-slit spectroscopic study \citep{1995ApJS...98..171V}. Pa$\alpha$ emission-line region is concentrated at the center of the galaxy.\\

\indent {\bf 3. NGC 232 (IRAS F00402$-$2349; VV 830; AM 0040-234):}
This has a companion galaxy (NGC 235) at a distance of 2$\arcmin$. It is a barred spiral (SBa; HyperLeda) classified as an H{\sc ii} galaxy \citep{2003ApJ...583..670C,1995ApJS...98..171V}. H{\sc ii} blobs, suggested as optical debris, between these two galaxies are detected in H$\alpha$ \citep[e.g.,][]{2002ApJS..143...47D, 1994AJ....107...99R}, but the Pa$\alpha$ data is not enough deep to detect the blobs. While NGC 232 has bright FIR emission, NGC 235 has no FIR flux though it has bright H{\sc ii} blobs \citep{1994AJ....107...99R}.
\clearpage
\onecolumngrid

\setcounter{figure}{2}
\begin{figure}[htb]
 \epsscale{0.8}
  \begin{center}
   \plottwo{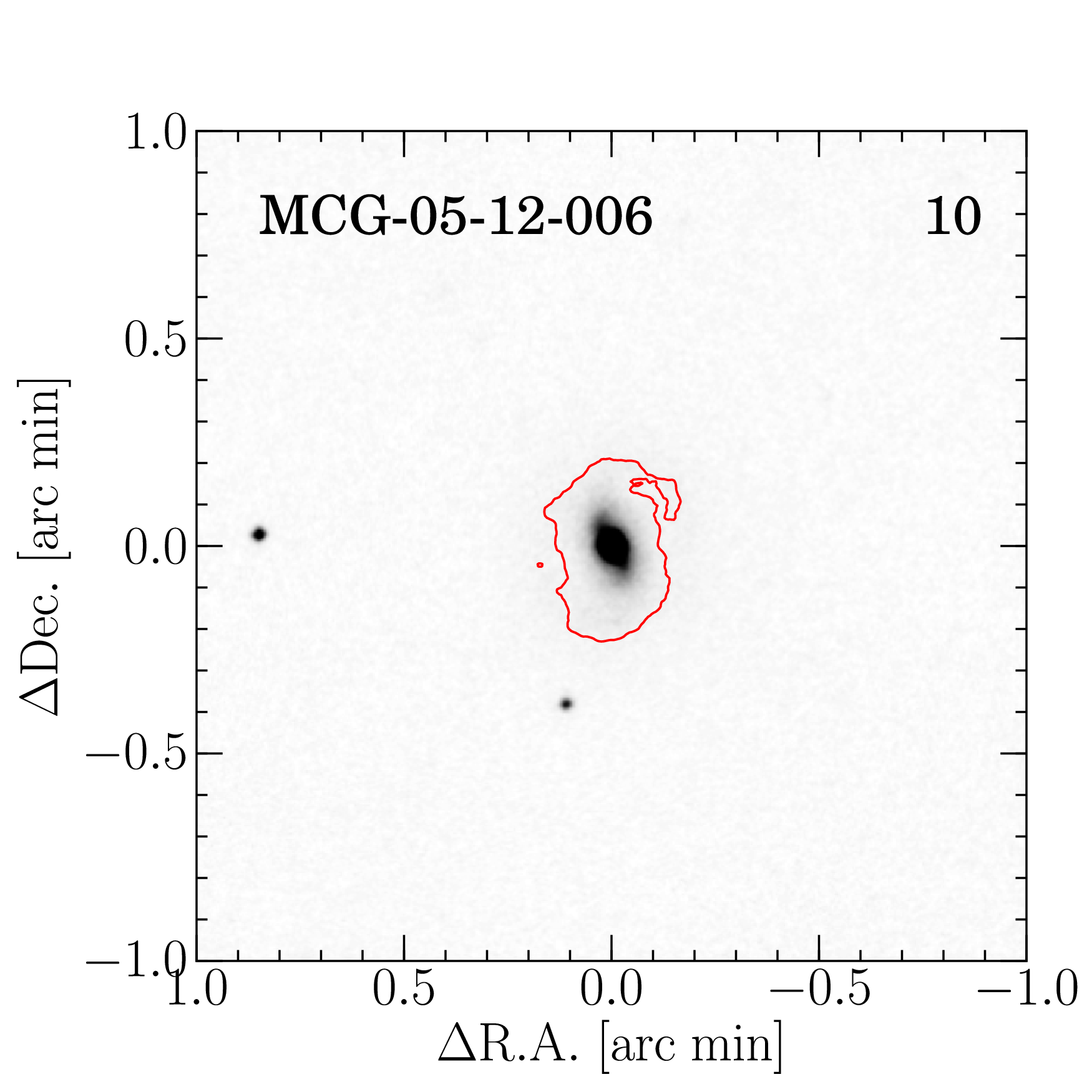}{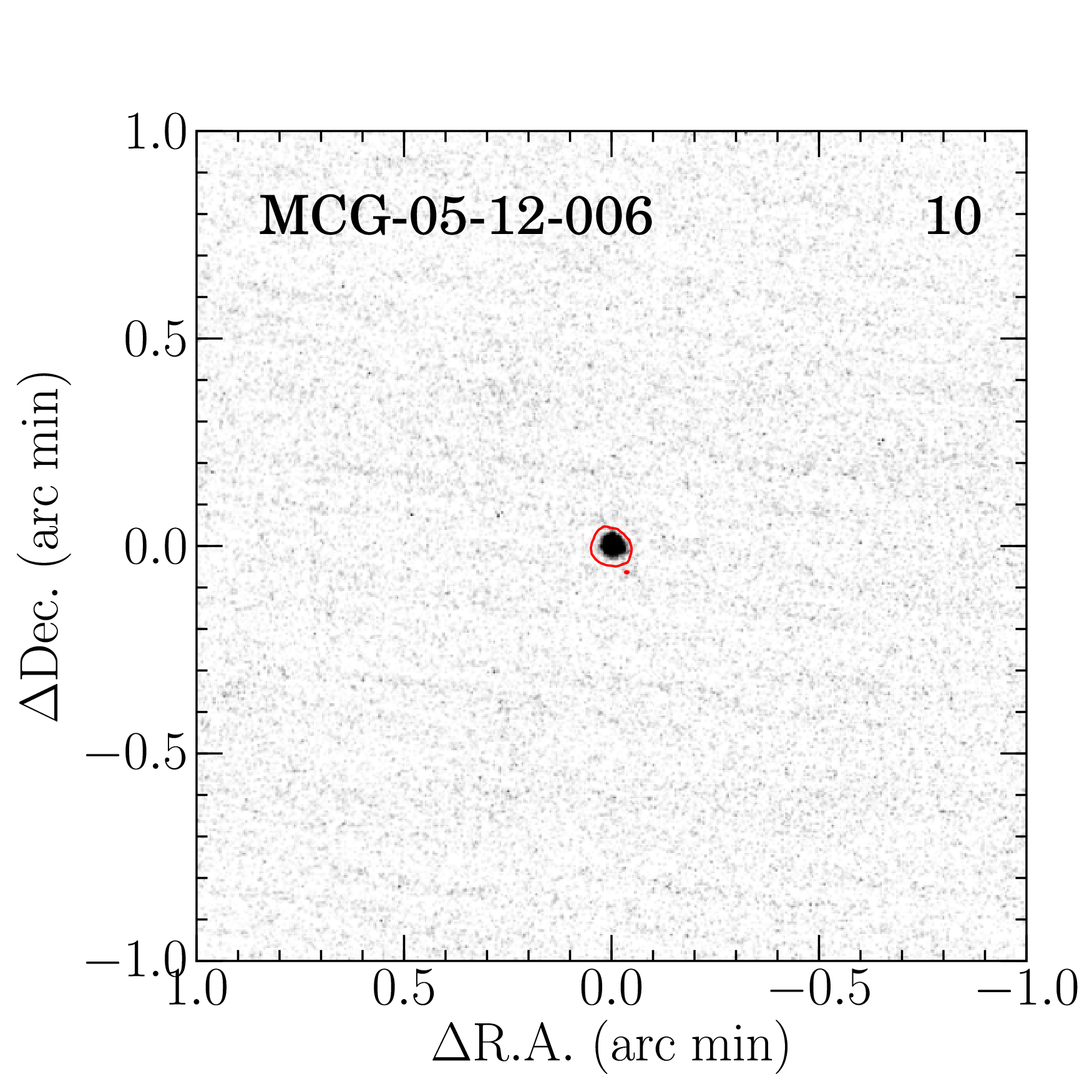}\\
   \plottwo{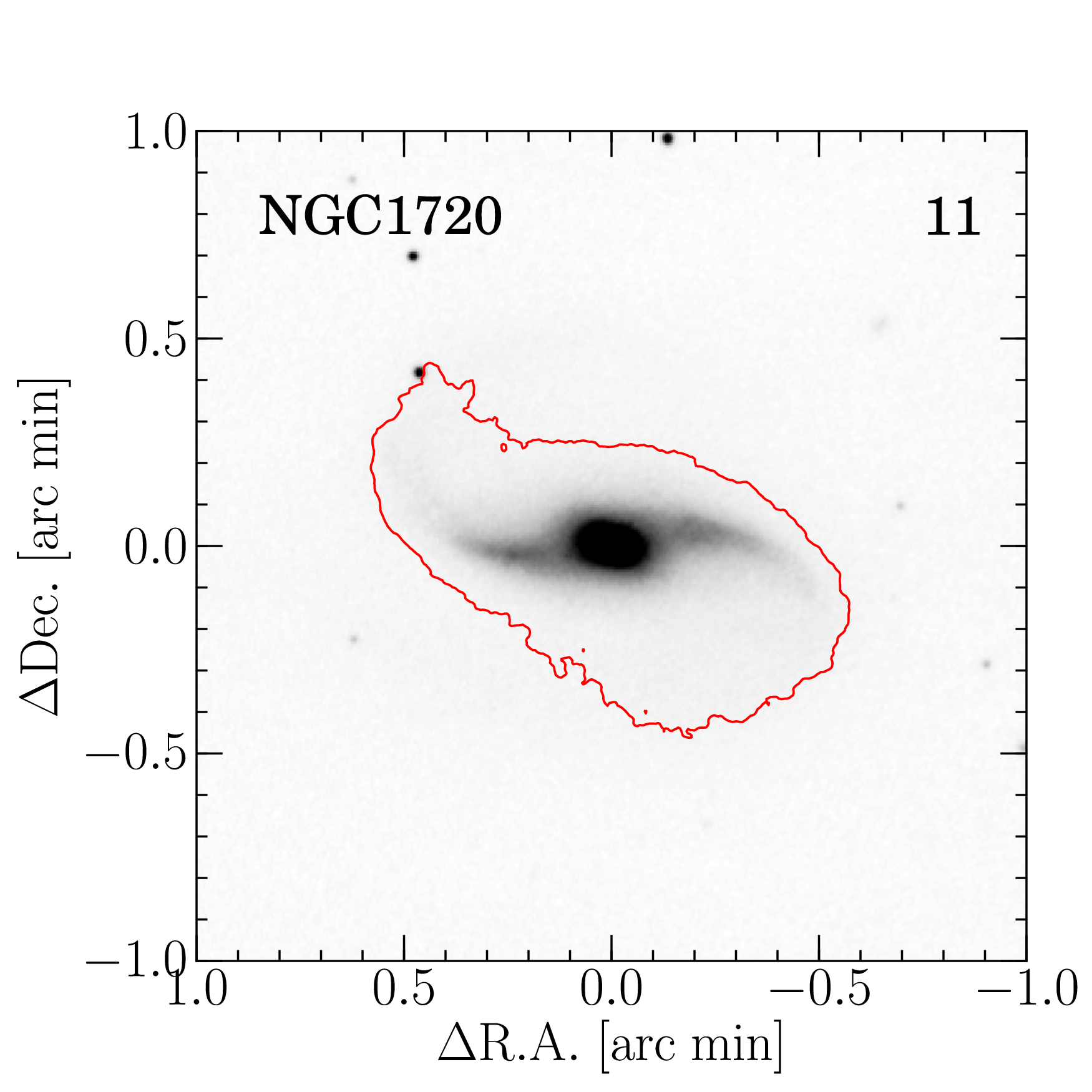}{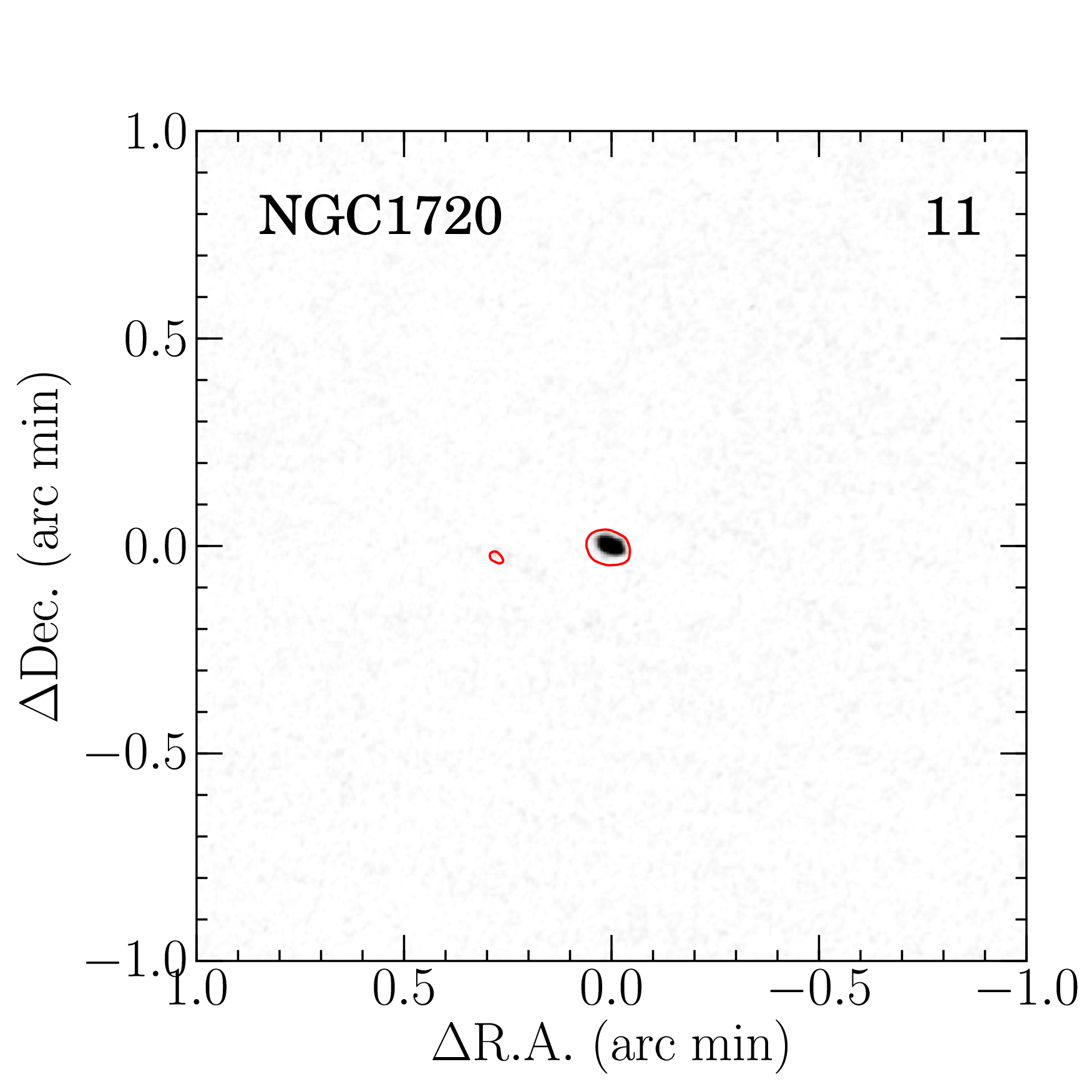}\\
   \plottwo{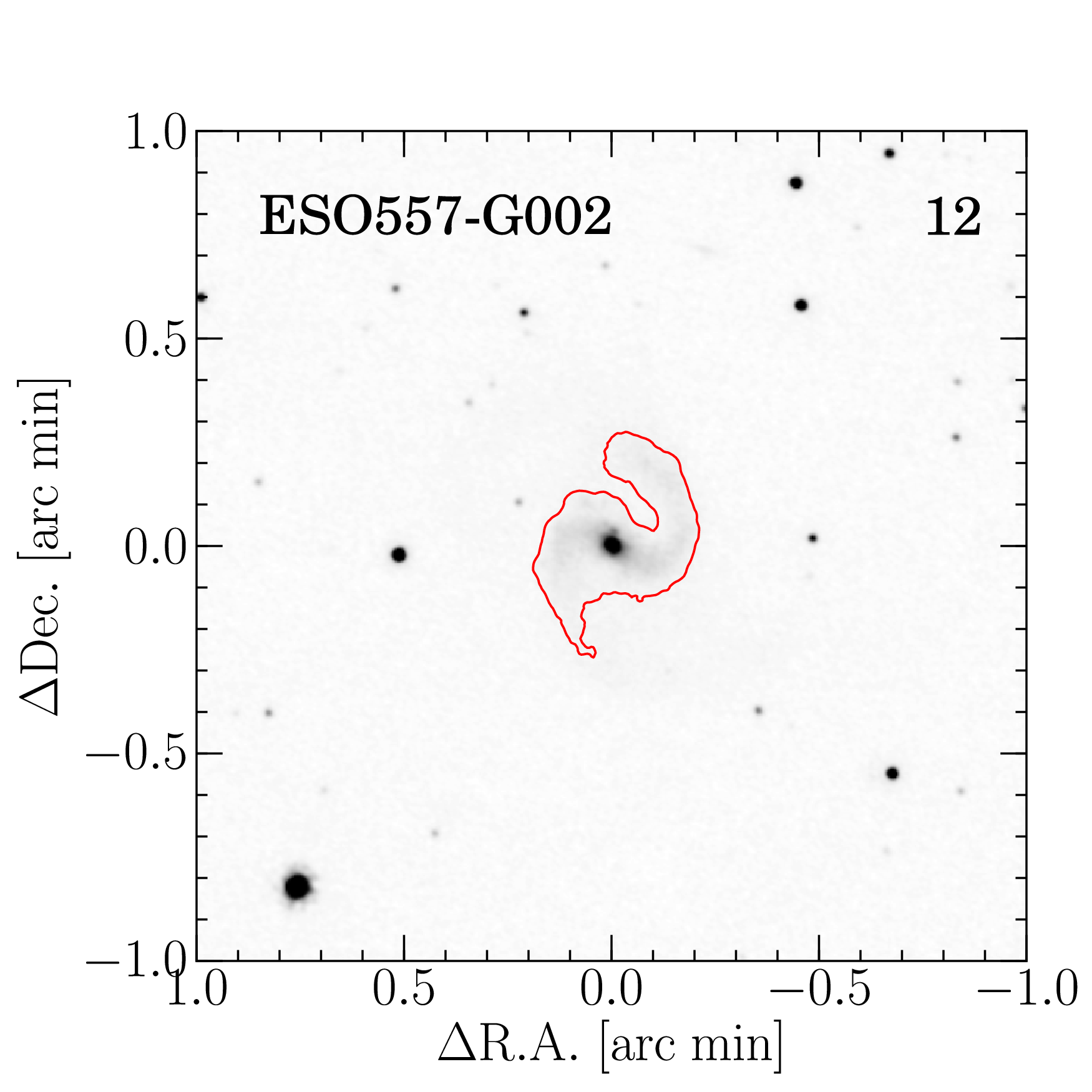}{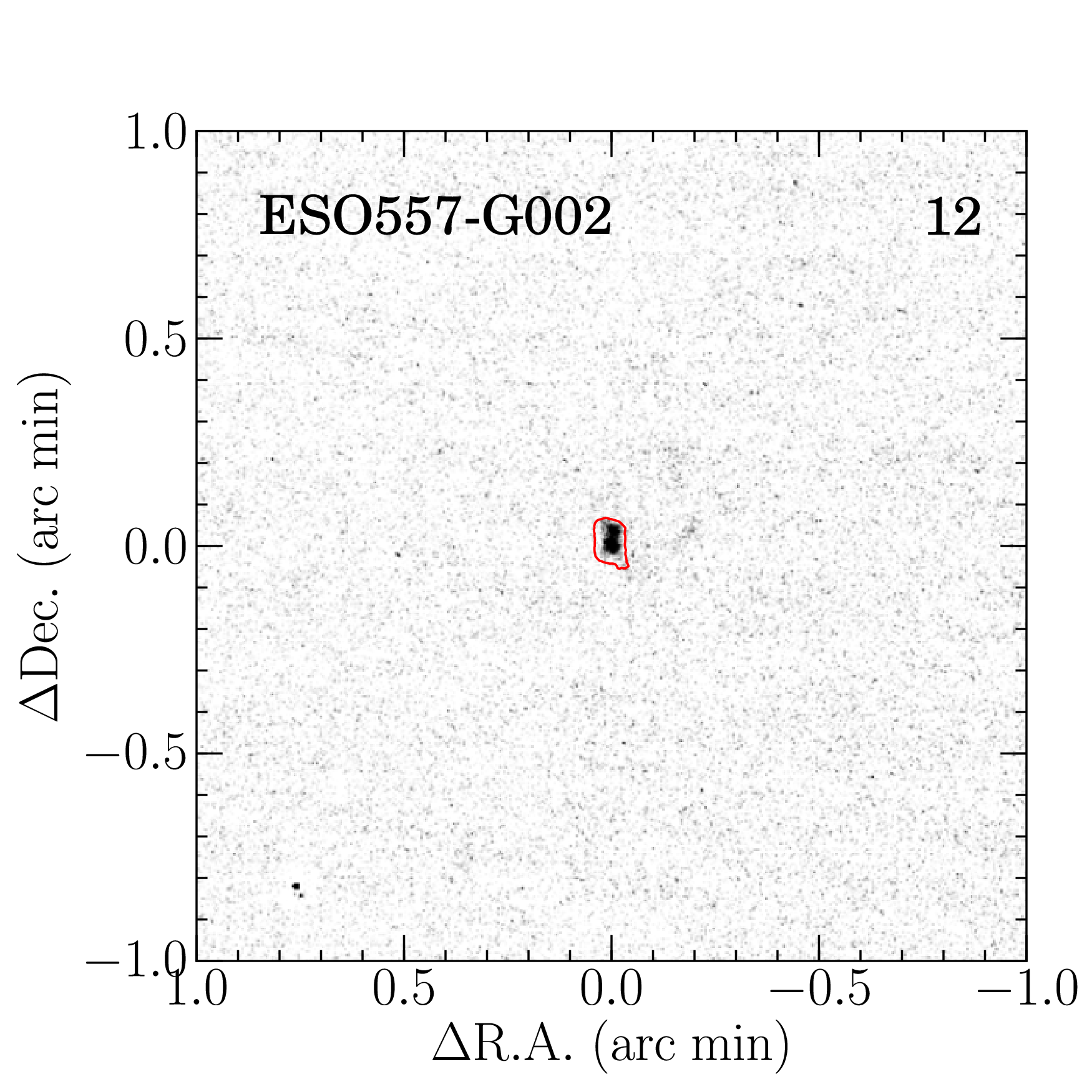}\\
   \end{center}
   \caption{Continued}
 \end{figure}

\twocolumngrid

\indent {\bf 4. IC 1623A/B (IRAS F00402$-$2349; VV 114; Arp 236):}
This system is known as a merger at its middle or late stage, and consists of two galaxies, IC 1623B and IC 1623A. An obscured AGN in IC 1623B is revealed by MIR spectroscopy and X-ray observations \citep[e.g.,][]{2002AJ....124..166A,2006ApJ...648..310G} suggesting that both starburst and AGN activities might be triggered by the ongoing merger. These galaxies are classified as an H{\sc ii} galaxy \citep{2003ApJ...583..670C,1995ApJS...98..171V}. The diffuse component of Pa$\alpha$ emission is distributed between the two galaxies over 10 kpc \citep{2013PASJ...65L...7I,2013ASPC..476..287S}.\\

\indent {\bf 5. ESO 244-G012 (IRAS F01159$-$4443; VV 827; AM 0115-444):}
This system is on-going interacting paired galaxies \citep{2000AJ....119...94A}, separated by 17$\arcsec$. Both are spirals (Sc; HyperLeda), and the northern galaxy is classified as an H{\sc ii} galaxy \citep{2003ApJ...583..670C}, while the class of the southern galaxy is ambiguous \citep{2003ApJ...583..670C}. The northern galaxy has bright con-
\clearpage
\onecolumngrid

\setcounter{figure}{2}
\begin{figure}[htb]
 \epsscale{0.8}
  \begin{center}
   \plottwo{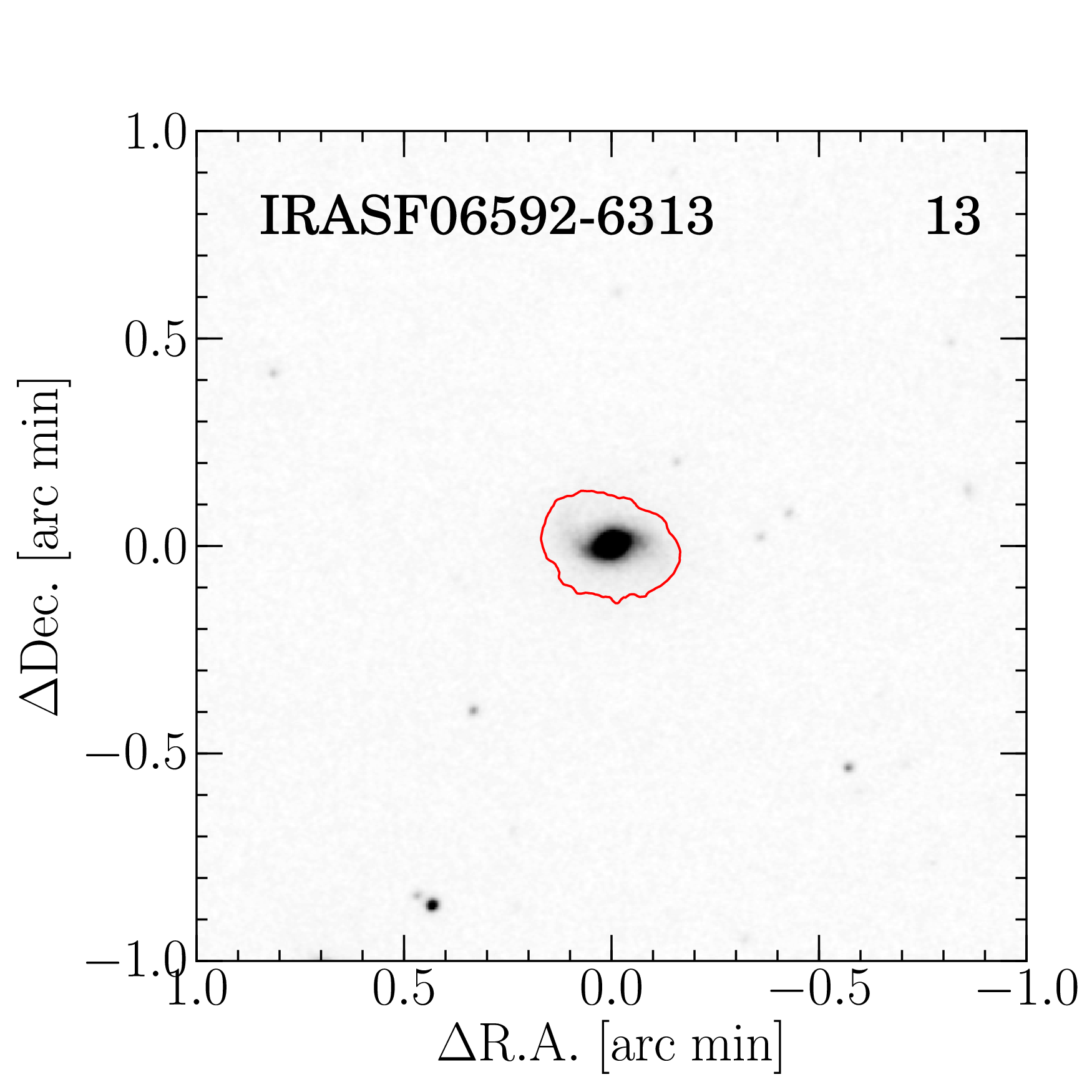}{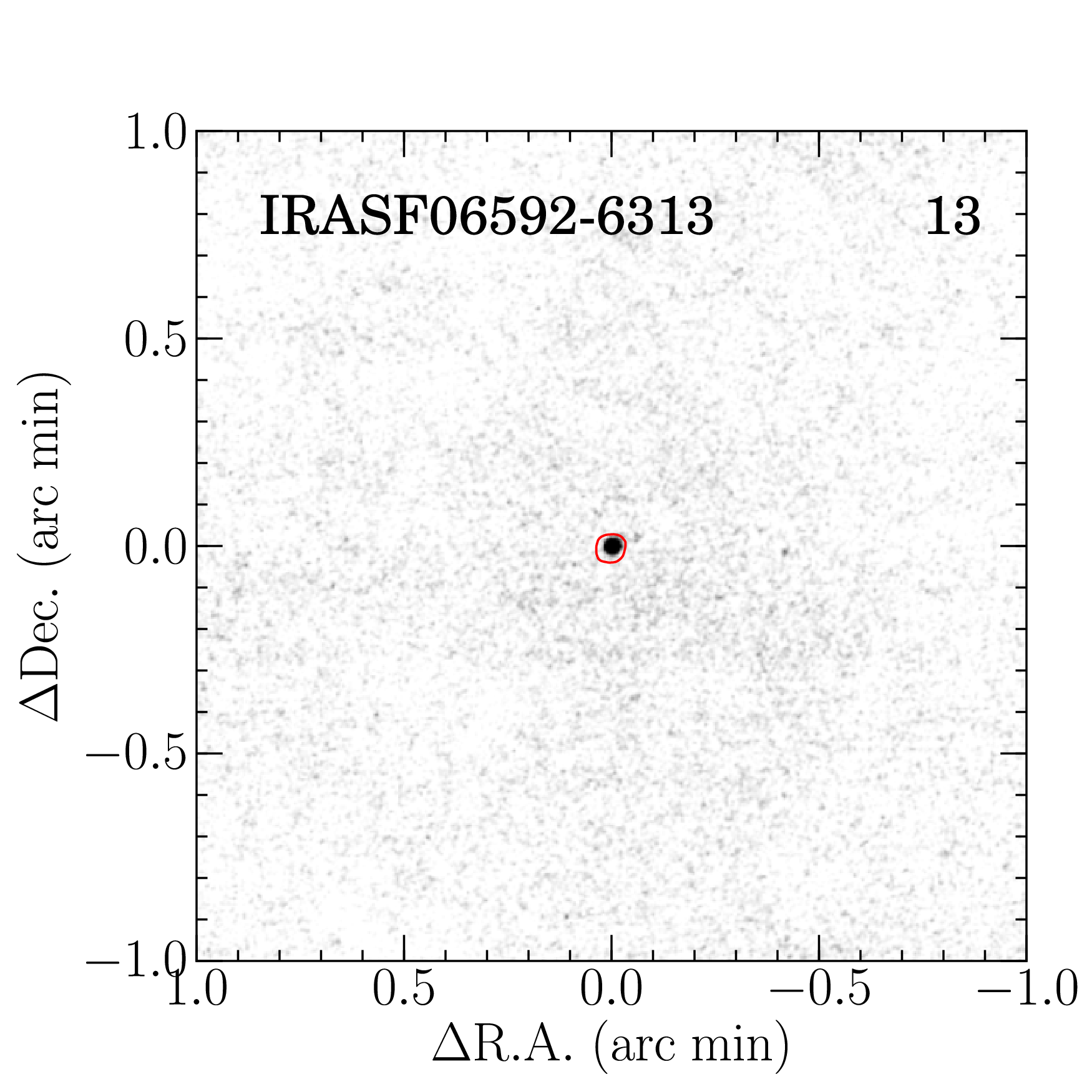}\\
   \plottwo{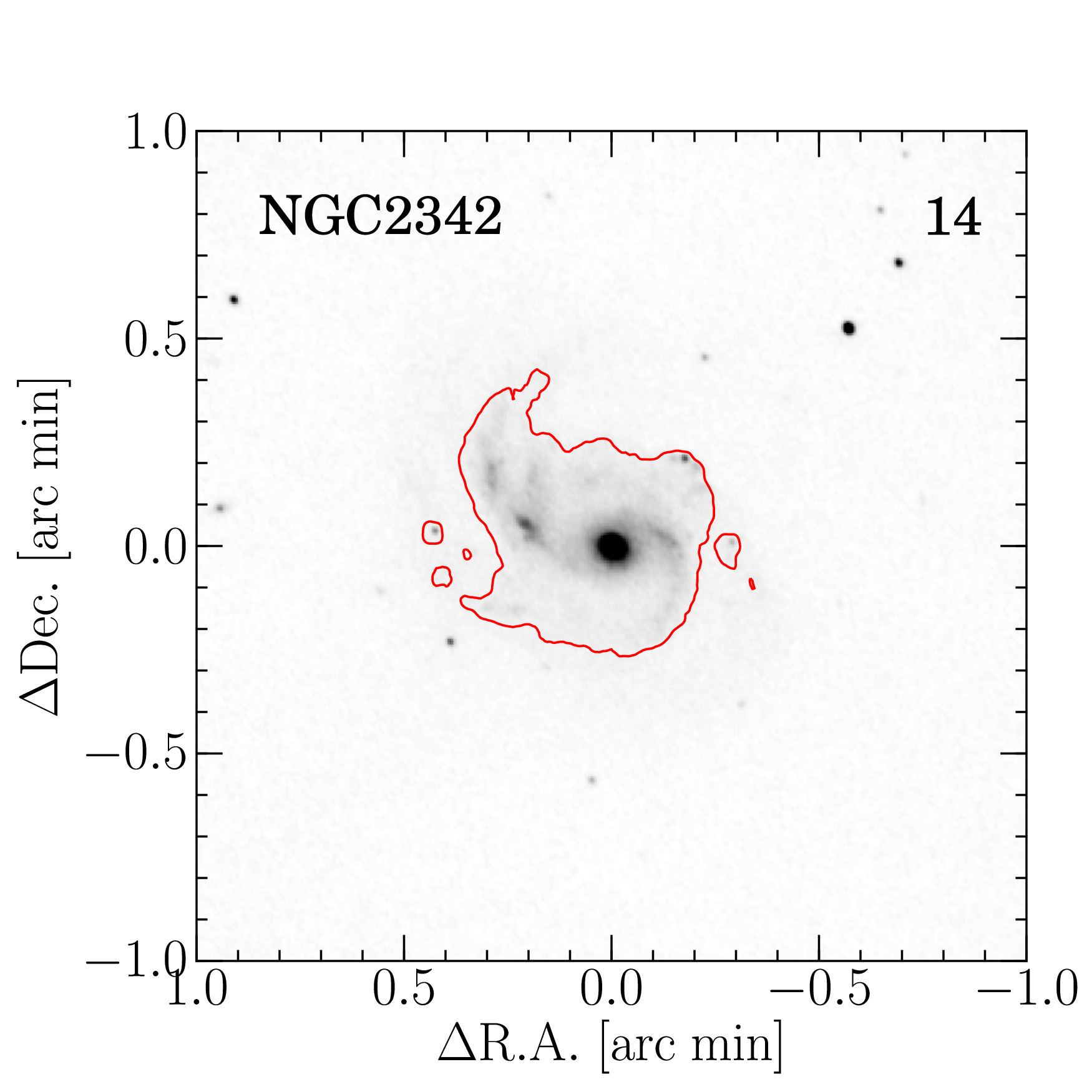}{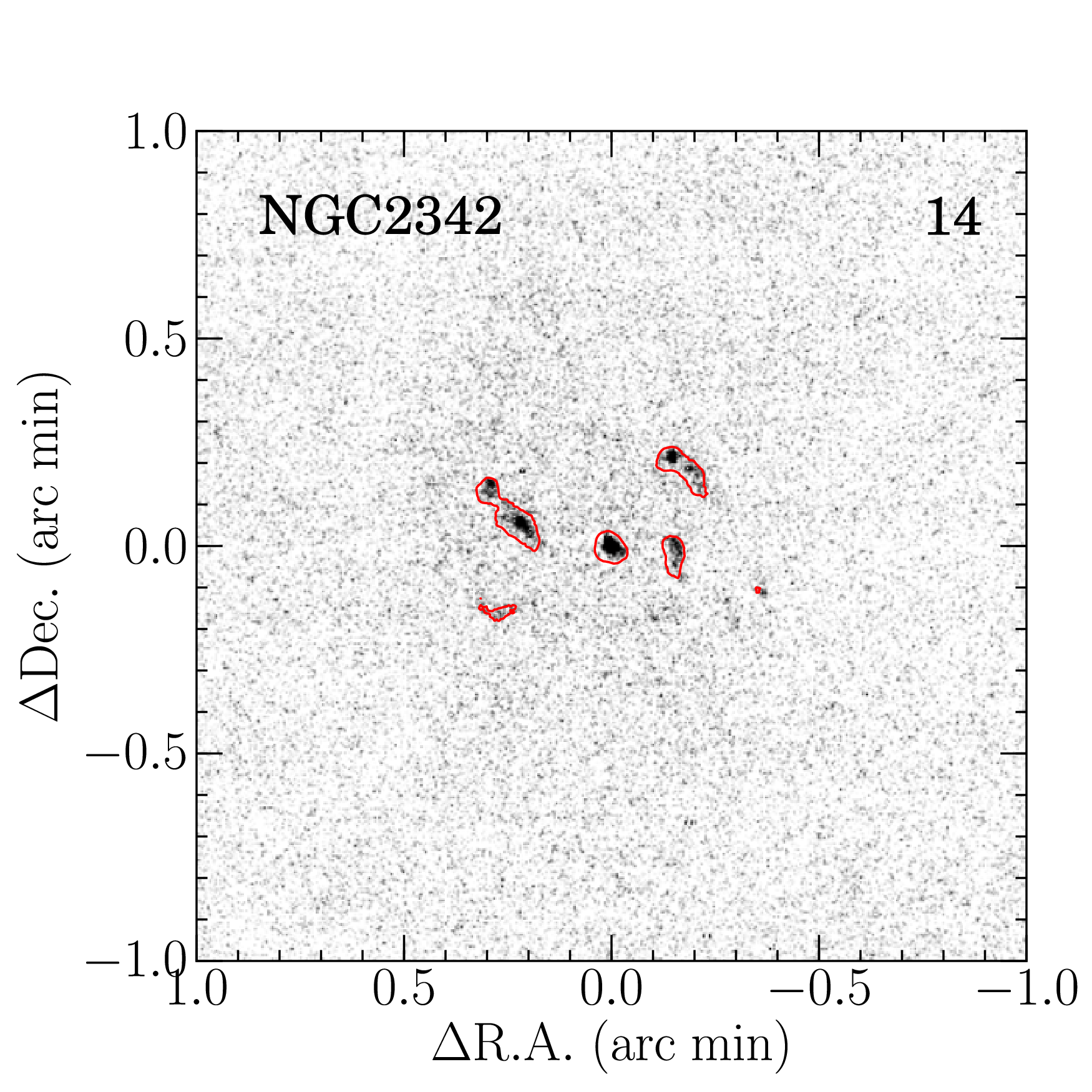}\\
   \plottwo{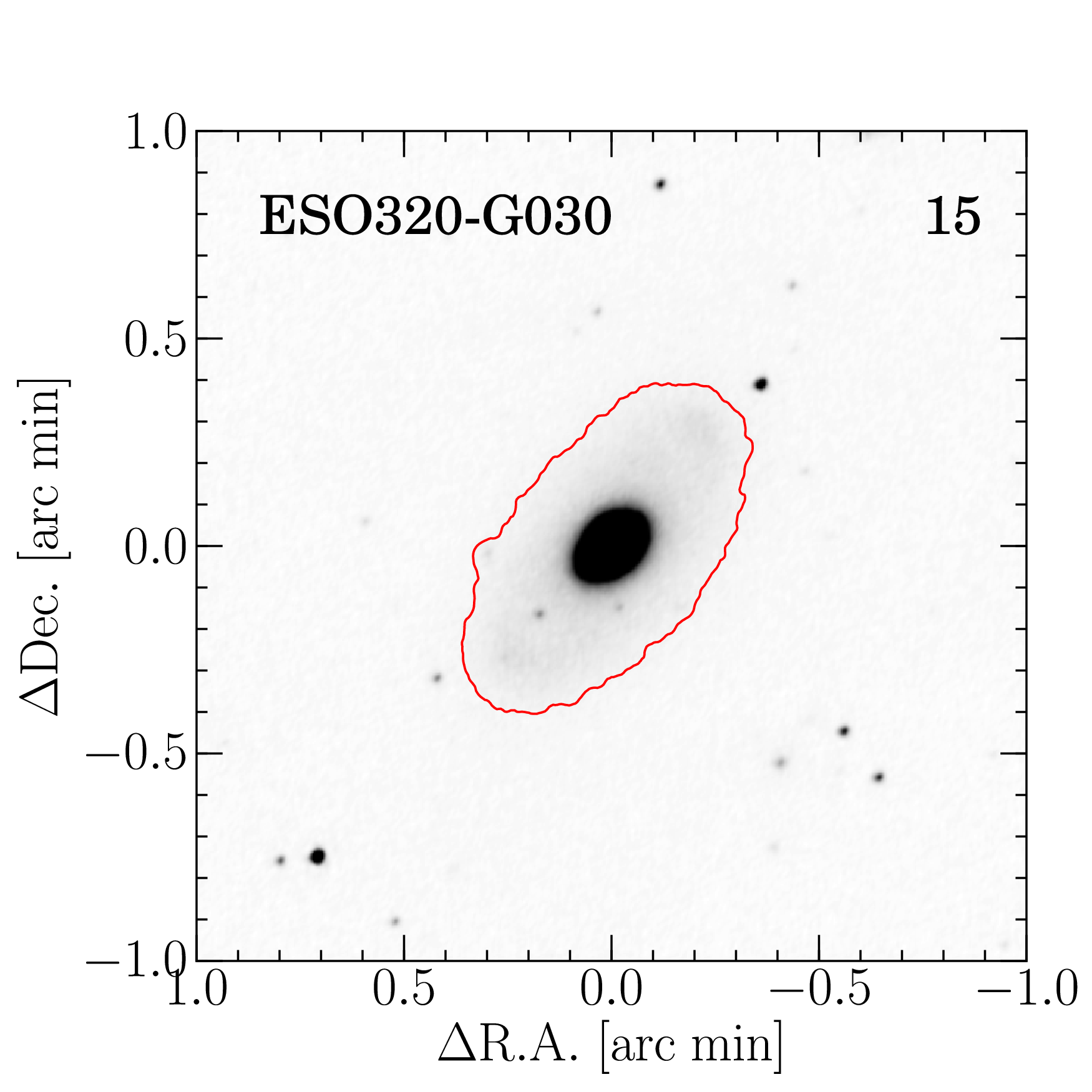}{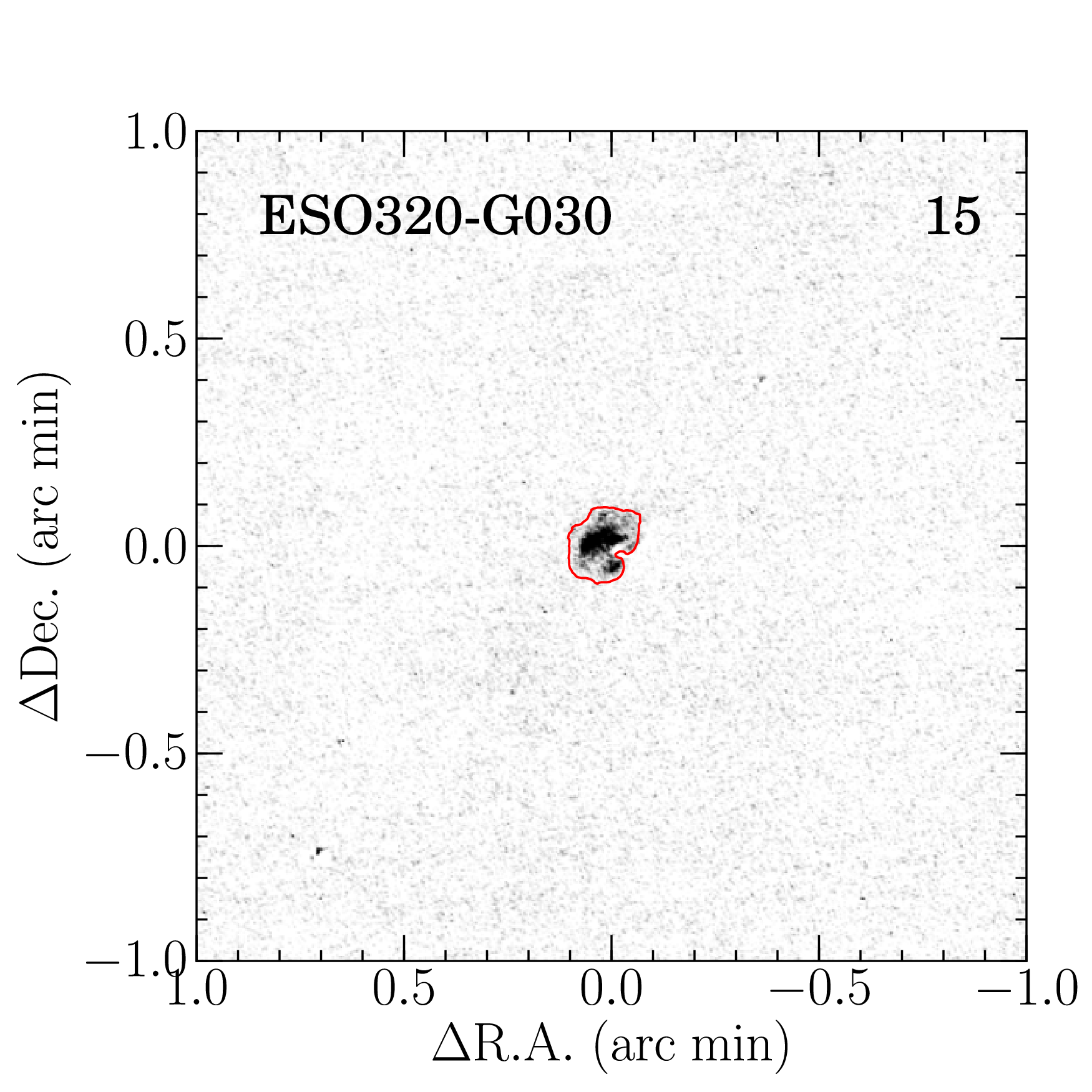}\\
   \end{center}
   \caption{Continued}
 \end{figure}

\twocolumngrid

\noindent centrated Pa$\alpha$ emission at its center, while the southern source is faint.\\

\indent {\bf 6. UGC 2238 (IRAS F02435$+$1253):}
This object is known as a merger remnant, and it is questionable whether it is undergoing or has undergone any amount of violent relaxation \citep{2004AJ....128.2098R}. It is an edge-on disk galaxy (Sm; HyperLeda) classified as a LINER \citep{1995ApJS...98..171V}. Strong Pa$\alpha$ emission are detected not only from the central region, but also from the disk component.\\

\indent {\bf 7. IRAS F02437$+$2122:}
This is an elliptical galaxy (E; HyperLeda) classified as a LINER \citep{1995ApJS...98..171V}. X-ray emission is detected (S/N=2.7, $L_\mathrm{14-195 keV} < 10^{42.9}$ (erg s$^{-1}$)) by Swift/BAT \citep{2013ApJ...765L..26K}, but no high-quality X-ray data is obtained. The Pa$\alpha$ morphology is compact and concentrated at the central region. 
\clearpage
\onecolumngrid

\setcounter{figure}{2}
\begin{figure}[htb]
 \epsscale{0.8}
  \begin{center}
   \plottwo{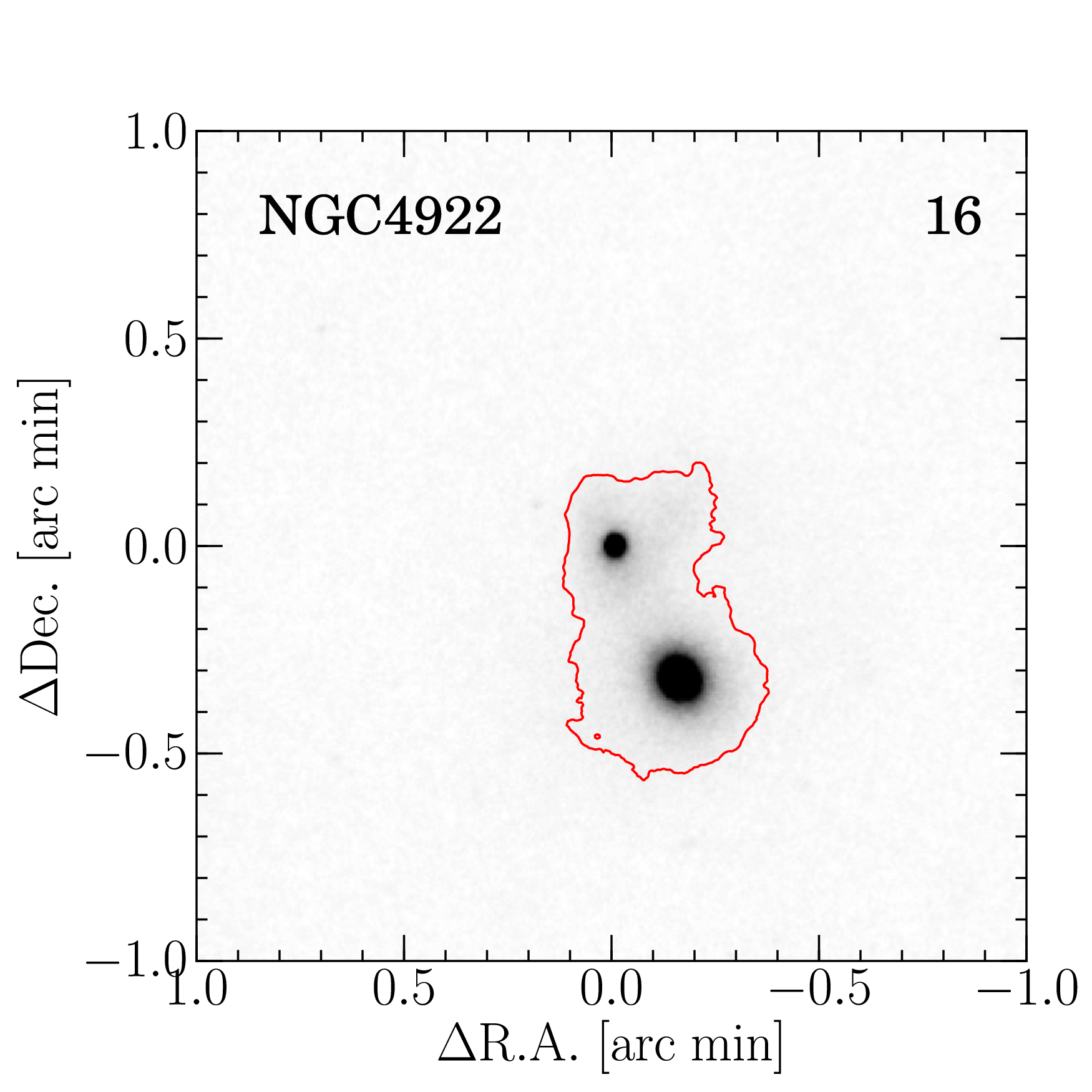}{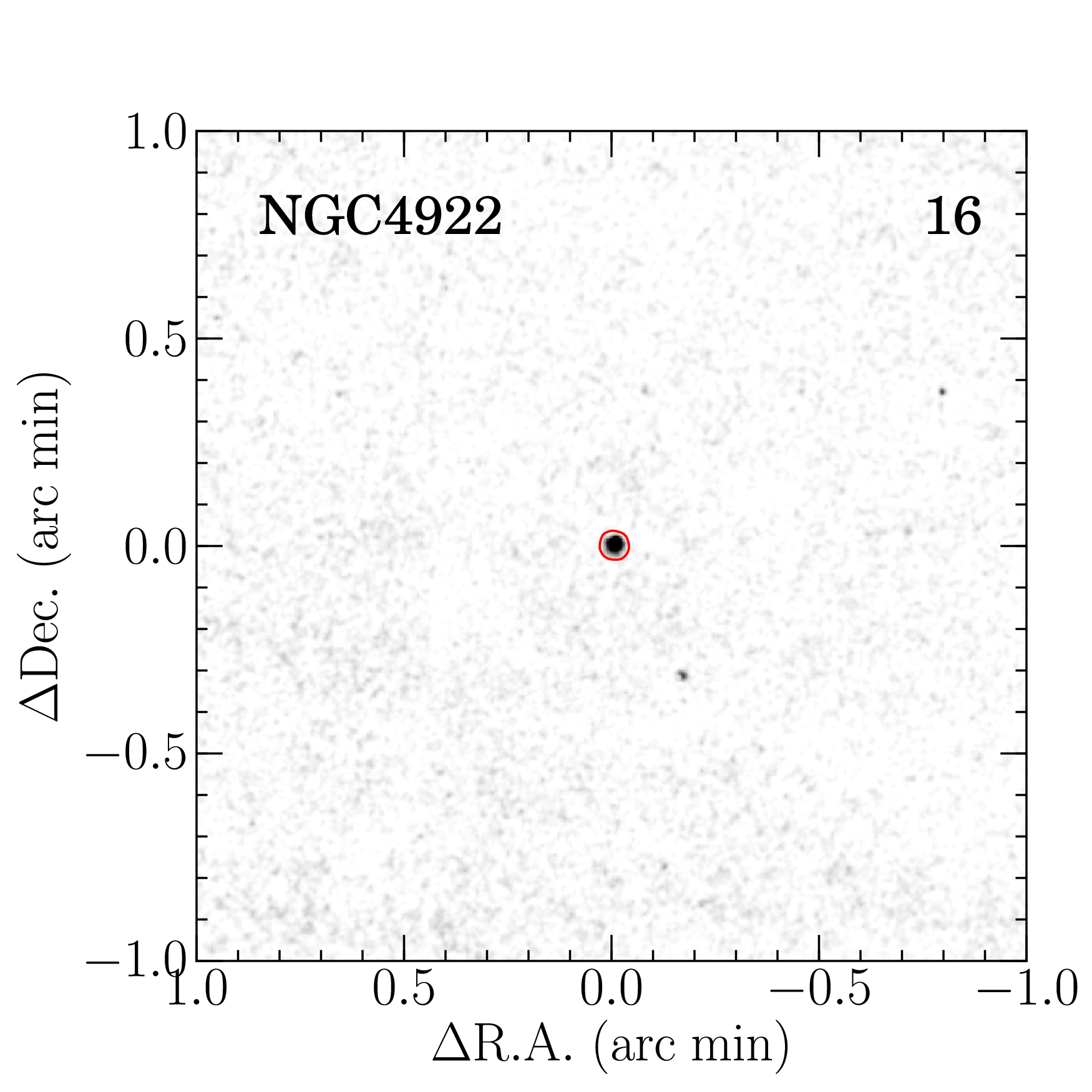}\\
   \plottwo{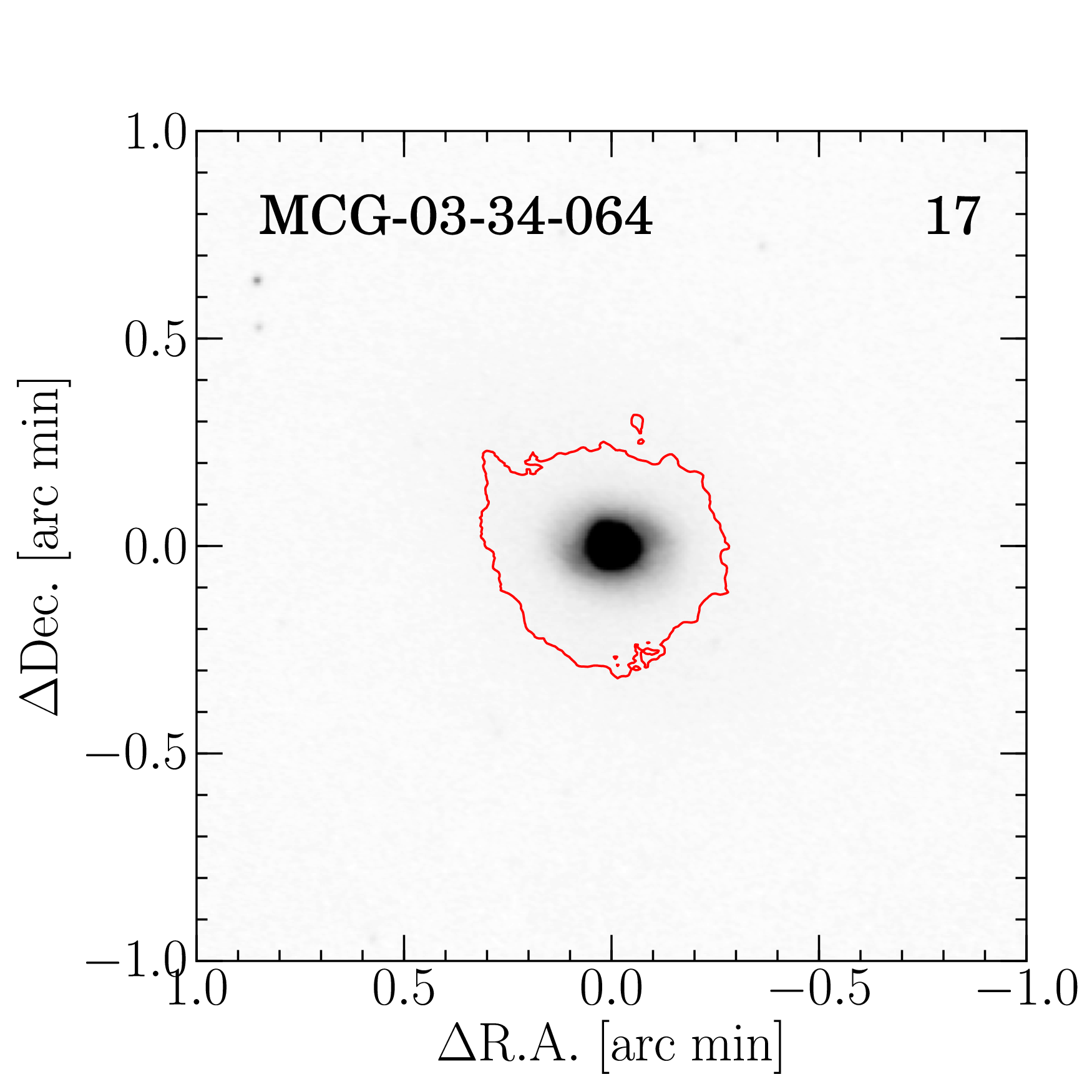}{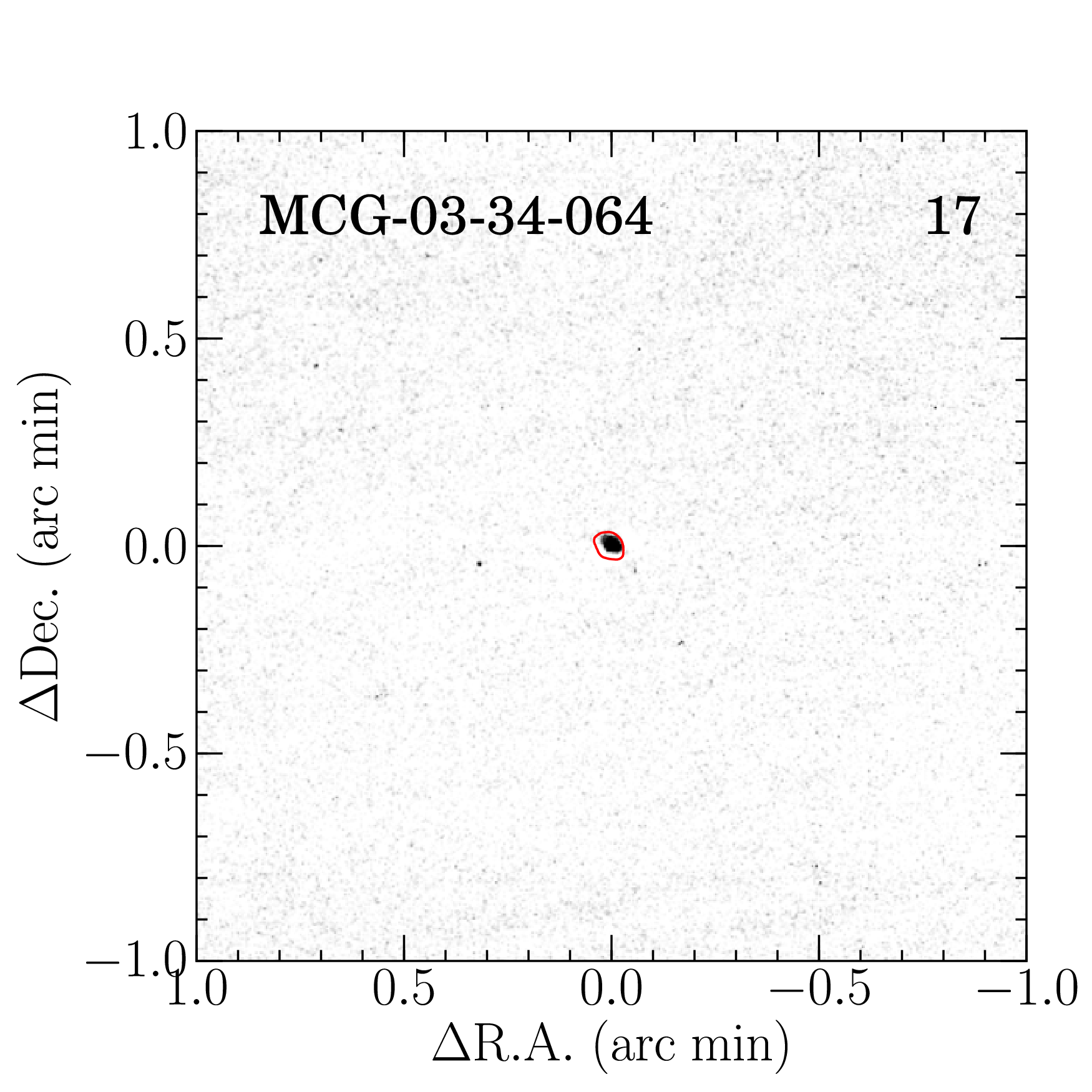}\\
   \plottwo{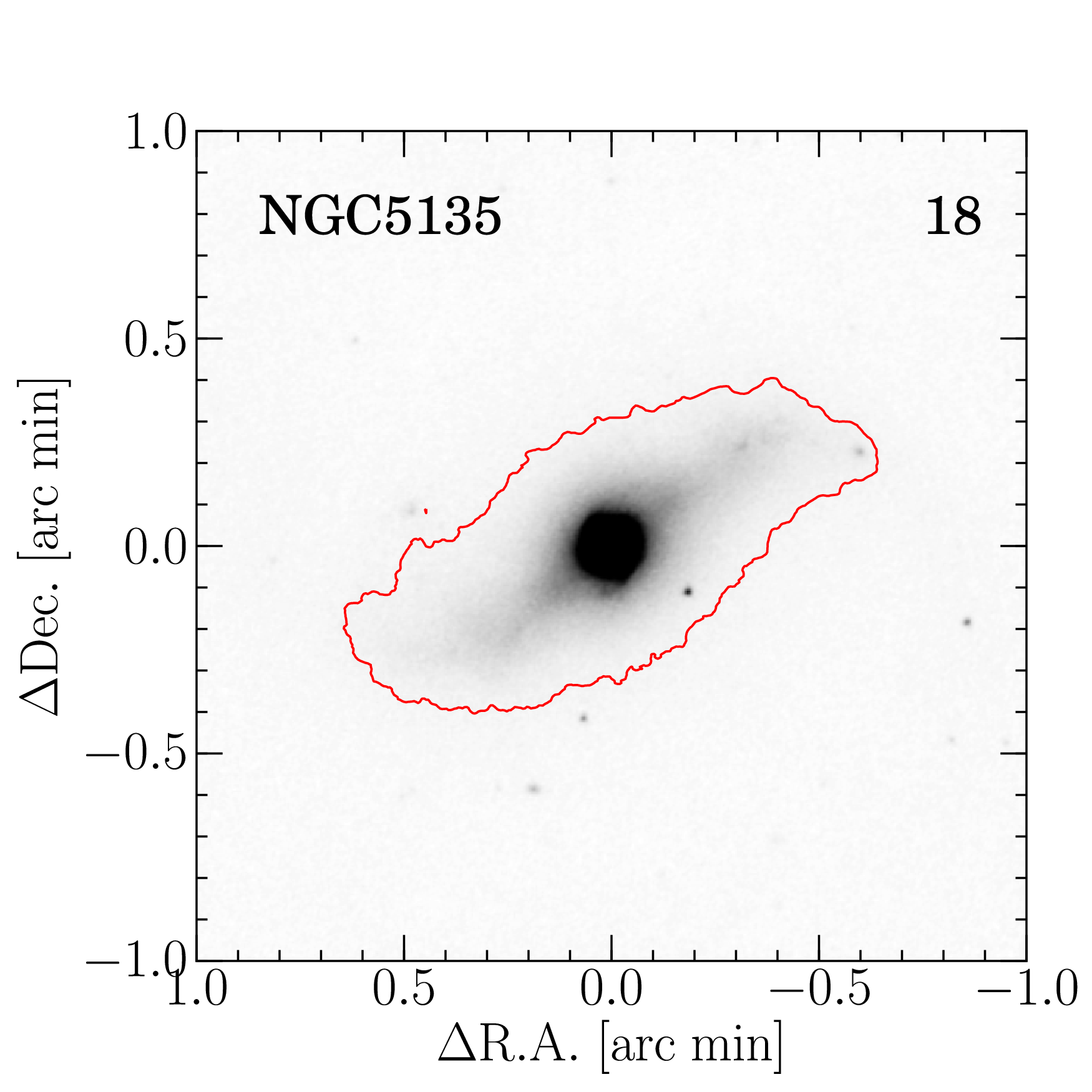}{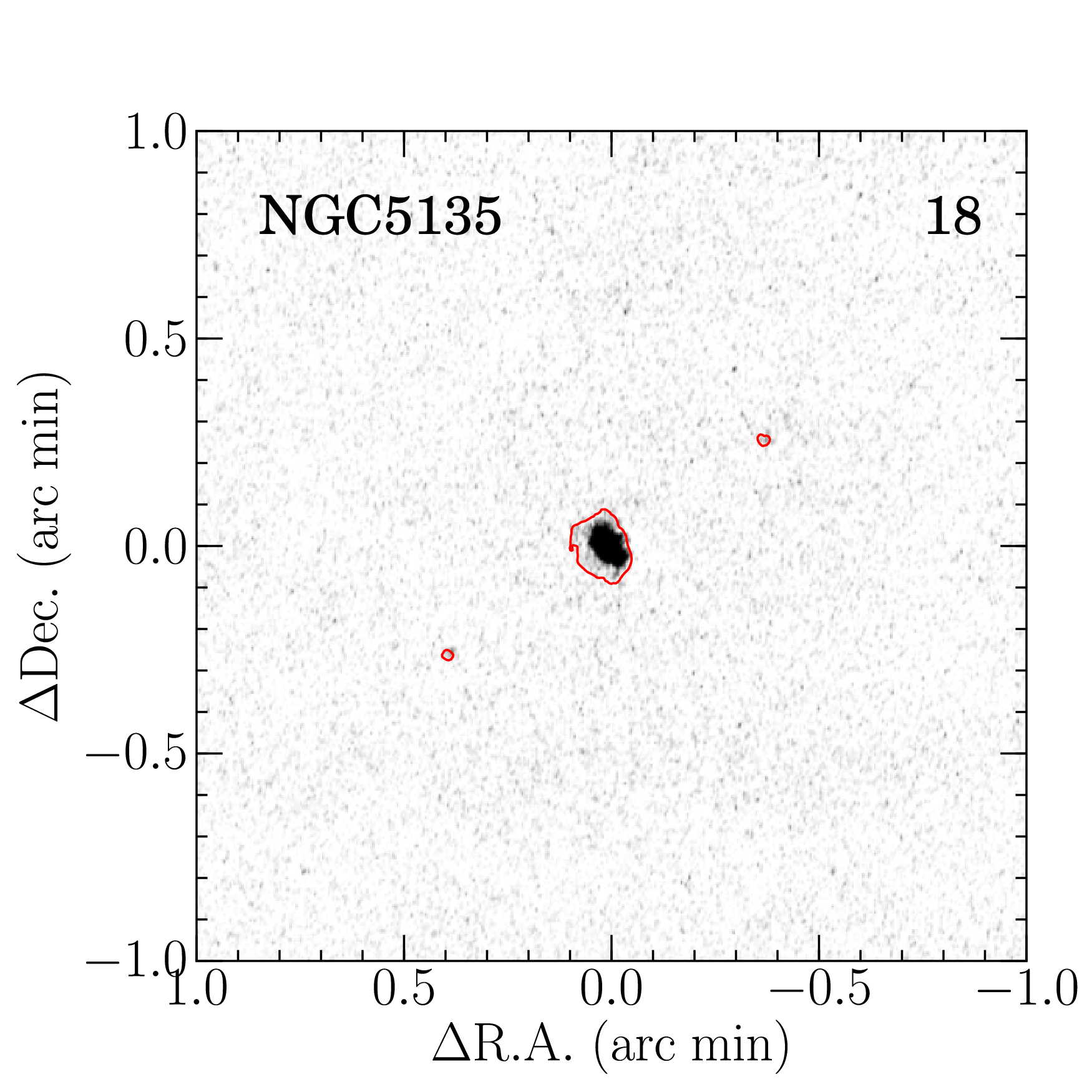}\\
   \end{center}
   \caption{Continued}
 \end{figure}

\twocolumngrid

\indent {\bf 8. UGC 2982 (IRAS F04097$+$0525):}
This is a barred spiral (SABa; HyperLeda) classified as an H{\sc ii} galaxy by a long-slit spectroscopic study \citep{1995ApJS...98..171V}. It is a diffuse isolated system having extended H{\sc ii} gas \citep{1991PASP..103...35C} and 850 $\mu$m emission extends to the periphery of a optical disk \citep{2004MNRAS.351..362T}. Pa$\alpha$ emission-line region is also extended out to the disk with clumpy blobs.\\

\indent {\bf 9. NGC 1614 (IRAS F04315$-$0840; ARP 186; Mrk 617; IIZW 015):}
This is a well known merger at its late stage, and found to be a minor merger system with a mass ratio of 5:1$\sim$3:1 \citep{2012MNRAS.420.2209V}. It is a barred spiral (SBc; HyperLeda) classified as an H{\sc ii} galaxy \citep{1995ApJS...98..171V,2001ApJ...546..952A,2003ApJ...583..670C}. A tidal tail can be seen in our continuum image, which is consistent with other broad-band images \citep{2002ApJS..143...47D,2011A&A...527A..60R}. A ring-like structure surrounding a nuclear region is discovered in a Pa$\alpha$ image with $HST$/NICMOS
\clearpage
\onecolumngrid

\setcounter{figure}{2}
\begin{figure}[htb]
 \epsscale{0.8}
  \begin{center}
   \plottwo{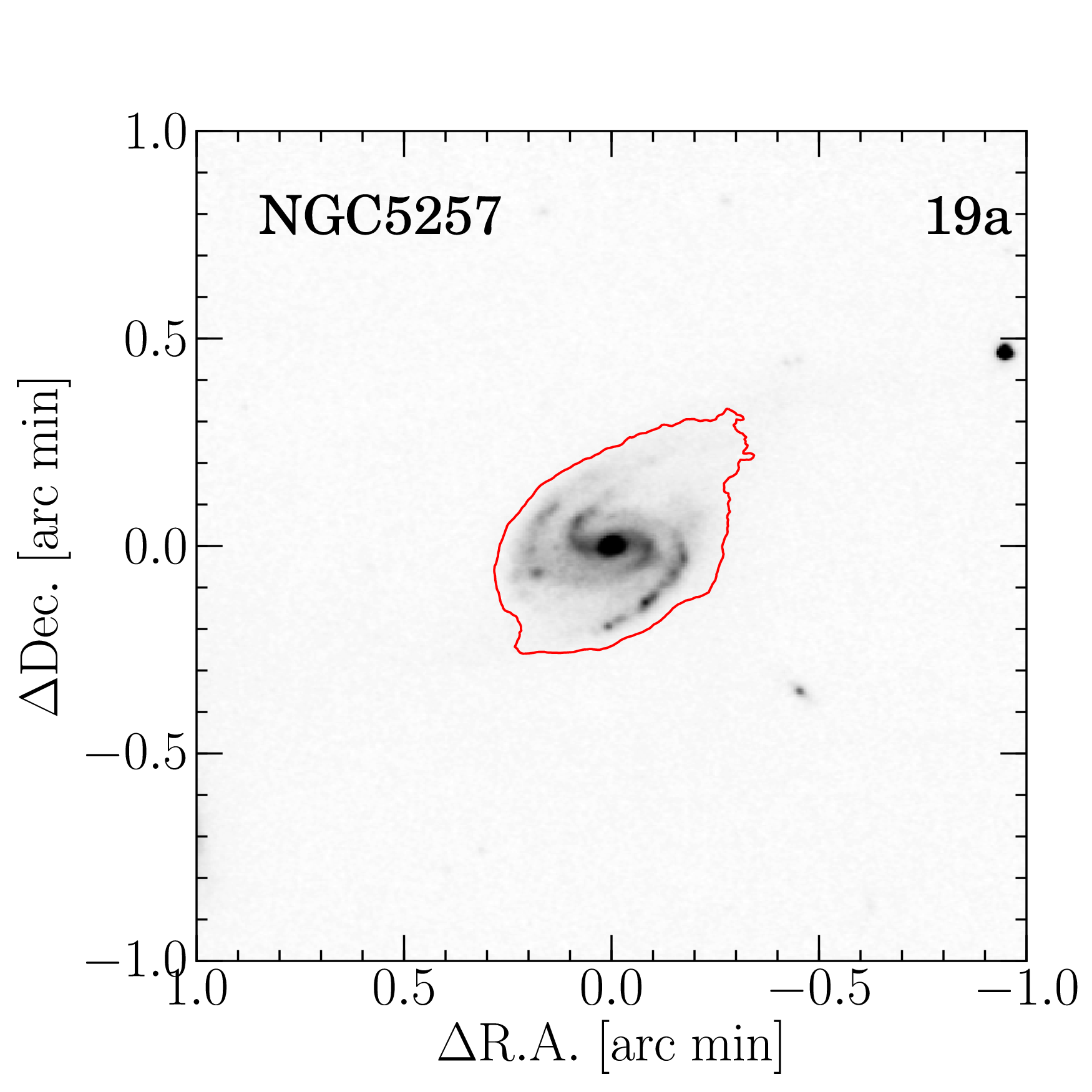}{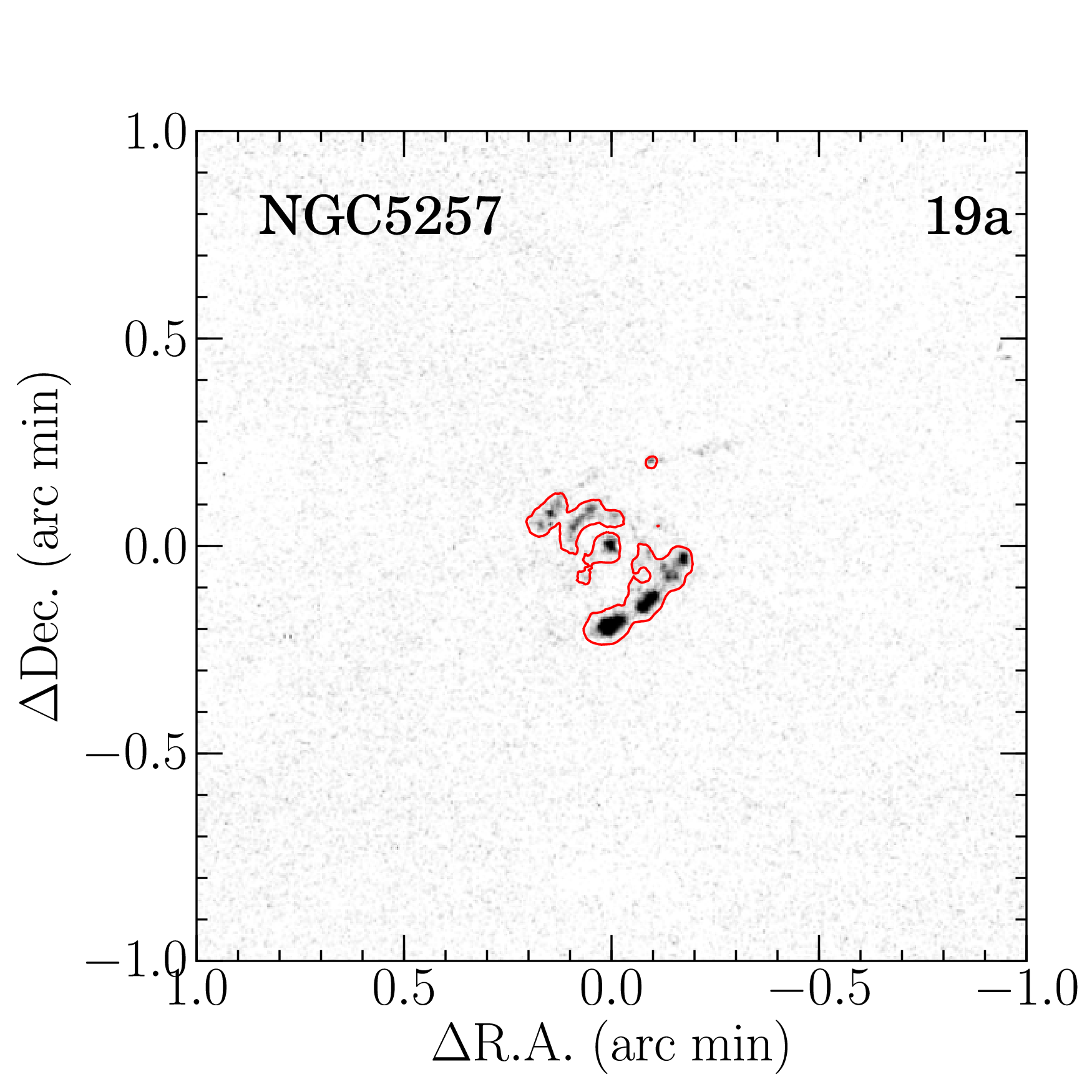}\\
   \plottwo{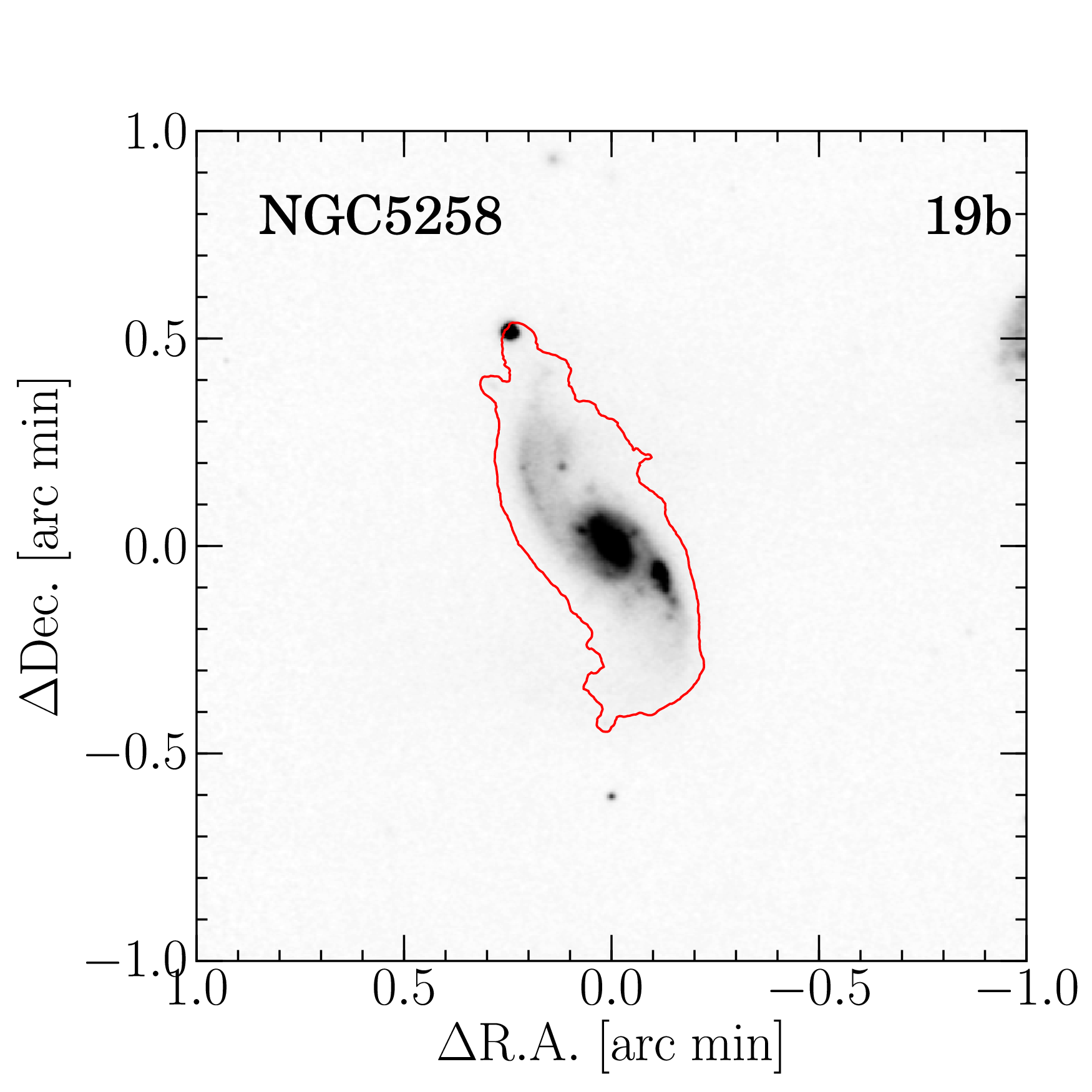}{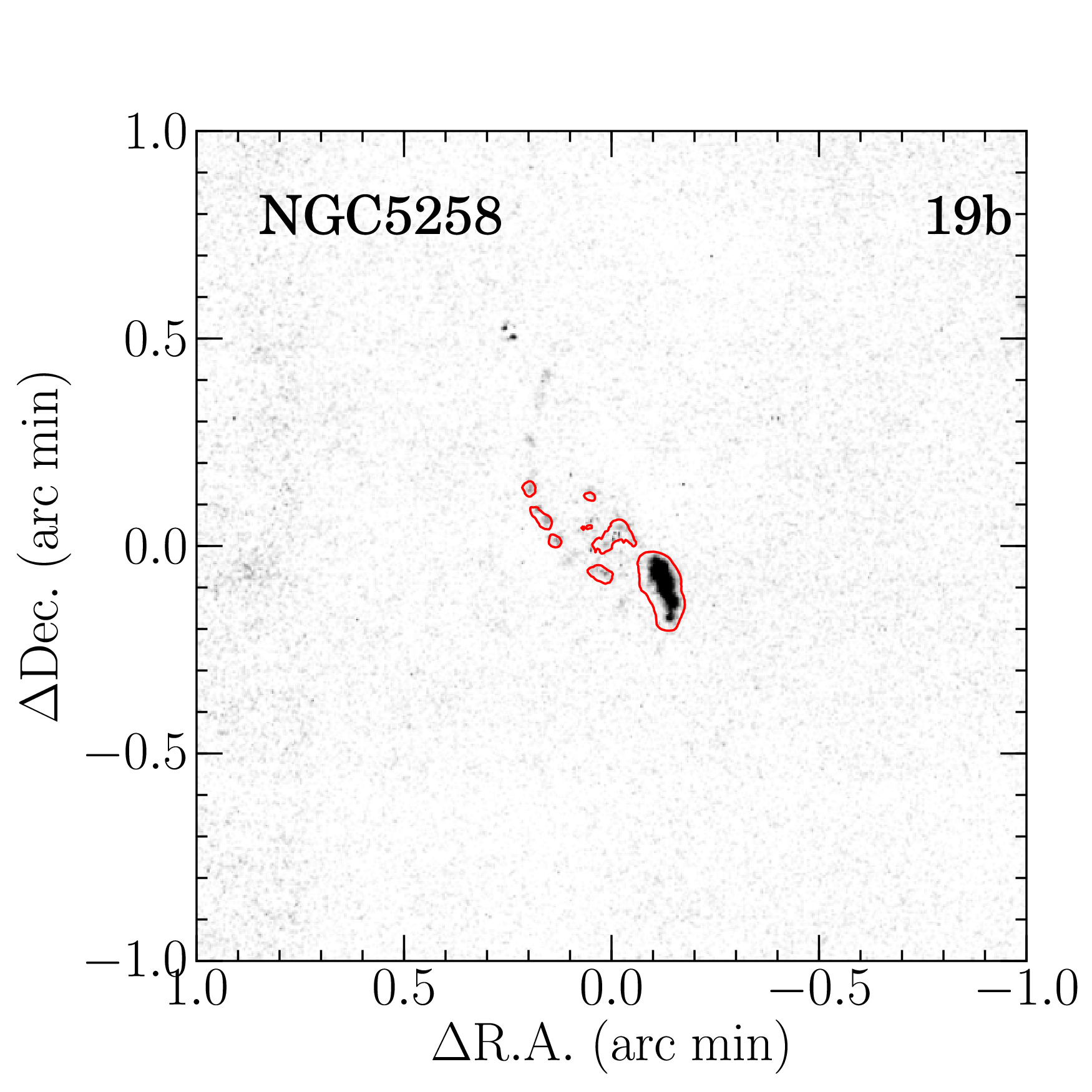}\\
   \plottwo{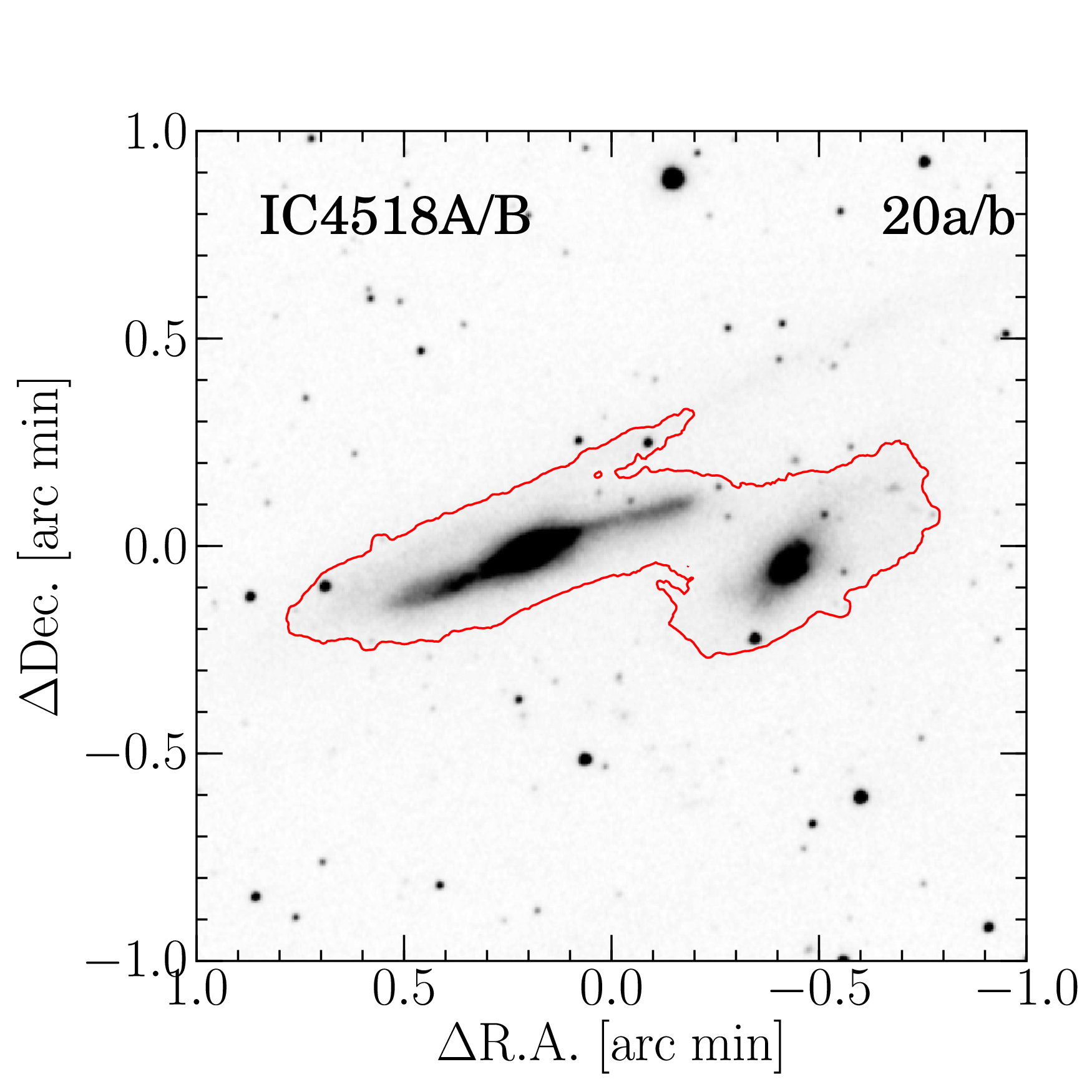}{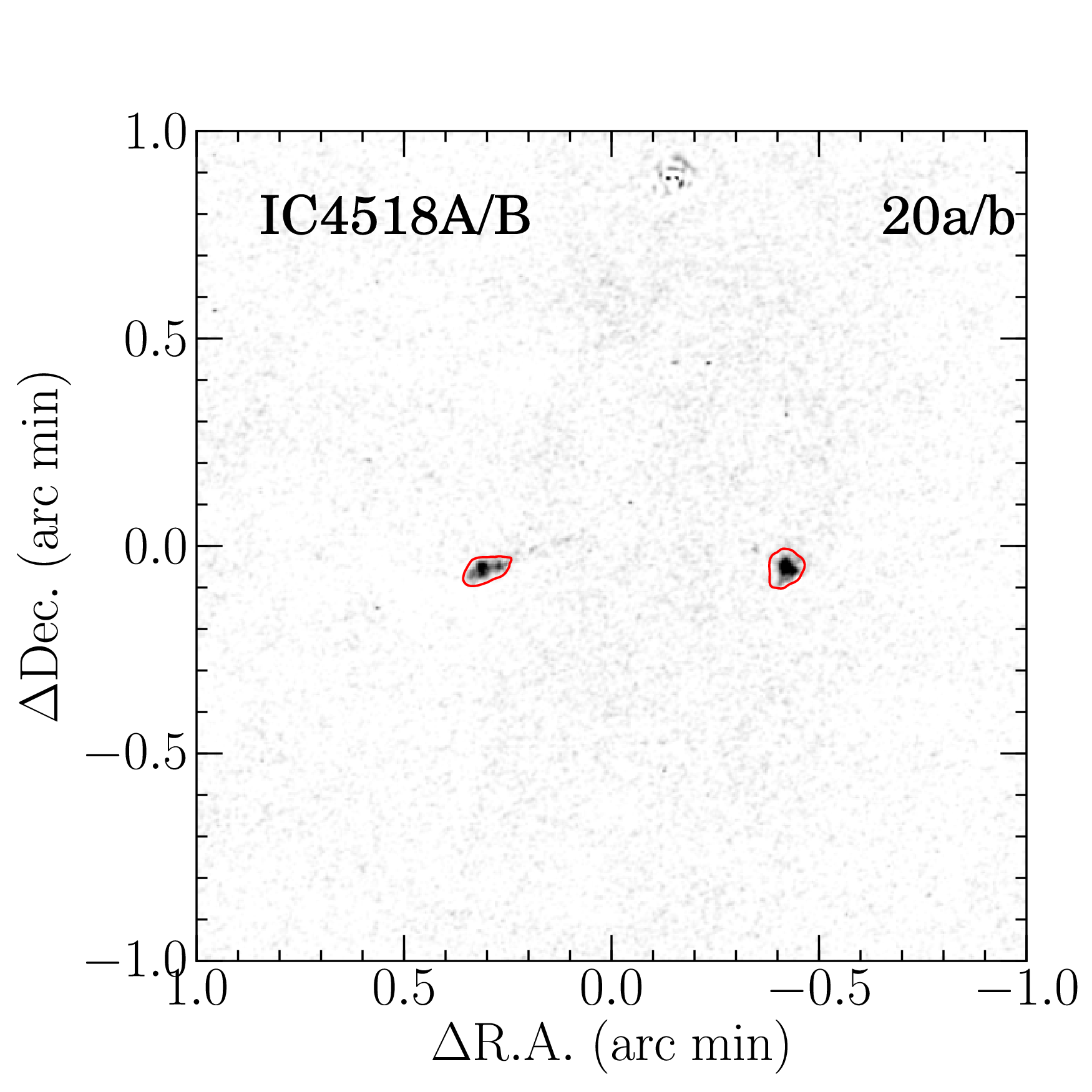}\\
   \end{center}
   \caption{Continued}
 \end{figure}

\twocolumngrid

\noindent \citep{2001ApJ...546..952A}. An extended star-forming region is detected in the Pa$\alpha$ image, but the ring-like structure can not be detected in the $0\farcs8$ spatial resolution of ANIR. \citet{2001ApJ...546..952A} describes that the growth of the ring structure is $``$inside-out$"$, but \citet{2010A&A...513A..11O} suggests that the ring is the result of a resonance.\\

\indent {\bf 10. MCG $-$05-12-006 (IRAS F04502$-$3304):}
This is an isolated barred spiral (SBb; HyperLeda) with a tidal tail \citep{2010ApJ...709..884Y} and classified as an H{\sc ii} galaxy \citep{2010ApJ...709..884Y}. Pa$\alpha$ emission is compact and concentrated at the center of the galaxy.\\

\indent {\bf 11. NGC 1720 (IRAS F04569$-$0756):}
This is a paired galaxy with NGC 1726 at a distance of 8$\farcs$2. It is a barred spiral (SBab; HyperLeda), but its energy source is not identified. The Pa$\alpha$ emission is concentrated at the center of the galaxy, and little emission can be seen at the spiral arm.
\clearpage
\onecolumngrid

\setcounter{figure}{2}
\begin{figure}[htb]
 \epsscale{0.8}
  \begin{center}
   \plottwo{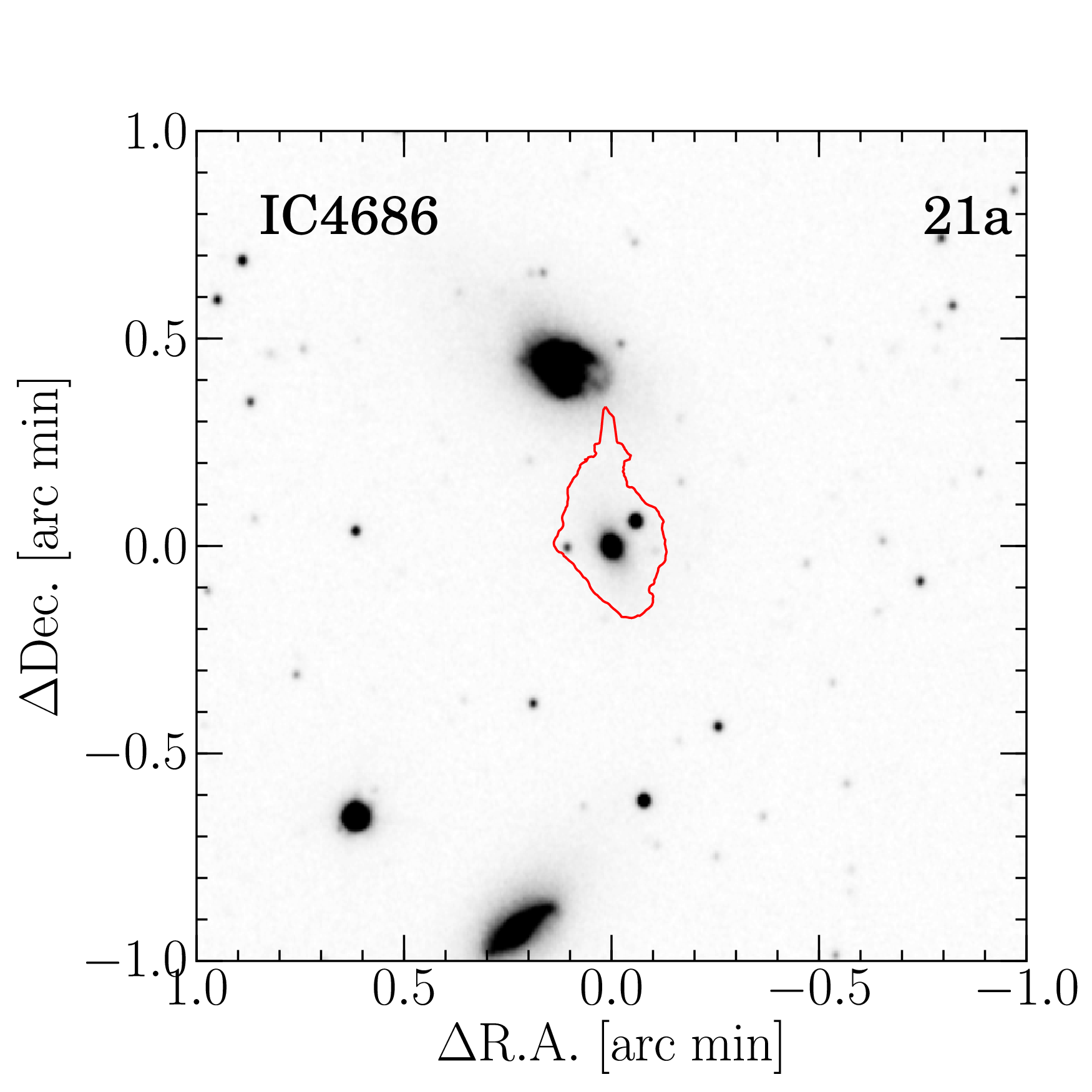}{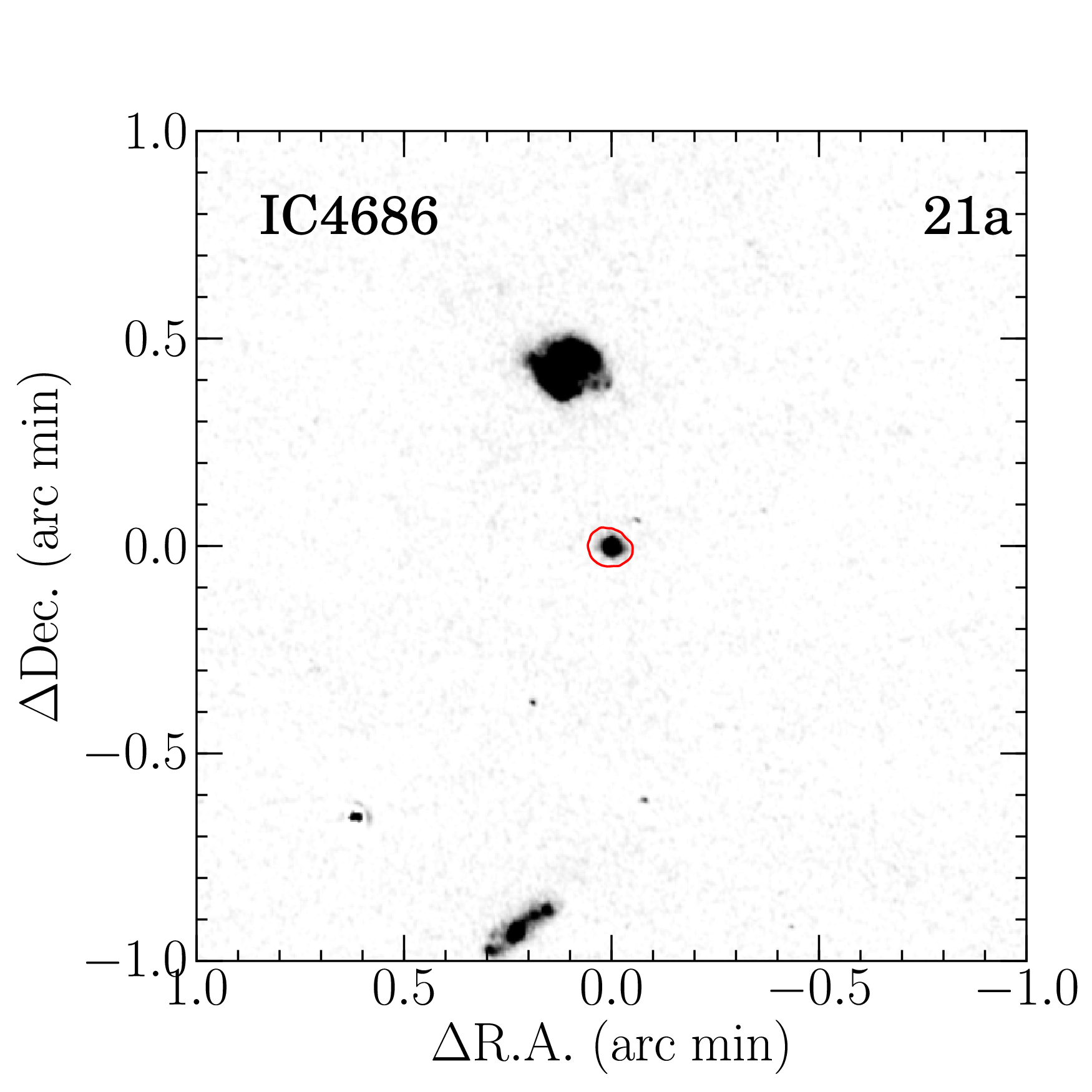}\\
   \plottwo{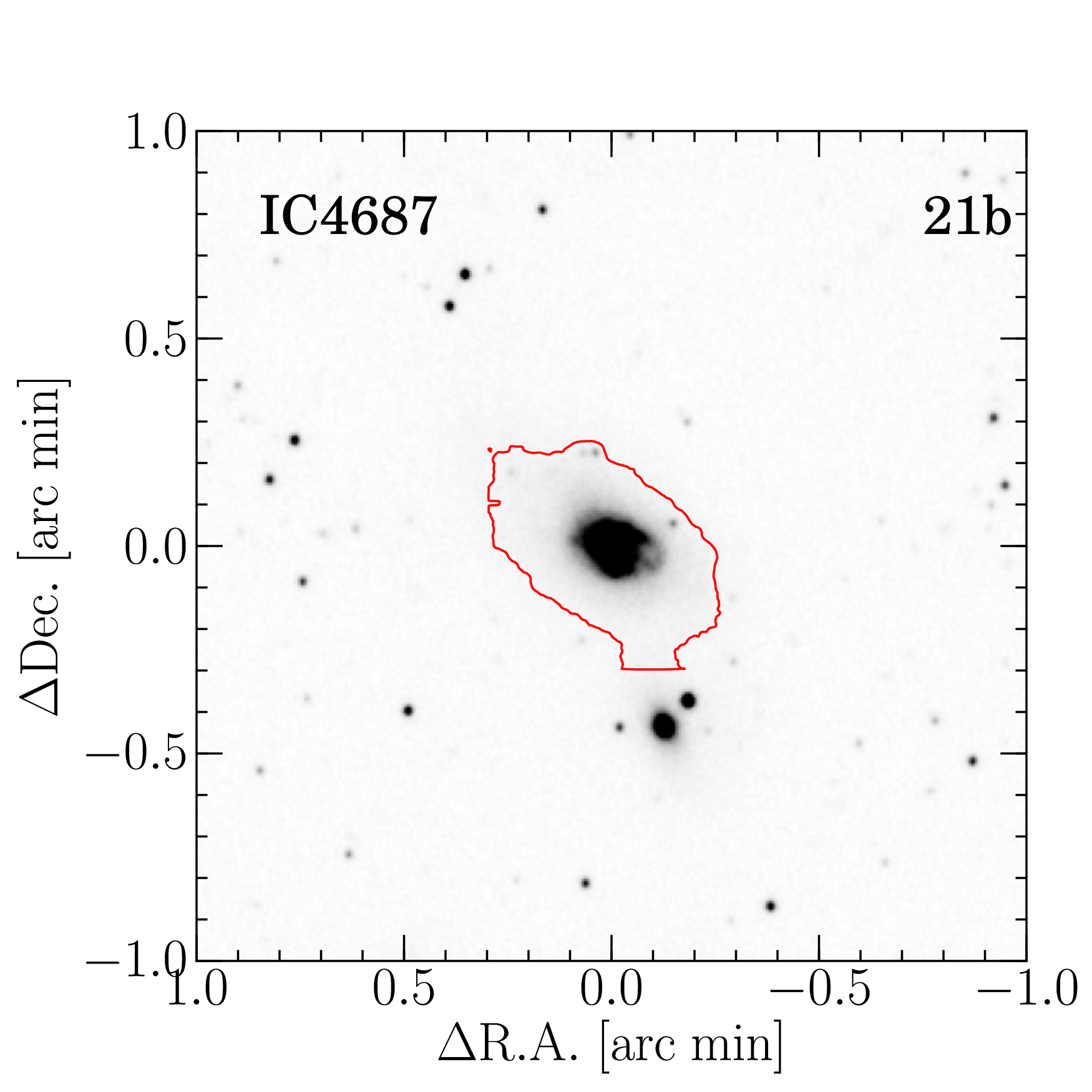}{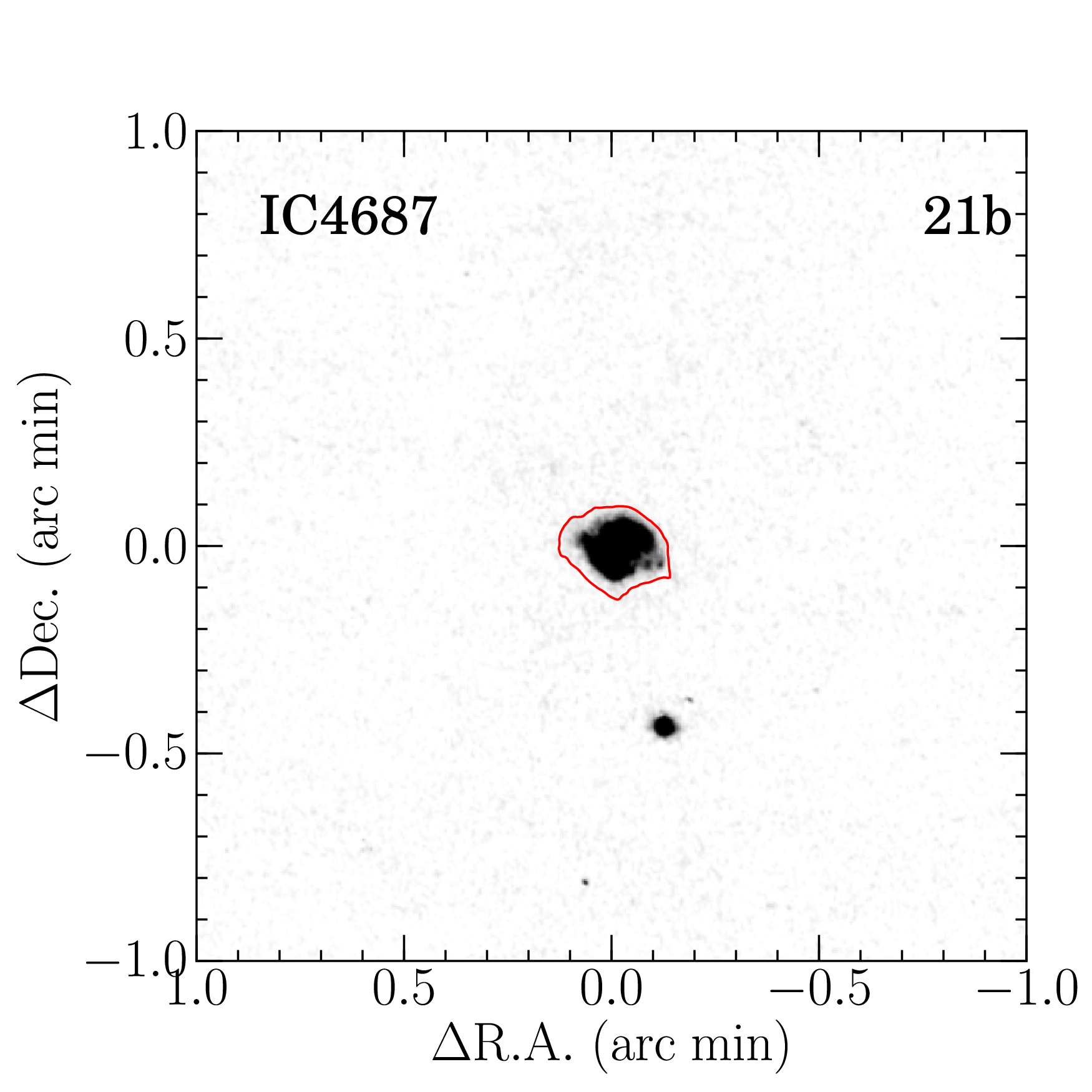}\\
   \plottwo{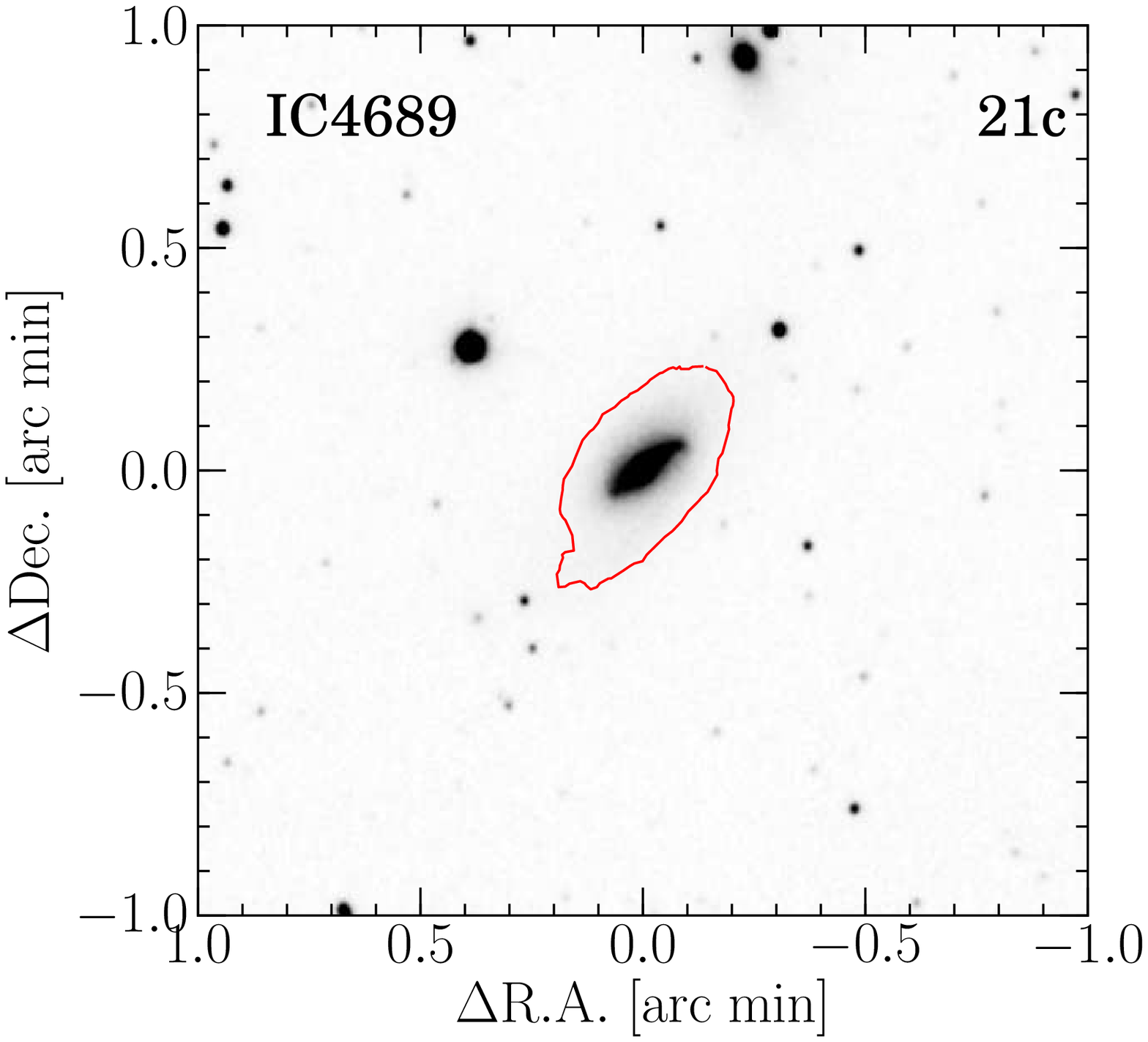}{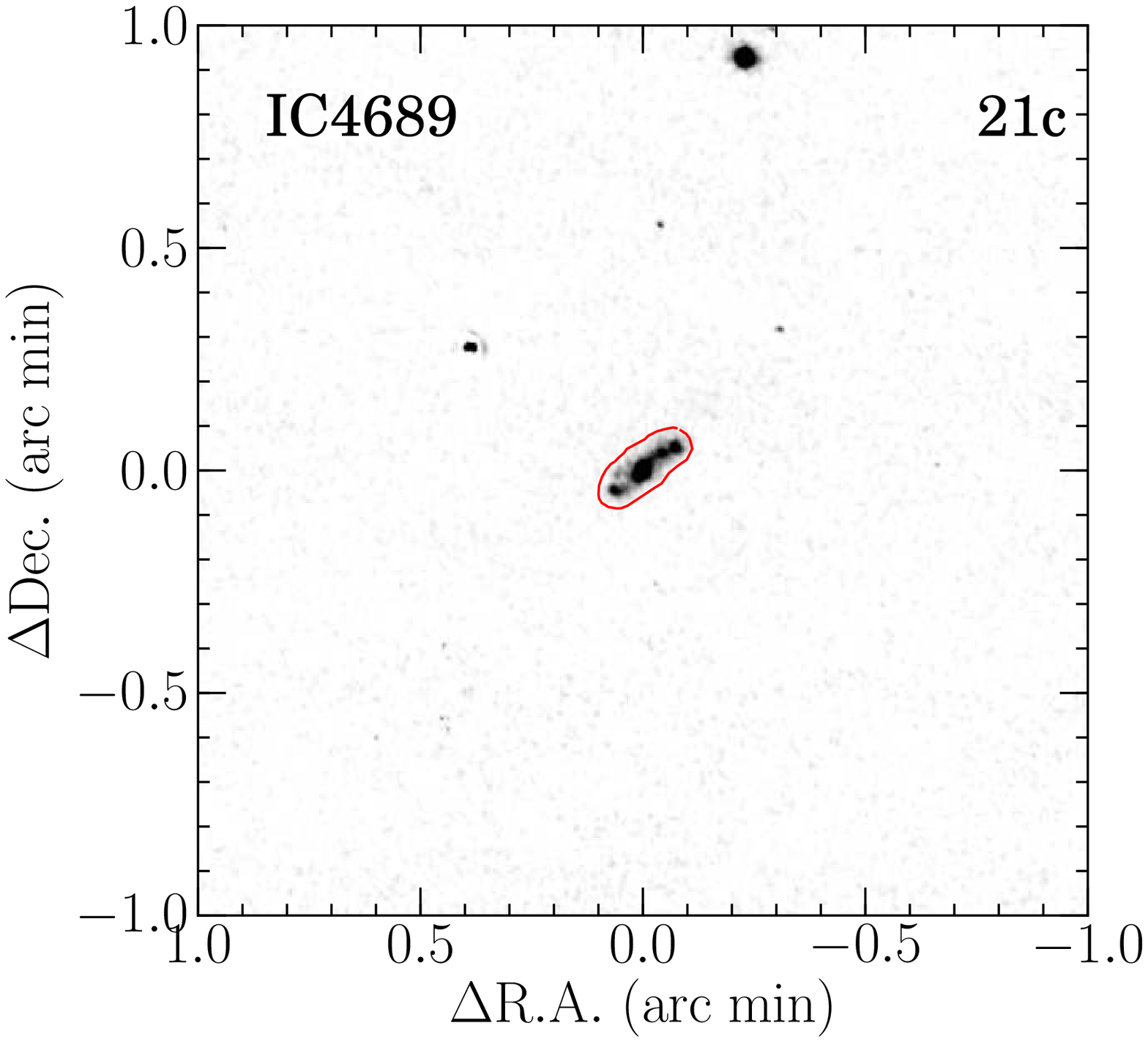}\\
   \end{center}
   \caption{Continued}
 \end{figure}

\twocolumngrid

\indent {\bf 12. ESO 557-G002 (IRAS F06295$-$1735):}
This galaxy has a companion galaxy (ESO 557-G001) at a distance of 1$\farcm$6 towards the south. It is a barred spiral (SBbc; HyperLeda) classified as an H{\sc ii} galaxy by a long-slit spectroscopic study \citep{2003ApJ...583..670C}. Ultra-hard X-ray (14-195 keV) emission cannot be detected by Swift/BAT \citep{2013ApJ...765L..26K}. The companion shows tidal distortion in the $R$-band and an H$\alpha$ image \citep{2002ApJS..143...47D} but does not show clearly in the $K_{\mathrm{s}}$-band and the Pa$\alpha$ image. Two strong concentrated peaks at the center of the galaxy is detected on the Pa$\alpha$ image.\\

\indent {\bf 13. IRAS F06592$-$6313:}
This is an isolated barred spiral (SABb; HyperLeda) with a tidal tail \citep{2010ApJ...709..884Y} classified as an H{\sc ii} galaxy \citep{2010ApJ...709..884Y}. It has H$\alpha$ condensation outside its main body at 7$\farcs$0 towards the north \citep{2011A&A...527A..60R}, of which the Pa$\alpha$ image is not enough deep to detect. Distribution of Pa$\alpha$ emission is concentrated at the center of the galaxy.
\clearpage
\onecolumngrid

\setcounter{figure}{2}
\begin{figure}[htb]
 \epsscale{0.8}
  \begin{center}
   \plottwo{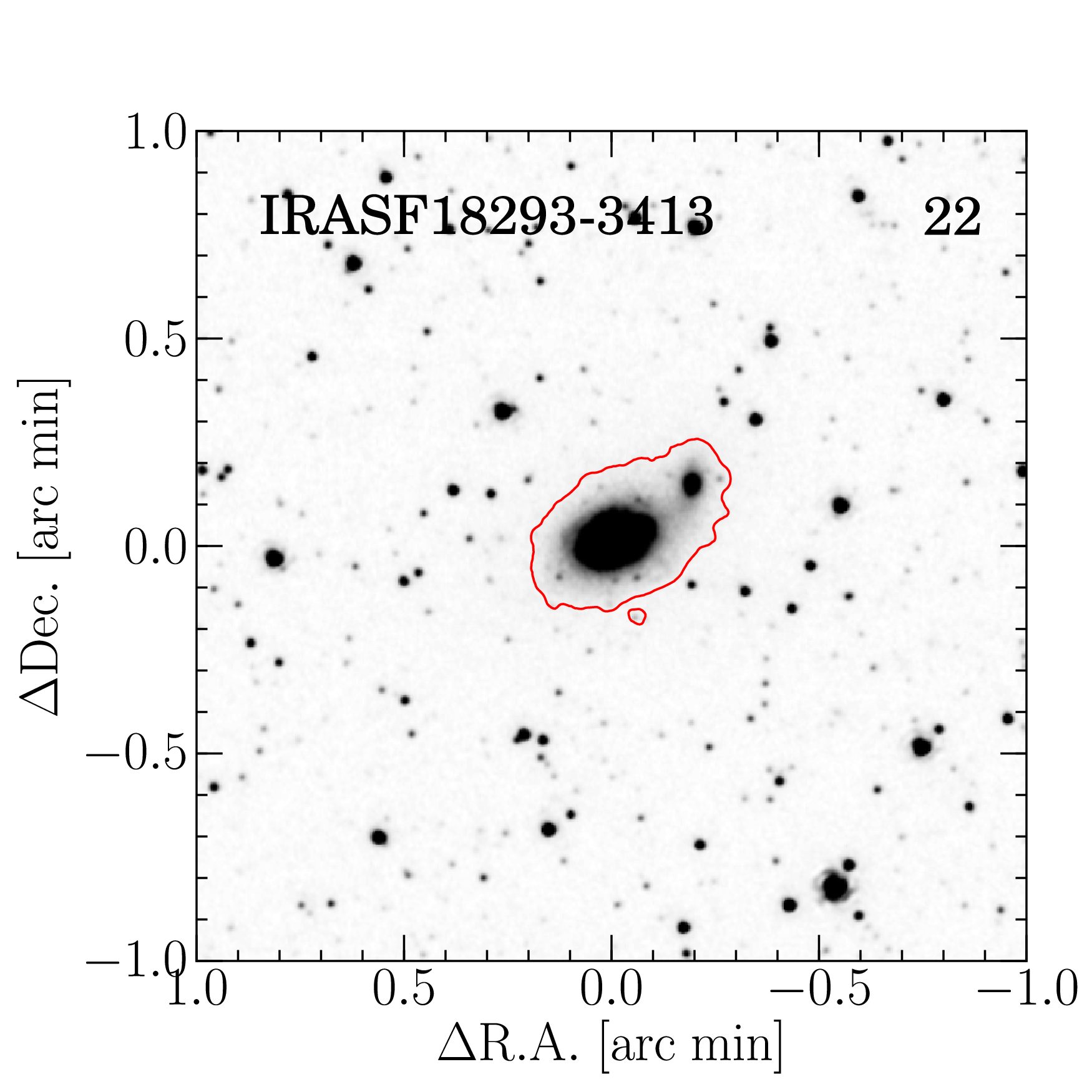}{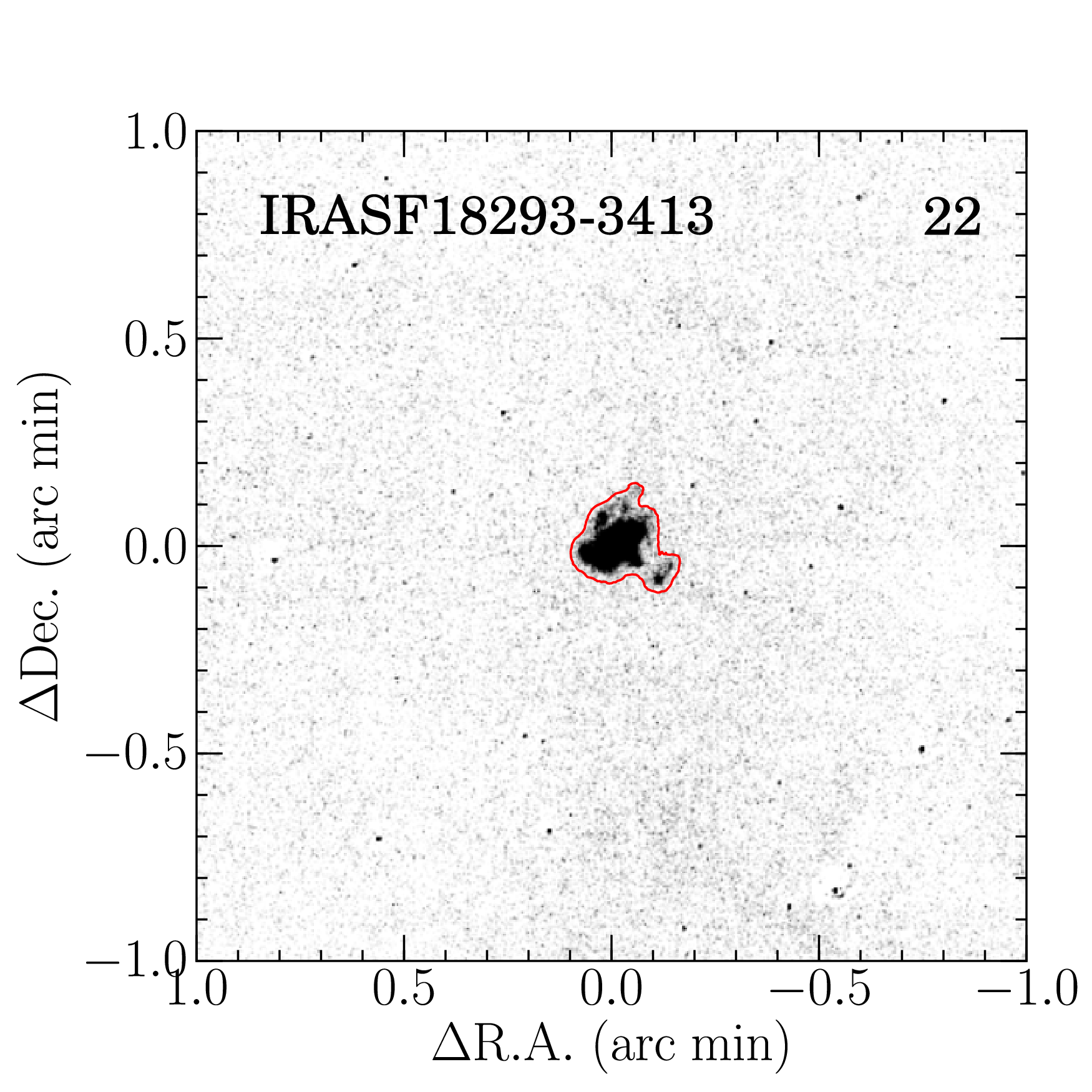}\\
   \plottwo{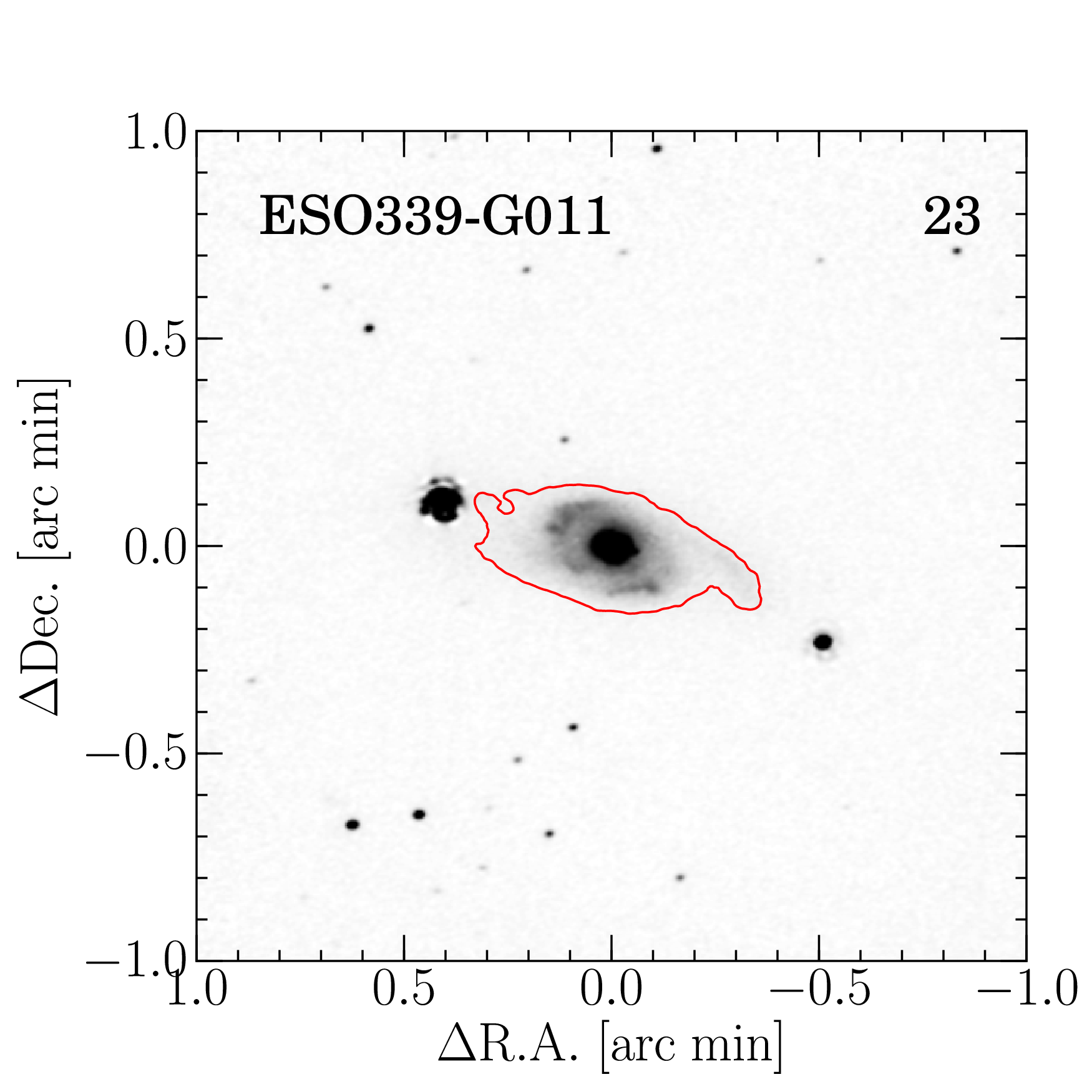}{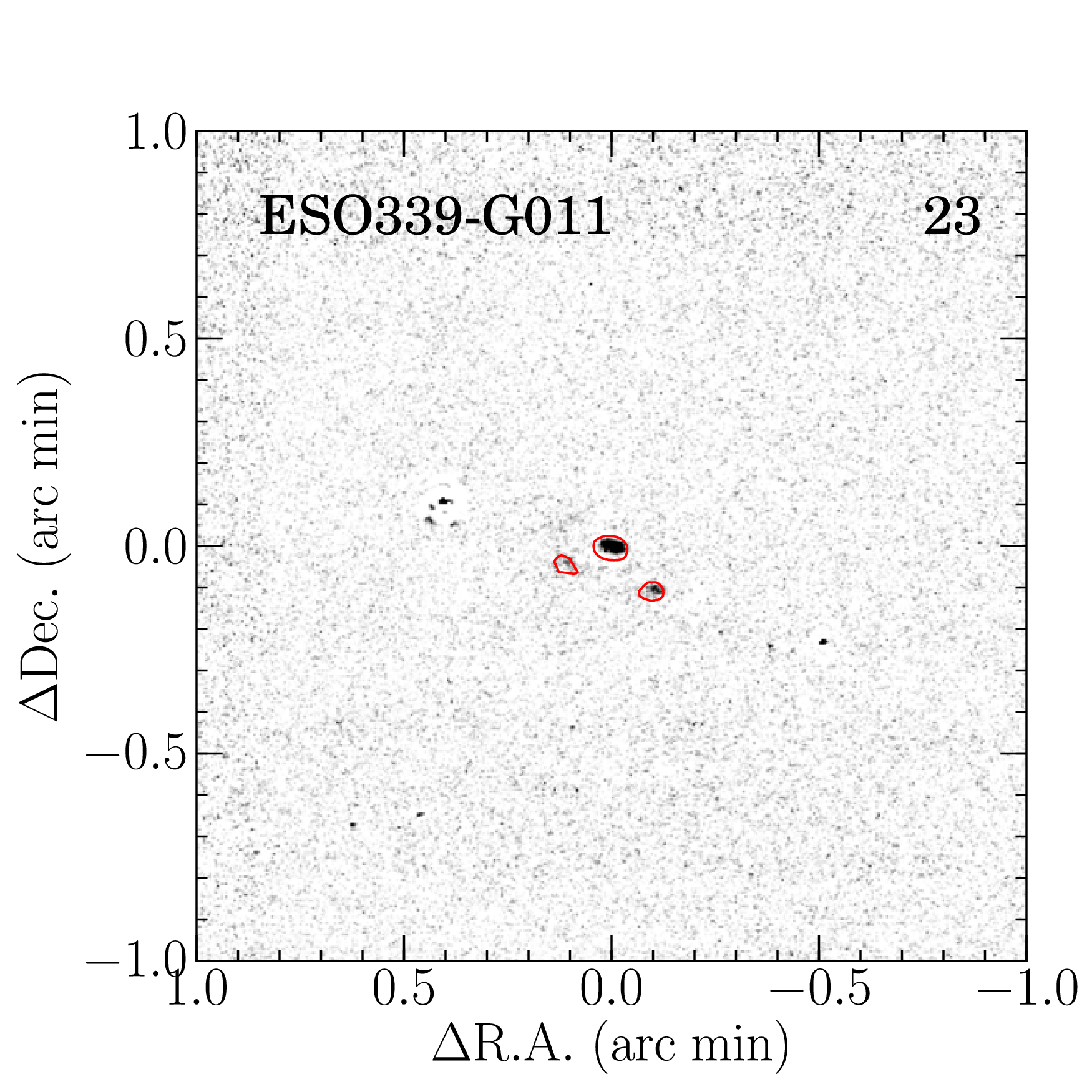}\\
   \plottwo{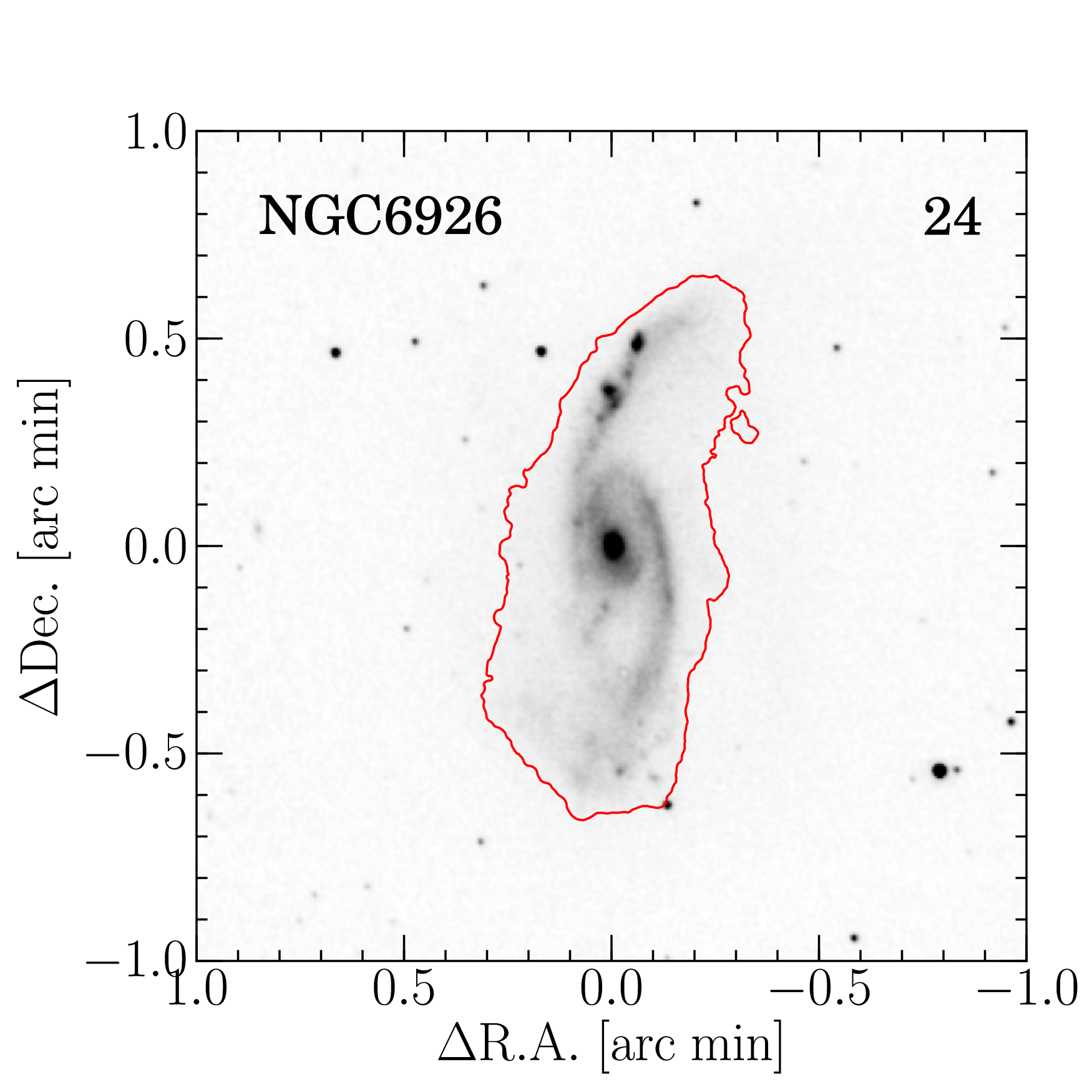}{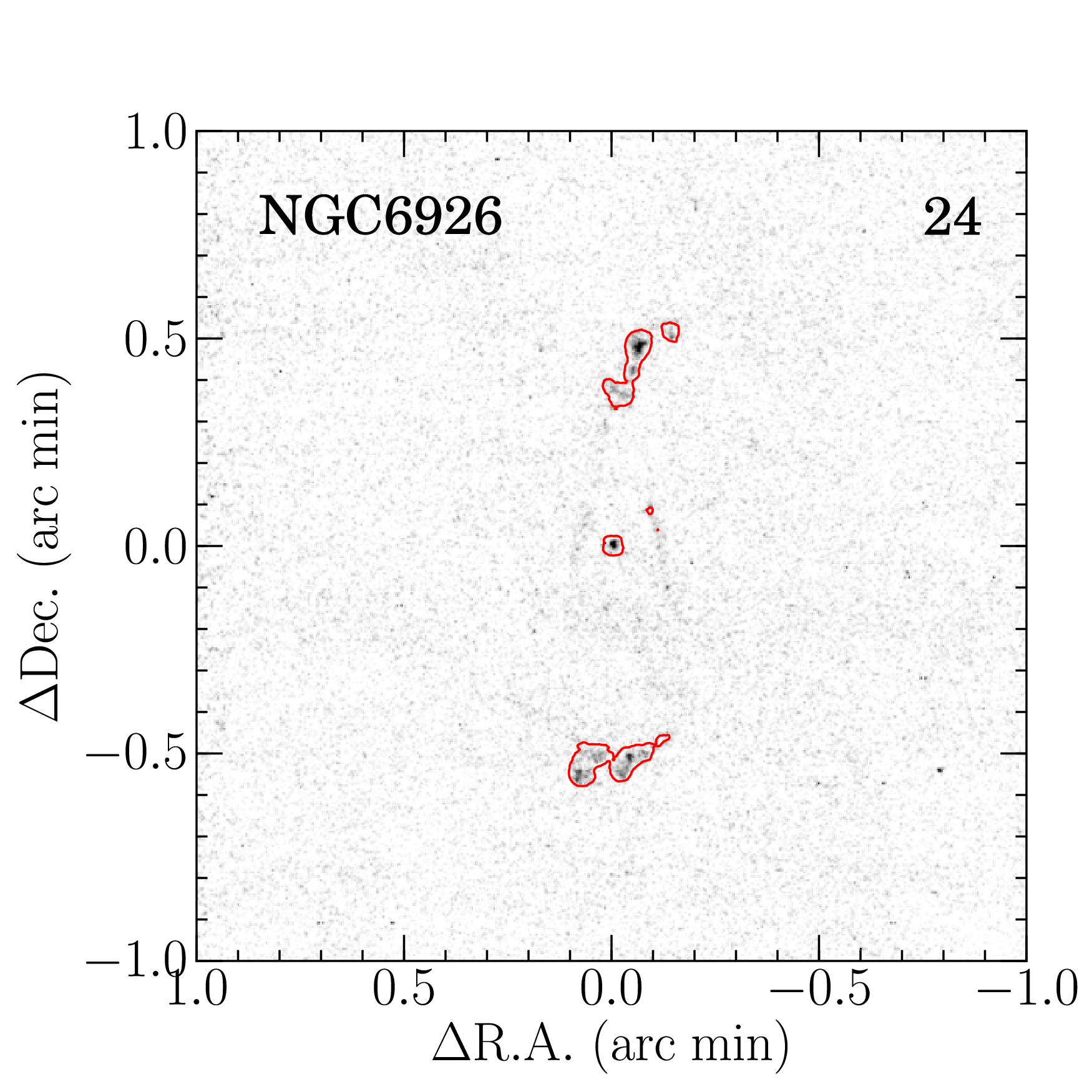}\\
   \end{center}
   \caption{Continued}
 \end{figure}

\twocolumngrid

\indent {\bf 14. NGC 2342 (IRAS 07063$+$2043):}
This is a spiral (Sc; HyperLeda) classified as an H{\sc ii} galaxy \citep{1997ApJS..112..315H} without an AGN activity at any wavelength, having a paired galaxy (NGC 2341; \citealt{2006ApJ...650..835A}) at a distance of 2$\farcm$5. This is not a well studied pair \citep{2005MNRAS.357..109J}. The both galaxies have high IR luminosities of $\log(L_\mathrm{FIR}/L_{\sun}) = 10.8$ \citep{2003AJ....126.1607S}. Pa$\alpha$ emission is extended along the spiral arms over 10 kpc.\\

\indent {\bf 15. ESO 320-G030 (IRAS F11506$-$3851):}
This is a barred spiral (SBb; HyperLeda) classified as an H{\sc ii} galaxy \citep{1991A&AS...91...61V}. VLT-VIMOS/H$\alpha$ \citep{2011A&A...527A..60R} and $HST$/NICMOS Pa$\alpha$ \citep{2006ApJ...650..835A} observations report starburst regions distributed in a ring-like shape, while they can not be seen in continuum images. In the Pa$\alpha$ image, the same ring structure can be seen, and the emission line region is extended beyond the ring.
\clearpage
\onecolumngrid

\setcounter{figure}{2}
\begin{figure}[htb]
 \epsscale{0.8}
  \begin{center}
   \plottwo{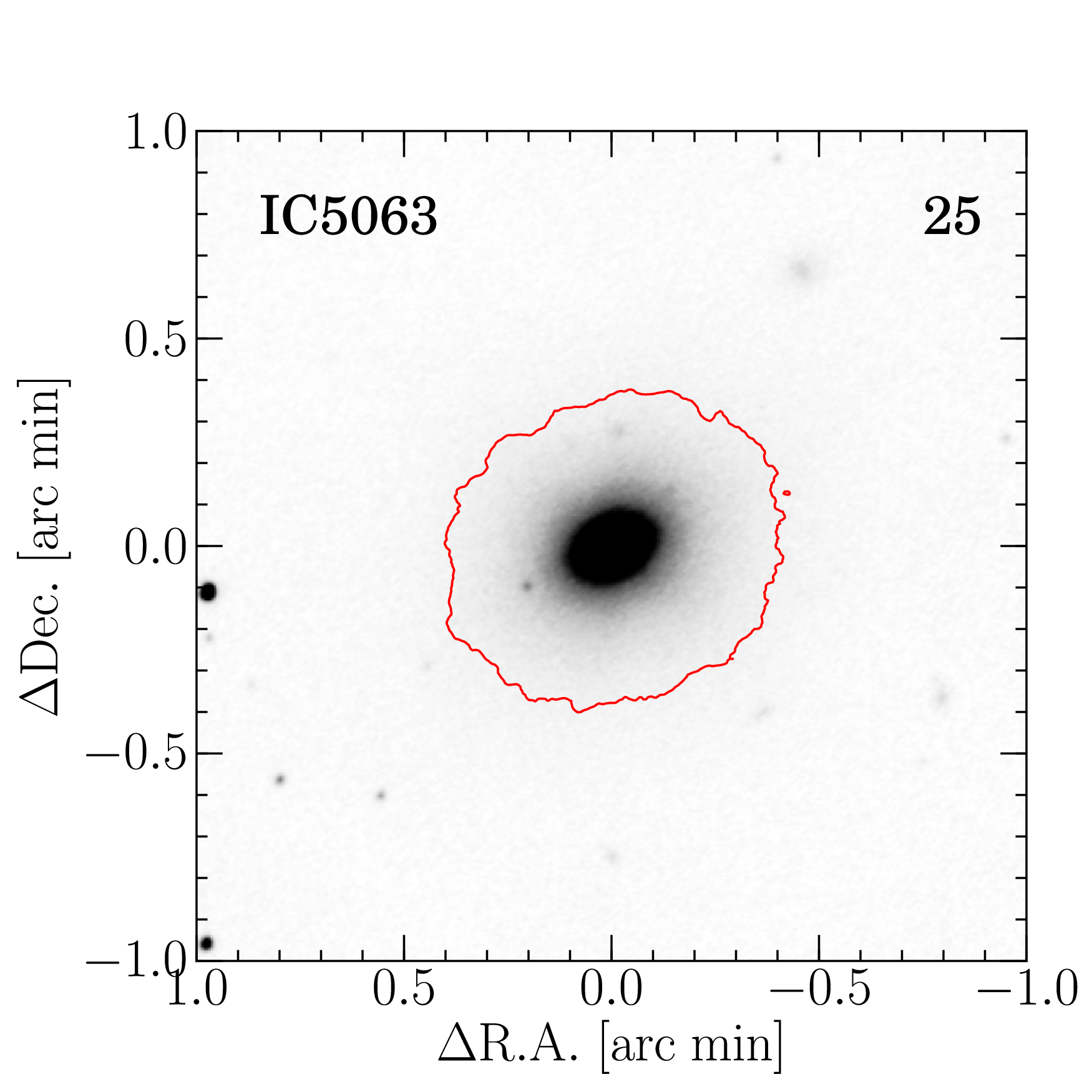}{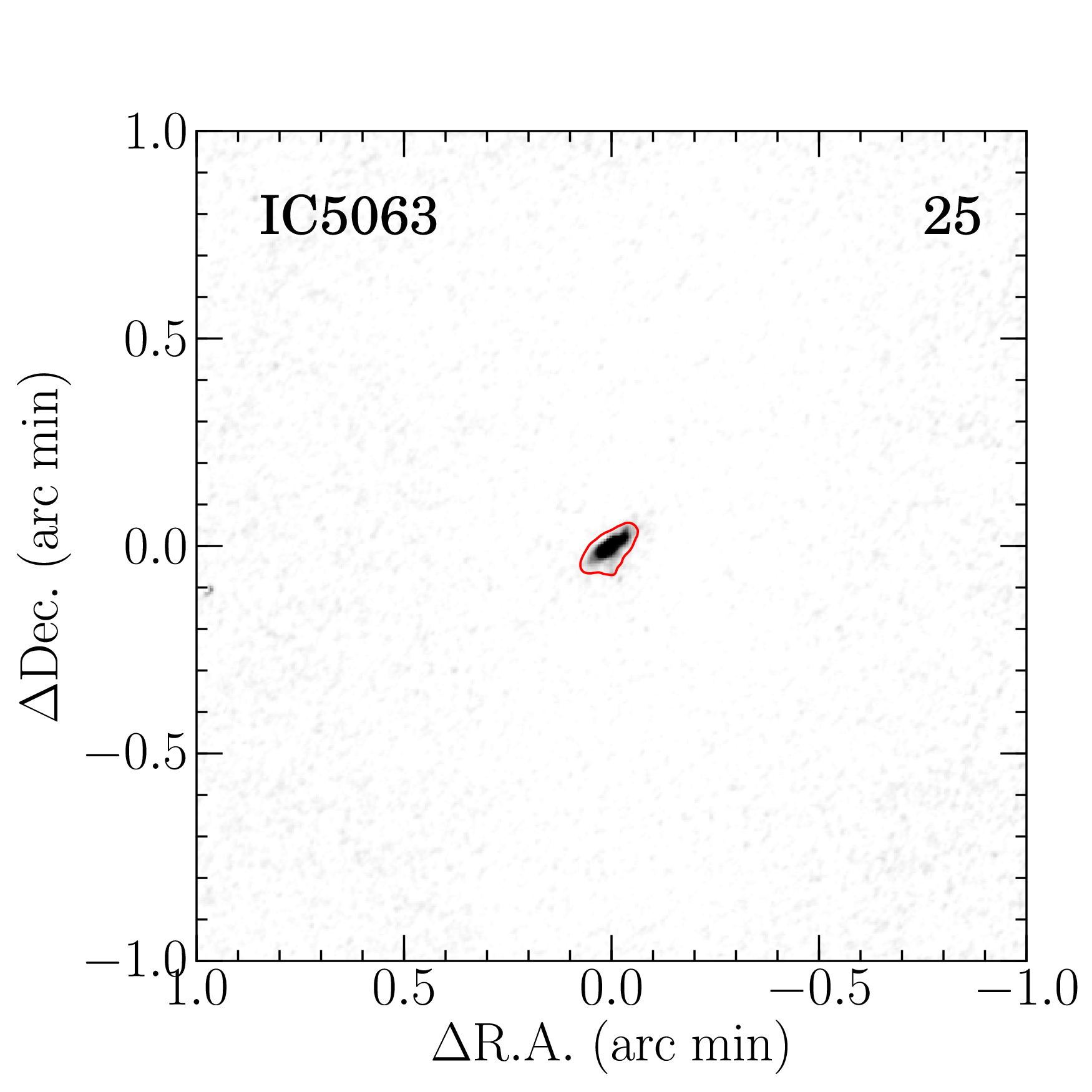}\\
   \plottwo{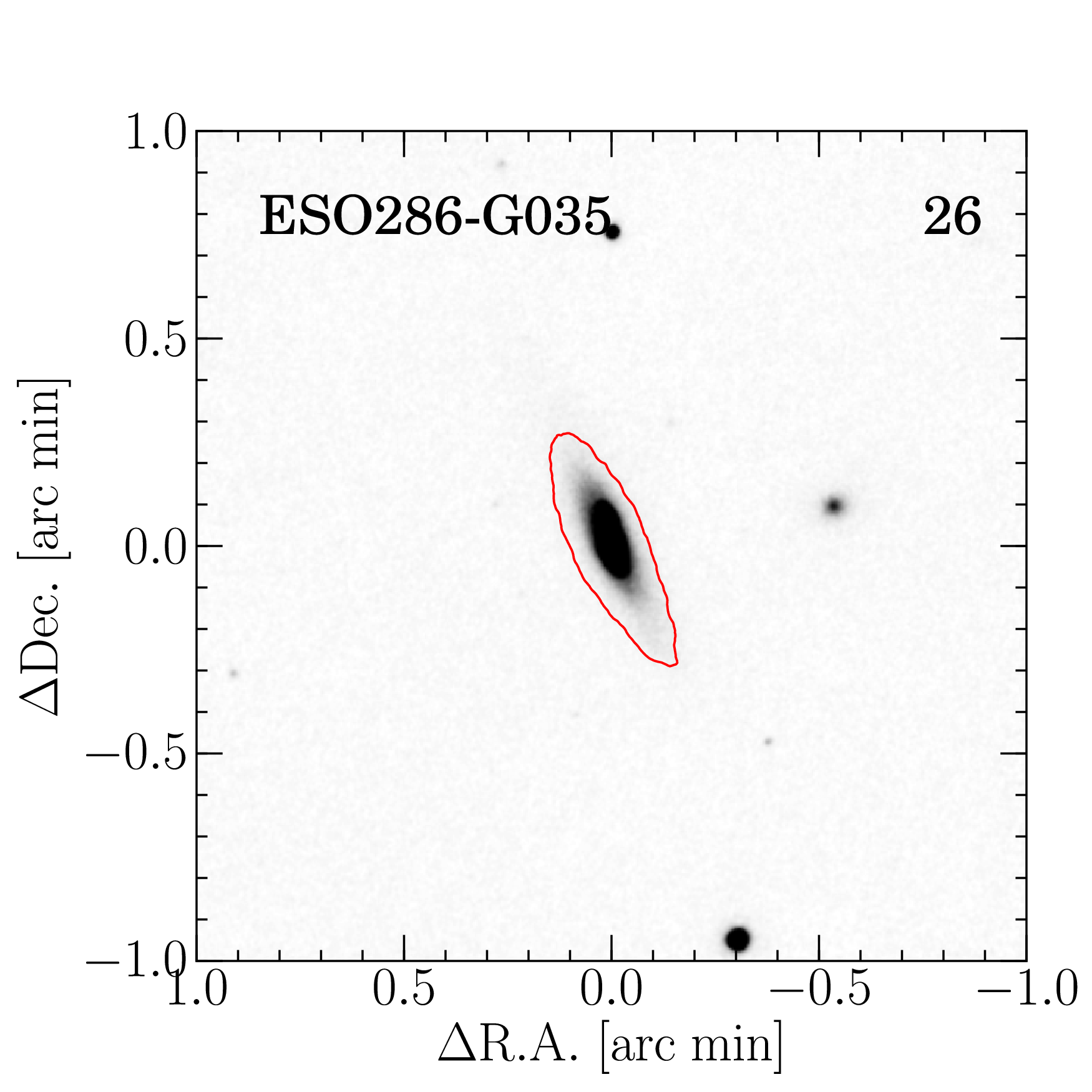}{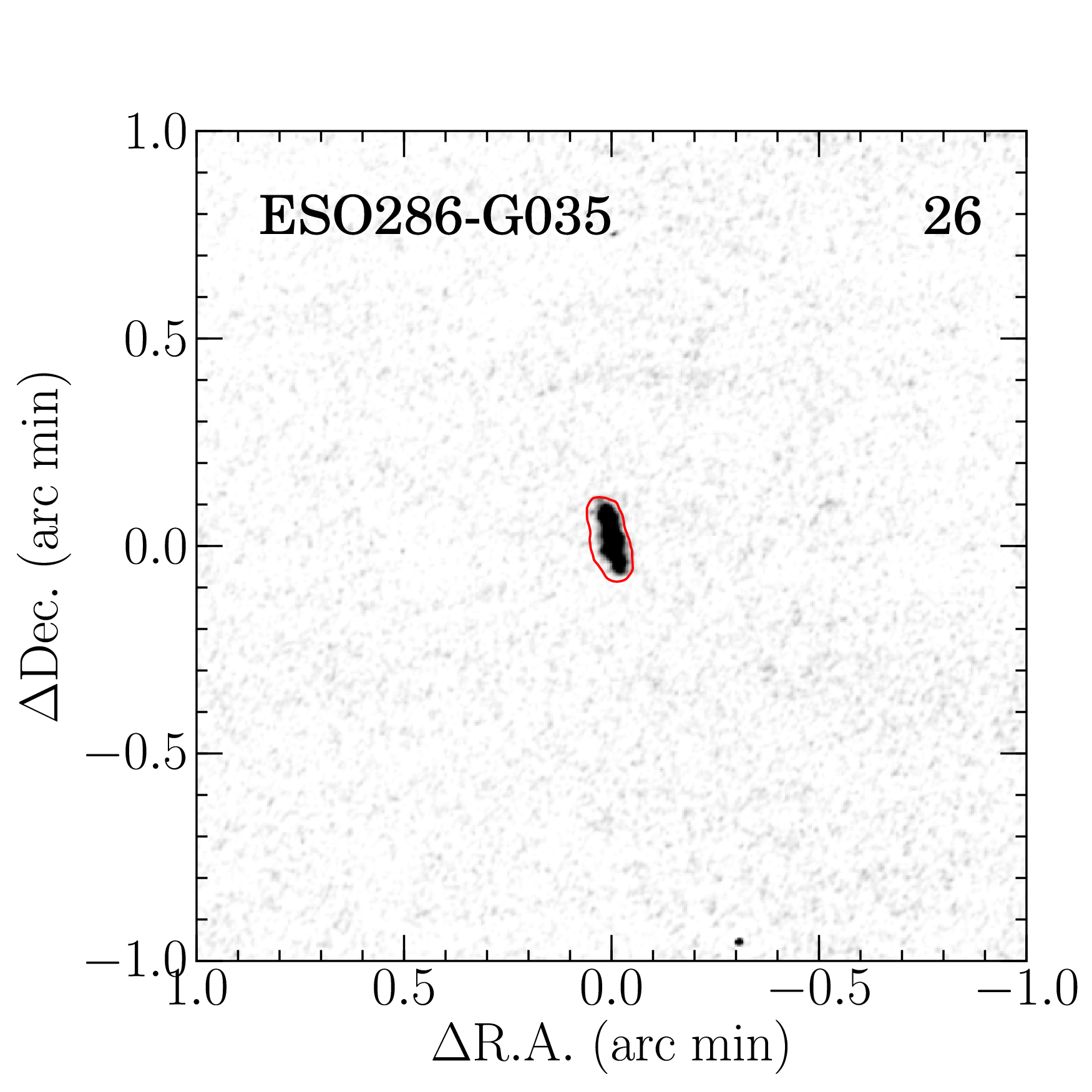}\\
   \plottwo{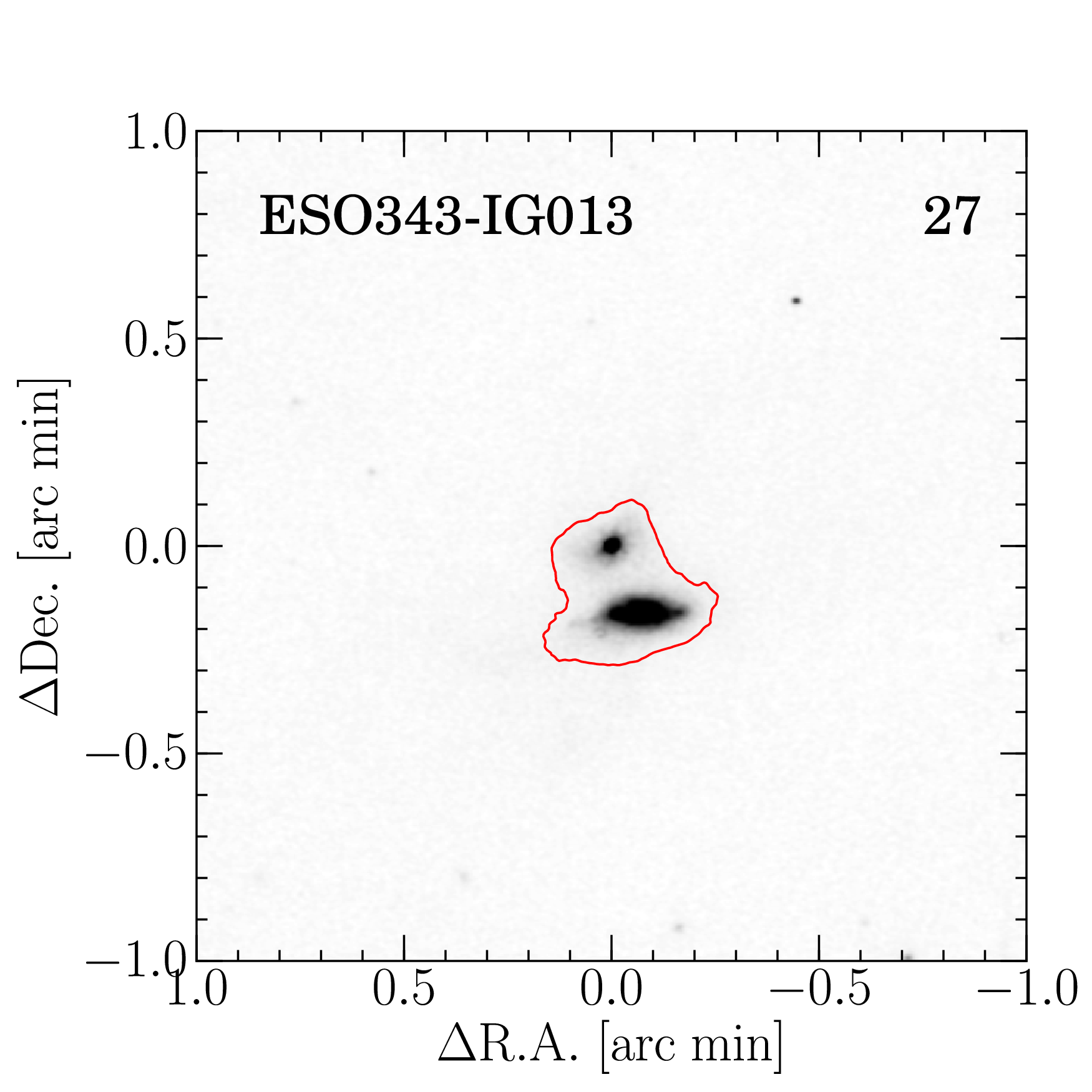}{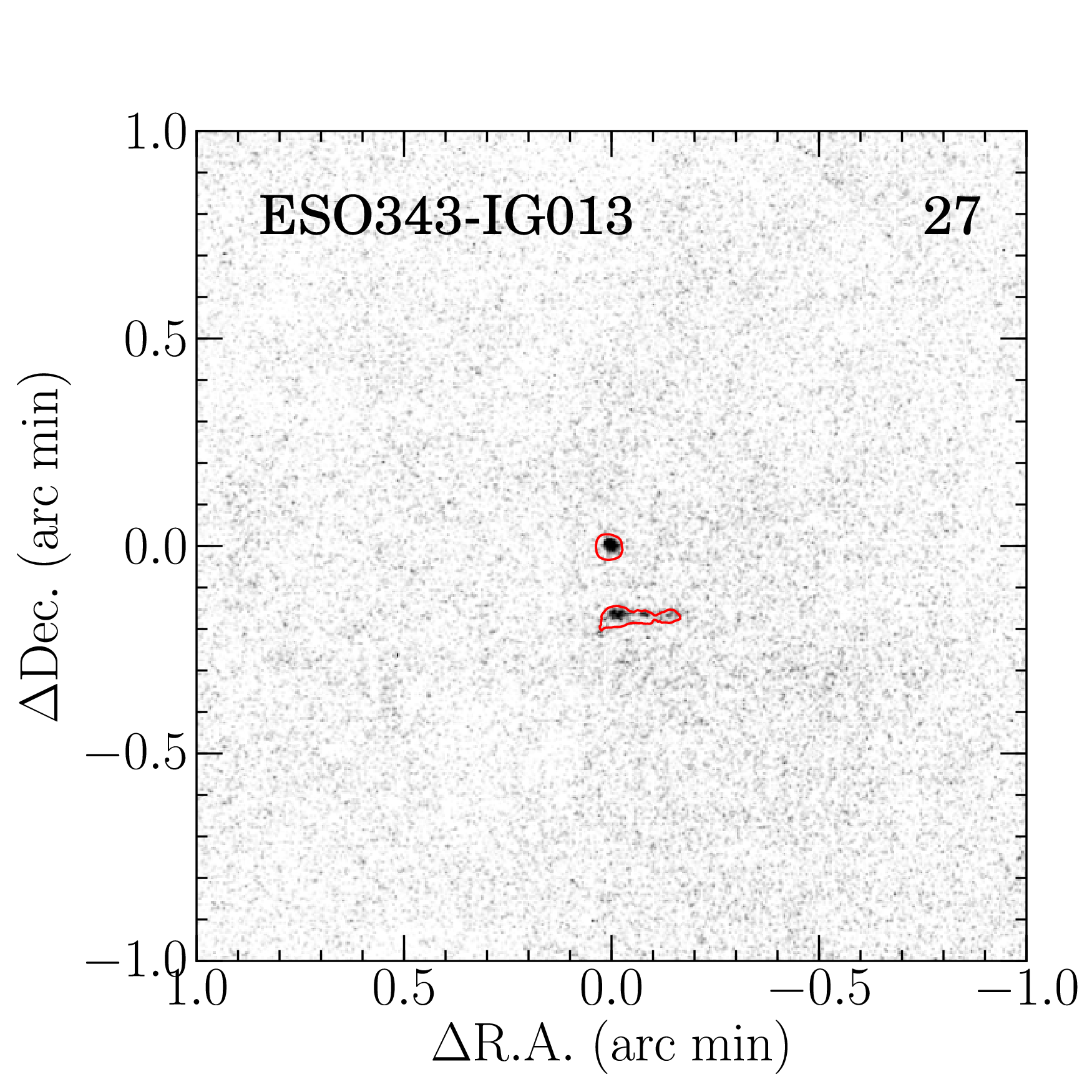}\\
   \end{center}
   \caption{Continued}
 \end{figure}

\twocolumngrid

\indent {\bf 16. NGC 4922 (IRAS F12590$+$2934; VV 609; KPG 363A/B):}
This is known as a post-merger between an early-type and a spiral, located at the outskirts of the Coma cluster \citep{2009AJ....138.1911S}. It is an elliptical (E; HyperLeda) classified as a LINER \citep{1991A&AS...91...61V}, and shows extended soft X-ray emission not originated from an AGN but possibly related to the on-going star formation \citep{1999MNRAS.302..561A} in the northern galaxy, especially. In the Pa$\alpha$ image, the concentrated northern nucleus has strong and extended emission, but the southern one is faint.\\

\indent {\bf 17. MCG $-$03-34-064 (IRAS F13197$-$1627):}
This galaxy forms a wide binary system with MCG $-$30-34-063 \citep{2010ApJ...709..884Y}, and is an S0/a \citep{1995MNRAS.274.1107N} classified as a Seyfert 1 spectroscopically \citep{2006A&A...455..773V}. Pa$\alpha$ emission is compact and concentrated at the center of the galaxy. In addition, a paired galaxy located 1$\farcm$8 away are found have a compact Pa$\alpha$ source.
\clearpage
\onecolumngrid

\setcounter{figure}{2}
\begin{figure}[htb]
 \epsscale{0.8}
  \begin{center}
   \plottwo{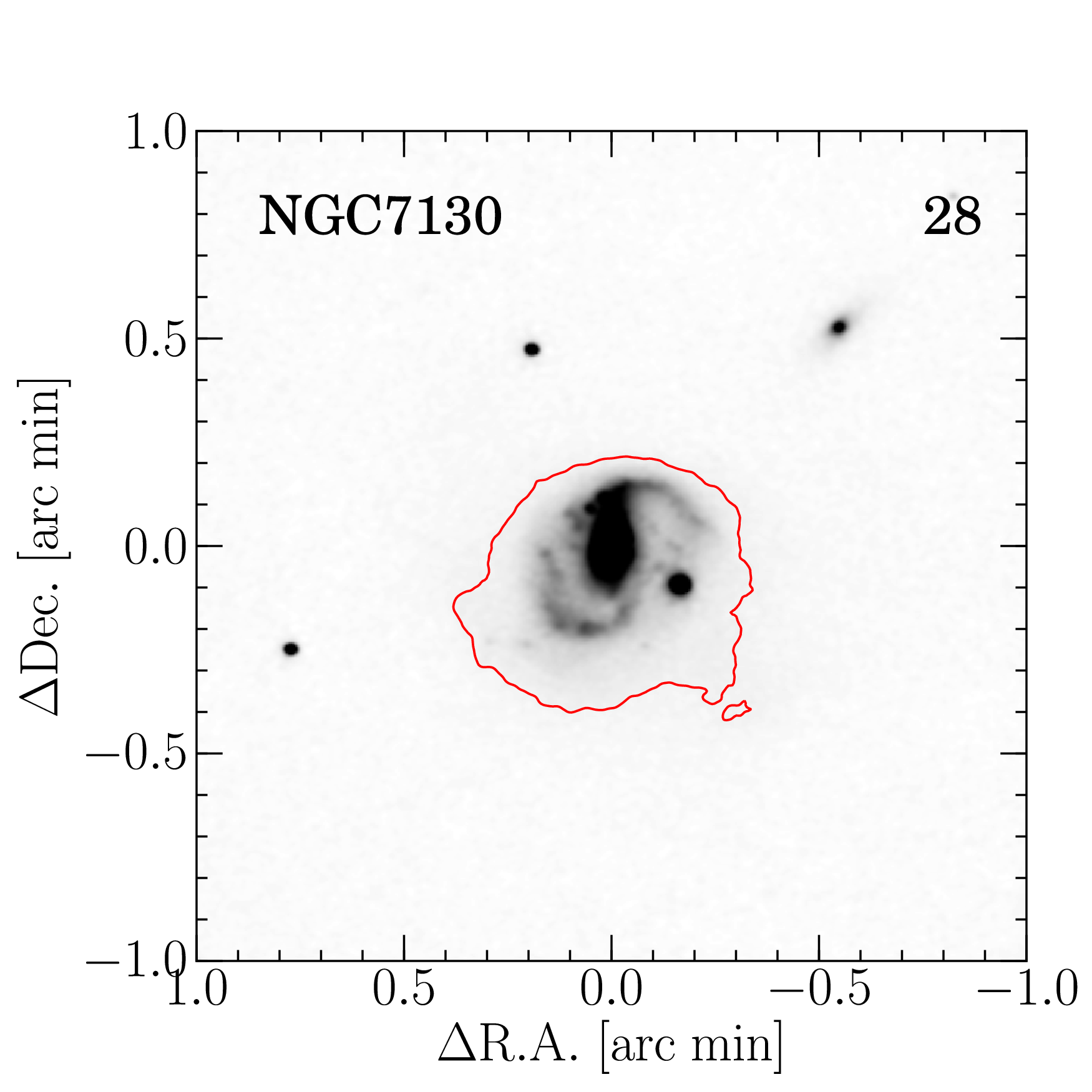}{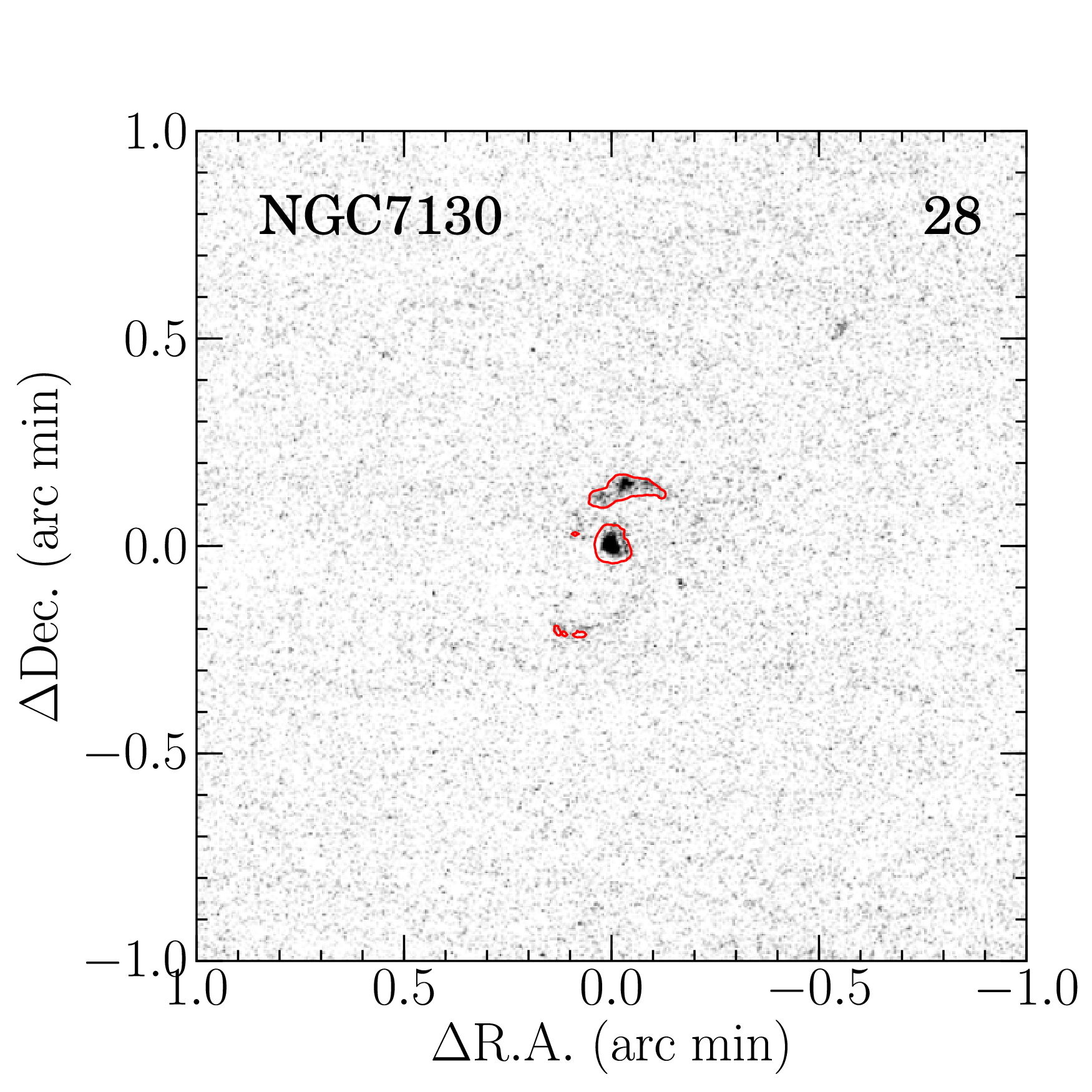}\\
   \plottwo{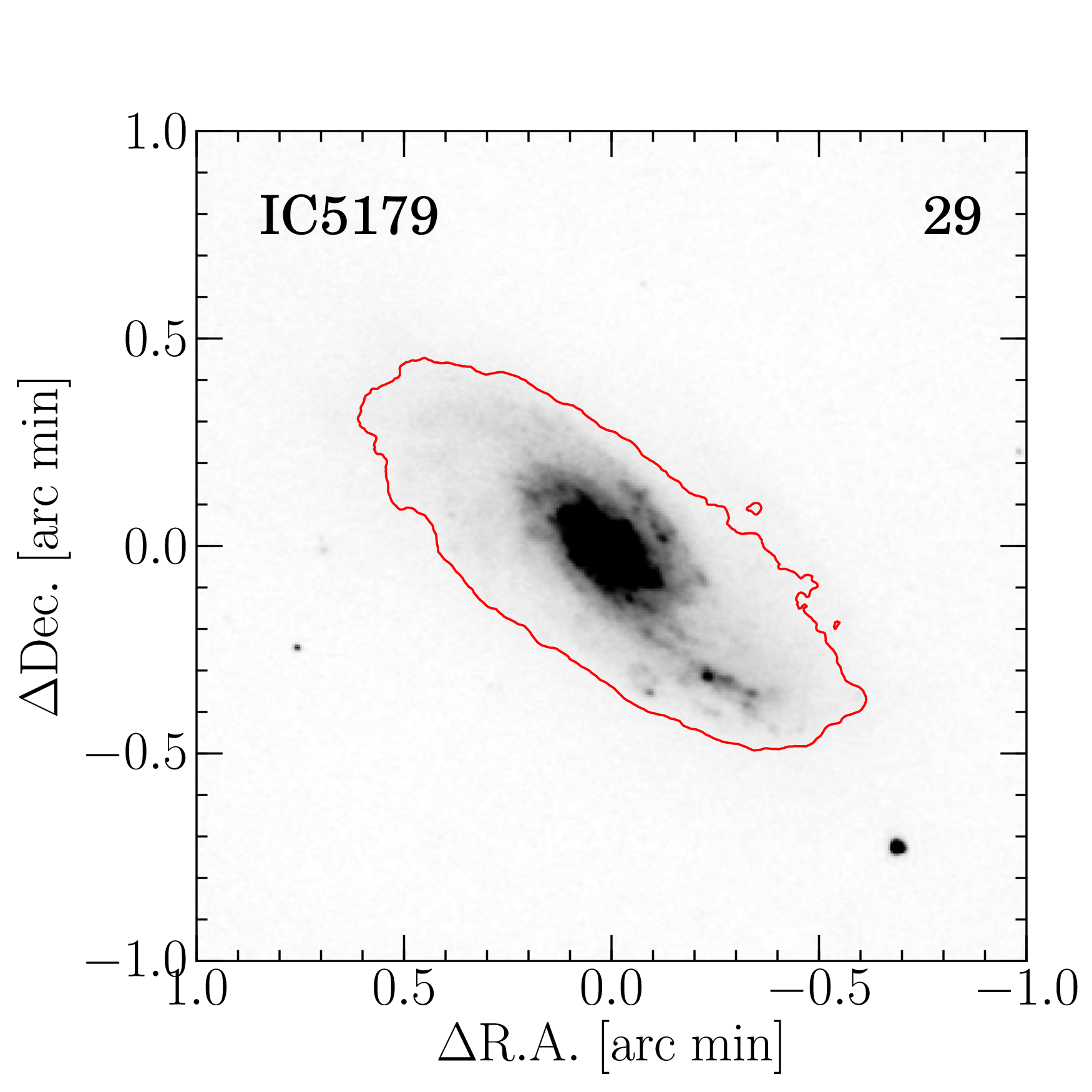}{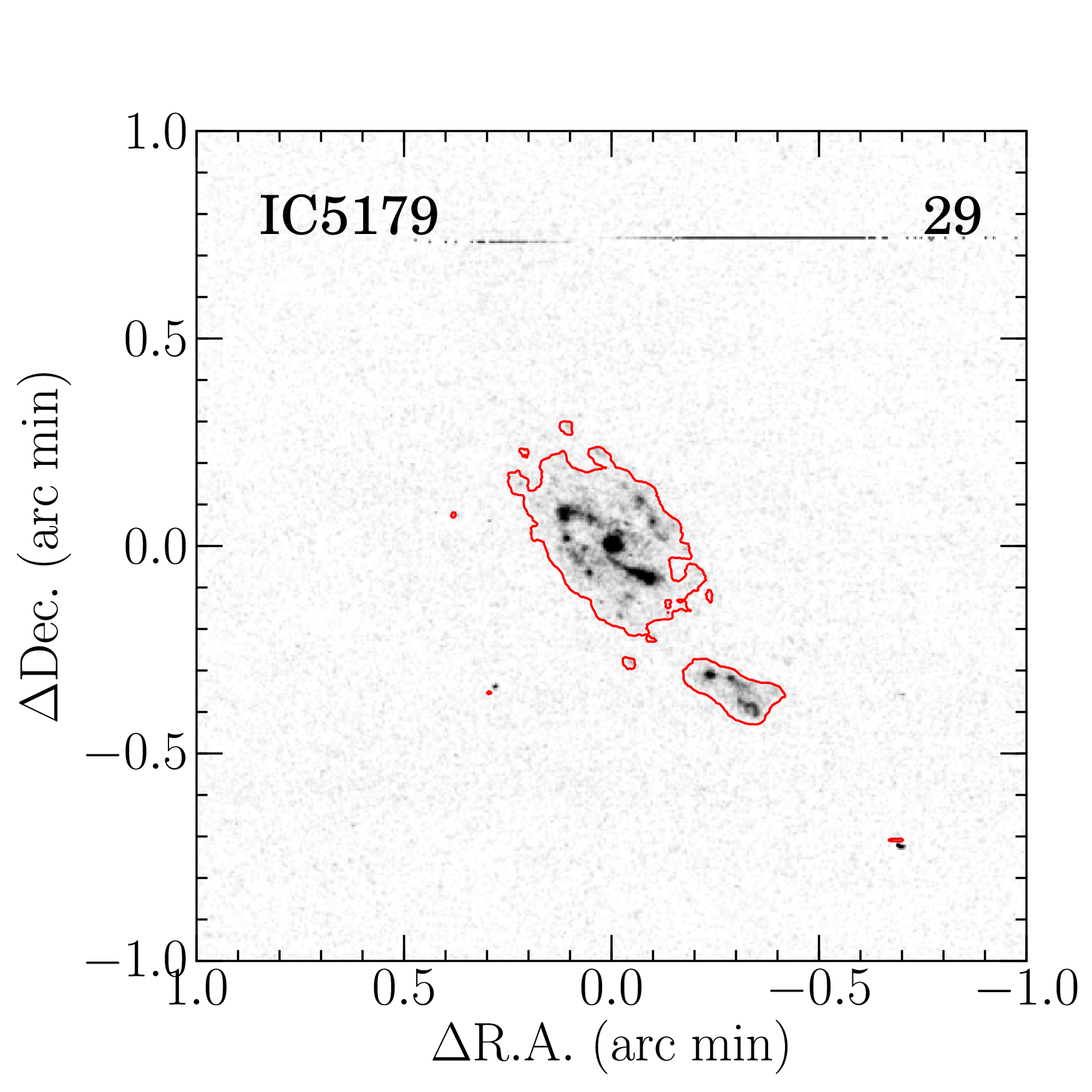}\\
   \plottwo{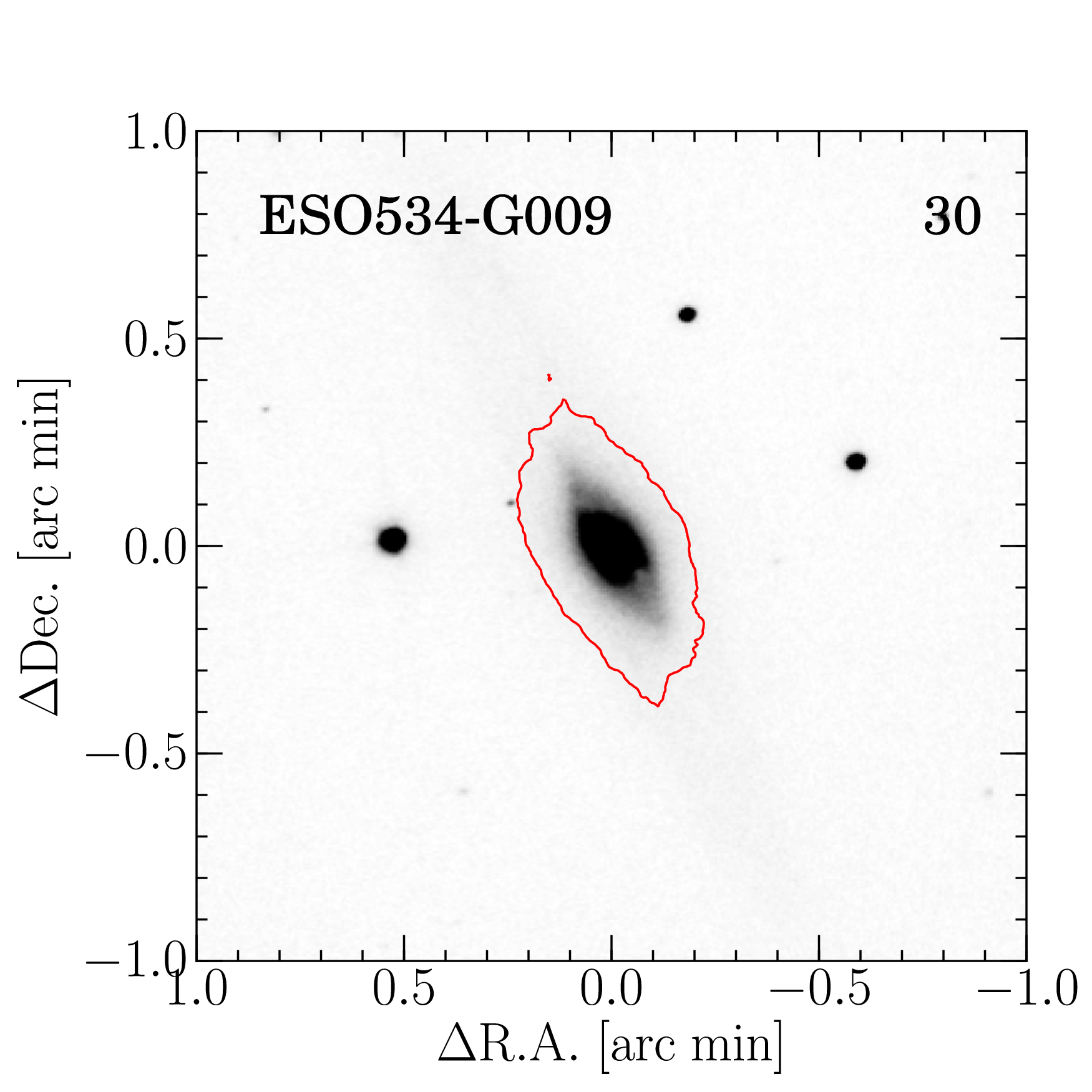}{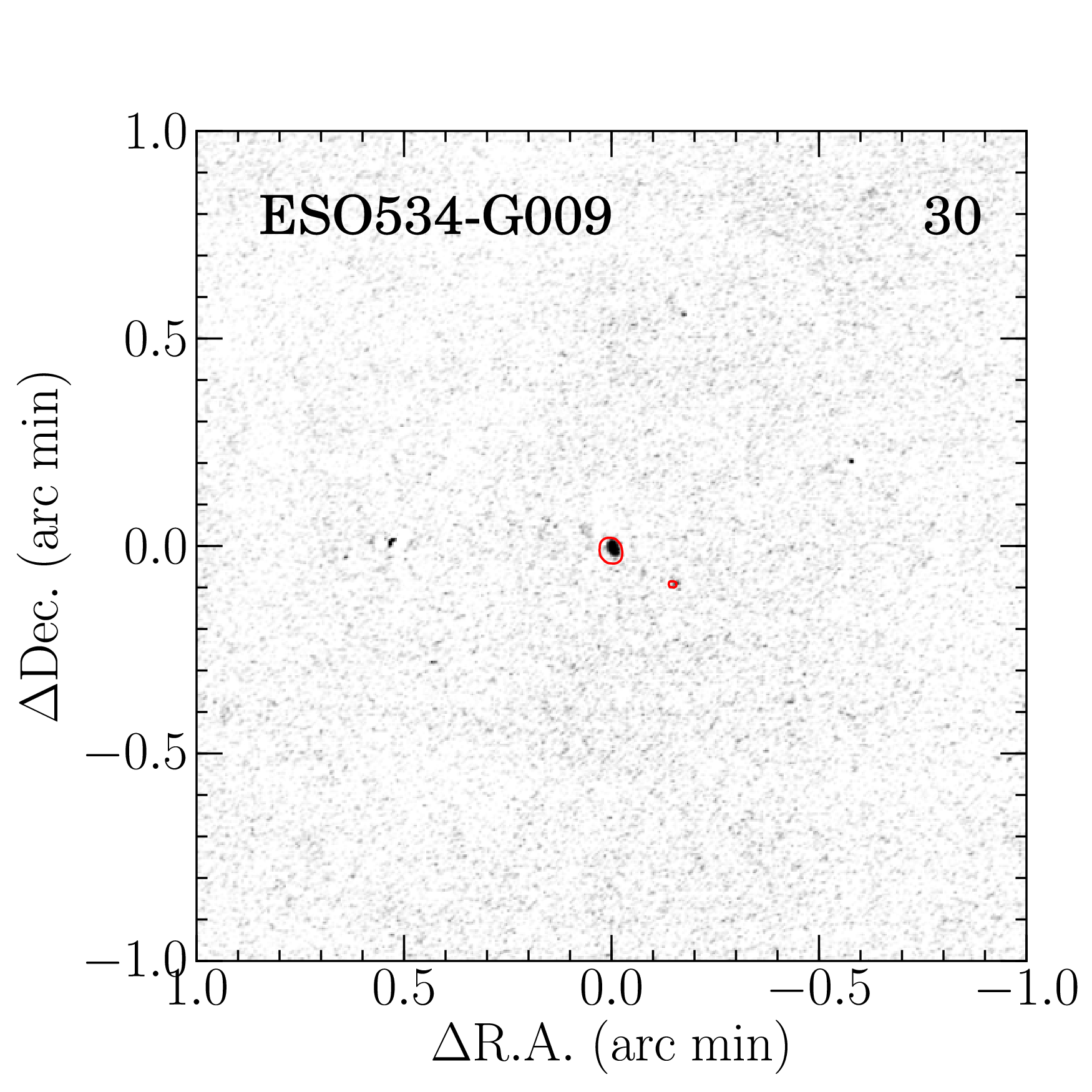}\\
   \end{center}
   \caption{Continued}
 \end{figure}

\twocolumngrid

\indent {\bf 18. NGC 5135 (IRAS F13229$-$2934): }
This galaxy is an isolated system \citep{2010ApJ...709..884Y}, and is an SB(s)ab \citep{1991S&T....82Q.621D} classified as a Seyfert 2 \citep{2003ApJ...583..670C}. In the Pa$\alpha$ image, clumpy blobs are detected at the central region which have been already reported in previous H$\alpha$ and Pa$\alpha$ imaging observations \citep[e.g.,][]{2011A&A...527A..60R,2006ApJ...650..835A}. Also, a mini-spiral structure can be seen in our Pa$\alpha$ image.\\

\indent {\bf 19. NGC 5257/8 (IRAS F13373$+$0105 NW/SE; Arp 240; VV 055; KPG 389):}
These galaxies form an interacting pair with a separation of 1$\farcm$3. NGC 5258 is an SA(s)b; peculiar, and NGC 5257 is an SAB(s)b; peculiar \citep{1991S&T....82Q.621D} both are classified as H{\sc ii} galaxies \citep{2003ApJ...583..670C,1995ApJS...98..171V}. While a tidal tail between these galaxies can be seen in the continuum image, Pa$\alpha$ emission is not detected there, which is consistent with H$\alpha$ imaging observations \citep{2002ApJS..143...47D}. NGC 5257 has Pa$\alpha$ blobs along the
\clearpage
\onecolumngrid

\setcounter{figure}{2}
\begin{figure}[htb]
 \epsscale{0.8}
  \begin{center}
   \plottwo{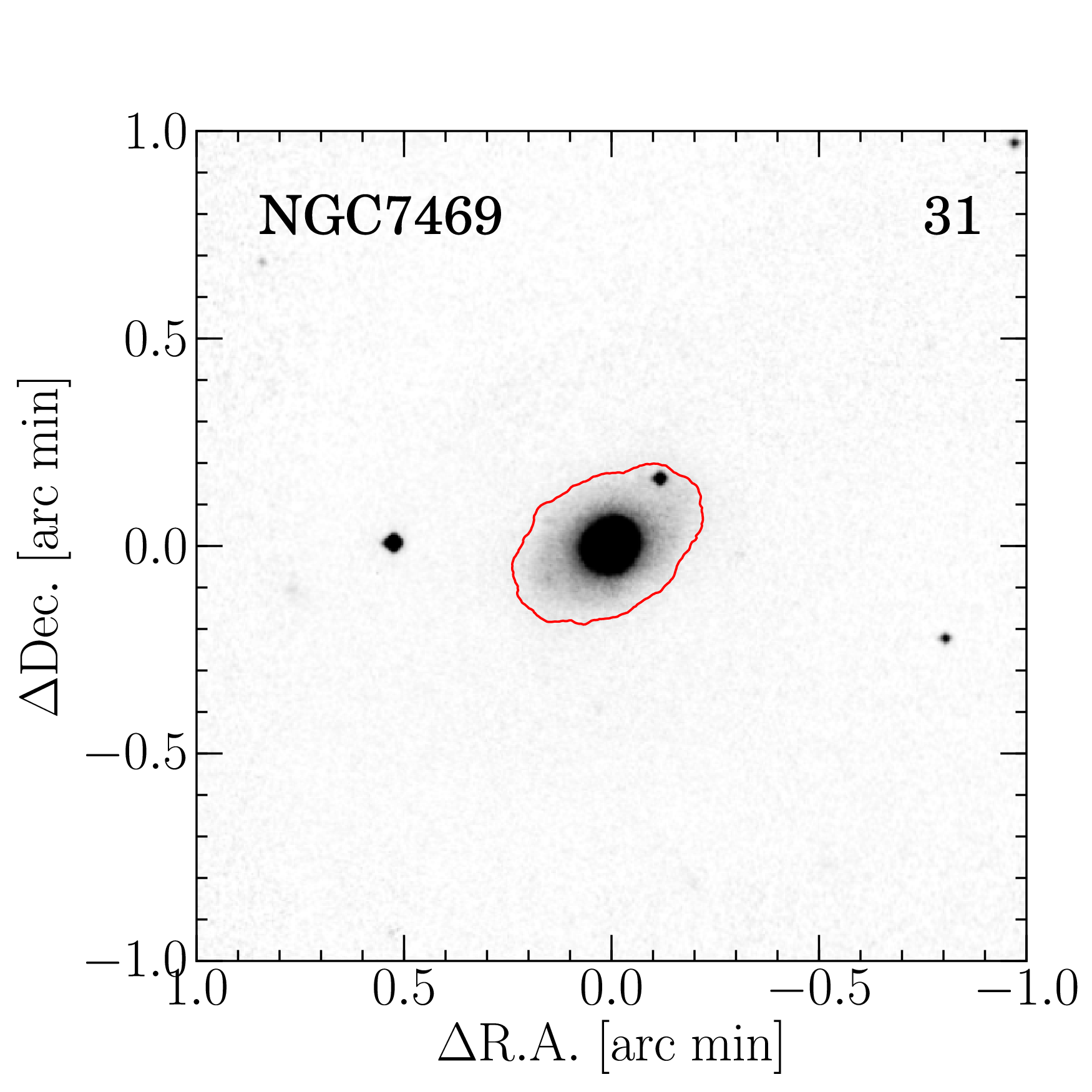}{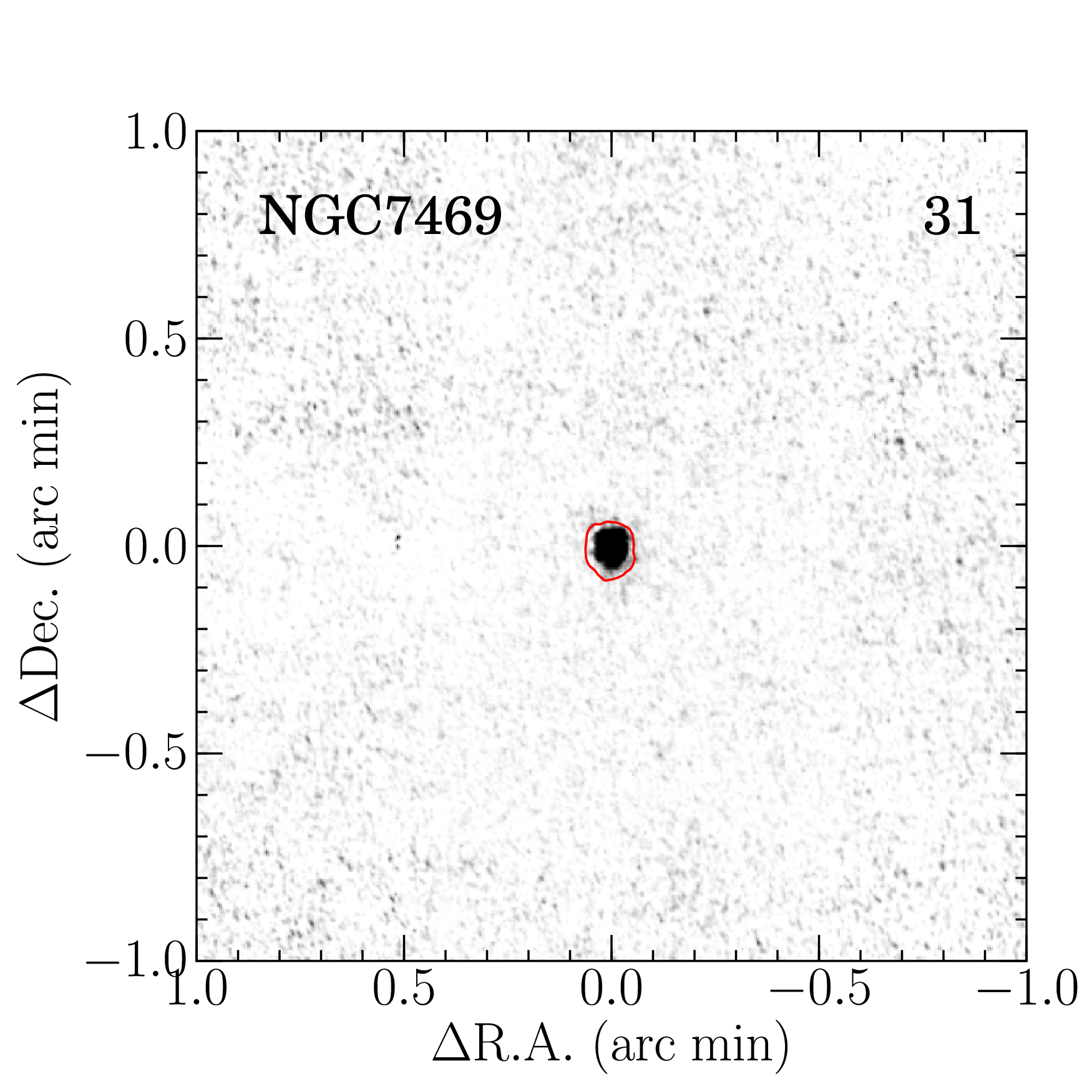}\\
   \plottwo{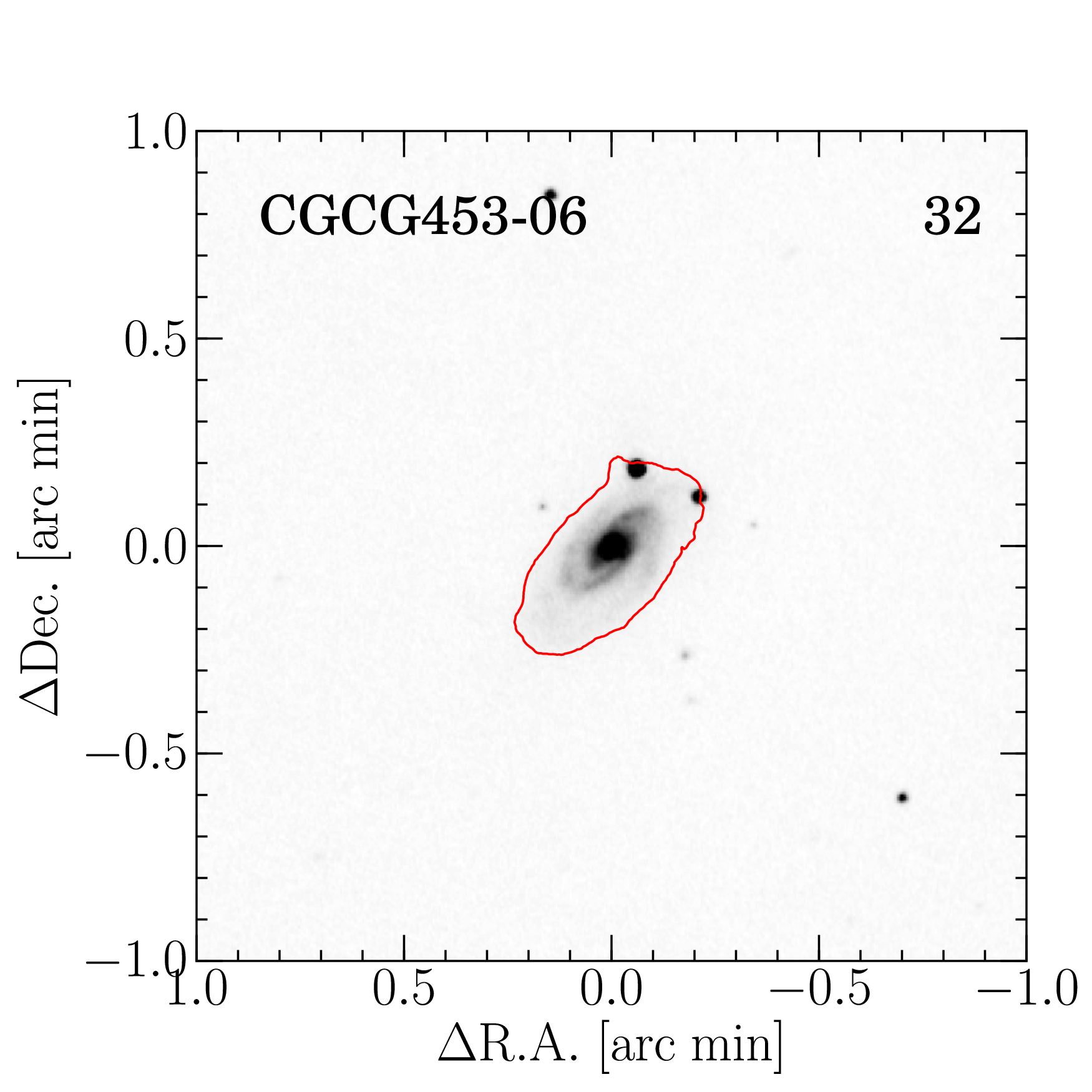}{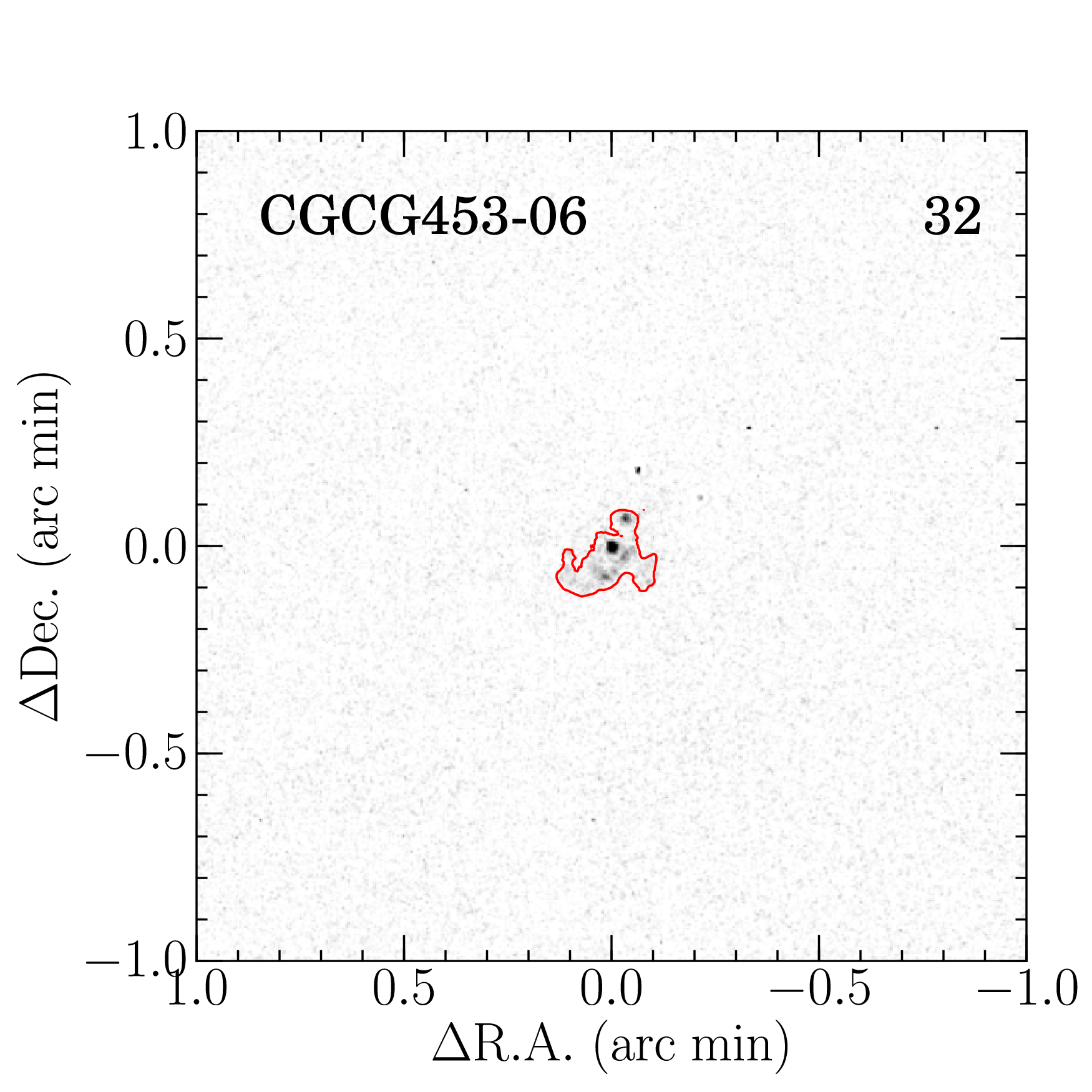}\\
   \plottwo{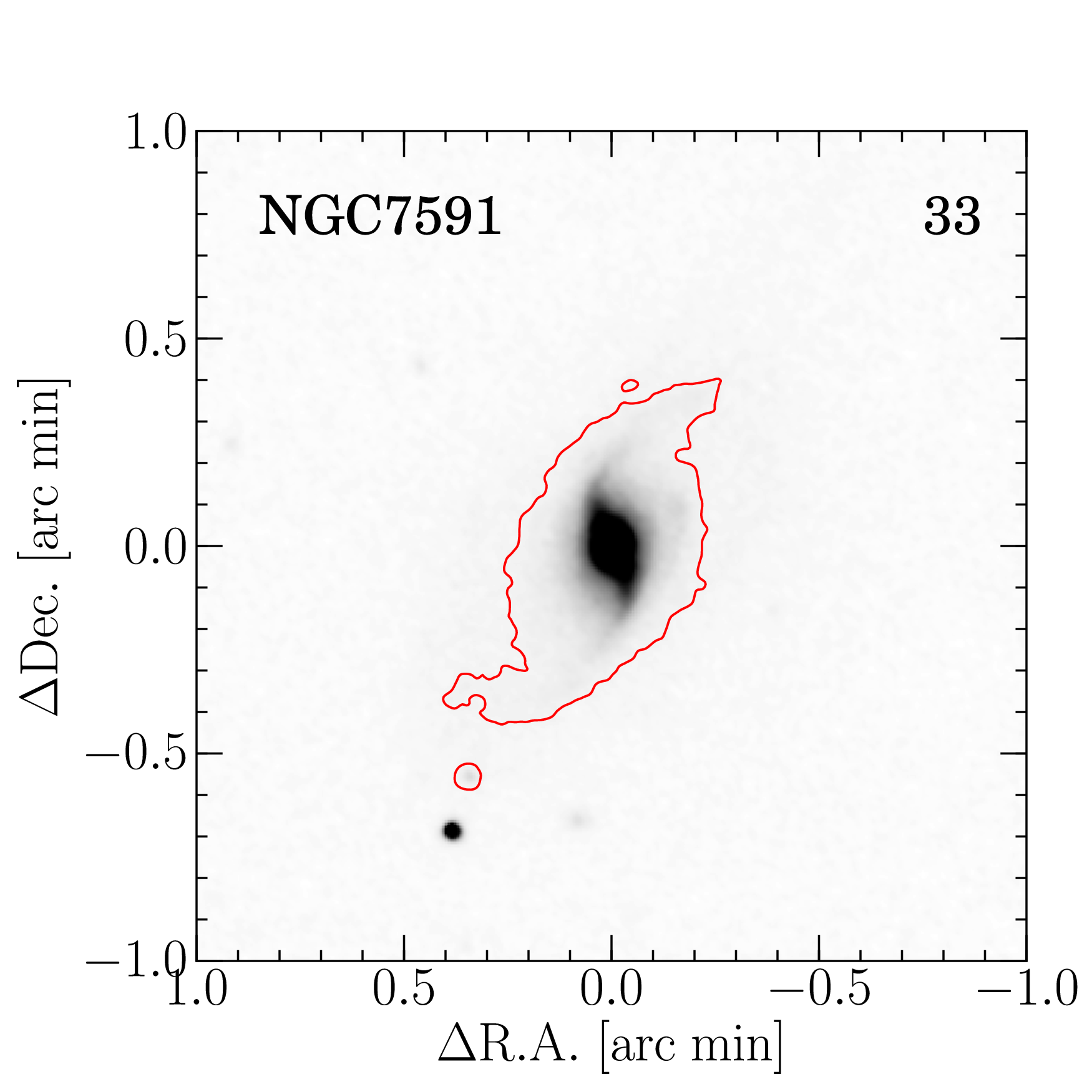}{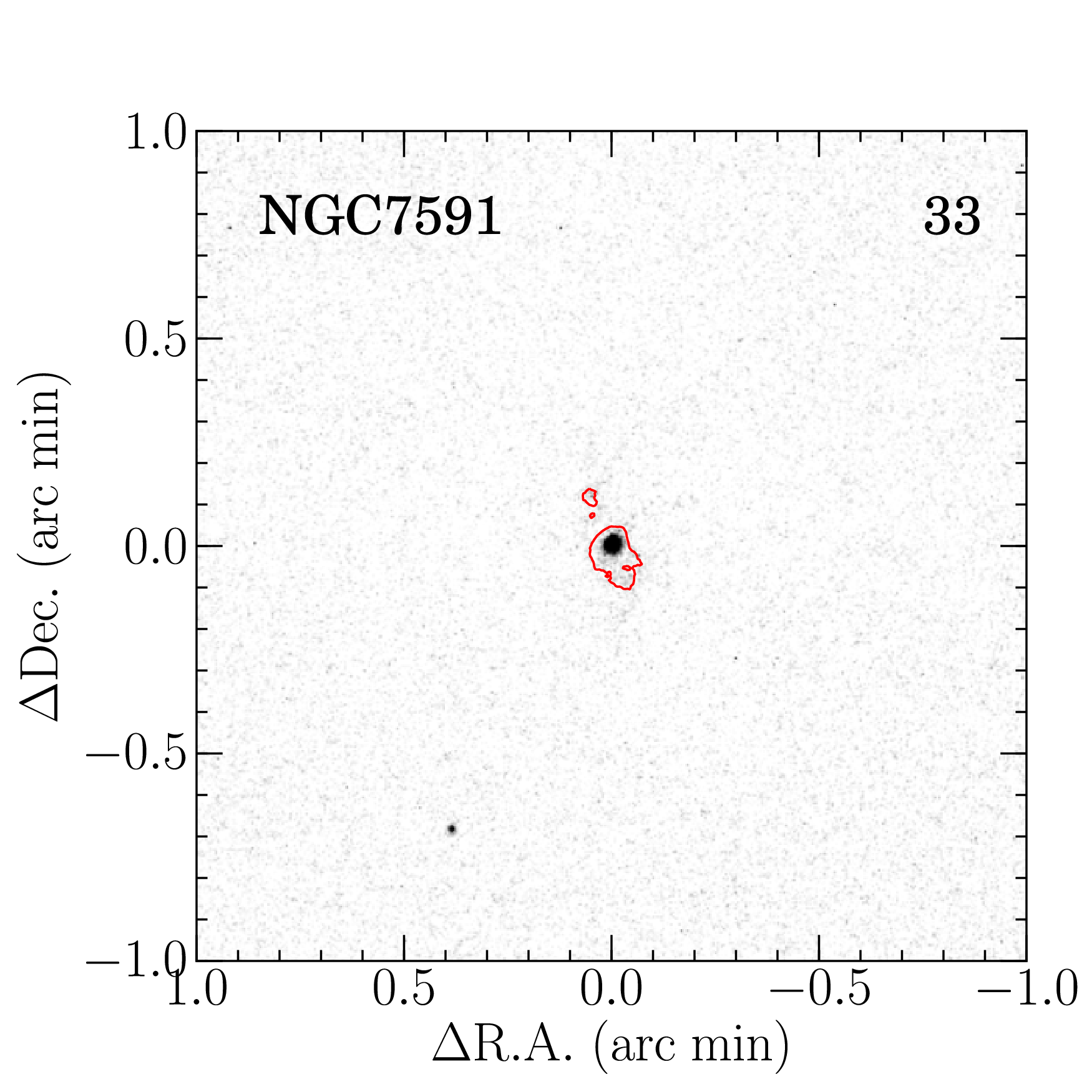}\\
   \end{center}
   \caption{Continued}
 \end{figure}

\twocolumngrid

\noindent spiral arms and the southern arm has stronger and bigger clumps of emission. NGC 5258 also has Pa$\alpha$ blobs along spiral arms, with the southern arm having stronger Pa$\alpha$ emission where larger amount of CO($1-0$) emission is detected \citep{2005ApJS..158....1I}.\\

\indent {\bf 20. IC 4518A/B (IRAS F14544$-$4255; VV 780; AM 1454$-$425):}
This is a strongly interacting pair of galaxies with a separation of 37$\farcs$6. Both are spirals (Sc; HyperLeda) classified as Seyfert 2 \citep{2003ApJ...583..670C}. Although tidal tails exist not only between these galaxies but also at spiral arms of each galaxy as can be seen in the continuum image, the Pa$\alpha$ emission is not detected there, being consistent with H$\alpha$ imaging observations \citep{2002ApJS..143...47D}. While the western galaxy has a strong and concentrated Pa$\alpha$ emission at the central region, the eastern galaxy has extended Pa$\alpha$ emission not only at its center but also along its spiral arms.\\

\clearpage
\onecolumngrid

\setcounter{figure}{2}
\begin{figure}[htb]
 \epsscale{0.8}
  \begin{center}
   \plottwo{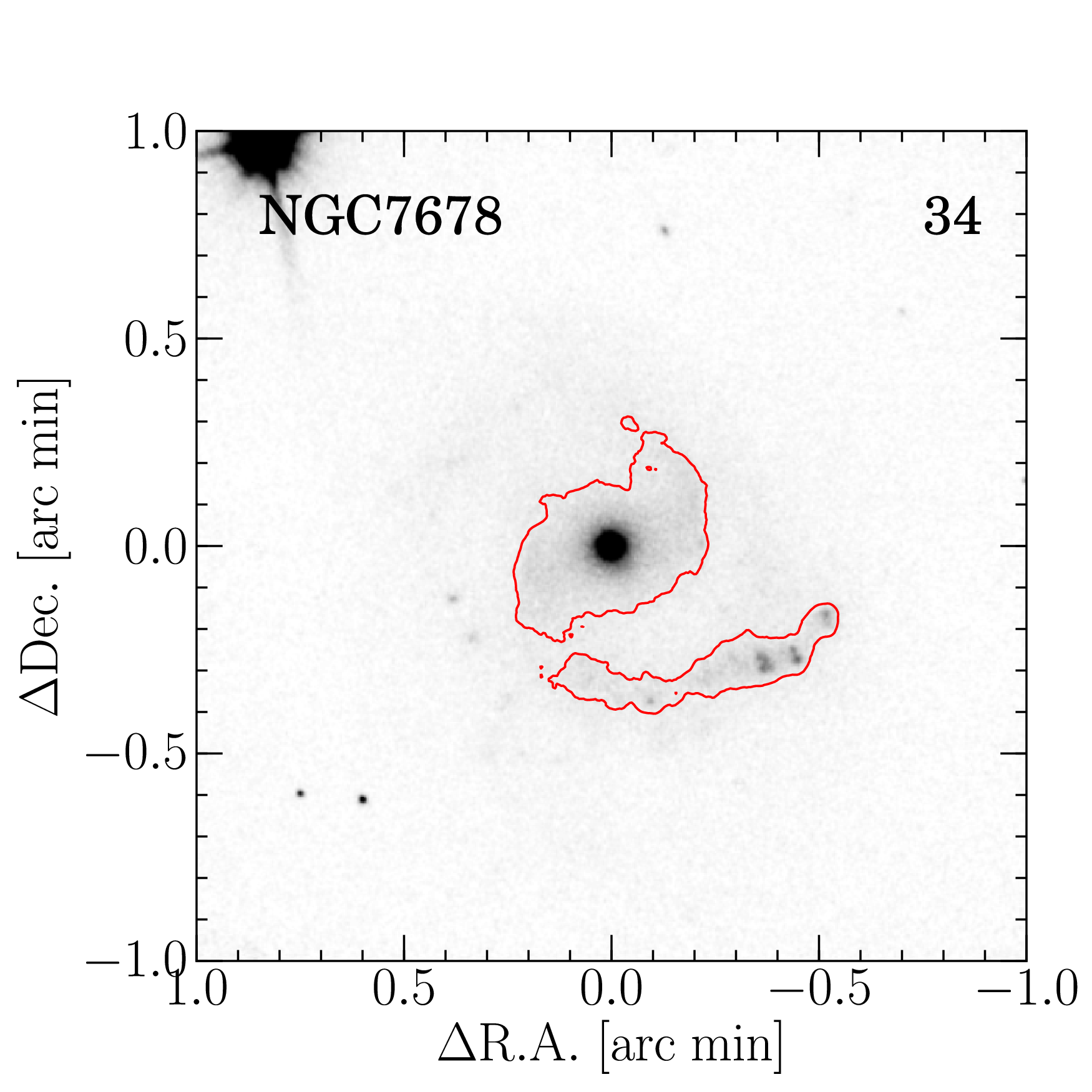}{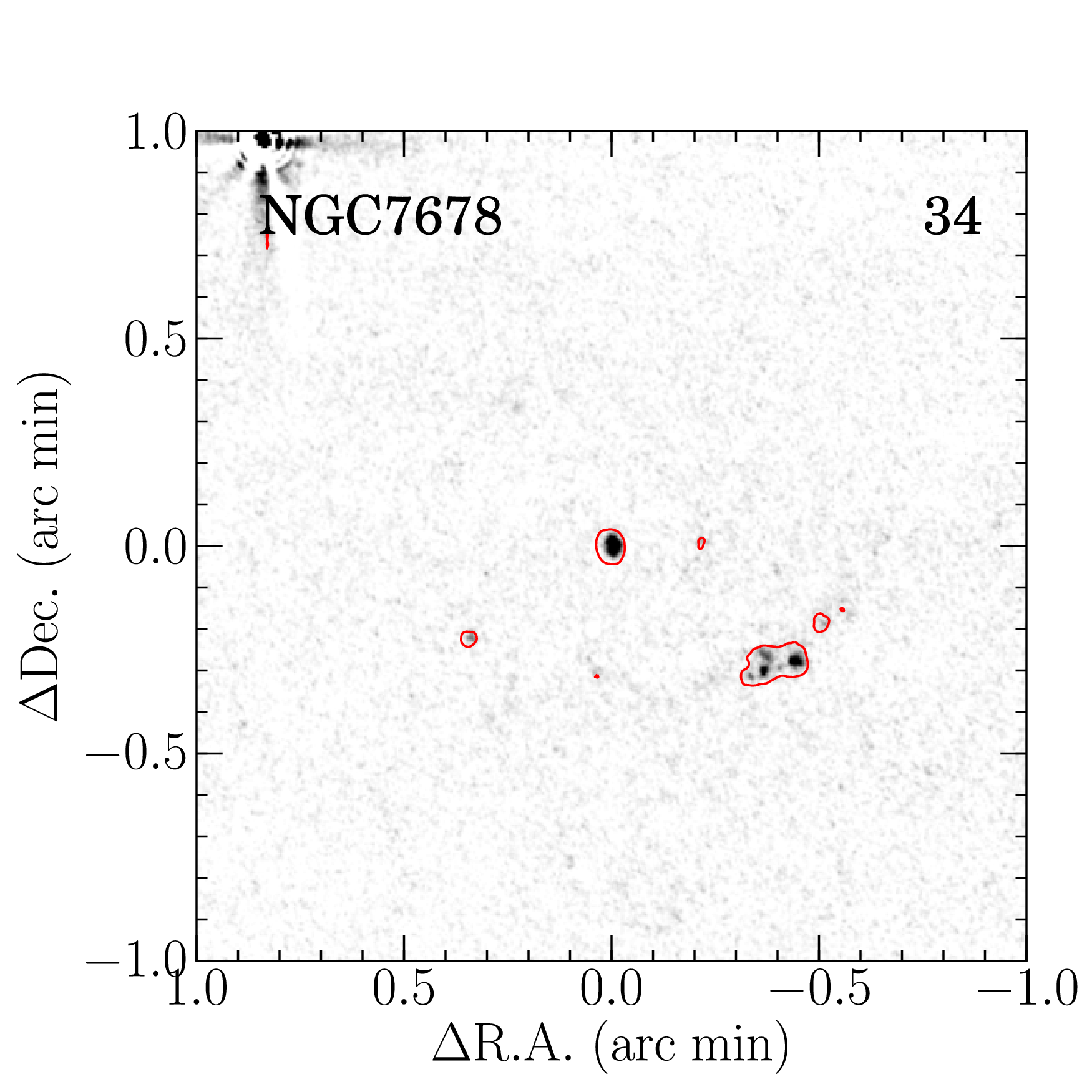}\\
   \plottwo{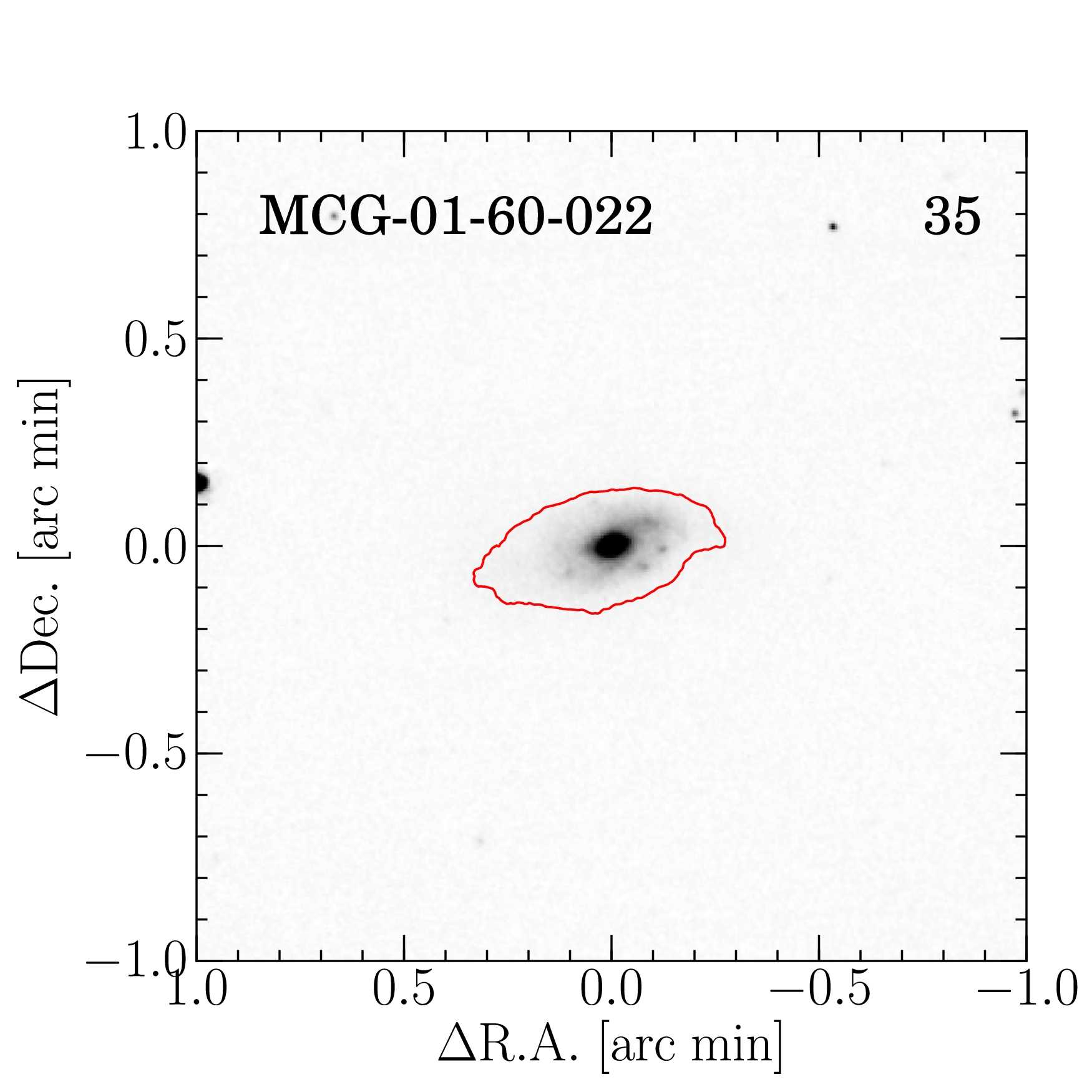}{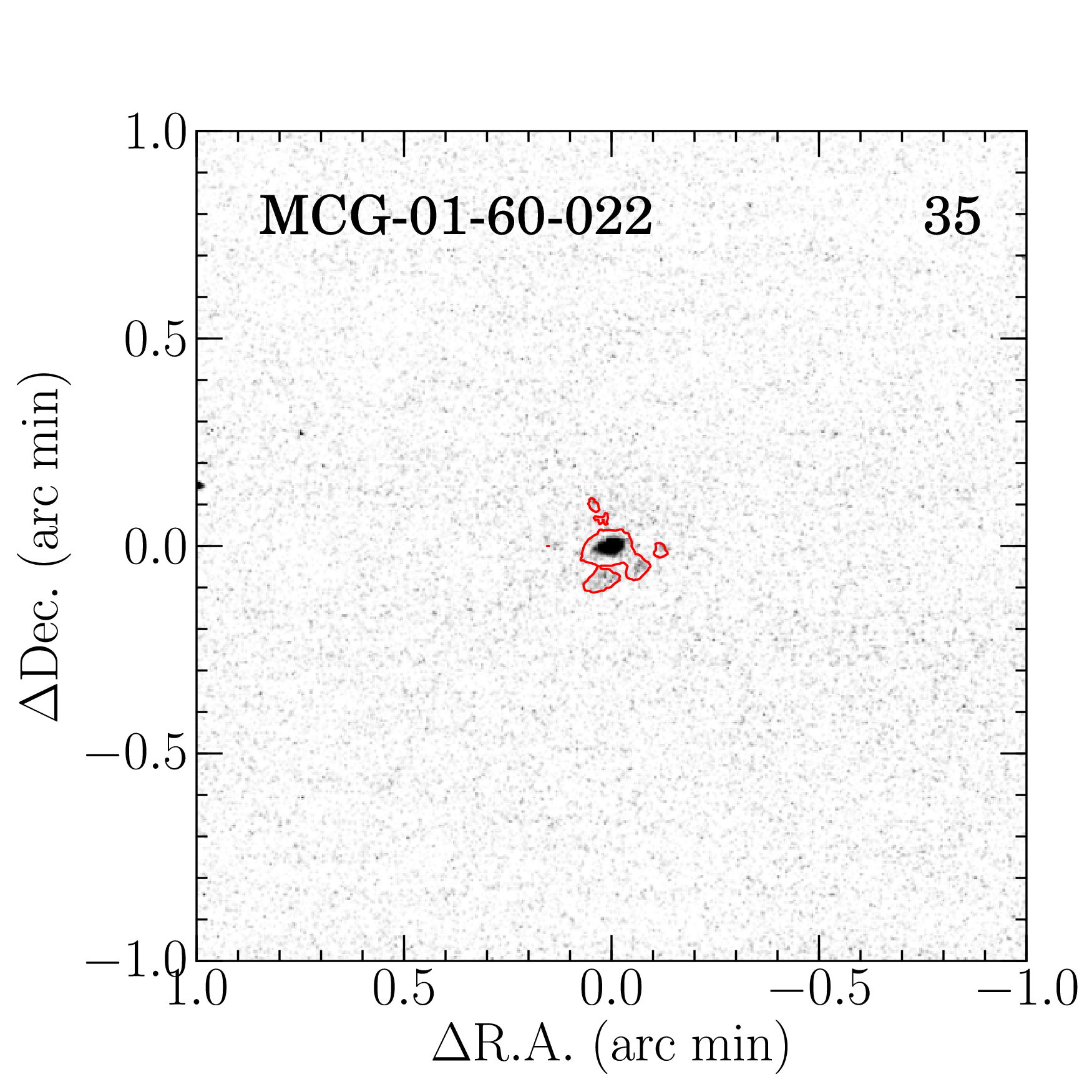}\\
   \plottwo{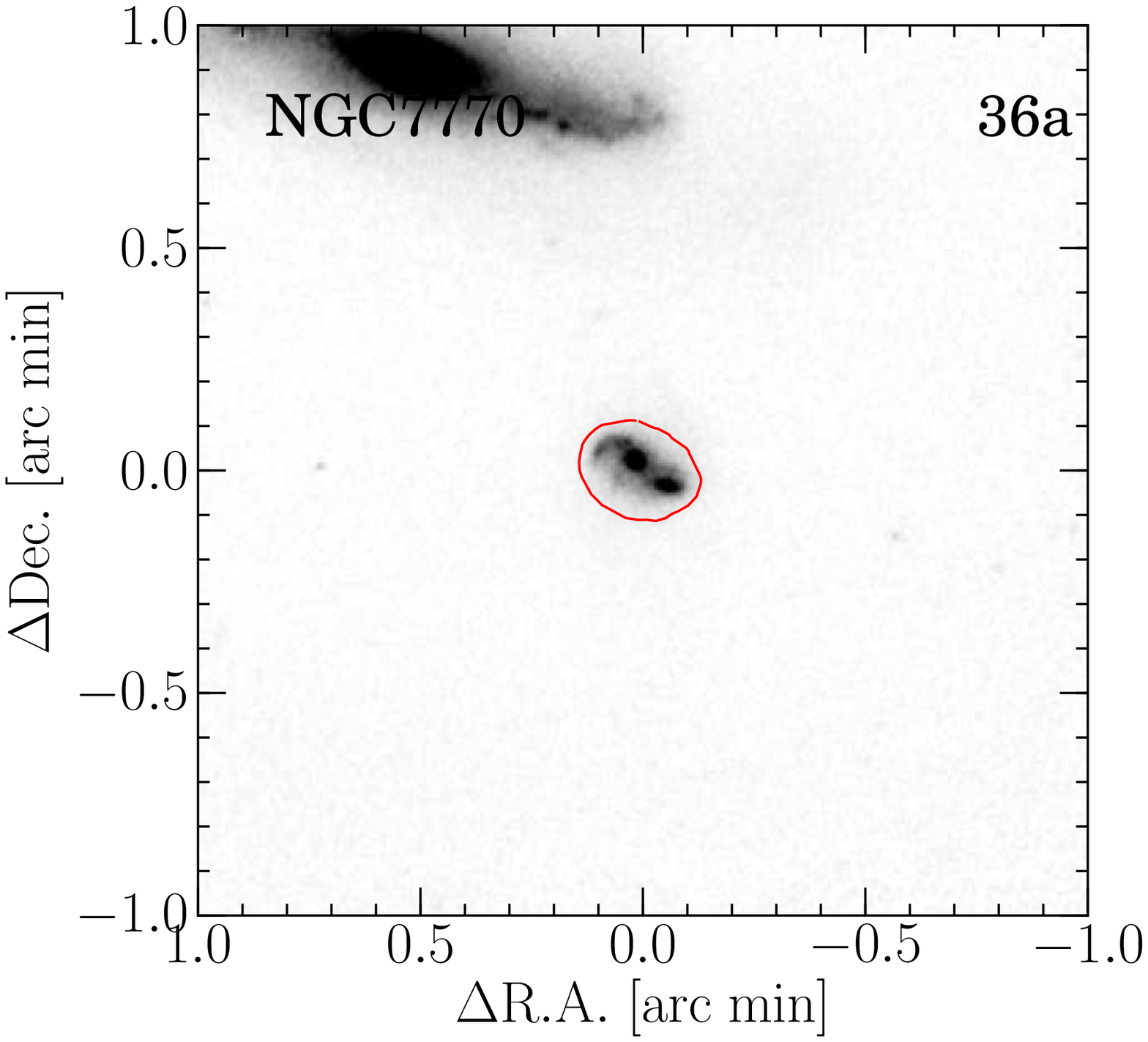}{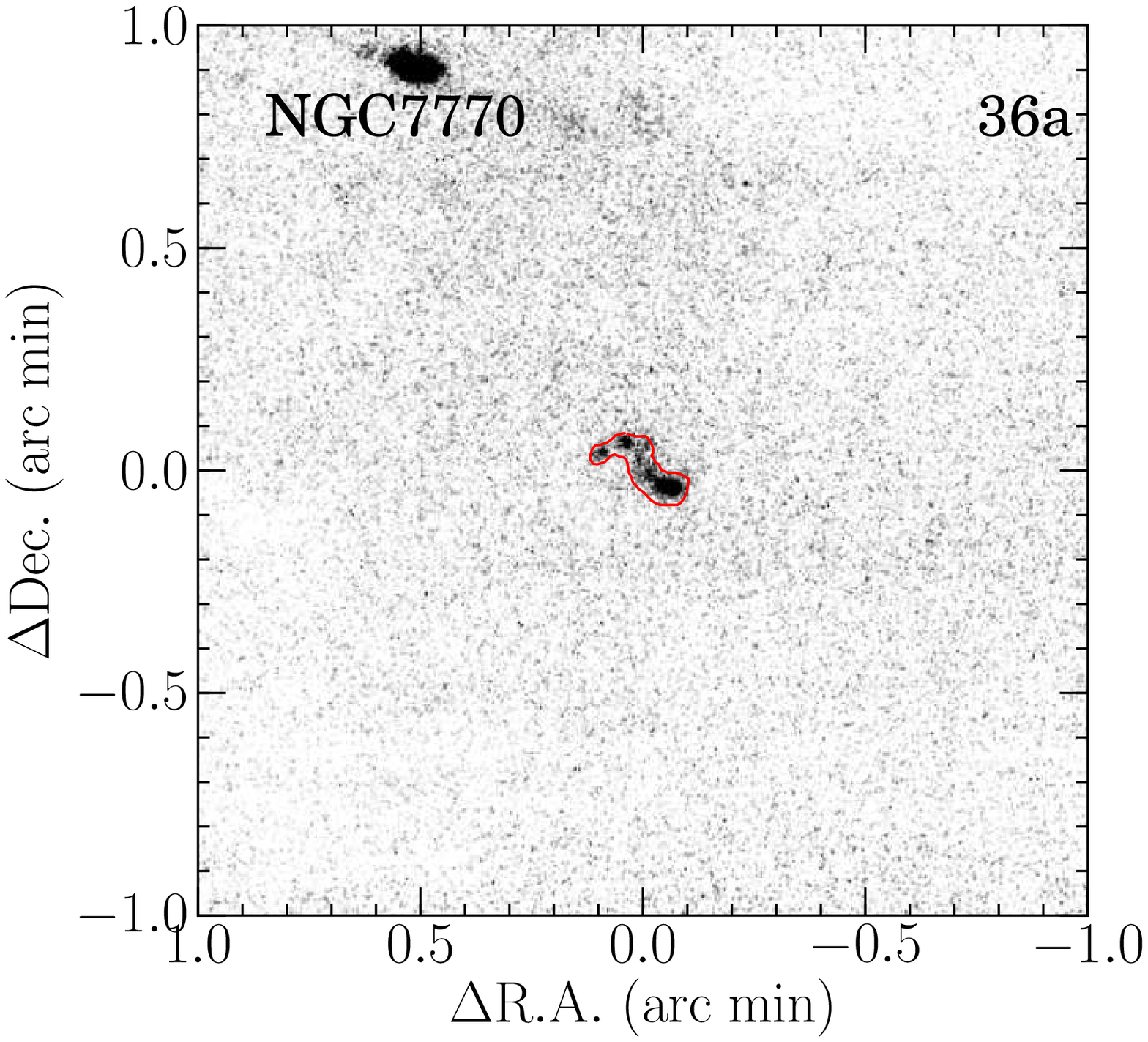}\\
   \end{center}
   \caption{Continued}
 \end{figure}

\twocolumngrid

\indent {\bf 21. IC 4687/6 (IRAS F18093$-$5744; AM 1809$-$574):}
This is a system of three interacting galaxies (IC 4687; IC 4686; IC 4689) \citep{1976A&A....46..327W}. The separation between the northern (IC 4687) and center (IC 4686) galaxy is 27$\farcs$8, and between the center and the southern (IC 4689) galaxy is 56$\farcs$8. The northern galaxy is a barred spiral (SABb; HyperLeda), the center is an elliptical (E; HyperLeda), and the southern is a barred spiral (SABa; HyperLeda), all classified as an H{\sc ii} galaxies \citep{2010ApJ...709..884Y,1995ApJS...98..171V}. These galaxies show strong Pa$\alpha$ emission; the northern galaxy has disturbed blobs like a ring starburst, which is consistent with a Pa$\alpha$ image of $HST$/NICMOS \citep{2006ApJ...650..835A}, the central galaxy has concentrated Pa$\alpha$ emission at the center, and the southern galaxy has extended emission regions along the spiral arms. We can not see the Pa$\alpha$ emission between these galaxies, which is consistent with H$\alpha$ imaging observations \citep{2011A&A...527A..60R,2002ApJS..143...47D}.

\clearpage
\onecolumngrid

\setcounter{figure}{2}
\begin{figure}[htb]
 \epsscale{0.8}
  \begin{center}
   \plottwo{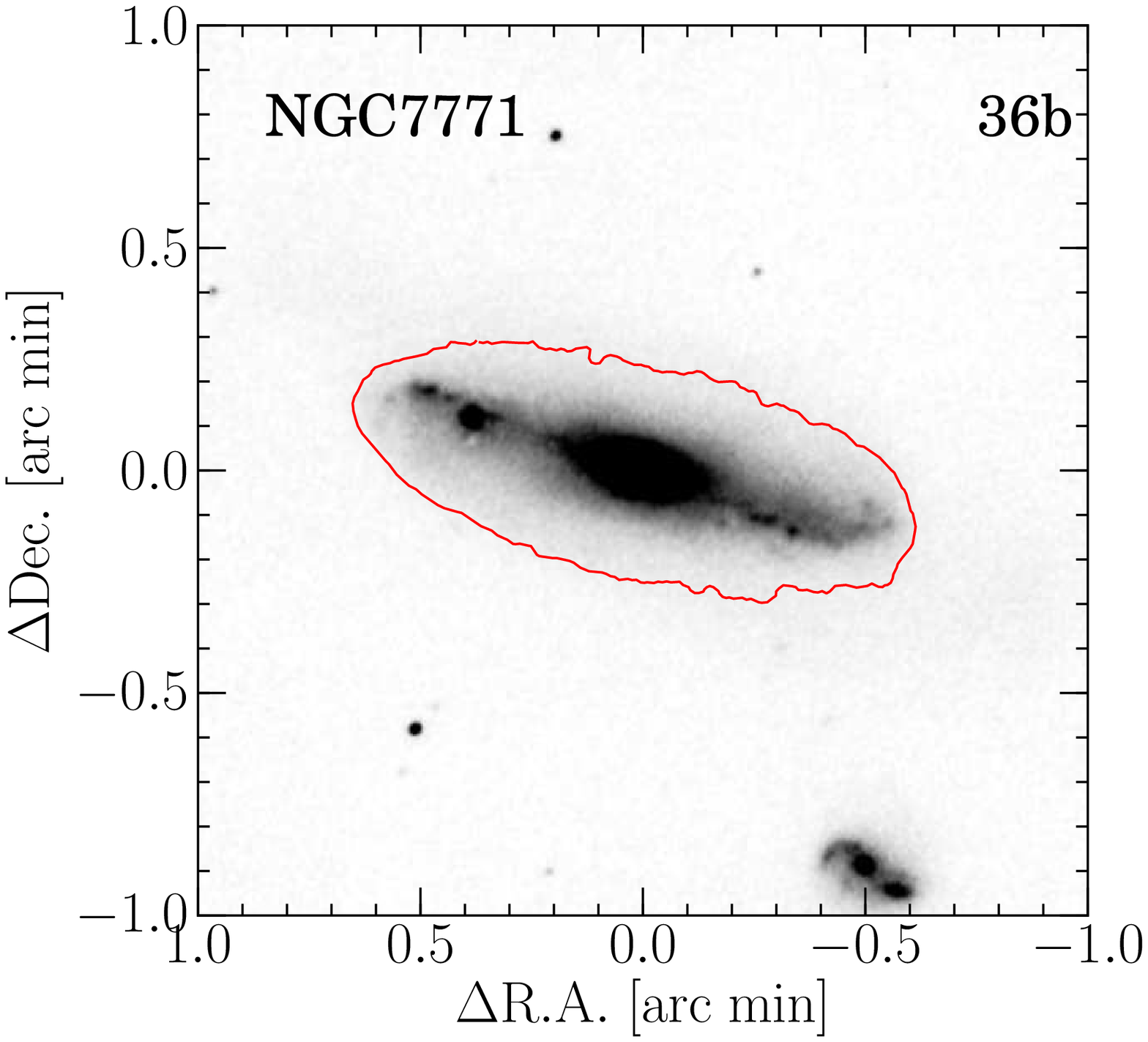}{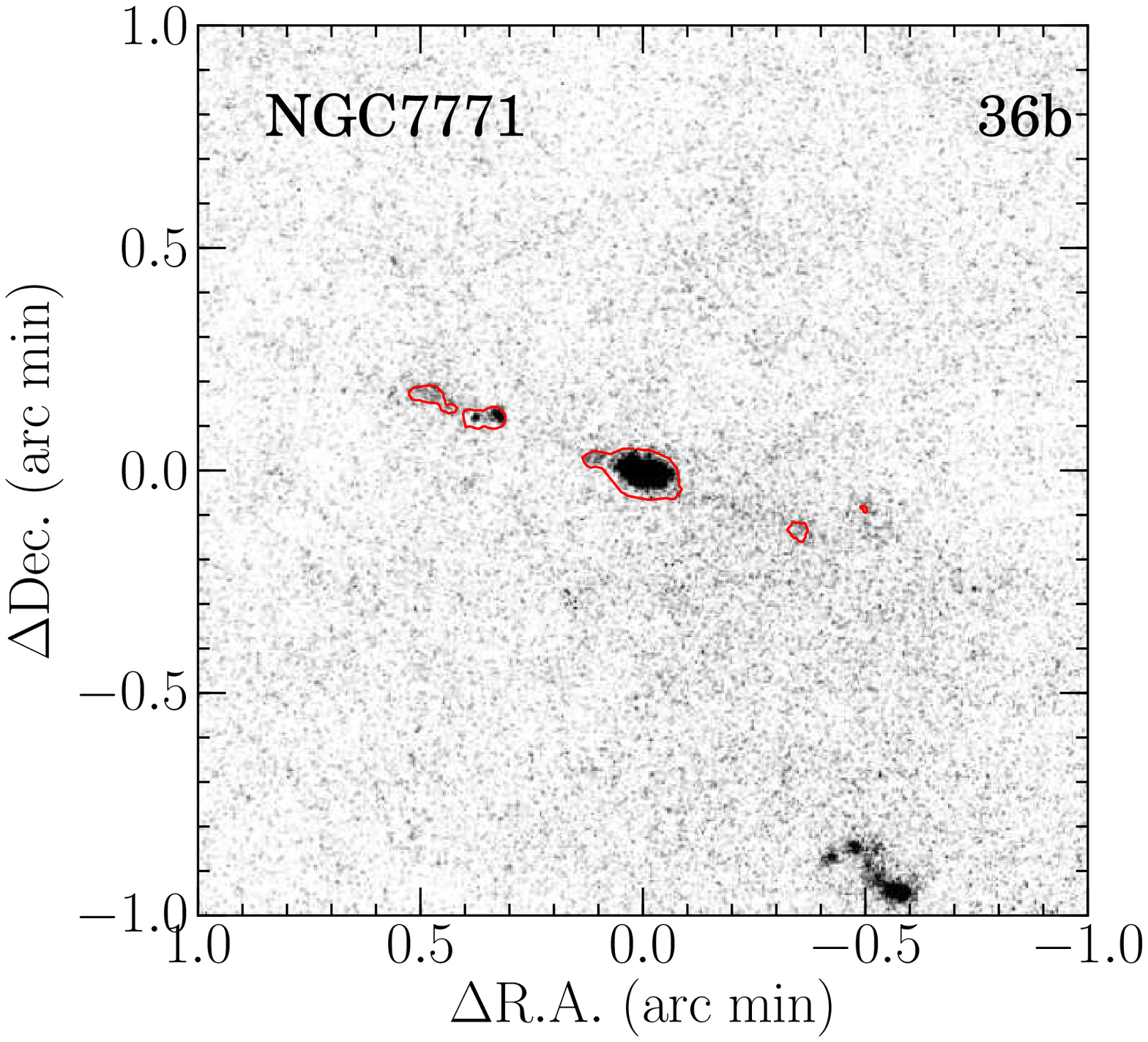}\\
   \plottwo{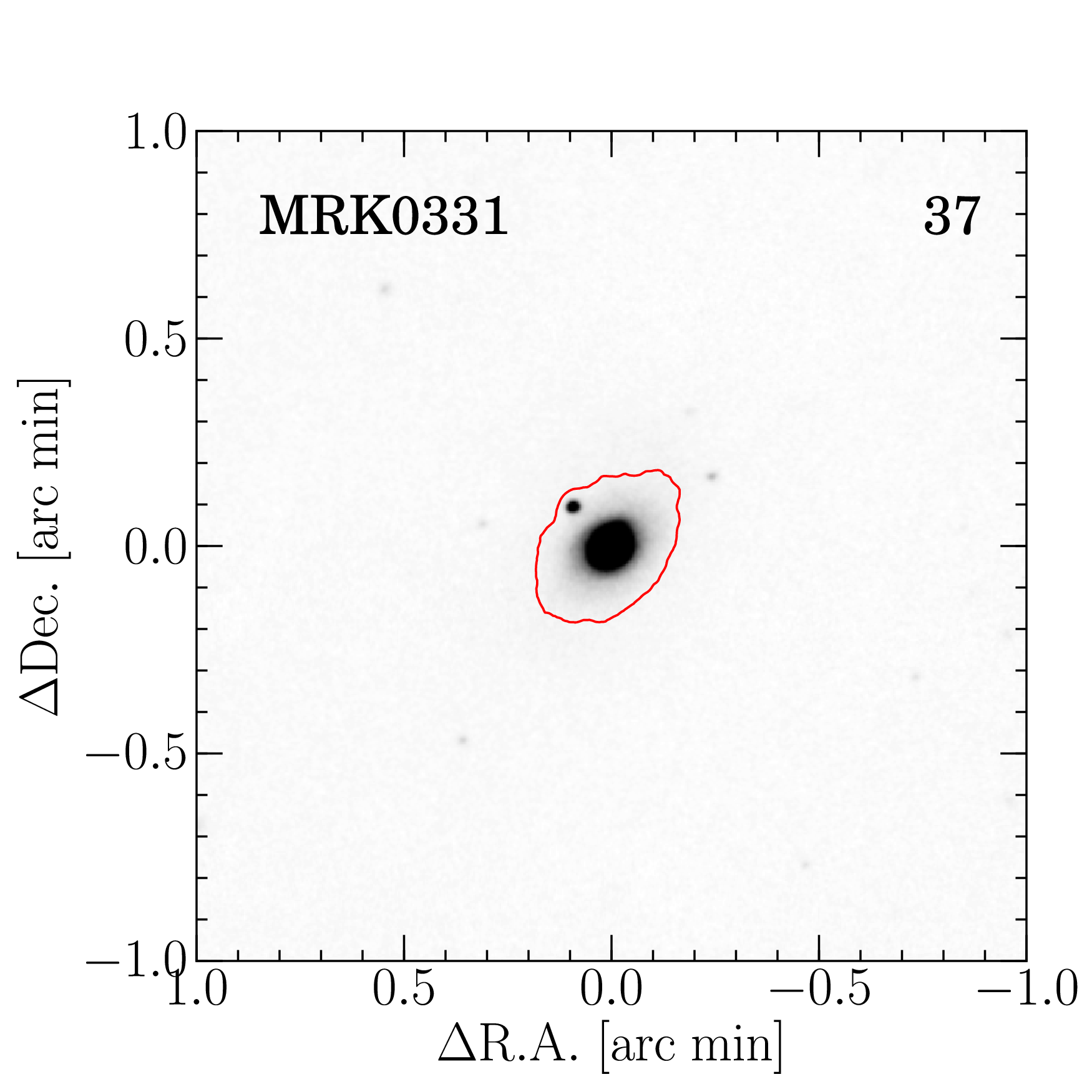}{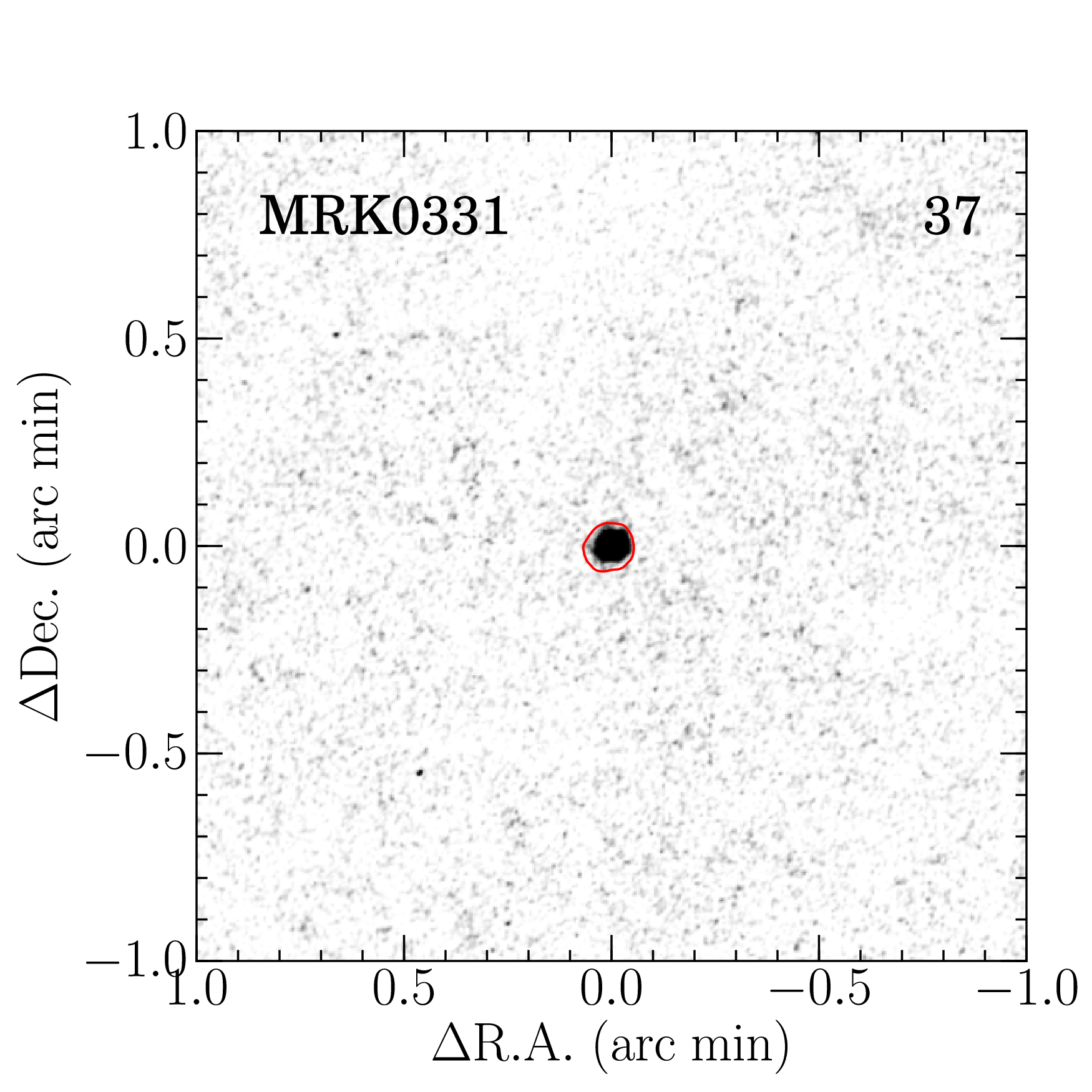}\\
   \plottwo{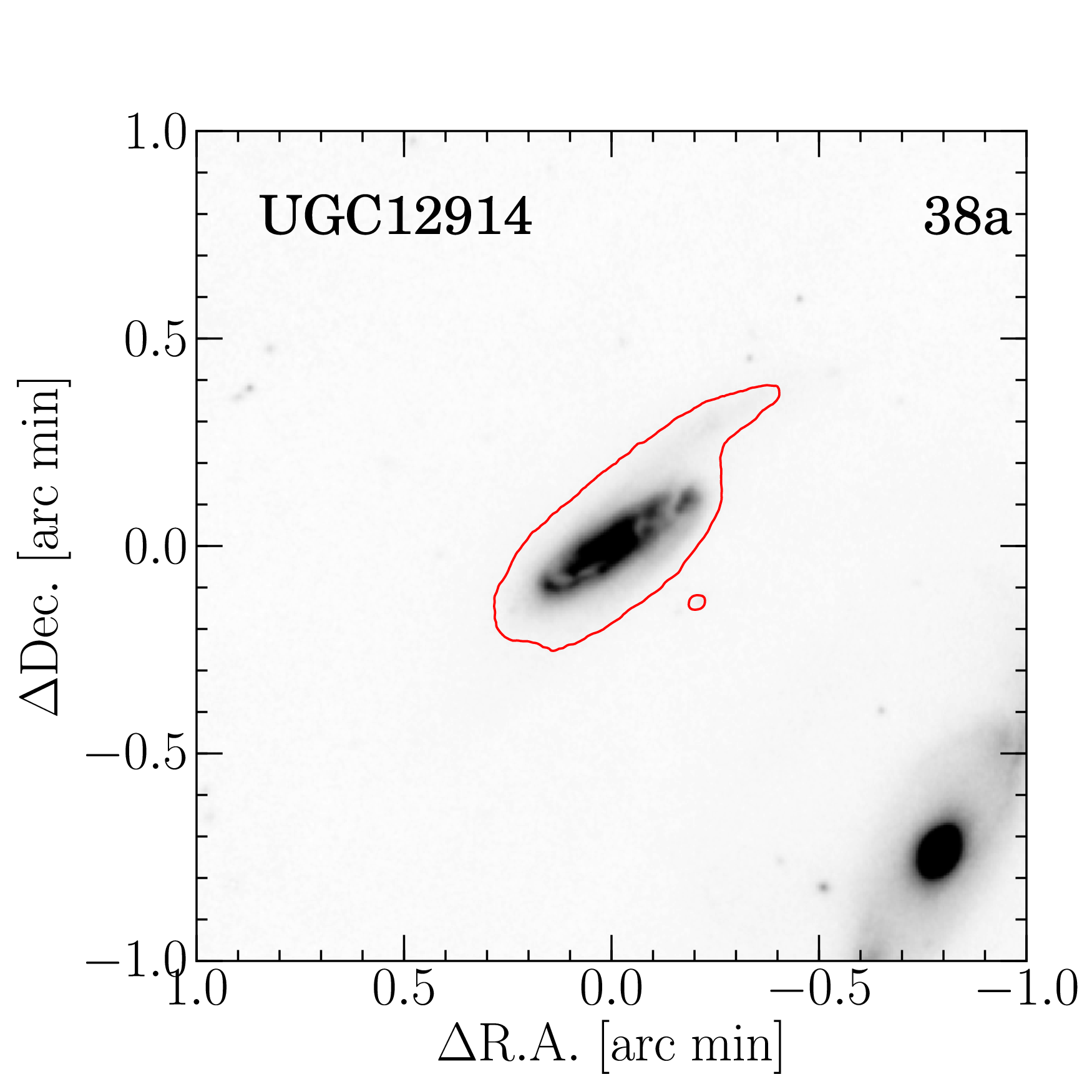}{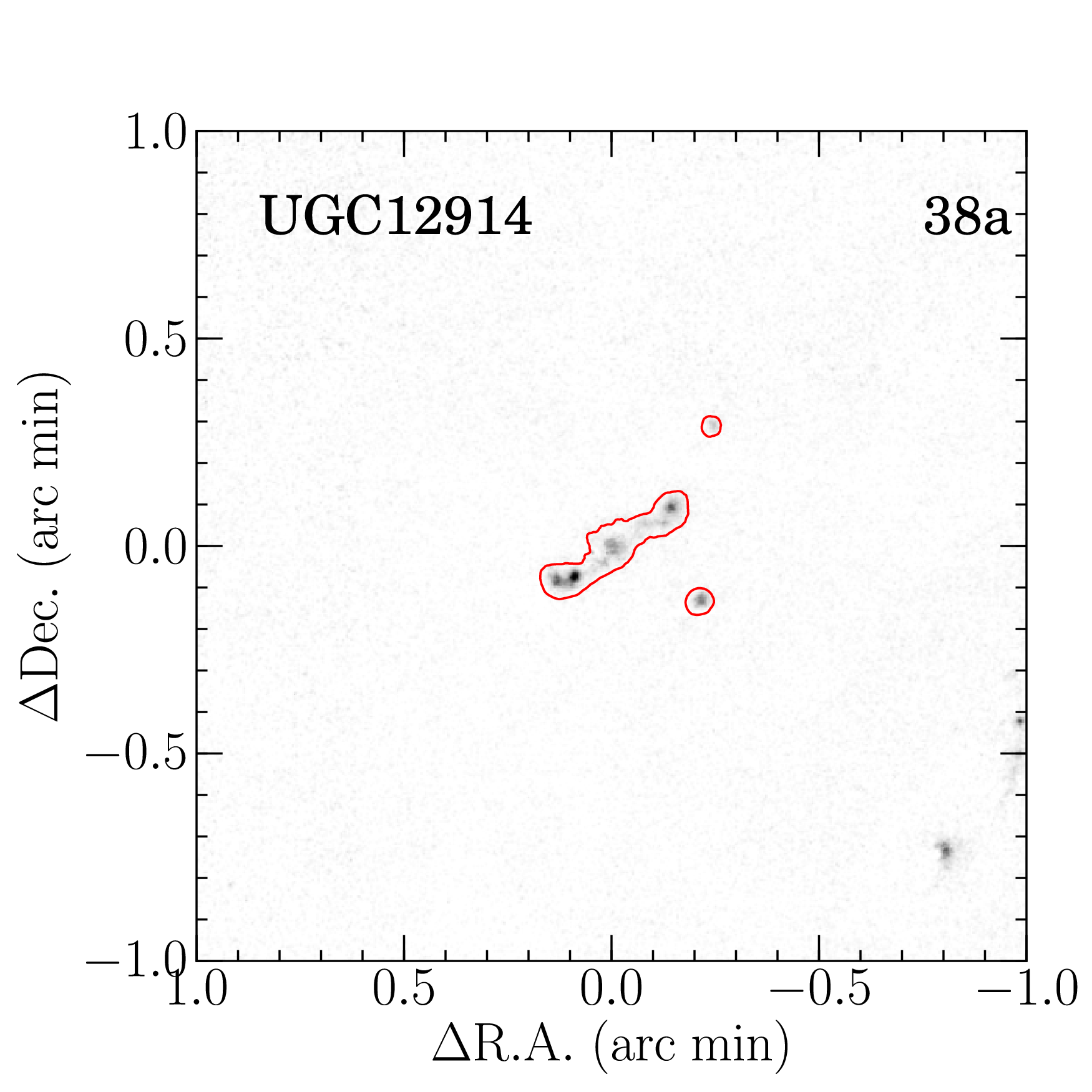}\\
   \end{center}
   \caption{Continued}
 \end{figure}

\twocolumngrid

\indent {\bf 22. IRAS F18293$-$3413:}
This galaxy is an S0/a (HyperLeda) classified as an H{\sc ii} galaxy \citep{1995ApJS...98..171V}. It shows a concentrated Pa$\alpha$ emission region at the center and diffuse region along the disk.\\

\indent {\bf 23. ESO 339-G011 (IRAS F19542$-$3804):}
This is an isolated barred galaxy (SBb, HyperLeda) with a tidal tail \citep{2010ApJ...709..884Y} classified as a Seyfert 2 \citep{2010ApJ...709..884Y}. It has disturbed Pa$\alpha$ emission at the center and some blobs can be detected along the disk. It seems to have a companion with a separation of 14\farcs3 with a tidal tail. The Pa$\alpha$ emission region possibly were induced by merger, but the emission can not be detected in the companion.\\

\indent {\bf 24. NGC 6926 (IRAS F20305$-$0211; VV 621):}
This is a barred galaxy (Sc; HyperLeda) and has a companion of dwarf elliptical (NGC 6929) at a distance of $\sim$ 4$'$ towards east. The starburst activity of this galaxy has presumably been triggered by a M51-type density wave
\clearpage
\onecolumngrid

\setcounter{figure}{2}
\begin{figure}[htb]
 \epsscale{0.8}
  \begin{center}
   \plottwo{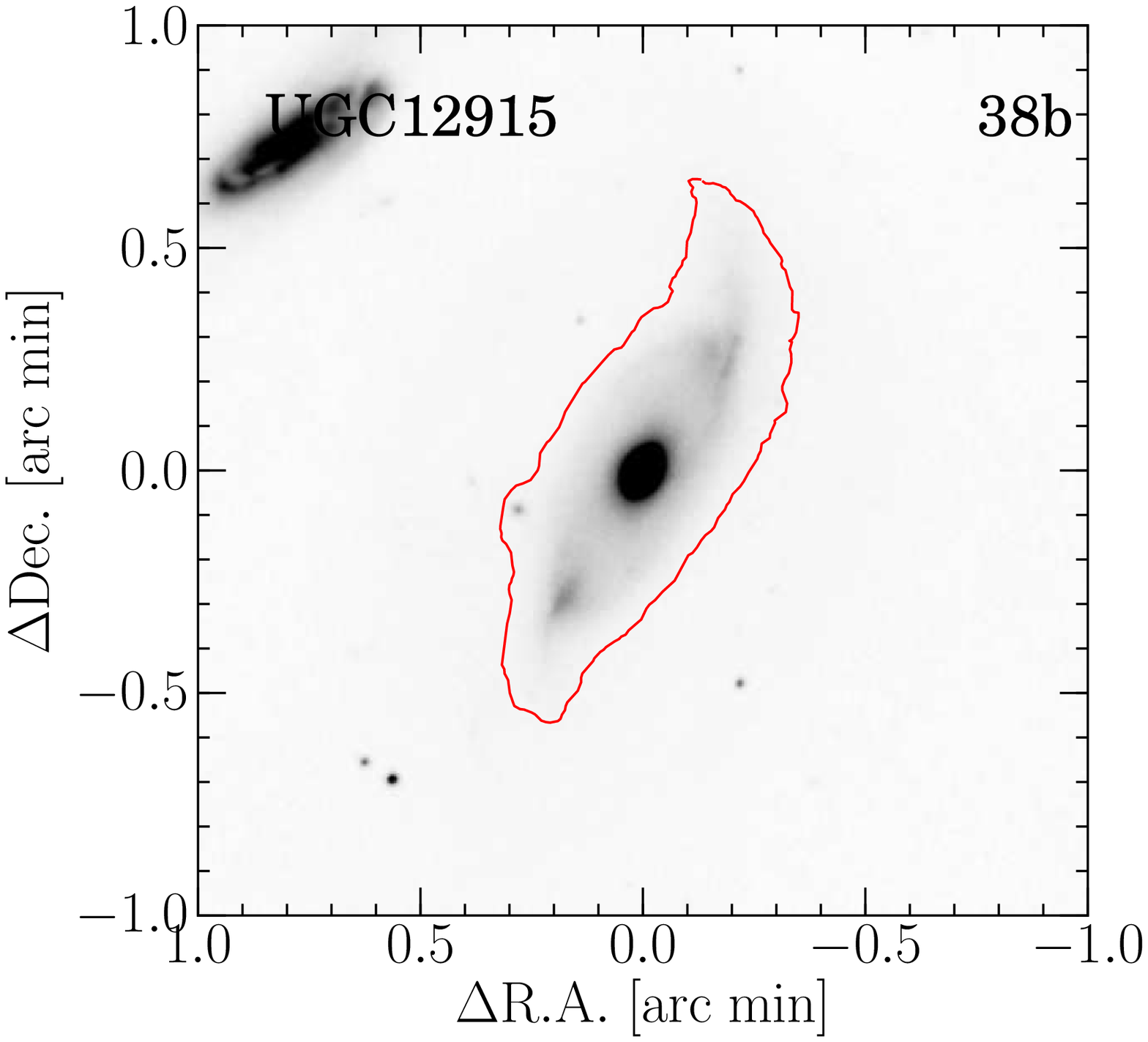}{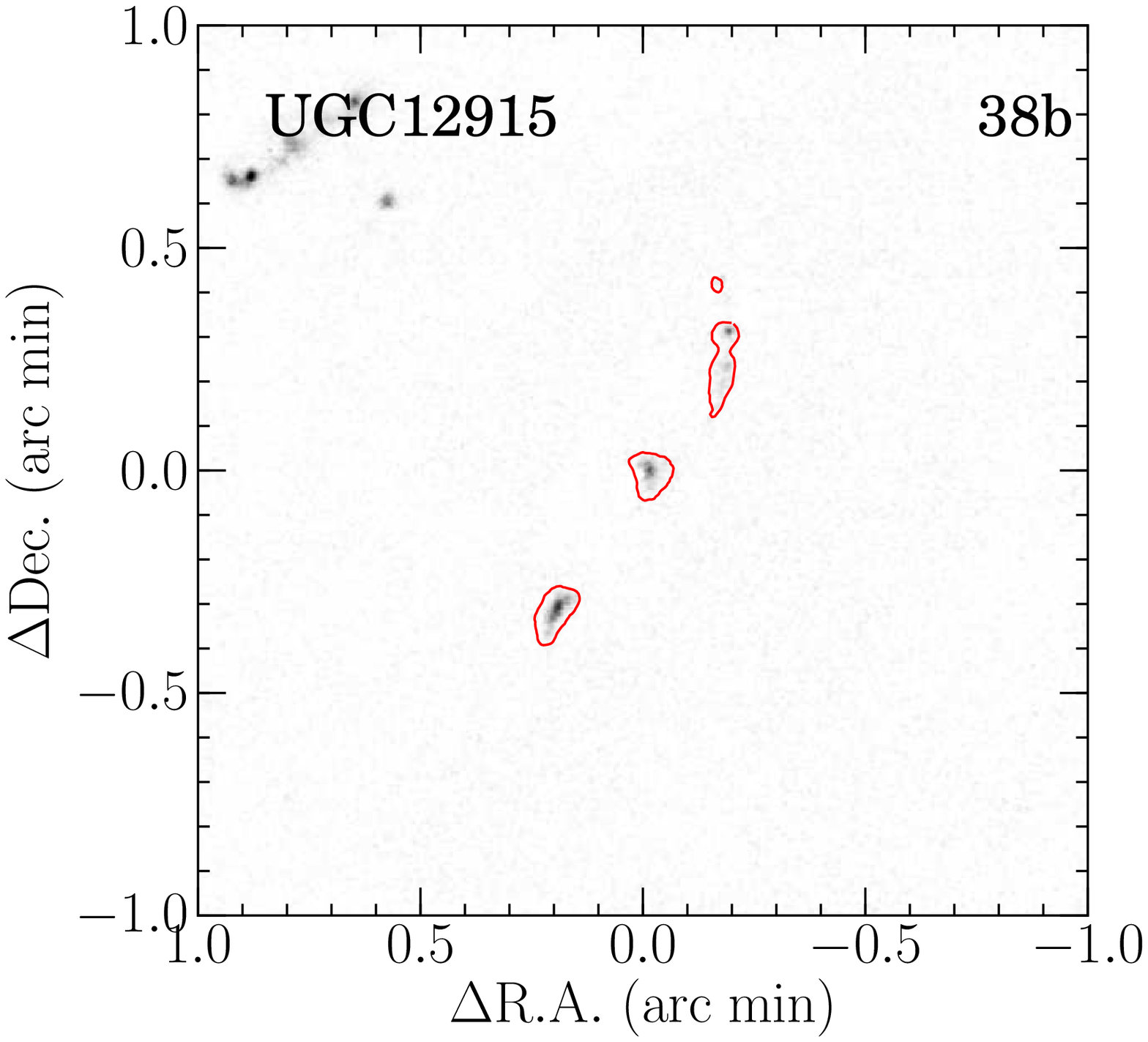}\\
   \end{center}
   \caption{Continued}
 \end{figure}

\twocolumngrid

\noindent induced by the companion \citep{1991IAUS..146..312L}. This is classified as a Seyfert 2 \citep{1995ApJS...98..171V} or an H{\sc ii} galaxy \citep{2003ApJ...583..670C}. While strong H$\alpha$ emissions are seen along the spiral arms especially in the northern part, no emission is detected in the nucleus \citep{2002ApJS..143...47D}. However, bright Pa$\alpha$ emission is detected not only at the spiral arms but also in the nucleus.\\

\indent {\bf 25. IC 5063 (IRAS F20481$-$5715; AM2048$-$571):}
This is an S0/a (HyperLeda) considered as a merger remnant \citep{1991ApJ...370..102C} and classified as a Seyfert 2. It shows high polarization in the near IR continuum \citep{1987MNRAS.224.1013H} and strong broad H$\alpha$ emission in polarized flux \citep{1993MNRAS.263..895I}. These results suggest that there is an obscured hidden broad-line region (HBLR) \citep[e.g.,][]{2001ApJ...554L..19T,2001MNRAS.327..459L,1985ApJ...297..621A} and the broad emission line is scattered into our line of sight by scatters outside the obscured regions \citep{2007A&A...476..735M}. Also, this object is the first galaxy where a fast gas outflow has been discovered \citep{1998AJ....115..915M}. A Pa$\alpha$ emission is strongly concentrated at the center and extended towards NW-SE direction, which may consistent with the direction of extend NLR reported by \citet{1991ApJ...370..102C}.\\

\indent {\bf 26. ESO 286-G035 (IRAS F21008$-$4347):}
This is a spiral (Sc; HyperLeda) classified as an H{\sc ii} galaxy \citep{1995ApJS...98..171V}. In the Pa$\alpha$ image, there is a strong emission line region at the center extended along the disk.\\

\indent {\bf 27. ESO 343-IG013 (IRAS F21330$-$3846; VV 714; AM 2133$-$384):}
This is a barred spiral (Sbc; HyperLeda) classed as an H{\sc ii} galaxy \citep{2010ApJ...709..884Y,1995ApJS...98..171V}. It is known as a strongly interacting pair \citep{1974MCG...C05....0V} with a separation between the two galactic nuclei of 10$\farcs$9 in the $K_{\mathrm{s}}$-band image. The northern part of the galaxy has concentrated Pa$\alpha$ emission at its center, while the southern part has strong Pa$\alpha$ emission at the its center and extended region along spiral arms.\\

\indent {\bf 28. NGC 7130 (IRAS F21453$-$3511; AM 2145$-$351; NED 02):}
This is a spiral (Sc; HyperLeda) classified as LINER/Seyfert 1 \citep{2003ApJ...583..670C,1995ApJS...98..171V}. It is known as a peculiar/disturbed spiral \citep{1982euse.book.....L} and a starburst/AGN composite galaxy \citep{2005ApJ...618..167L}. In the Pa$\alpha$ image, strong emission exists at the central region, as well as along the spiral arms, especially at the northern arm which is consistent with an H$\alpha$ image \citep{2011A&A...527A..60R}, and a Pa$\alpha$ image by $HST$/NICMOS \citep{2006ApJ...650..835A}.\\

\indent {\bf 29. IC 5179 (IRAS F22132$-$3705; AM 2213$-$370):}
This is a barred spiral (Sbc; HyperLeda) classified as an H{\sc ii} galaxy \citep{2010ApJ...709..884Y}. Type Ia SN 1999ee and the Type Ib/c SN 1999ex have been discovered in this galaxy \citep{2002AJ....124.2100S}. Pa$\alpha$ emission is distributed along the spiral arms as clumpy knots. These H{\sc ii} knots, which are widely spread over the entire galaxy, is not be covered by the field of view of VLT/VIMOS \citep{2011A&A...527A..60R} and $HST$/NICMOS \citep{2006ApJ...650..835A} observations.\\

\indent {\bf 30. ESO 534-G009 (IRAS F22359$-$2606):}
This is a spiral (Sab; HyperLeda) classified as a LINER \citep{1995ApJS...98..171V}. There is a strong Pa$\alpha$ emission at the central region.\\

\indent {\bf 31. NGC 7469 (IRAS F23007$+$0836; Arp 298; NED 01; Mrk 1514; KPG 575A):}
This is a spiral (Sa; HyperLeda) classified as a Seyfert 1 \citep{1995ApJS...98..171V}. It is known to have a circumnuclear ring with a diameter on scales of 1$\farcs$5--2$\farcs$5 \citep{2004ApJ...602..148D}, which is observed in the radio \citep{1991ApJ...381...79W,1991ApJ...378...65C,2001ApJ...553L..19C}, in the optical \citep{1994A&A...285...44M}, in the mid-infrared \citep{1994ApJ...425L..37M,2003AJ....126..143S}, and in the near-infrared \citep{1995ApJ...444..129G,1999ESOC...56..555L,2000AJ....119..991S} wavelengths. In Pa$\alpha$ image, the ring structure can not be resolved by the seeing size of the ANIR data.\\

\indent {\bf 32. CGCG 453-062 (IRAS F23024$+$1916):}
This is a spiral (Sab; HyperLeda) classified as a LINER \citep{1991S&T....82Q.621D}. In Pa$\alpha$ image, it has extended
\clearpage
\onecolumngrid

%\begin{turnpage}
%\begin{landscape}
\begin{deluxetable}{ccccccccc}
\tabletypesize{\tiny}
%\tabletypesize{\scriptsize}
%\rotate
\tablecolumns{10}
\tablewidth{0pc}
%\rotate
\tablecaption{Pa$\alpha$ luminosities and derived star formation rates. \label{table_sfr}}
%\tablewidth{0pt}

\tablehead{
\colhead{ID} & \colhead{Galaxy}& \colhead{$\mathrm{Size}(\mathrm{Pa}\alpha)$}& \colhead{$L(\mathrm{Pa}\alpha)$}& \colhead{$SFR(\mathrm{Pa}\alpha)$}& \colhead{$SFR(\mathrm{Pa}\alpha)_\mathrm{corr}$}& \colhead{$SFR(\mathrm{IR})$}
& \colhead{} & \colhead{}\\
 & \colhead{Name}& \colhead{(kpc)}& \colhead{(ergs s$^{-1}$)}& \colhead{($M_{\odot}$ yr$^{-1}$)}& \colhead{($M_{\odot}$ yr$^{-1}$)}& \colhead{($M_{\odot}$ yr$^{-1}$)}
& \colhead{$E(B-V)$} & \colhead{Ref.}\\
\colhead{(1)} & \colhead{(2)} & \colhead{(3)}& \colhead{(4)}& \colhead{(5)}& \colhead{(6)}& \colhead{(7)}& \colhead{(8)}& \colhead{(9)}
}
\startdata
1 & NGC 23\dotfill & 1.9 & 1.17$\times$10$^{41}$ & 7.5 & 11.2 & 14.5 & 0.74 & 1 \\ 
2 & NGC 34\dotfill & 1.1 & 3.30$\times$10$^{41}$ & 21.1 & 65.0 & 35.4 & 2.08 & 1 \\ 
3 & NGC 232\dotfill & 0.9 & 7.98$\times$10$^{40}$ & 5.1 & 10.4 & 26.5 & 1.31 & 1 \\ 
4 & IC 1623A/B\dotfill & 9.8 & 6.02$\times$10$^{41}$ & 38.5 & 44.3 & 58.9 & 0.26 & 1 \\ 
5 & ESO 244-G012\dotfill & 1.8 & 4.33$\times$10$^{41}$ & 27.7 & 59.3 & 32.6 & 1.41 & 3 \\ 
6 & UGC 2238\dotfill & 3.3 & 3.88$\times$10$^{41}$ & 24.8 & 83.8 & 23.0 & 2.25 & 1 \\ 
7 & IRAS F02437$+$2122\dotfill & 0.5 & 2.46$\times$10$^{40}$ & 1.6 & 5.0 & 17.6 & 2.13 & 1 \\ 
8 & UGC 2982\dotfill & 4.7 & 2.67$\times$10$^{41}$ & 17.1 & 47.5 & 17.0 & 1.89 & 1 \\ 
9 & NGC 1614\dotfill & 1.1 & 5.97$\times$10$^{41}$ & 38.2 & 74.7 & 49.0 & 1.24 & 1,2,3 \\ 
10 & MCG $-$05-12-006\dotfill & 0.8 & 1.28$\times$10$^{41}$ & 8.2 & [14.5] & 15.8 & [1.05] & -- \\ 
11 & NGC 1720\dotfill & 0.9 & 3.32$\times$10$^{40}$ & 2.1 & [3.8] & 8.5 & [1.05] & -- \\ 
12 & ESO 557-G002\dotfill & 1.2 & 9.02$\times$10$^{40}$ & 5.8 & 7.3 & 18.6 & 0.43 & 2,3 \\ 
13 & IRAS F06592-6313\dotfill & 0.7 & 7.39$\times$10$^{40}$ & 4.7 & 9.4 & 17.0 & 1.28 & 3 \\ 
14 & NGC 2342\dotfill & 10.9 & 1.55$\times$10$^{41}$ & 9.9 & [17.5] & 27.0 & [1.05] & -- \\ 
15 & ESO 320-G030\dotfill & 1.5 & 4.26$\times$10$^{40}$ & 2.7 & 5.1 & 20.3 & 1.16 & 3 \\ 
16 & NGC 4922\dotfill & 1.2 & 5.50$\times$10$^{40}$ & 3.5 & [6.2] & 22.7 & [1.05] & -- \\ 
17 & MCG $-$03-34-064\dotfill & 0.3 & 7.45$\times$10$^{40}$ & 4.8 & [8.4] & 20.4 & [1.05] & -- \\ 
18 & NGC 5135\dotfill & 1.4 & 1.72$\times$10$^{41}$ & 11.0 & 14.2 & 20.0 & 0.46 & 3 \\ 
19 & NGC 5257/8\dotfill & 9.5/10.6 & 2.65$\times$10$^{41}$/1.93$\times$10$^{41}$ & 16.9/12.4 & 20.8/15.2 & 37.5 & 0.38 & 1,2 \\ 
20 & IC 4518A/B\dotfill & 0.7/4.6 & 4.51$\times$10$^{+40}$/2.86$\times$10$^{40}$ & 2.9/1.8 & 3.7/2.3 & 13.4 & 0.43 & 2,3 \\ 
21 & IC 4686/87\dotfill & 0.6/2.5 & 8.81$\times$10$^{40}$/5.41$\times$10$^{41}$ & 5.6/34.6 & 8.2/45.6 & 38.2 & 0.69 & 2,3 \\ 
21c & IC 4689\dotfill & 2.2 & 1.42$\times$10$^{41}$ & 9.1 & 16.0 & -- & [1.05] & - \\ 
22 & IRAS F18293-3413\dotfill & 1.8 & 6.29$\times$10$^{41}$ & 40.3 & 104.3 & 70.6 & 1.76 & 1 \\ 
23 & ESO 339-G011\dotfill & 1.3 & 7.37$\times$10$^{40}$ & 4.7 & [8.3] & 14.8 & [1.05] & -- \\ 
24 & NGC 6926\dotfill & 24.5 & 1.16$\times$10$^{41}$ & 7.4 & 14.3 & 22.2 & 1.22 & 1,2 \\ 
25 & IC 5063\dotfill & 0.8 & 2.64$\times$10$^{40}$ & 1.7 & [3.0] & 7.8 & [1.05] & -- \\ 
26 & ESO 286-G035\dotfill & 2.1 & 1.76$\times$10$^{41}$ & 11.3 & [19.9] & 19.0 & [1.05] & -- \\ 
27 & ESO 343-IG013\dotfill & 1.6 & 7.09$\times$10$^{40}$ & 4.5 & 7.5 & 13.7 & 0.93 & 1,2 \\ 
28 & NGC 7130\dotfill & 1.0 & 1.35$\times$10$^{41}$ & 8.6 & 14.2 & 26.5 & 0.92 & 1,2,3 \\ 
29 & IC 5179\dotfill & 5.4 & 1.62$\times$10$^{41}$ & 10.4 & 19.0 & 17.5 & 1.11 & 1,3 \\ 
30 & ESO 534-G009\dotfill & 0.5 & 6.53$\times$10$^{39}$ & 0.4 & 0.6 & 5.4 & 0.63 & 1 \\ 
31 & NGC 7469\dotfill & 1.0 & 3.71$\times$10$^{41}$ & 23.7 & 23.7 & 50.0 & 0.00 & 1 \\ 
32 & CGCG 453-062\dotfill & 4.1 & 1.23$\times$10$^{41}$ & 7.9 & [13.9] & 27.7 & [1.05] & -- \\ 
33 & NGC 7591\dotfill & 5.4 & 6.79$\times$10$^{40}$ & 4.3 & 9.2 & 13.9 & 1.39 & 1 \\ 
34 & NGC 7678\dotfill & 18.8 & 3.41$\times$10$^{40}$ & 2.2 & [3.8] & 7.4 & [1.05] & -- \\ 
35 & MCG $-$01-60-022\dotfill & 1.3 & 1.61$\times$10$^{41}$ & 10.3 & 21.9 & 20.9 & 1.40 & 2 \\ 
36 & NGC 7770/1\dotfill & 2.3/3.4 & 8.75$\times$10$^{40}$/1.77$\times$10$^{+41}$ & 5.6/11.3 & 7.8/34.0 & 28.6 & 2.03 & 1 \\ 
37 & Mrk 331\dotfill & 1.6 & 2.14$\times$10$^{41}$ & 13.7 & [24.2] & 32.4 & [1.05] & -- \\ 
38 & UGC 12914/15\dotfill & 6.8/12.2 & 6.34$\times$10$^{40}$/3.85$\times$10$^{40}$ & 4.1/2.5 & [7.2]/[4.4] & 10.5 & [1.05] & -- 
\enddata
%% Text for table notes should follow after the \enddata but before
%% the \end{deluxetable}. Make sure there is at least one \tablenotemark
%% in the table for each \tablenotetext.
\tablecomments{Column (1): Galaxy ID in this paper. Column (2): Galaxy name. Column (3): Size of Pa$\alpha$ emission line regions in diameter defined to be an elliptical (major axis) contained 50\% flux within the Petrosian radius \citep{2012ApJ...751...11P}. Column (4): Observed Pa$\alpha$ luminosities. Column (5): Pa$\alpha$-derived star formation rates. Column (6): Pa$\alpha$-derived star formation rates with dust extinction correction. Brackets values are dust-extinction corrected SFRs by using assumed $E(B-V)$ = 1.05, the average value of all our objects. Column (7): IR-drived star formation rates. Column (8): Amount of dust extinction derived from balmer decrement using the flux ratio of H$\alpha$ and H$\beta$. Column (9): References for the dust extinction. 1; \citet{1995ApJS...98..171V}, 2; \citet{2002ApJS..143...47D}, 3; \citet{2011A&A...527A..60R}}
%\tablenotetext{a}{Sample footnote for table~\ref{tbl-1} that was generated with the deluxetable environment}
%\tablenotetext{b}{Another sample footnote for table~\ref{tbl-1}}
\end{deluxetable}
%\end{landscape}
%\end{sidewaystable}
%\clearpage
%\end{turnpage}

\twocolumngrid

\noindent star-forming region distributed not only at the central region but also along the disk.\\

\indent {\bf 33. NGC 7591 (IRAS F23157$+$0618):}
This is an isolated \citep{2010ApJ...709..884Y} barred spiral (SBbc; HyperLeda) classified as a LINER \citep{1995ApJS...98..171V}. In Pa$\alpha$ image, a strong emission peak is at the central region, and diffuse emission regions distributed along the spiral arms and the barred structure.\\

\indent {\bf 34. NGC 7678 (IRAS F23259$+$2208; Arp 028; VV 359):}
This is a barred spiral (Sc; HyperLeda) classified as an H{\sc ii} galaxy \citep{1996JKAS...29..255A}. It has a massive spiral arm in the southern part of the galaxy, and is considered to have experienced strong interaction which induced active star formation in the nucleus and in the southern arm \citep{1996JKAS...29..255A}. In Pa$\alpha$ image, strong emission line regions are at the central region, and at the southern arm. Extended diffuse emission can be seen along the spiral arms.\\

\indent {\bf 35. MCG $-$01-60-022 (IRAS F23394$-$0353; VV 034a; Arp 295B):}
This is a merging spiral (Sb; HyperLeda) classified as an H{\sc ii} galaxy \citep{2003ApJ...583..670C}. It is known as an interacting galaxy \citep{1974MCG...C05....0V,2007RMxAA..43..179R} paired with MCG $-$01-60-021, separated by 4$\farcm$.5 in the $K_{\mathrm{s}}$-band image.   In Pa$\alpha$ image, a strong emission is detected not only at the central region but also along the disk, while a Pa$\alpha$ emission can not be detected in the tidal tail between Arp 295A and Arp 295B.\\

\indent {\bf 36. NGC 7770/1 (IRAS F23488$+$19489;  Mrk 9006; KPG 592B):}
This appears to be in the early stage of an interaction with NGC 7770 where a separation between the two galactic nuclei is 1$'$ in the $K_\mathrm{s}$-band image. NGC 7770 is a spiral (S0-a; HyperLeda), and NGC 7771 is a barred spiral (Sa; HyperLeda) classified as an H{\sc ii} galaxy \citep{1995ApJS...98..171V}. Both NGC 7771 and NGC 7770 show a strong Pa$\alpha$ emission. NGC 7770 has strong emission in the spiral arm rather than at the central region, while NGC 7771 has strong emission at the central region rather than in the disk. The morphology of central Pa$\alpha$ emission in NGC 7771 seems to be a ring starburst. No Pa$\alpha$ emission is detected between these galaxies.\\

\indent {\bf 37. Mrk 331 (IRAS F23488$+$1949; KPG 593B):}
This is a member of a group of three galaxies; an irregular separated by 1$\farcm$.4 and a spiral by 2$'$ \citep{1983ApJ...270L..35M}. It is a spiral (Sa; HyperLeda) classified as an H{\sc ii}/Seyfert 2 (\citealt{2001AJ....122.1213S}). In Pa$\alpha$ image, an emission line appears to be concentrated at the central region.\\

\indent {\bf 38. UGC 12914/5 (IRAS F23591$+$2312; VV 254; III Zw 125; KPG 603; TaffyI):}
These are barred galaxies (SBc; HyperLeda). It is an interacting system with an extended shock-induced synchrotron radio emission connecting the two galaxies \citep{1993AJ....105.1730C,2012ApJ...751...11P}. The steepening of the radio spectral index at the bridge indicates a face-on collision which occurred only 20 Myr ago \citep{1993AJ....105.1730C}. There are many H{\sc ii} blobs in a Pa$\alpha$ image, and it is considered that many of them are induced by the collision \citep{2012ApJ...757..138K}.

%-------------
\begin{figure}[t]
\begin{center}
\plotone{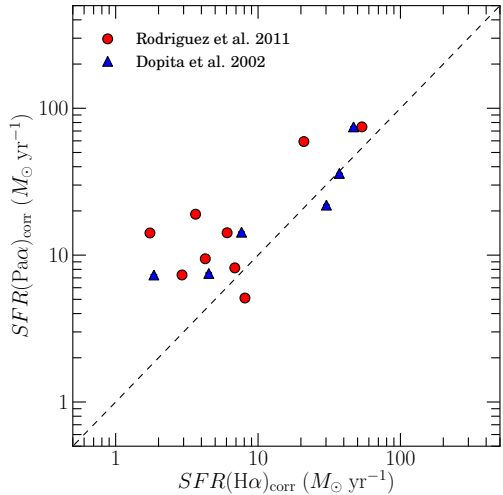}
\caption{Comparison of SFRs derived from H$\alpha$ and Pa$\alpha$ luminosities, both corrected for dust extinction. H$\alpha$ luminosities are taken from \citet{2011A&A...527A..60R} (circles) and \citet{2002ApJS..143...47D} (triangles). The dashed line shows the one-to-one relation.}\label{fig:sfr_paha}
\end{center}
\end{figure}
%-------------

%%%==========================================================================%%%
%%%   Section 5 Discussion
%%%==========================================================================%%%
\section{Discussion}

%%%___Subsection 5.1 Dust Extinction___________________________________
\subsection{The Effect of Dust-extinction for Pa$\alpha$ Flux}
Even though the Pa$\alpha$ emission line is less sensitive against dust extinction, we can not ignore its effect. Typical amount of dust extinction in LIRGs is $A_{V}$ $\sim$ 4 mag \citep{2006ApJ...650..835A} as derived by the ``Balmer Decrement Method'' \citep{2000ApJ...533..682C} using the flux ratio of H$\alpha$ and Pa$\alpha$. To correct for the dust-extinction in our sample, we adopt the extinction curve of \citet{2000ApJ...533..682C} with $R_{V} = 4.05$, which results in $A_V$/$A_{\mathrm{Pa}\alpha}$ = 6.97, and $A_{\mathrm{H}\alpha}$/$A_{\mathrm{Pa}\alpha}$ = 5.68, meaning that Pa$\alpha$ is typically attenuated by 0.57 mag in LIRGs. In Table \ref{table_sfr}, color excess ($E(B-V)$) derived by balmer decrement method using ratios of H$\alpha$ and H$\beta$ taken from \citet{1995ApJS...98..171V}, \citet{2002ApJS..143...47D}, and \citet{2011A&A...527A..60R} are listed. 26 galaxies, with H$\alpha$ and H$\beta$ data in our sample have $A_{V}$ = 4.25 mag ($E(B-V)$ = 1.05) on average, and we adopt this value for the rest of the sample.

%%%___Subsection 5.2 Star Formation Rates______________________________
\subsection{Star Formation Rates}
SFRs are derived from Pa$\alpha$ luminosity using the following relation, assuming a relation on SFR-H$\alpha$ luminosity assuming a Kroupa IMF \citep{2009ApJ...703.1672K} and a flux ratio of H$\alpha$/Pa$\alpha$ $=$ 8.6 ($T=1000\ \mathrm{K}$, $n_e=10^4\ \mathrm{cm^{-3}}$, \citealt{1987MNRAS.224..801H}) in starburst galaxies;

\begin{eqnarray}
SFR(\mathrm{Pa}\alpha)(M_{\sun}\ \mathrm{yr^{-1}}) \equiv 6.4 \times 10^{-41}L(\mathrm{Pa\alpha})\mathrm{(erg\ s^{-1})}
\end{eqnarray}
where $L(\mathrm{Pa}\alpha)$ is the luminosity of Pa$\alpha$. Dust-extinction uncorrected star formation rates ($SFR(\mathrm{Pa}\alpha)$) obtained from $L(\mathrm{Pa}\alpha)$ and those corrected for dust-extinction ($SFR(\mathrm{Pa}\alpha)_\mathrm{corr}$) are shown in Table \ref{table_sfr}.

Figure \ref{fig:sfr_paha} shows the comparision between $SFR(\mathrm{Pa}\alpha)_\mathrm{corr}$ and SFR derived from the dust-extinction corrected H$\alpha$ luminosity ($SFR(\mathrm{H}\alpha)_\mathrm{corr}$). The dust-extinction corrected H$\alpha$ luminosities are taken from \citet{2011A&A...527A..60R} and \citet{2002ApJS..143...47D}, which are obtained by narrow-band imaging or integral field spectroscopy. Their correlation is generally good, but $SFR(\mathrm{Pa}\alpha)_\mathrm{corr}$ is larger than $SFR(\mathrm{H}\alpha)_\mathrm{corr}$ systematically, while both are expected to be the same value because it is emitted from a same region by a same mechanism. \citet{2013ApJ...778L..41L} have found that correcting dust extinction using H$\alpha$/Pa$\beta$ gives much larger star formation rate than using H$\alpha$/H$\beta$ (Balmer decrement) in star formation regions of M83. Considering this study, the difference between $SFR(\mathrm{Pa}\alpha)_\mathrm{corr}$ and $SFR(\mathrm{H}\alpha)_\mathrm{corr}$ suggests that Pa$\alpha$ can see star-forming activity through a more dusty region than H$\alpha$.

FIR and bolometric infrared luminosities are also good indicators for star formation in dusty starburst galaxies \citep[e.g.,][]{1998ARA&A..36..189K, 2002AJ....124.3135K, 2003A&A...410...83H}, because MIR to FIR emission in these galaxies arises from re-radiation of dust-absorbed shorter-wavelength photons. Bolometric IR luminosities ($L(\mathrm{IR})$) are derived from the $IRAS$ ADDSCAN/SCANPI 12 $\mu \mathrm{m}$, 25 $\mu \mathrm{m}$, 60 $\mu \mathrm{m}$, and 100 $\mu \mathrm{m}$ data \citep{1993ApJS...89....1R,2003AJ....126.1607S} compiled by \citet{2003AJ....126.1607S} and the following relation \citep{1996ARA&A..34..749S};

\begin{eqnarray}
L(\mathrm{IR}) &=& 4\pi D_L^2 \times (1.8\times10^{-14}(13.48 f_{12}\nonumber\\
&&+ 5.16 f_{25} + 2.58 f_{60} + f_{100})),\ 
\end{eqnarray}
where $D_L$ is the luminosity distance in meters, and $f_{12}$, $f_{25}$, $f_{60}$, and $f_{100}$ are the IRAS flux densities in Jy at 12, 25, 60, and 100 $\mu \mathrm{m}$, respectively. Then, $SFR(\mathrm{IR})$ is derived using the following relation \citep{2013arXiv:1208.2997v1};
\begin{equation}
SFR(\mathrm{IR})\ (M_{\sun}\ \mathrm{yr^{-1}}) \equiv 2.8 \times 10^{-44}
\hspace{0.3em}L(\rm IR)\ \mathrm{(erg\ s^{-1})},
\end{equation}
\clearpage

%-------------
\begin{figure}[t]
\begin{center}
\plotone{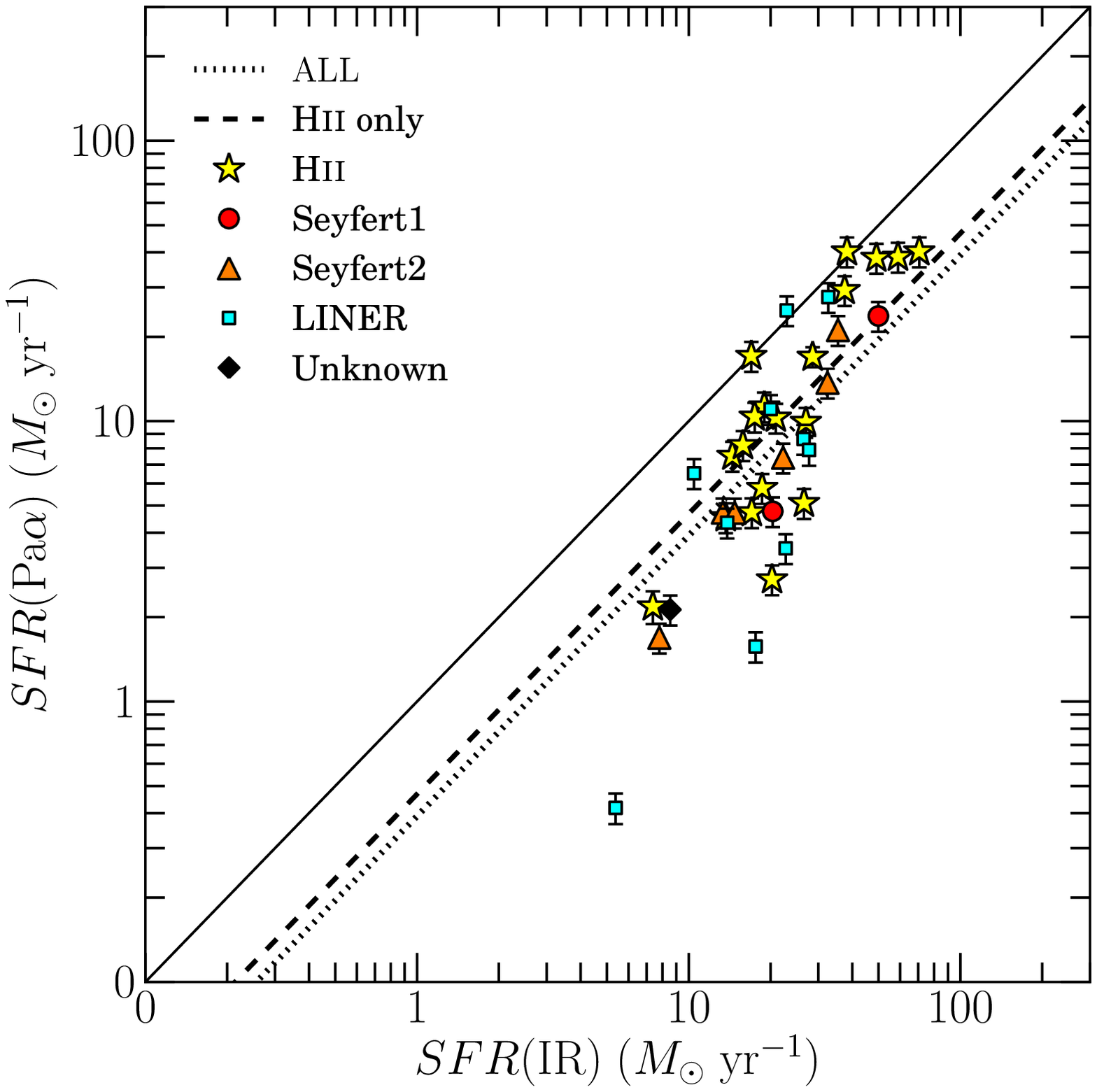}
\caption{Comparison of SFRs derived from bolometric infrared luminosities calculated using $IRAS$ ADDSCAN FIR luminosities of 12 $\mu \mathrm{m}$, 25 $\mu \mathrm{m}$, 60 $\mu \mathrm{m}$, and 100 $\mu \mathrm{m}$, and those derived from Pa$\alpha$ luminosities. The dotted line represents the best fit relation for all galaxies assuming a slope of one while the dashed line represents for the H{\sc ii} galaxies. The solid line shows the one-to-one relation. Stars represent H{\sc ii} galaxies, circles Seyfert Is, triangles Seyfert IIs, squares LINERs, and diamonds galaxies without any classification. The error bars are the total uncertainties on the measurement of the Pa$\alpha$ flux ($\sigma_\mathrm{total}$; see text).}\label{fig:plot_sfr1}
\end{center}
\end{figure}
%-------------

%-------------
\begin{figure}[h]
\begin{center}
\plotone{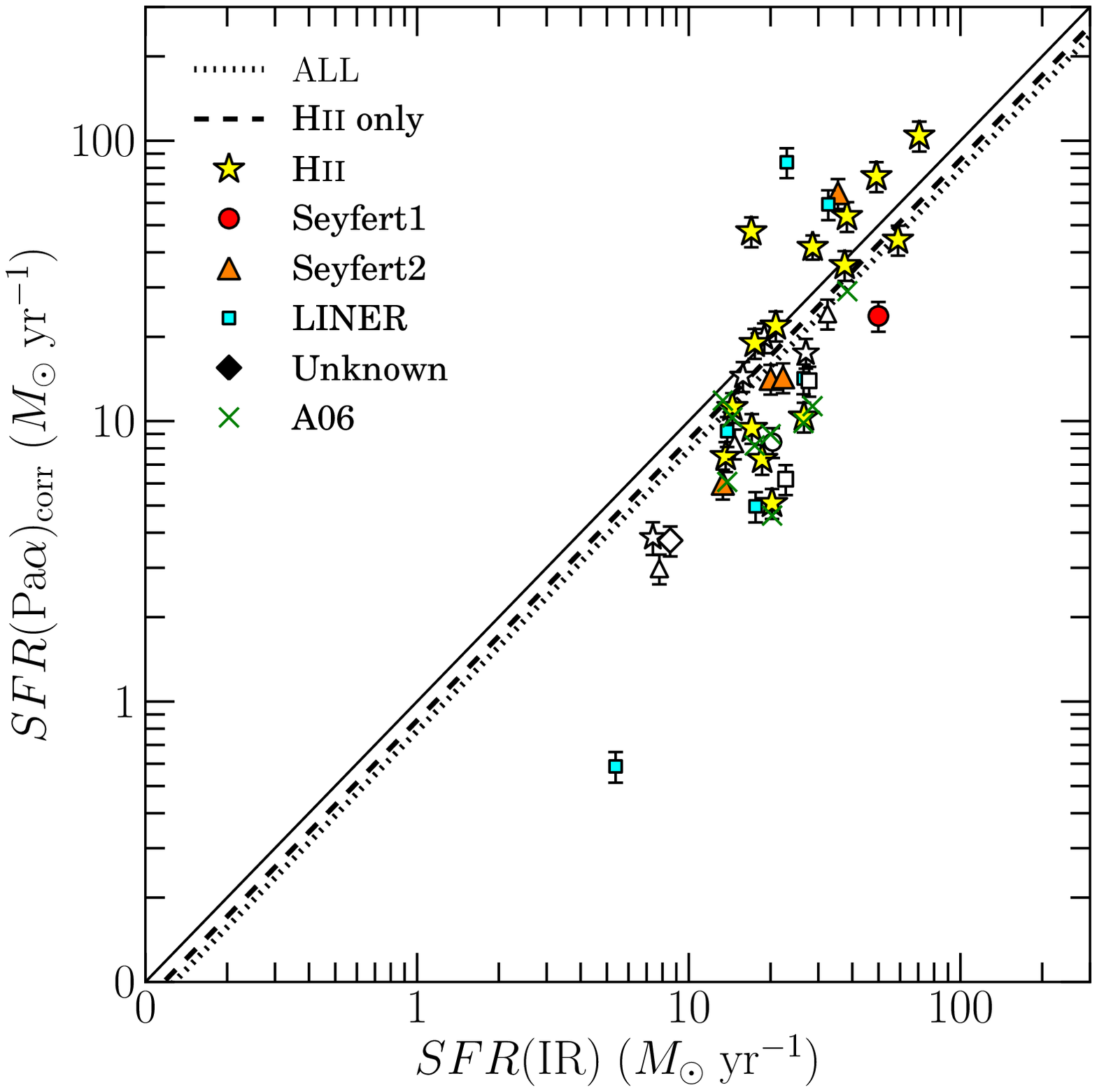}
\caption{Same as Figure \ref{fig:plot_sfr1}, but plot for SFRs derived from dust-extinction corrected Pa$\alpha$ luminosities. Filled symbols represent galaxies whose extinction is corrected using Balmer decrement method while open symbols assuming $E(B-V)$=1.05. Crosses show galaxies whose Pa$\alpha$ data are observed by $HST$/NICMOS and its extinction is corrected using Balmer decrement method (\citealt{2006ApJ...650..835A}, A06). The error bars are the total uncertainties on the measurement of the Pa$\alpha$ flux ($\sigma_\mathrm{total}$; see text).}\label{fig:plot_sfr2}
\end{center}
\end{figure}
%-------------

\noindent The results are listed in Table \ref{table_sfr}.

Comparison of $SFR(\mathrm{IR})$ and $SFR(\mathrm{Pa}\alpha)$ is shown in Figure \ref{fig:plot_sfr1}. The solid line shows the one-to-one relation between $SFR(\mathrm{IR})$ and $SFR(\mathrm{Pa}\alpha)$.
The dotted line represents the best fit relation for all galaxies assuming a slope of one and the dashed line represents the same but using only H{\sc ii} galaxies. Both indicate that $SFR(\mathrm{Pa}\alpha)$ are systematically offset by $-0.3$ dex from $SFR(\mathrm{IR})$.

Figure \ref{fig:plot_sfr2} is comparison of $SFR(\mathrm{IR})$ and $SFR(\mathrm{Pa}\alpha)_\mathrm{corr}$. Some samples have no H$\alpha$ data, therefore we correct for dust-extinction assuming $E(B-V) = 1.05\ (A_{V} = 4.3)$ which is an average value of our sample. In this figure, we also plot a sample of galaxies from \citet{2006ApJ...650..835A} where $SFR(\mathrm{Pa}\alpha)$ are derived in the same way as described above, but we do not use this sample in our best-fitting relations. The offset between $SFR(\mathrm{Pa}\alpha)_\mathrm{corr}$ and $SFR(\mathrm{IR})$ is $-0.07$ dex within a scatter for $0.27$ dex.

%%%___Subsection 5.3 Selection Bias____________________________________
\subsubsection{Malmquist Bias}
In Figure \ref{fig:plot_sfr1} and \ref{fig:plot_sfr2}, it seems that the slopes of the distribution of H{\sc ii} galaxies are larger than that of one. Actually, the slopes are 1.8 in Figure \ref{fig:plot_sfr1} and 2.1 in Figure \ref{fig:plot_sfr2}. This may be caused by $``$Malmquist bias$"$ \citep{1925Obs....48..142M}, where brightness of a sample falls off quickly until the brightness falls below observational threshold, because the luminosity range of our sample is spread over only 1 order of magnitude. Therefore we have estimated this effect in our sample with the following a simple test.

We first create a model set of 38 galaxies with SFR of $``X"$ ($M_\odot\ \mathrm{yr}^{-1}$) within the same IR luminosity range as our sample. This $X$ for each galaxy is converted into $``Y"$ by adding Gaussian-random $``$noise$"$ with $\sigma$ = 0.3 dex, which is the same value as the dispersion between $\log(SFR(\mathrm{Pa}\alpha)_{\mathrm{corr}})$ and $\log(SFR(\mathrm{IR}))$. Then, a best-fit relation is obtained for the plot of $X$ and $Y$. In this fitting, two kinds of relations are used; one is 1 free-parameter function ($\log(Y) = \log(X) + b$) and the other is two free-parameters function ($\log(Y) = a\cdot\log(X) + b$). We have carried out the above procedure 1000 times.

In the 2 free-parameter fitting, we find the slope $a$ to be 2.0 $\pm$ 0.3, which is larger than the intrinsic value of 1 and consistent with the observed value of $a = 2.1$. In the 1 free-parameter fitting, we find that the offset is $b$ = 0 $\pm$ 0.05. The Large slope of the distributions in Figure \ref{fig:plot_sfr1} and in Figure \ref{fig:plot_sfr2} come from the Malmquist bias. To remove the Malmquist bias it is necessary to expand the luminosity range. The offset of $-0.07$ dex in Figure \ref{fig:plot_sfr2} with 1 free-paramter fitting is within the statistical error.

%%%___Subsection 5.3 Selection Bias____________________________________
\subsubsection{Comparison Between IR SFR and Pa$\alpha$ SFR}
From our simple test, the vertical offsets in Figure \ref{fig:plot_sfr1} and in Figure \ref{fig:plot_sfr2} are insensitive to the Malmquist bias. To explain the offset of $-0.3$ dex in Figure \ref{fig:plot_sfr1} by dust-extinction, $A_V$ = 5.7 is required. \citet{2006ApJ...650..835A} suggested that the spectroscopic dust-extinction derived from an H$\alpha$ and Pa$\alpha$ is $A_V$ =  4.1 on average, which is almost the same value as our result but slightly smaller.
Figure \ref{fig:plot_sfr2} shows that there is a relatively good coincidence between the $SFR(\mathrm{Pa}\alpha)_\mathrm{corr}$ and the $SFR(\mathrm{IR})$ on average.

Although our result shows that there is a good correlation between $SFR(\mathrm{Pa}\alpha)_\mathrm{corr}$ and $SFR(\mathrm{IR})$, some of the galaxies have offset from it. For example, ESO 534-G009, having lowest $SFR(\mathrm{Pa}\alpha)_\mathrm{corr}$ in our sample, is a late-type spiral (Sab) classified as a LINER, and its Pa$\alpha$ emission is emitted from a unresolved compact central region. This galaxy shows an order of magnitude lower $SFR(\mathrm{Pa}\alpha)_\mathrm{corr}$ than $SFR(\mathrm{IR})$.
The galaxies with smaller $SFR(\mathrm{Pa}\alpha)_\mathrm{corr}$ than $SFR(\mathrm{IR})$ may have larger
\clearpage

%-------------
\begin{figure}[t]
\begin{center}
\plotone{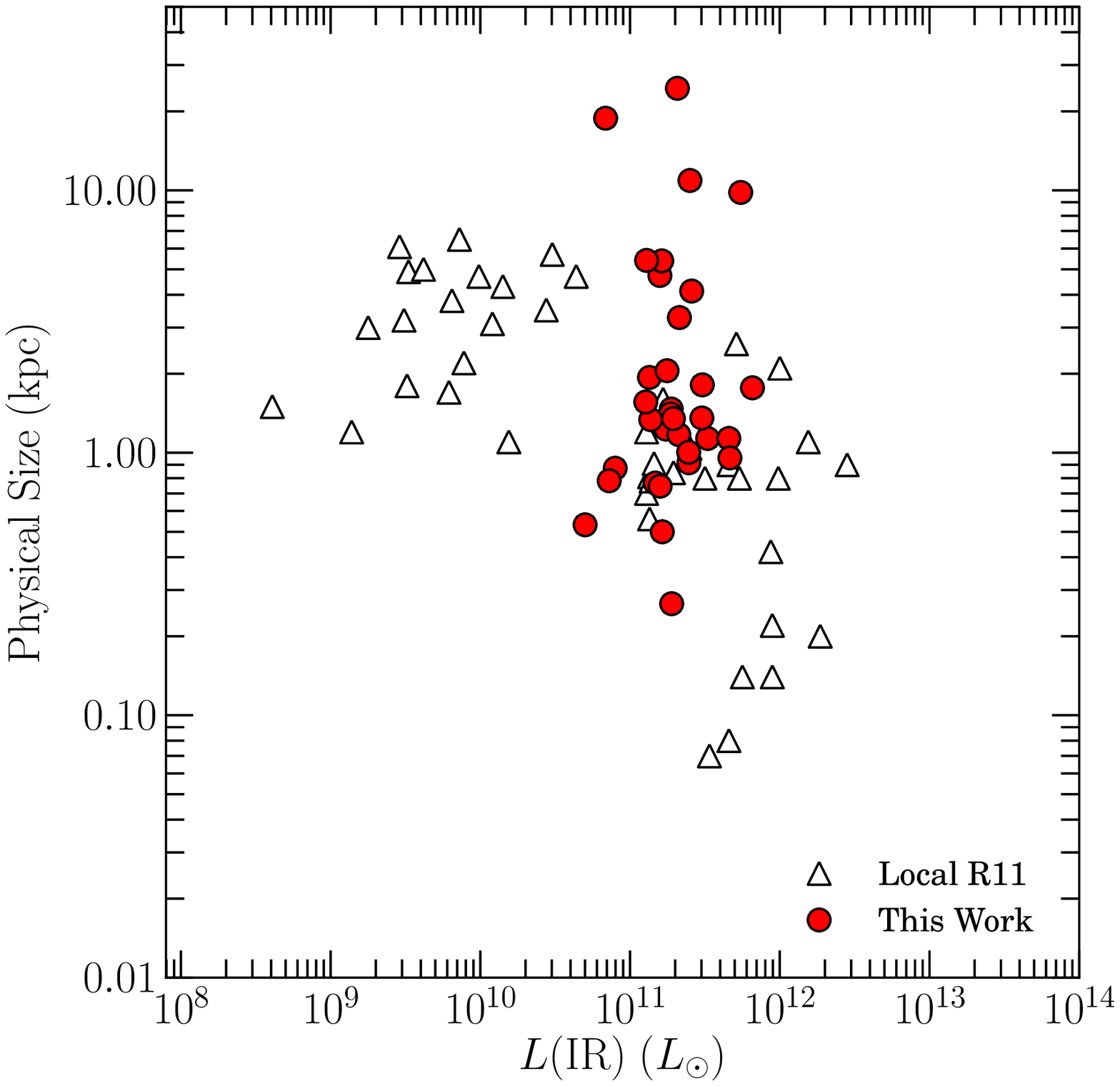}
\caption{Comparison of $L(\mathrm{IR})$ and physical sizes (in diameter) of star-forming regions. To compare with local sample in R11, the area of star-forming region is measured in Pa$\alpha$ images convolved with a gaussian function with $\sigma$ = 4 kpc in physical scale. Triangles represent normal galaxies and U/LIRGs derived from R11.
}\label{fig:rujoplot1_iras}
\end{center}
\end{figure}
%-------------

%-------------
\begin{figure}[t]
\begin{center}
\plotone{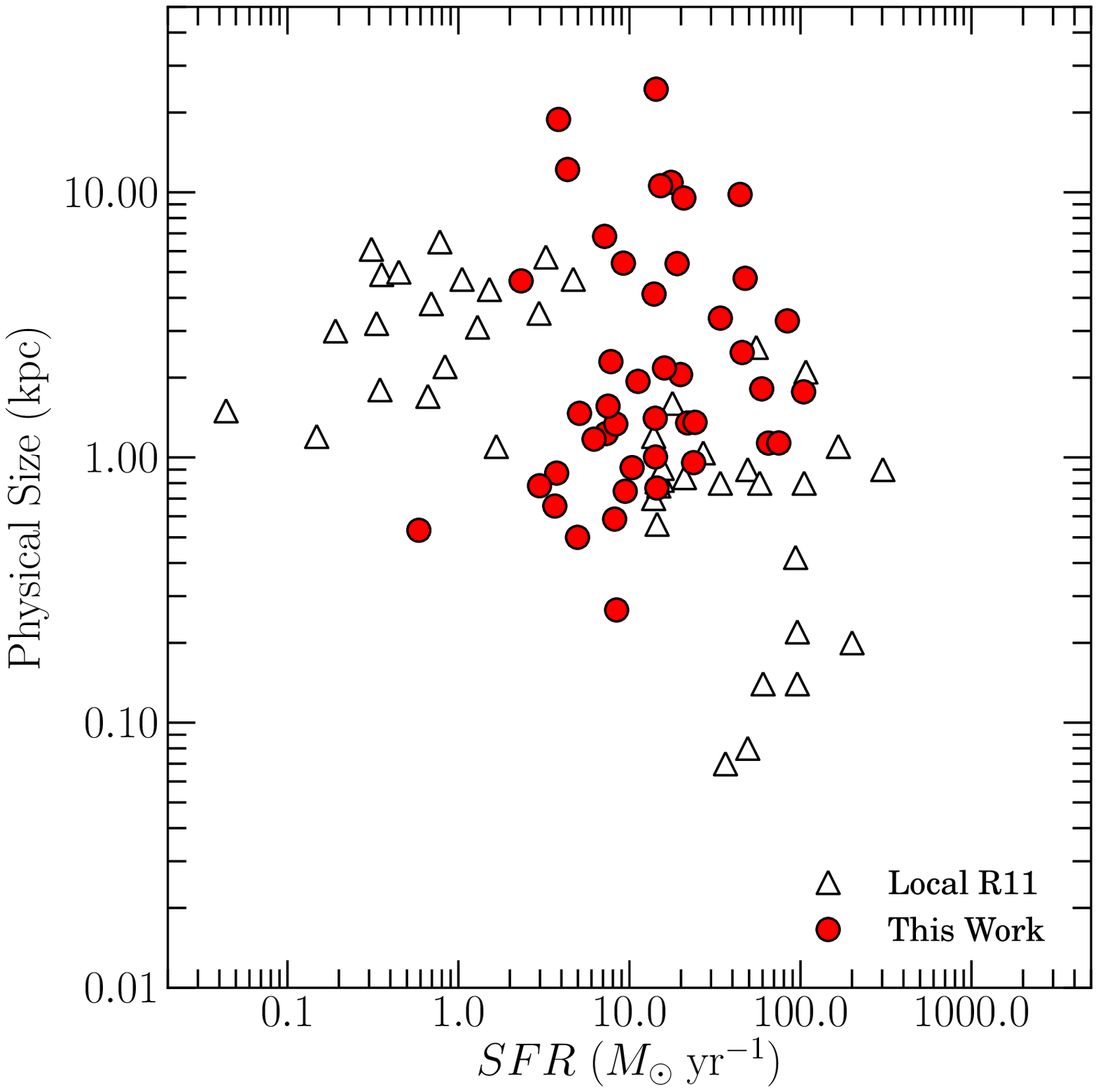}
\caption{Comparison of $SFR$ and physical sizes (in diameter) of star-forming regions. Same as Figure \ref{fig:rujoplot1_iras}, the area of star-forming region is measured in Pa$\alpha$ images with being convolved. SFRs are derived not only from Pa$\alpha$ but also from MIR luminosity in R11.
}\label{fig:rujoplot1}
\end{center}
\end{figure}
%-------------

\noindent dust extinction than measured; indeed, \citet{2013arXiv1304.0894P} shows that dust-extinction of central region of LIRGs are estimated to be $A_V$ = 5 $\sim$ 13 mag by using infrared indicators, larger than the value obtained using optical. Another possibility is contribution to IR luminosity by dust particles heated by more evolved stars, called the ``IR cirrus'' component \citep[e.g.,][]{1998ARA&A..36..189K,2009ApJ...703.1672K}, which may overestimate $SFR(\mathrm{IR})$.

%%%---Subsection 4.5 Surface Densities of Star Formation Rates
\subsection{Surface Densities of IR Luminosity and Star Formation Rate}

The size of star-forming regions in LIRGs at high IR luminosity end and in ULIRGs are compact while the normal galaxies show extended star-forming region over a few kilo-parsecs along the spiral arm \citep[e.g.,][]{2000AJ....119..509S,2010ApJ...723..993D,2012ApJ...744....2A}. R11 find that the size of star-forming regions changes drastically at the LIRG luminosity, suggesting that these differences are caused by strong interaction \citep{2011PASJ...63.1181T}. However, there are few sample at the $``$transition$"$ point. To investigate this relationship between the size and star-forming activity, we measure sizes of star-forming regions in our large sample of LIRGs with Pa$\alpha$ emission line images, which fill the gap between normal galaxies and ULIRGs in R11.

We define size of a star-forming region of a galaxy to be an elliptical diameter (major axis) containing 50\% flux within a Petrosian radius \citep{1976ApJ...209L...1P} defined in the Pa$\alpha$ line image convolved with a Gaussian function with $\sigma$ = 4 kpc in physical scale as done in R11 and list it in column-(3) of Table \ref{table_sfr}. Figure \ref{fig:rujoplot1_iras} shows the comparison between the IR luminosities and the sizes of the star-forming regions except for the 5 galaxies (NGC 5257/8, IC 4518A/B, IC 4686/7, NGC 7770/1, UGC 12914/5) which are paired galaxies and the size of SFR can not be defined. We find that the sizes of the LIRGs in our sample is distributed from 0.3 kpc or less to 25 kpc. Especially, IC 1623A/B, NGC 2342, NGC 7678, NGC 6926 have a large ($>$ 9.5 kpc) star-forming regions along their spiral arms. Figure \ref{fig:rujoplot1} shows the comparison between the SFRs derived from Pa$\alpha$ and the sizes of the star-forming regions including 5 galaxies (NGC 5257/8, IC 4518A/B, IC 4686/7, NGC 7770/1, UGC 12914/5) removed from Figure \ref{fig:rujoplot1_iras}. Our results fill the blank parameter space between normal galaxies and ULIRGs in R11.

\begin{figure}[t]
\begin{center}
\plotone{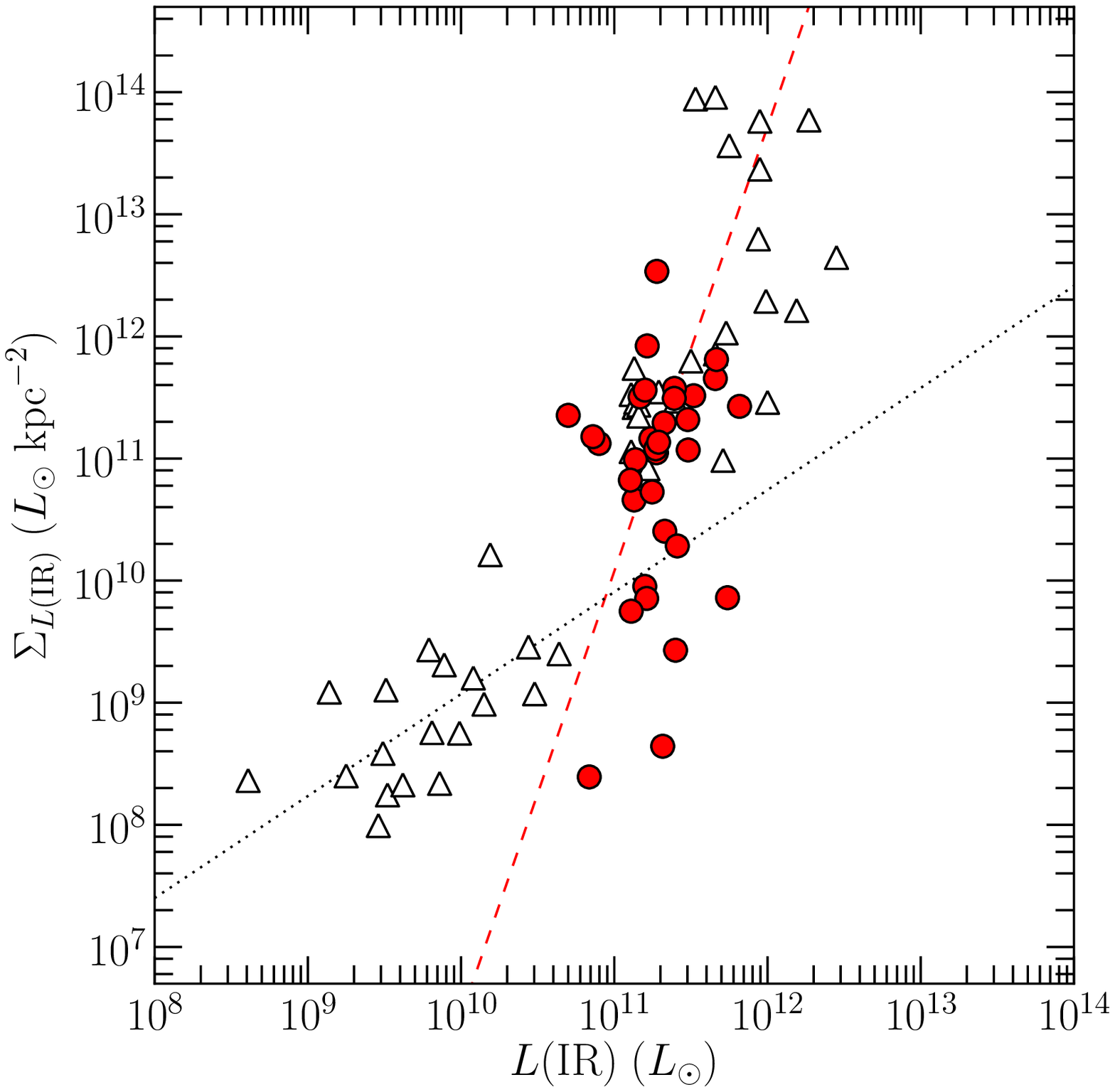}
\caption{Comparison of $L(\mathrm{IR})$ and $L(\mathrm{IR})$ surface densities. The dotted line represent the sequence of normal galaxies (R11), while dashed line the sequence of U/LIRGs, which is the best fit relation using the sample from this work and R11.}\label{fig:rujoplot2_iras}
\end{center}
\end{figure}

Figure \ref{fig:rujoplot2_iras} shows a comparison of IR luminosites and surface densities of IR luminosity ($\Sigma_{L(\mathrm{IR})}$) derived by dividing a IR by area of a star-forming region;
\begin{eqnarray}
\Sigma_{L(\mathrm{IR})}\ \equiv L(\mathrm{IR})/(\pi r_m^2),
\end{eqnarray}
where $r_m$ is the Petrosian radius listed in Table \ref{table_sfr}. In this figure, the dotted line represents the sequence of normal galaxies derived by best fitting of local normal galaxies and high-z star-forming galaxies in R11 and \citet{2013ApJ...767...73R}. R11 shows that the sequence of local U/LIRGs is different from that of local normal galaxies. Therefore we obtained the sequence of U/LIRGs by fitting sub-LIRGs ($L(\mathrm{IR}) \geq 10^{10}\ L_{\sun}$) and ULIRGs of both ours and those in R11 with the dashed line, that is $\log_{10}(\Sigma_{L(\mathrm{IR})}) = 3.6\times\log_{10}(L(\mathrm{IR})) - 30.0$.

Figure \ref{fig:rujoplot1_iras} and \ref{fig:rujoplot2_iras} show that $\Sigma_{L(\mathrm{IR})}$ become higher and the size of the star-forming region become more compact as the $L(\mathrm{IR})$ increase, and the transition point from the sequence of normal galaxies and U/LIRGs is at $L(\mathrm{IR})=8\times10^{10}\ L_{\sun}$, which is consistent with the result of R11.

\begin{figure}[t]
\begin{center}
\plotone{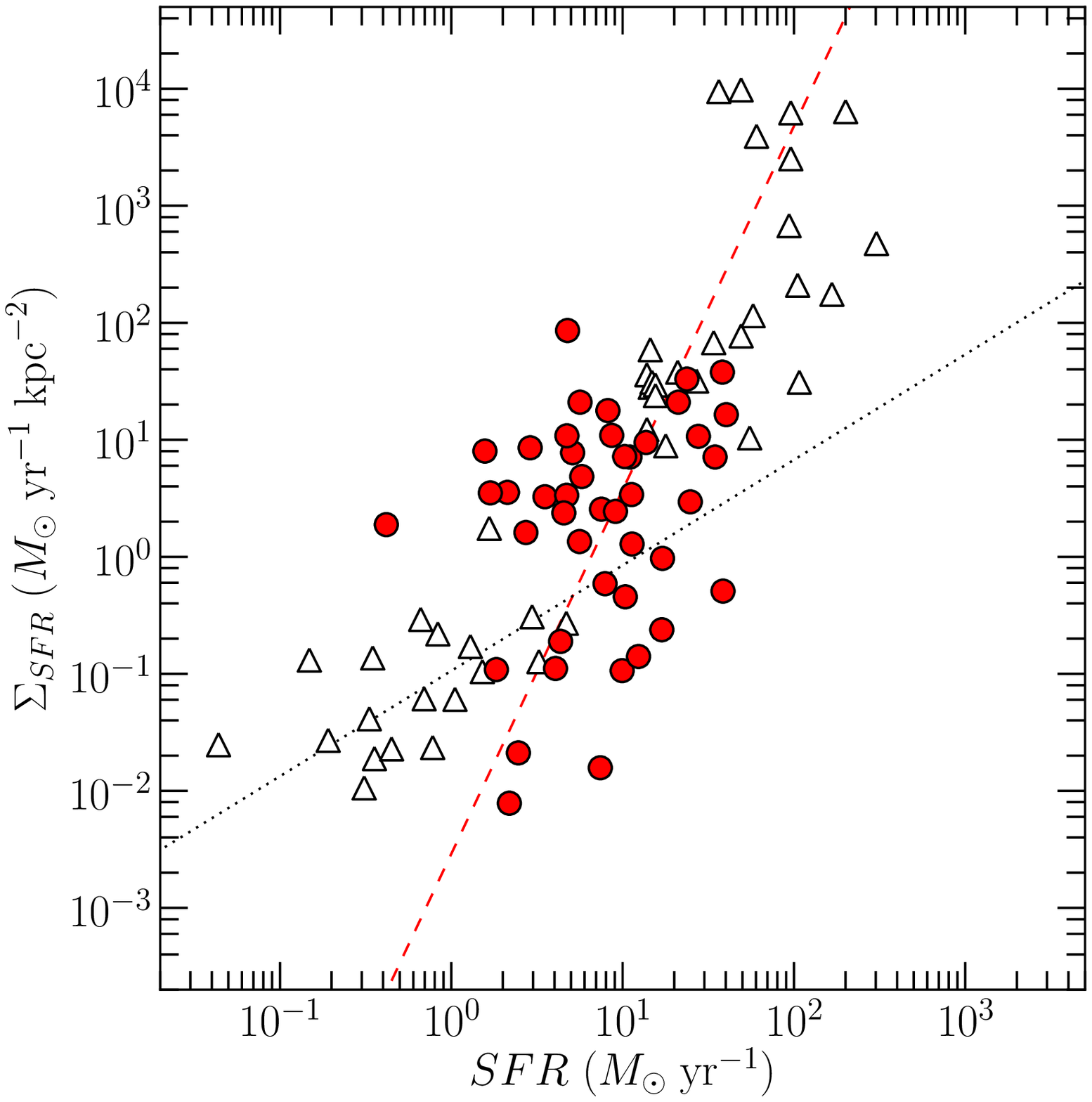}
\caption{Comparison of SFRs derived from Pa$\alpha$ luminosities and SFR surface densities. The dotted line represent the sequence of normal galaxies (R11; \citealt{2013ApJ...767...73R}), while dashed line the sequence of U/LIRGs, which is the best fit relation using the sample from this work and R11.}\label{fig:rujoplot2}
\end{center}
\end{figure}

Figure \ref{fig:rujoplot2} shows a comparison of SFRs derived from Pa$\alpha$ and surface densities of SFR ($\Sigma_{\mathrm{SFR}}$) derived by dividing a SFR by area of a star-forming region;
\begin{eqnarray}
\Sigma_{SFR}\ \equiv \mathrm{SFR}/(\pi r_m^2).
\end{eqnarray}
The dashed line is the best fit linear relation for U/LIRGs and sub-LIRGs ($SFR \geq 1.1\ M_{\sun}\ \mathrm{yr^{-1}}$) including both ours and those in R11, where $\log_{10}(\Sigma_{\mathrm{SFR}}) = 3.4\times\log_{10}(\mathrm{SFR}) - 3.2$. In this plot, our results fill the blank
\clearpage

%-------------
%-------------

%-------------
%-------------

\noindent between the sequence of normal galaxies and U/LIRGs, and the transition point is at $SFR$ = 7 $M_{\sun}\ \mathrm{yr}^{-1}$, which is a consistent value with that in the $\Sigma_{L(\mathrm{IR})}$--$L(\mathrm{IR})$ relation.

Also we find that there is a large scatter in both figures, different from those of normal galaxies and U/LIRGs in R11. Some simulations of galaxy formation suggest that while merging or interacting galaxies in their early stages have extended star-forming regions along their collision interfaces and disks \citep{2009PASJ...61..481S}, late-stage mergers have compact and concentrated star-forming regions at their centers \citep{1996ApJ...471..115B}. Our results show that while LIRGs have a wide range in sizes of the star-forming regions from less than 1 kpc to over 10 kpc, almost all LIRGs at high luminosity end and ULIRGs have compact star-forming region around 1 kpc or less. Considering the fact that all the ULIRGs are mergers \citep{2000AJ....119..509S}, there is a possibility that ULIRGs and LIRGs at high luminosity end with compact star-forming regions are at the late-stage in the merging history. On the other hand, from sub-LIRGs and LIRGs at low luminosity end with extended star-forming regions may be at first-stage in merging history.

%%%==========================================================================%%%
%%%   Section 6 Summary
%%%==========================================================================%%%
\section{Summary}
We have observed 38 galaxy system listed in $IRAS$ RBGS catalog in Pa$\alpha$ with narrow-band imaging ($cz = 2800$--$8100\ ${km~s$^{-1}$}, $L(\mathrm{IR}) = 4.5\times 10^{10}$--$6.5\times 10^{11}\ L_{\sun}$) using near infrared camera (ANIR) for the University of Tokyo Atacama Observatory 1m telescope, installed at the summit of Co. Chajnantor in northern Chile.
 
We have estimated the Pa$\alpha$ fluxes from the narrow-band images with our newly developed flux calibration method. We find that $SFR(\mathrm{Pa}\alpha)_\mathrm{corr}$ which is SFR obtained from Pa$\alpha$ luminosity corrected for the effect of dust extinction with balmer decrement method (H$\beta$/H$\alpha$) shows good agreement with $SFR(\mathrm{IR})$, which is star formation estimated from total infrared luminoisity. This result suggests that Pa$\alpha$ with dust-correction is sufficient for estimating SFR of whole the galaxy. However, some galaxies have large differences between the $SFR(\mathrm{Pa}\alpha)_\mathrm{corr}$ and the $SFR(\mathrm{IR})$, which may be caused by effect of AGNs, strong dust-extinction, or IR cirrus component.

We also obtain surface densities of $L(\mathrm{IR})$ ($\Sigma_{L(\mathrm{IR})}$) and SFR ($\Sigma_{\mathrm{SFR}}$) for individual galaxies by measuring physical sizes of star-forming regions. The range of SFR in our sample (from 0.6 to $\sim$ 104 $M_\odot$ yr$^{-1}$) fill the blank of the range of SFR in the previous work. We find that most of the samples follow a sequence of local U/LIRGs on the $L(\mathrm{IR})$-$\Sigma_{L(\mathrm{IR})}$ and SFR-$\Sigma_{\mathrm{SFR}}$ plane. We confirm that a transition of the sequence from normal galaxies to U/LIRGs is seen at $L(\mathrm{IR})=8\times10^{10}$. Also, we find that there is a large scatter in physical size, different from those of normal galaxies or ULIRGs. Considering the fact that most of U/LIRGs are merging or interacting galaxies, these scatters may be caused by the strong external factors or the differences of their merging stage.

%%%==========================================================================%%%
%%%   Acknowledgments
%%%==========================================================================%%%
\acknowledgments
We deeply thank the referee (Dr. Guilin Liu) for useful comments and suggestions that helped improve the quality of the paper, and we also thank M. Malkan (UCLA) and S. Howard (CfA) for enlightening discussions on this topic. This work is supported by Ministry of Education, Culture, Sports, Science and Technology of Japan, Grant-in-Aid for Scientific Research (17104002, 20040003, 20041003, 21018003, 21018005, 21684006, 22253002, 22540258,and 23540261) from the Japan Society for the Promotion of Science (JSPS). Operation of ANIR on the miniTAO 1m telescope is also supported by NAOJ Research Grant for Universities and Optical \& Near-Infrared Astronomy Inter-University Cooperation Program, supported by the MEXT of Japan. Part of this work has been supported by the Institutional Program for Young Researcher Overseas Visits operated by JSPS. The Image Reduction and Analysis Facility (IRAF) used in this paper is distributed by the National Optical Astronomy Observatories, which are operated by the Association of Universities for Research in Astronomy, Inc., under cooperative agreement with the National Science Foundation. We acknowledge the usage of the HyperLeda database (http://leda.univ-lyon1.fr).

%%%==========================================================================%%%
%%%   Reference
%%%==========================================================================%%%

%%%==========================================================================%%%
%%%   Appendix. Paα Flux Calibration Method
%%%==========================================================================%%%
\appendix

\section{Derivation of Pa$\alpha$ Flux Affected by Atmospheric Absorptions}
%-------------
\begin{figure}[b]
\begin{center}
\plotone{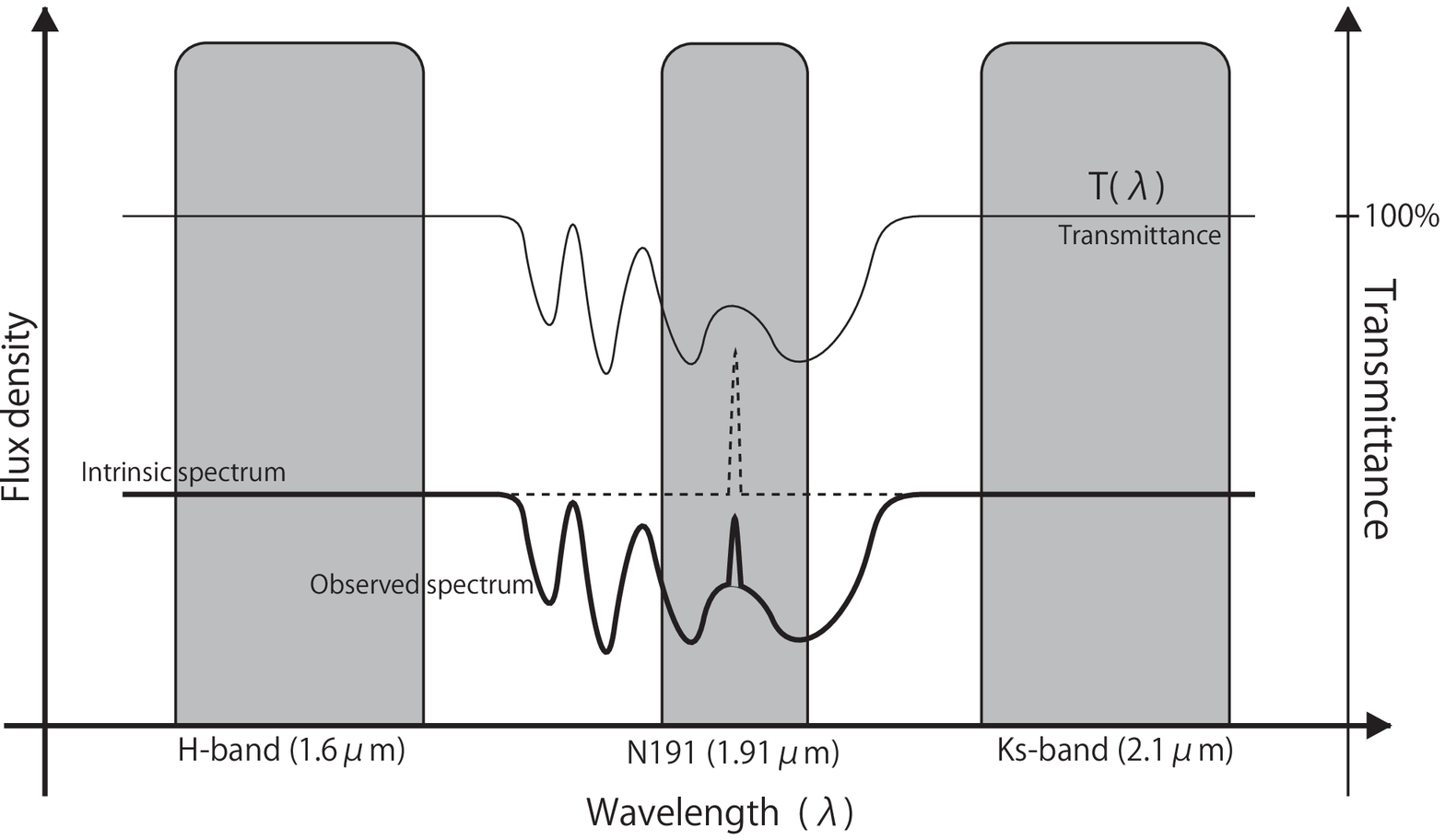}
\caption{Cartoon of a model spectrum around the wavelength of Pa$\alpha$ emission line, affected by the atmospheric absorption. $H$ and $K_\mathrm{s}$ are the broad-band filters, and $N191$ is a narrow-band filter installed used for our observations. The thin curve represents the atmospheric transmittance, the dashed curve the intrinsic spectrum of the target, and the thick line the observed spectrum.}\label{fig:transmodel_mono}
\end{center}
\end{figure}
%-------------

The TAO site is a suitable place for ground-based Pa$\alpha$ observation thanks its low PWV. Still, there are many atmospheric absorption features within the wavelength range of the narrow-band filter (Figure \ref{fig:transmodel}) which vary temporally due to change of PWV, and it is difficult to obtain the emission-line flux accurately. We then estimate the intrinsic Pa$\alpha$ flux as follows. Figure \ref{fig:transmodel_mono} represents a cartoon of a spectrum affected by the atmospheric absorption. The thin-solid curve represents the atmospheric transmittance, the dashed curve an intrinsic spectrum of a target, bold-solid an observed spectrum affected by the atmosphere, and the shaded areas the wavelength ranges of the $H$-, $N191$- and $K_\mathrm{s}$-band filters.

The intrinsic flux in the narrow-band filter ($F_{\mathrm{int}}^{N191}$) can be divided into two components; one is continuum ($F_\mathrm{int}^{c}$) and the other is emission line ($F_\mathrm{int}^{l}$). Then, the intrinsic flux can be written as
\begin{eqnarray}
F_{\mathrm{int}}^{N191} &=& F_\mathrm{int}^{c} + F_\mathrm{int}^{l}.
\end{eqnarray}
We also assume that the wavelength dependence of the continuum within the $N191$ filter is negligible. To derive $F_\mathrm{int}^{l}$ it is necessary to estimate the effect of instrumental absorption and atmospheric absorption due to PWV. The throughput ($\zeta_\mathrm{ANIR}$) can be written as
\begin{eqnarray}
\zeta_\mathrm{ANIR} &=& T^{N191}_\mathrm{Tel} \times T^{N191}_\mathrm{ANIR} \times T^{N191}_\mathrm{filter}
\end{eqnarray}
where $T^{N191}_\mathrm{Tel}$, $T^{N191}_\mathrm{ANIR}$, and $T^{N191}_\mathrm{filter}$ represent transmittance of the telescope, the instrument except the filter, and the N191 filter, described in K14. The atmospheric transmittance within the wavelength of the $N191$ narrow-band filter ($T^{\mathrm{PWV},N191}_\mathrm{atm})^X$) can be written as
\begin{eqnarray}
(T^{\mathrm{PWV},N191}_\mathrm{atm})^X &=& \frac{\int^{\lambda_2}_{\lambda_1}T_\mathrm{atm}(\lambda)d\lambda}{\int^{\lambda_2}_{\lambda_1}d\lambda},
\end{eqnarray}
$\lambda_1$ and $\lambda_2$ represent cut-on and cut-off wavelength of the $N191$ narrow-band filter. Then the observed flux density of $N191$ can be written as follows,
\begin{eqnarray}
F_{\mathrm{obs}}^{N191} &=& F_\mathrm{int}^{c} \zeta_\mathrm{ANIR} (T^{\mathrm{PWV},N191}_\mathrm{atm})^X + F_\mathrm{int}^{l} \zeta_\mathrm{ANIR} T_\mathrm{line}\\
F_\mathrm{int}^{l} &=& \frac{1}{\zeta_\mathrm{ANIR} T_\mathrm{line}}(F_{\mathrm{obs}}^{N191}-F_\mathrm{int}^{c}\zeta_\mathrm{ANIR}(T^{\mathrm{PWV},N191}_\mathrm{atm})^X).
\end{eqnarray}
In a narrow-band filter observation, an observed flux is the averaged flux of emission line within the wavelength range from $\lambda_1$ to $\lambda_2$, and the flux are calibrated by using the flux of 2MASS stars. The relation between observed flux and calibrated flux ($f_\mathrm{cal}^{N191}$) is as follows,
\begin{eqnarray}
F_\mathrm{obs}^{N191} &=& \zeta_\mathrm{ANIR}(T^{\mathrm{PWV},N191}_\mathrm{atm})^X f_\mathrm{cal}^{N191}\Delta\lambda.
\end{eqnarray}
We can derive a calibrated flux of continuum ($f_\mathrm{cal}^c$) by the interpolated flux of $H$ and $K_\mathrm{s}$ broad-band filter ($f_\mathrm{cal}^{H-K_\mathrm{s}}$), which are not affected by the atmospheric absorption ($f_\mathrm{cal}^c \equiv f_\mathrm{cal}^{H-K_\mathrm{s}}$). The relation between intrinsic flux of continuum ($F_\mathrm{int}^c$) and calibrated flux is as follows,
\begin{eqnarray}
F_\mathrm{int}^c &=& \zeta_\mathrm{ANIR}f_\mathrm{cal}^{c}\Delta\lambda\\
&=& \zeta_\mathrm{ANIR}f_\mathrm{cal}^{H-K_\mathrm{s}}\Delta\lambda.
\end{eqnarray}
Then, we can obtain the following equation by using (A5), (A6) and (A8),
\begin{eqnarray}
F_{\mathrm{int}}^l &=& \frac{(T^{\mathrm{PWV},N191}_{\mathrm{atm}})^X}{T_{\mathrm{line}}}(f_{\mathrm{cal}}^{N191}-f_{\mathrm{cal}}^{H-K_\mathrm{s}})\Delta\lambda.
\end{eqnarray}


\begin{thebibliography}{50}
\expandafter\ifx\csname natexlab\endcsname\relax\def\natexlab#1{#1}\fi


\bibitem[Alonso-Herrero et al.(1999)]{1999MNRAS.302..561A} Alonso-Herrero, A., Ward, M.~J., Aragon-Salamanca, A.,\& Zamorano, J.\ 1999, \mnras, 302, 561 

\bibitem[Alonso-Herrero et al.(2001)]{2001ApJ...546..952A} Alonso-Herrero, A., Engelbracht, C.~W., Rieke, M.~J., Rieke, G.~H., \& Quillen, A.~C.\ 2001, \apj, 546, 952 

\bibitem[Alonso-Herrero et al.(2002)]{2002AJ....124..166A} Alonso-Herrero, A., Rieke, G.~H., Rieke, M.~J., \& Scoville, N.~Z.\ 2002, \aj, 124, 166

\bibitem[Alonso-Herrero et al.(2006)]{2006ApJ...650..835A} Alonso-Herrero, A., Rieke, G.~H., Rieke, M.~J., et al.\ 2006, \apj, 650, 835

\bibitem[Alonso-Herrero et al.(2009)]{2009A&A...506.1541A} Alonso-Herrero, A., Garc{\'{\i}}a-Mar{\'{\i}}n, M., Monreal-Ibero, A., et al.\ 2009, \aap, 506, 1541 

\bibitem[Alonso-Herrero et al.(2012)]{2012ApJ...744....2A} Alonso-Herrero, A., Pereira-Santaella, M., Rieke, G.~H., \& Rigopoulou, D.\ 2012, \apj, 744, 2 

\bibitem[Ann \& Kim(1996)]{1996JKAS...29..255A} Ann, H.-B., \& Kim, J.-M.\ 1996, Journal of Korean Astronomical Society, 29, 255

\bibitem[Antonucci \& Miller(1985)]{1985ApJ...297..621A} Antonucci, R.~R.~J., \& Miller, J.~S.\ 1985, \apj, 297, 621 

\bibitem[Ag{\"u}ero et al.(2000)]{2000AJ....119...94A} Ag{\"u}ero, E.~L., Paolantonio, S., G{\"u}nthardt, G.\ 2000, \aj, 119, 94

\bibitem[Barnes \& Hernquist(1996)]{1996ApJ...471..115B} Barnes, J.~E., \& Hernquist, L.\ 1996, \apj, 471, 115 

\bibitem[Calzetti et al.(2000)]{2000ApJ...533..682C} Calzetti, D., Armus, L., Bohlin, R.~C., et al.\ 2000, \apj, 533, 682

\bibitem[Calzetti et al.(2007)]{2007ApJ...666..870C} Calzetti, D., Kennicutt, R.~C., Engelbracht, C.~W., et al.\ 2007, \apj, 666, 870 

\bibitem[Calzetti (2013)]{2013arXiv:1208.2997v1} Calzetti, D., \ 2013, arXiv:1208.2997v1, in press

\bibitem[Cardelli et al.(1989)]{1989ApJ...345..245C} Cardelli, J.~A., Clayton, G.~C., \& Mathis, J.~S.\ 1989, \apj, 345, 245

\bibitem[Caputi et al.(2007)]{2007ApJ...660...97C} Caputi, K.~I., Lagache, G., Yan, L., et al.\ 2007, \apj, 660, 97 

\bibitem[Chaboyer \& Vader(1991)]{1991PASP..103...35C} Chaboyer, B., \& Vader, J.~P.\ 1991, \pasp, 103, 35 

\bibitem[Colina et al.(2001)]{2001ApJ...553L..19C} Colina, L., Alberdi, A., Torrelles, J.~M., Panagia, N., \& Wilson, A.~S.\ 2001, \apjl, 553, L19 

\bibitem[Colina et al.(1991)]{1991ApJ...370..102C} Colina, L., Sparks, W.~B., \& Macchetto, F.\ 1991, \apj, 370, 102 

\bibitem[Condon et al.(1991)]{1991ApJ...378...65C} Condon, J.~J., Huang, Z.-P., Yin, Q.~F., \& Thuan, T.~X.\ 1991, \apj, 378, 65 

\bibitem[Condon et al.(1993)]{1993AJ....105.1730C} Condon, J.~J., Helou, G., Sanders, D.~B., \& Soifer, B.~T.\ 1993, \aj, 105, 1730 

\bibitem[Corbett et al.(2003)]{2003ApJ...583..670C} Corbett, E.~A., Kewley, L., Appleton, P.~N., et al.\ 2003, \apj, 583, 670 

\bibitem[Daddi et al.(2010)]{2010ApJ...714L.118D} Daddi, E., Elbaz, D., Walter, F., et al.\ 2010, \apjl, 714, L118 

\bibitem[Davies et al.(2004)]{2004ApJ...602..148D} Davies, R.~I., Tacconi, L.~J., \& Genzel, R.\ 2004, \apj, 602, 148 

\bibitem[de Vaucouleurs et al.(1991)]{1991S&T....82Q.621D} de Vaucouleurs, G., de Vaucouleurs, A., Corwin, H.~G., Jr., et al.\ 1991, \skytel, 82, 621
\bibitem[D{\'{\i}}az-Santos et al.(2010)]{2010ApJ...723..993D} D{\'{\i}}az-Santos, T., Charmandaris, V., Armus, L., et al.\ 2010, \apj, 723, 993 
\bibitem[Dopita et al.(2002)]{2002ApJS..143...47D} Dopita, M.~A., Pereira, M., Kewley, L.~J., \& Capaccioli, M.\ 2002, \apjs, 143, 47
\bibitem[Falcke et al.(1998)]{1998ApJ...494L.155F} Falcke, H., Rieke, M.~J., Rieke, G.~H., Simpson, C., \& Wilson, A.~S.\ 1998, \apjl, 494, L155 
\bibitem[Fern{\'a}ndez et al.(2010)]{2010AJ....140.1965F} Fern{\'a}ndez, X., van Gorkom, J.~H., Schweizer, F., \& Barnes, J.~E.\ 2010, \aj, 140, 1965 
\bibitem[Genzel et al.(1995)]{1995ApJ...444..129G} Genzel, R., Weitzel, L., Tacconi-Garman, L.~E., et al.\ 1995, \apj, 444, 129 
\bibitem[Goto et al.(2010)]{2010A&A...514A...6G} Goto, T., Takagi, T., Matsuhara, H., et al.\ 2010, \aap, 514, A6 
\bibitem[Grimes et al.(2006)]{2006ApJ...648..310G} Grimes, J.~P., Heckman, T., Hoopes, C., et al.\ 2006, \apj, 648, 310 
\bibitem[Hill et al.(1996)]{1996ApJ...462..163H} Hill, G.~J., Goodrich, R.~W., \& Depoy, D.~L.\ 1996, \apj, 462, 163 
\bibitem[Hirashita et al.(2003)]{2003A&A...410...83H} Hirashita, H., Buat, V., \& Inoue, A.~K.\ 2003, \aap, 410, 83 
\bibitem[Hough et al.(1987)]{1987MNRAS.224.1013H} Hough, J.~H., Brindle, C., Axon, D.~J., Bailey, J., \& Sparks, W.~B.\ 1987, \mnras, 224, 1013 
\bibitem[Ho et al.(1997)]{1997ApJS..112..315H} Ho, L.~C., Filippenko, A.~V., \& Sargent, W.~L.~W.\ 1997, \apjs, 112, 315 
\bibitem[Hopkins \& Beacom(2006)]{2006ApJ...651..142H} Hopkins, A.~M., \& Beacom, J.~F.\ 2006, \apj, 651, 142
\bibitem[Hummer \& Storey(1987)]{1987MNRAS.224..801H} Hummer, D.~G., \& Storey, P.~J.\ 1987, \mnras, 224, 801 
\bibitem[Inglis et al.(1993)]{1993MNRAS.263..895I} Inglis, M.~D., Brindle, C., Hough, J.~H., et al.\ 1993, \mnras, 263, 895 
\bibitem[Iono et al.(2005)]{2005ApJS..158....1I} Iono, D., Yun, M.~S., \& Ho, P.~T.~P.\ 2005, \apjs, 158, 1
\bibitem[Iono et al.(2013)]{2013PASJ...65L...7I} Iono, D., Saito, T., Yun, M.~S., et al.\ 2013, \pasj, 65, L7
\bibitem[Jenkins et al.(2005)]{2005MNRAS.357..109J} Jenkins, L.~P., Roberts, T.~P., Ward, M.~J., \& Zezas, A.\ 2005, \mnras, 357, 109 
\bibitem[Kennicutt(1998)]{1998ARA&A..36..189K} Kennicutt, R.~C., Jr.\ 1998, \araa, 36, 189
\bibitem[Kennicutt et al.(2007)]{2007ApJ...671..333K} Kennicutt, R.~C., Jr., Calzetti, D., Walter, F., et al.\ 2007, \apj, 671, 333
\bibitem[Kennicutt et al.(2009)]{2009ApJ...703.1672K} Kennicutt, R.~C., Jr., Hao, C.-N., Calzetti, D., et al.\ 2009, \apj, 703, 1672 
\bibitem[Kewley et al.(2002)]{2002AJ....124.3135K} Kewley, L.~J., Geller, M.~J., Jansen, R.~A., \& Dopita, M.~A.\ 2002, \aj, 124, 3135 
\bibitem[Kim et al.(1998)]{1998ApJ...508..627K} Kim, D.-C., Veilleux, S., \& Sanders, D.~B.\ 1998, \apj, 508, 627 
\bibitem[Kim et al.(2010)]{2010ApJ...724..386K} Kim, D., Im, M., \& Kim, M.\ 2010, \apj, 724, 386 
\bibitem[Komugi et al.(2005)]{2005PASJ...57..733K} Komugi, S., Sofue, Y., Nakanishi, H., Onodera, S., \& Egusa, F.\ 2005, \pasj, 57, 733 
\bibitem[Komugi et al.(2012)]{2012ApJ...757..138K} Komugi, S., Tateuchi, K., Motohara, K., et al.\ 2012, \apj, 757, 138
\bibitem[Konishi et al.(2014)]{2013K} Konishi, M., Motohara, K., Tateuchi, K., et al.\ 2013, PASJ, accepted
\bibitem[Koss et al.(2013)]{2013ApJ...765L..26K} Koss, M., Mushotzky, R., Baumgartner, W., et al.\ 2013, \apjl, 765, L26 
\bibitem[Lada et al.(2012)]{2012ApJ...745..190L} Lada, C.~J., Forbrich, J., Lombardi, M., \& Alves, J.~F.\ 2012, \apj, 745, 190
\bibitem[Lai et al.(1999)]{1999ESOC...56..555L} Lai, O., Rouan, D., \& Alloin, D.\ 1999, European Southern Observatory Conference and Workshop Proceedings, 56, 555 
\bibitem[Lauberts(1982)]{1982euse.book.....L} Lauberts, A.\ 1982, Garching: European Southern Observatory (ESO), 1982,  
\bibitem[Levenson et al.(2005)]{2005ApJ...618..167L} Levenson, N.~A., Weaver, K.~A., Heckman, T.~M., Awaki, H., \& Terashima, Y.\ 2005, \apj, 618, 167  
\bibitem[Liu et al.(2013a)]{2013ApJ...772...27L} Liu, G., Calzetti, D., Kennicutt, R.~C., Jr., et al.\ 2013, \apj, 772, 27 
\bibitem[Liu et al.(2013b)]{2013ApJ...778L..41L} Liu, G., Calzetti, D., Hong, S., et al.\ 2013, \apjl, 778, LL41
\bibitem[Lord (1992)]{Lord...1992} Lord, S. D.\ 1992,\ NASA Technical Memorandum, 103957
\bibitem[Lumsden et al.(2001)]{2001MNRAS.327..459L} Lumsden, S.~L., Heisler, C.~A., Bailey, J.~A., Hough, J.~H., \& Young, S.\ 2001, \mnras, 327, 459
\bibitem[Lutz(1991)]{1991IAUS..146..312L} Lutz, D.\ 1991, Dynamics of Galaxies and Their Molecular Cloud Distributions, 146, 312 
\bibitem[Malmquist(1925)]{1925Obs....48..142M} Malmquist, K.~G.\ 1925, The Observatory, 48, 142
\bibitem[Mauder et al.(1994)]{1994A&A...285...44M} Mauder, W., Weigelt, G., Appenzeller, I., \& Wagner, S.~J.\ 1994, \aap, 285, 44 
\bibitem[Miles et al.(1994)]{1994ApJ...425L..37M} Miles, J.~W., Houck, J.~R., \& Hayward, T.~L.\ 1994, \apjl, 425, L37 
\bibitem[Mirabel(1983)]{1983ApJ...270L..35M} Mirabel, I.~F.\ 1983, \apjl, 270, L35 
\bibitem[Minezaki et al.(2010)]{2010SPIE.7733E.163M} Minezaki, T., Kato, D., Sako, S., et al.\ 2010, \procspie, 7733, 
\bibitem[Morganti et al.(1998)]{1998AJ....115..915M} Morganti, R., Oosterloo, T., \& Tsvetanov, Z.\ 1998, \aj, 115, 915 
\bibitem[Morganti et al.(2007)]{2007A&A...476..735M} Morganti, R., Holt, J., Saripalli, L., Oosterloo, T.~A., \& Tadhunter, C.~N.\ 2007, \aap, 476, 735 
\bibitem[Motohara et al.(2008)]{2008SPIE.7012E.141M} Motohara, K., Aoki, T., Sako, S., et al.\ 2008, \procspie, 7012,  
\bibitem[Motohara et al.(2008)]{2008SPIE.7014E..94M} Motohara, K., Mitani, N., Sako, S., et al.\ 2008, \procspie, 7014, 
\bibitem[Motohara et al.(2010)]{2010SPIE.7735E.120M} Motohara, K., Konishi, M., Toshikawa, K., et al.\ 2010, \procspie, 7735,
\bibitem[Motohara et al.(2011)]{2011RMxAC..41...83M} Motohara, K.~M., Aoki, T., Asano, K., et al.\ 2011, Revista Mexicana de Astronomia y Astrofisica Conference Series, 41, 83
\bibitem[Murphy et al.(1999)]{1999ApJ...525L..85M} Murphy, T.~W., Jr., Soifer, B.~T., Matthews, K., Kiger, J.~R., \& Armus, L.\ 1999, \apjl, 525, L85 
\bibitem[Naim et al.(1995)]{1995MNRAS.274.1107N} Naim, A., Lahav, O., Buta, R.~J., et al.\ 1995, \mnras, 274, 1107 
\bibitem[Olsson et al.(2010)]{2010A&A...513A..11O} Olsson, E., Aalto, S., Thomasson, M., \& Beswick, R.\ 2010, \aap, 513, A11
\bibitem[\protect\citeauthoryear{Paturel et al.}{2003}]{2003A&A...412...45P} Paturel G., Petit C., Prugniel P., Theureau G., Rousseau J., Brouty M., Dubois P., Cambr{\'e}sy L., 2003, A\&A, 412, 45 
\bibitem[Peterson et al.(2012)]{2012ApJ...751...11P} Peterson, B.~W., Appleton, P.~N., Helou, G., et al.\ 2012, \apj, 751, 11 
\bibitem[Petrosian(1976)]{1976ApJ...209L...1P} Petrosian, V.\ 1976, \apjl, 209, L1 
\bibitem[Rieke et al.(2009)]{2009ApJ...692..556R} Rieke, G.~H., Alonso-Herrero, A., Weiner, B.~J., et al.\ 2009, \apj, 692, 556 
\bibitem[Piqueras L{\'o}pez et al.(2013)]{2013arXiv1304.0894P} Piqueras L{\'o}pez, J., Colina, L., Arribas, S., 
\& Alonso-Herrero, A.\ 2013, arXiv:1304.0894 
\bibitem[Richter et al.(1994)]{1994AJ....107...99R} Richter, O.-G., Sackett, P.~D., \& Sparke, L.~S.\ 1994, \aj, 107, 99  
\bibitem[Roche(2007)]{2007RMxAA..43..179R} Roche, N.~D.\ 2007, RMXAA, 43, 179 
\bibitem[Rodr{\'{\i}}guez-Zaur{\'{\i}}n et al.(2011)]{2011A&A...527A..60R} Rodr{\'{\i}}guez-Zaur{\'{\i}}n, J., Arribas, S., Monreal-Ibero, A., et al.\ 2011, \aap, 527, A60 
\bibitem[Rothberg \& Joseph(2004)]{2004AJ....128.2098R} Rothberg, B., \& Joseph, R.~D.\ 2004, \aj, 128, 2098 
\bibitem[Rujopakarn et al.(2010)]{ruj10} Rujopakarn, W., Eisenstein, D.~J., Rieke, G.~H., et al.\ 2010, \apj, 718, 1171
\bibitem[Rujopakarn et al.(2011)]{2011ApJ...726...93R} Rujopakarn, W., Rieke, G.~H., Eisenstein, D.~J., \& Juneau, S.\ 2011, \apj, 726, 93 (R11)
\bibitem[Rujopakarn et al.(2013)]{2013ApJ...767...73R} Rujopakarn, W., Rieke, G.~H., Weiner, B.~J., et al.\ 2013, \apj, 767, 73
\bibitem[Rush et al.(1993)]{1993ApJS...89....1R} Rush, B., Malkan, M.~A., \& Spinoglio, L.\ 1993, \apjs, 89, 1 
\bibitem[Saito et al.(2013)]{2013ASPC..476..287S} Saito, T., Iono, D., Yun, M., et al.\ 2013, Astronomical Society of the Pacific Conference Series, 476, 287 
\bibitem[Saitoh et al.(2009)]{2009PASJ...61..481S} Saitoh, T.~R., Daisaka, H., Kokubo, E., et al.\ 2009, \pasj, 61, 481 
\bibitem[Sanders \& Mirabel(1996)]{1996ARA&A..34..749S} Sanders, D.~B., \& Mirabel, I.~F.\ 1996, \araa, 34, 749 
\bibitem[Sanders et al.(2003)]{2003AJ....126.1607S} Sanders, D.~B., Mazzarella, J.~M., Kim, D.-C., Surace, J.~A., \& Soifer, B.~T.\ 2003, \aj, 126, 1607
\bibitem[Scoville et al.(2000)]{2000AJ....119..991S} Scoville, N.~Z., Evans, A.~S., Thompson, R., et al.\ 2000, \aj, 119, 991 
\bibitem[Scoville et al.(2001)]{2001AJ....122.3017S} Scoville, N.~Z., Polletta, M., Ewald, S., et al.\ 2001, \aj, 122, 3017 
\bibitem[Schweizer \& Seitzer(2007)]{2007AJ....133.2132S} Schweizer, F., \& Seitzer, P.\ 2007, \aj, 133, 2132 
\bibitem[Sheen et al.(2009)]{2009AJ....138.1911S} Sheen, Y.-K., Jeong, H., Yi, S.~K., et al.\ 2009, \aj, 138, 1911 
\bibitem[Skrutskie et al.(2006)]{2006AJ....131.1163S} Skrutskie, M.~F., Cutri, R.~M., Stiening, R., et al.\ 2006, \aj, 131, 1163 
\bibitem[Stritzinger et al.(2002)]{2002AJ....124.2100S} Stritzinger, M., Hamuy, M., Suntzeff, N.~B., et al.\ 2002, \aj, 124, 2100 
\bibitem[Spinoglio et al.(1995)]{1995ApJ...453..616S} Spinoglio, L., Malkan, M.~A., Rush, B., Carrasco, L., \& Recillas-Cruz, E.\ 1995, \apj, 453, 616 
\bibitem[Spergel et al.(2003)]{2003ApJS..148..175S} Spergel, D.~N., Verde, L., Peiris, H.~V., et al.\ 2003, \apjs, 148, 175 
\bibitem[Soifer et al.(2000)]{2000AJ....119..509S} Soifer, B.~T., Neugebauer, G., Matthews, K., et al.\ 2000, \aj, 119, 509
\bibitem[Soifer et al.(2001)]{2001AJ....122.1213S} Soifer, B.~T., Neugebauer, G., Matthews, K., et al.\ 2001, \aj, 122, 1213 
\bibitem[Soifer et al.(2003)]{2003AJ....126..143S} Soifer, B.~T., Bock, J.~J., Marsh, K., et al.\ 2003, \aj, 126, 143
\bibitem[Tanab{\'e} et al.(2013)]{2013PASJ...65...55T} Tanab{\'e}, T., Motohara, K., Tateuchi, K., et al.\ 2013, \pasj, 65, 55
\bibitem[Tateuchi et al.(2012a)]{2012PKAS...27..297T} Tateuchi, K., Motohara, K., Konishi, M., et al.\ 2012a, Publication of Korean Astronomical Society, 27, 297 
\bibitem[Tateuchi et al.(2012b)]{2012SPIE.8446E..7DT} Tateuchi, K., Motohara, K., Konishi, M., et al.\ 2012b, \procspie, 8446, 
\bibitem[Tateuchi et al.(2013)]{2013ASPC..476..301T} Tateuchi, K., Motohara, K., Konishi, M., et al.\ 2013, Astronomical Society of the Pacific Conference Series, 476, 301 
\bibitem[Thomas et al.(2004)]{2004MNRAS.351..362T} Thomas, H.~C., Alexander, P., Clemens, M.~S., et al.\ 2004, \mnras, 351, 362 
\bibitem[Totani et al.(2011)]{2011PASJ...63.1181T} Totani, T., Takeuchi, T.~T., Nagashima, M., Kobayashi, M.~A.~R., \& Makiya, R.\ 2011, \pasj, 63, 1181 
\bibitem[Tran(2001)]{2001ApJ...554L..19T} Tran, H.~D.\ 2001, \apjl, 554, L19 
\bibitem[van den Broek et al.(1991)]{1991A&AS...91...61V} van den Broek, A.~C., van Driel, W., de Jong, T., et al.\ 1991, \aaps, 91, 61 
\bibitem[V{\"a}is{\"a}nen et al.(2012)]{2012MNRAS.420.2209V} V{\"a}is{\"a}nen, P., Rajpaul, V., Zijlstra, A.~A., Reunanen, J., \& Kotilainen, J.\ 2012, \mnras, 420, 2209 
\bibitem[Veilleux et al.(1995)]{1995ApJS...98..171V} Veilleux, S., Kim, D.-C., Sanders, D.~B., Mazzarella, J.~M., \& Soifer, B.~T.\ 1995, \apjs, 98, 171 
\bibitem[V{\'e}ron-Cetty \& V{\'e}ron(2006)]{2006A&A...455..773V} V{\'e}ron-Cetty, M.-P., \& V{\'e}ron, P.\ 2006, \aap, 455, 773 
\bibitem[Vorontsov-Vel'Yaminov \& Arkhipova(1974)]{1974MCG...C05....0V} Vorontsov-Vel'Yaminov, B.~A., \& Arkhipova, V.~P.\ 1974, Morphological catalogue of galaxies., 5 (1974), 0 
\bibitem[Wilson et al.(1991)]{1991ApJ...381...79W} Wilson, A.~S., Helfer, T.~T., Haniff, C.~A., \& Ward, M.~J.\ 1991, \apj, 381, 79 
\bibitem[West(1976)]{1976A&A....46..327W} West, R.~M.\ 1976, \aap, 46, 327 
\bibitem[Yoshii et al.(2010)]{2010SPIE.7733E...6Y} Yoshii, Y., Aoki, T., Doi, M., et al.\ 2010, \procspie, 7733,
\bibitem[Yuan et al.(2010)]{2010ApJ...709..884Y} Yuan, T.-T., Kewley, L.~J., \& Sanders, D.~B.\ 2010, \apj, 709, 884 
\end{thebibliography}
\end{document}